    \documentclass[11pt,twoside,openleft,mathbbm,a4paper]{book}               
    \usepackage{cite}
    \usepackage[english]{babel}
    \usepackage{graphicx,bbm,psfrag}
    \usepackage{subfigure,makeidx}
    \usepackage{epsf,amsfonts,hyperref}
    \usepackage{fancyhdr}
    \usepackage[body={6.0in, 8.2in},left=1.25in,right=1.25in]{geometry}
    \usepackage{amsmath,amssymb}             
    \usepackage{rotating}                    
    \usepackage{setspace}                    
    \usepackage{txfonts}                     

  \makeatletter  
  \renewcommand{\chaptermark}[1]{\markboth{\@chapapp\ \thechapter.\ #1}{}}
  \makeatother    

    \makeatletter
    \def\cleardoublepage{\clearpage\if@twoside \ifodd\c@page\else%
        \hbox{}%
        \thispagestyle{empty}
        \newpage%
        \if@twocolumn\hbox{}\newpage\fi\fi\fi}
    \makeatother
    
\newcommand{\be}{\begin{equation}}
\newcommand{\bea}{\begin{eqnarray}}
\newcommand{\eea}{\end{eqnarray}}
\newcommand{\ee}{\end{equation}}

\newcommand{\cc}{\mathbbm{C}}
\newcommand{\nn}{\mathbbm{N}}
\newcommand{\rr}{\mathbbm{R}}

\input{Qcircuit}
 
\begin{document}



          \begin{titlepage}
          \begin{center}

					\vspace*{\stretch{6}}

					{\bf {\Huge Entanglement, quantum
					    phase transitions and quantum algorithms}}

					\vspace*{\stretch{4}}
					
					{\bf {\Large Rom\'an \'Oscar Or\'us Lacort}}
					
					\vspace*{\stretch{1}}

					{\bf Barcelona, July, 2006}

					\vspace*{\stretch{3}}

%

					{\bf {\large {\it  Universitat de Barcelona}}}
					
					\vspace*{\stretch{1}}
					
					{\bf {\large {\it Departament d'Estructura i Constituents de la Mat\`eria}}}
					
					\begin{center}
						\includegraphics[scale=1.25]{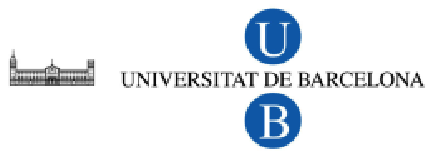}
					\end{center}

          \end{center}

          \end{titlepage}
					
					\pagebreak

					\chapter*{   }
					\thispagestyle{empty}					

					\begin{center}
					
					\vspace*{\stretch{1}}					

					{\bf {\Huge Entanglement, quantum
					    phase transitions and quantum algorithms}}
					
					\vspace*{\stretch{2}}					
					
					Memoria de la tesis presentada 

					por Rom\'an \'Oscar Or\'us Lacort para optar

					al grado de Doctor en Ciencias F\'\i sicas
					
					\vspace*{\stretch{1}}
					
					{Director de la tesis: Dr. Jos\'e
					Ignacio Latorre Sent\'\i s}
					
					\vspace*{\stretch{4}}
					
					Departament d'Estructura i Constituents de la Mat\`eria


					Programa de doctorado de {\it ``F\'\i sica avanzada''} 

					Bienio 2002-2004

					{\bf Universitat de Barcelona}
					
					\vspace*{\stretch{2}}

					\end{center}

					\pagebreak

					\chapter*{   }
					\thispagestyle{empty}

					\begin{flushright}
					
					\vspace*{\stretch{1}}
					
					\emph{A los que se fueron, y}
					
					\emph{a los que se quedaron,} 
					
					\emph{en especial a Mariano, Nati,}
							
					\emph{Mar\'{\i}a Mercedes y Ondiz}
					
					\vspace*{\stretch{4}}
					
					\end{flushright}

					\pagebreak					

					\chapter*{   }
					
					\thispagestyle{empty}

					\begin{flushright}
					
					\vspace*{\stretch{1}}
					
					\emph{Saber y saberlo demostrar es valer dos veces}
					
					-- Baltasar Graci\'an 
					
					\vspace{40pt}
										
					\emph{Las cuentas claras y el chocolate espeso}
					
					-- Refranero popular espa$\tilde{{\rm n}}$ol
					
					\vspace*{\stretch{4}}
					
					\end{flushright}

					\pagebreak

    \pagenumbering{roman} \setcounter{page}{0}

\chapter*{Agradecimientos}

He necesitado escribir m\'as de 150 p\'aginas llenas de ecuaciones raras y letras feas para poder saborear el delicioso momento de escribir los agradecimientos de mi tesis. Y es que dicen que de bien nacido es el ser agradecido, y tengo cuerda para rato, as\'{\i} que all\'a vamos. 

La primera persona a quien le he de agradecer muchas cosas es a Jos\'e Ignacio Latorre, quien se ofreci\'o a dirigirme una tesis doctoral en el mundillo este de la informaci\'on cu\'antica, de la cual yo no ten\'{\i}a ni idea hace cuatro a$\tilde{{\rm n}}$os y pico cuando entr\'e en su despacho cual pollito reci\'en licenciado y \'el me dijo aquello de ``hacer una tesis es como casarse, y si te casas, \emph{te casas}''. Memorables sentencias al margen, le agradezco la total y absoluta confianza que siempre ha depositado en m\'{\i} a lo largo de este tiempo, adem\'as de lo much\'{\i}simo que he aprendido con \'el benefici\'andome de su conocimiento multidisciplinar. Tampoco me olvido de alguna que otra cena  (aquella sopa de cebolla me hizo llorar de alegr\'{\i}a...), alguna que otra cata de vinos de la casa, y alg\'un que otro partidazo de basket o f\'utbol. Seguiremos en contacto. 

A Guifr\'e Vidal he de agradecerle su confianza en m\'{\i} casi desde el minuto cero. Ha sido un placer discutir de f\'{\i}sica contigo y he aprendido y disfrutado mucho. Seguiremos en Brisbane. 

Gracias a todas las personas con las que he discutido sobre lo divertida que es la mec\'anica cu\'antica. En la UB, los ``quantum boys'' Enric Jan\'e, Llu\'{\i}s Masanes, Enrique Rico (compa$\tilde{{\rm n}}$ero de batalla tantos a$\tilde{{\rm n}}$os, qu\'e grandes partidos de basket, qu\'e grandes sopas de cebolla), Joakim Bergli, Sofyan Iblisdir (Seigneur, pr\'evenez-moi \`a l'avance afin que j'y pr\'e-dispose mon syst\`eme digestif...), y los reci\'en llegados al campo Arnau Riera, Jos\'e Mar\'{\i}a Escart\'{\i}n y Pere Talavera. Tambi\'en a Pere Pascual por sus siempre buenos consejos e inter\'es por mi trabajo, a Rolf Tarrach y David Mateos por meterme entre ceja y ceja cuando hac\'{\i}a la carrera que lo de la cu\'antica era interesante, a N\'uria Barber\'an, Josep Tar\'on, y a los profesores visitantes Andy L$\ddot{{\rm u}}$tken y Krzystof Pilch. En la UAB, a Emili Bagan, Mariano Baig, Albert Bramon, John Casamiglia, Ram\'on Mu$\tilde{{\rm n}}$oz-Tapia, Anna Sanpera, y tambi\'en a \'Alex Monr\'as y Sergio Boixo (que pesadito estuve con QMA, ?`eh?). En el ICFO, a Toni Ac\'{\i}n le debo sabios consejos aqu\'{\i} y en alg\'un que otro pueblo perdido del pirineo. Gracias tambi\'en a Maciej Lewenstein, Joonwoo Bae, Miguel Navascu\'es (?`c\'omo hiciste el truco aqu\'el de la rata?), y Mafalda. 

Durante este tiempo he colaborado con gente en mis art\'{\i}culos a los que tambi\'en les agradezco lo mucho que he aprendido de ellos. Gracias a -- thanks to -- Jos\'e Ignacio Latorre, Jens Eisert, Marcus Cramer, MariCarmen Ba$\tilde{{\rm n}}$uls, Armando P\'erez (tenemos una paella pendiente), Pedro Ru\'{\i}z Femen\'{\i}a, Enrique Rico, Julien Vidal, Rolf Tarrach, Cameron Wellard, y Miguel \'Angel Mart\'{\i}n-Delgado. 

Al margen de la mec\'anica cu\'antica pura y dura, agradezco dentro de la UB y por diversos motivos a Dom\'enech Espriu, Joaquim Gomis, Joan Soto, Josep Mar\'{\i}a Pons, Llu\'{\i}s Garrido, Roberto Emparan, Artur Polls, Manel Barranco, Aurora Hern\'andez, y Marcel Porta, as\'{\i} como a toda la gente de la secretar\'{\i}a del departamento.  

Y lleg\'o la hora de volar. 

I want to thank Edward Farhi for inviting me to visit MIT. Thanks also to Jeffrey Goldstone, Sam Gutmann, Andrew Childs, Andrew Landahl, and Enrico Deotto: I had a really good time in Boston. Thanks to Julien Vidal for inviting Enrique and me to collaborate with him in Paris.
 A Ignacio Cirac he de agradecerle entre otras cosas sus buenos consejos y su confianza, as\'{\i} como el invitarme a visitar el grupo de Garching. Gracias a -- thanks to -- Mar\'{\i}a, Diego, Juanjo, Bel\'en, David (qu\'e gran disfraz el de aqu\'el d\'{\i}a), G\'eza, Susana, Michael, Christine, mi spanglish friend Elva, Stefan, Toby, Enrique Solano, Renate... hicisteis que Bavaria fuera como mi casa, y aprend\'{\i} mucho con vosotros. 
 Thanks to Daniel Gottesman and Debbie Leung for inviting me to visit Perimeter Institute and the University of Waterloo, and also to Mike Mosca, Lucien Hardy, Carlos Mochon, Mary Beth Ruskai and Frank Wilhem for their hospitality and for sharing interesting discussions about physics with me. I am grateful as well to David P. DiVincenzo for inviting me to visit  the IBM Watson Research Center, and thanks to Charles Bennett, John Smolin, Barbara Terhal, and Roberto Oliveira: it was wonderful in New York as well. 
 De nuevo, gracias a Guifr\'e Vidal por invitarme a visitar las ant\'{\i}podas y a comer carne de canguro, 
 and thanks to all the people that I met at the quantum information group and the physics' department of the University of Queensland for their hospitality and interesting discussions: Michael Nielsen, David Poulin, Alexei Gilchrist, Andrew Doherty, Kenny, Norma, Robert Spalek, Rolando Somma, \'Alvaro, Juliet, Aggie, Huan-Qiang Zhou, John Fjaerestad... I think we are going to meet again.  

Tambi\'en he conocido a much\'{\i}sima gente y he hecho amigos en congresos y escuelas, a quien en mayor o menor medida debo agradecerles lo bien que me lo he pasado haciendo f\'{\i}sica durante los \'ultimos cuatro a$\tilde{{\rm n}}$os y pico. El primer TAE en Pe$\tilde{{\rm n}}$\'{\i}scola alcanz\'o la categor\'{\i}a de genial: Olga, Ester, Carmen, Pedro... estuvo bien, ?`eh?
The time at the Les Houches summer school was memorable: thanks to all of you Les Houches guys, Elva, Carlos, Fabio, Derek, Alex, Silvia, Sara, Cameron, John, Neill, Andr\'e, Alessio, Luca, The-Russian-Guy, Toby, David, Maggie, Chris, Philippe, Ru-Fen, Augusto... the french national day will never be the same for me. Thanks also to Fabio Anselmi, Yasser Omar, Roberta Rodr\'{\i}guez, Jeremie Roland, Ver\'onica Cerletti (era ``posho'', ?`no?), Marcos Curty, Philipp Hyllus, Jiannis Pachos, Angelo Carollo, Almut Beige, Jonathan Oppenheim, Ivette Fuentes-Schuller, David Salgado, Ad\'an Cabello, Marcus Cramer, Shashank Virmani... and so many people that I met and who I can not remember right now, but to whom I am grateful too.  

Y aterrizamos en Barcelona. 

Una menci\'on especial se la merecen mis compa$\tilde{{\rm n}}$eros de departamento, hermanos de batalla cient\'{\i}fica en el arduo v\'{\i}a crucis del doctorado: Aleix, \'Alex, Toni Mateos, Toni Ram\'{\i}rez, Jan, Otger, Ernest (desgraciaaaaaaaaaaat!), M\'{\i}riam y su infinita paciencia, Xavi y sus pesas de buzo, Diego y sus acordes, Dani, Carlos y sus c\'omics, el \'{\i}nclito y maravilloso Luca, Jordi Garra, Joan Rojo, Jaume, Majo, Arnau Rios, Chumi alias Cristian, Jordi Mur, Sandro el revolucionario, Ignazio, Enrique, Llu\'{\i}s... sin vosotros me habr\'{\i}a aburrido como una ostra. Otra menci\'on especial a mis coleguillas de la carrera: sois tantos que no cab\'eis todos ni en 500 p\'aginas, pero ya sab\'eis quienes sois, as\'{\i} que daros por agradecidos. De todas formas, gracias en especial a Encarni (o actual se$\tilde{{\rm n}}$ora Pleguezuelos) por soportarme estoicamente a lo largo de innumerables caf\'es y crepes de queso de los menjadors, y tambi\'en a Alberto por aguantarme tantas paranoias. Qu\'e grandes partidos de basket con la gente de electr\'onica: gracias tambi\'en a vosotros. E igualmente gracias a la gente que he conocido por el IRC-Hispano, ya sabr\'eis quienes sois si os dais por aludidos al leer esto: me lo he pasado muy bien con vosotros tambi\'en. Tambi\'en he de dar gracias a Pablo, por comprender mi visi\'on del mundo y estar ah\'{\i} cuando hac\'{\i}a falta, entre Silvios y patxaranes. 

Gracias tambi\'en a Apple por inventar el PowerBook de 12'', a Google por encontrarme cada vez que me pierdo, a Donald Knuth por inventar el TeX, al caf\'e, al chocolate, a Haruko, a Guu, al Dr. Fleishman, y a Samantha Carter por ense$\tilde{{\rm n}}$arme a hacer explotar una estrella usando un stargate. Hab\'eis hecho mis \'ultimos cuatro a$\tilde{{\rm n}}$os y medio mucho m\'as agradables y llevaderos. 

Finalmente, gracias a mi familia extensa, el clan Or\'us Lacort y todas sus ramificaciones posibles en todos sus grados de consanguinidad y generaci\'on. Primos, sois de verdad mucho primo. Gracias a mis padres y a mi hermana por tantas cosas y tant\'{\i}simo apoyo incondicional. Gracias a Lidia por aguantarme en su casa de vez en cuando. Y gracias a mi peque$\tilde{{\rm n}}$o y particular desastre fraggel de ojos azules llamado Ondiz por estar siempre ah\'{\i} cuando le necesito. 

    \chapter*{Research papers}

This thesis is the conclusion of four and a half years of work at the \emph{Departament d'Estructura i Constituents de la Mat\`eria} of the \emph{University of Barcelona}. All along this time I have contributed in several research papers, most of them being the basis of the results that I present here. 

\bigskip

The papers on which this thesis is based, sorted in chronological order, are:

\begin{itemize}
\item{R. Or\'us, J. I. Latorre, J. Eisert, and M. Cramer. Half the entanglement in critical systems is distillable from a single specimen, 2005. quant-ph/0509023 (to appear in {\it Phys. Rev. A}).}
\item{ M. C. Ba$\tilde{\rm{n}}$uls, R. Or\'us, J. I. Latorre, A. P\'erez, and P. 
Ruiz-Femen{\' i}a. Simulation of many-qubit quantum computation with matrix 
product states. {\it Phys. Rev. A},  73:022344, 2006.}
\item{R. Or\'us. Entanglement and majorization in (1+1)-dimensional quantum
    systems. {\it Phys. Rev. A}, 71:052327, 2005.  Erratum-ibid  73:019904, 2006.}
\item{J. I. Latorre, R. Or\'us, E. Rico and J. Vidal. Entanglement entropy in the Lipkin-Meshkov-Glick model. {\it Phys. Rev. A}, 71:064101, 2005.}
\item{ R. Or\'us and J. I. Latorre. Universality of entanglement and quantum computation
    complexity. {\it Phys. Rev. A},  69:052308, 2004.}   
\item{J. I. Latorre and R. Or\'us. Adiabatic quantum computation and quantum phase
    transitions. {\it Phys. Rev. A}, 69:062302, 2004.}
\item{ R. Or\'us, J. I. Latorre, and M. A. Mart\'{\i}n-Delgado. Systematic analysis of majorization in quantum algorithms. {\it Eur. Phys. J. D},  29:119, 2004.}
\item{R. Or\'us,  J. I. Latorre, and M. A. Mart\'{\i}n-Delgado. Natural majorization of the quantum Fourier
    transformation in phase-estimation algorithms. {\it Quant. Inf. Proc.}, 4:283, 2003.}
\end{itemize}

\bigskip

Other papers in which I was involved, sorted in chronological order, are:

\begin{itemize}
\item{R. Or\'us. Two slightly-entangled NP-complete problems. {\it Quant. Inf. and Comp.}, 5:449, 2005.}    
\item{R. Or\'us and R. Tarrach. Weakly-entangled states are dense and robust. {\it Phys. Rev. A}, 70:050101, 2004.}
\item{C. Wellard and R. Or\'us. Quantum phase transitions in anti-ferromagnetic
planar cubic lattices. {\it Phys. Rev. A}, 70:062318, 2004.}
\end{itemize}


   \newpage
   \renewcommand{\contentsname}{Contents}  
    \begin{spacing}{1.2}  
    \tableofcontents
    \cleardoublepage
    \end{spacing}

    \pagenumbering{arabic} \setcounter{page}{1}

    \chapter{Introduction}

From the seminal ideas of Feynman \cite{Feynman82} and until now, quantum information and computation \cite{Nielsen-Chuang} has been a rapidly evolving field. While at the beginning, physicists looked at quantum mechanics as a theoretical framework to describe the fundamental processes that take place in Nature, it was during the $80$'s and $90$'s that people began to think about the intrinsic quantum behavior of our world as a tool to eventually develop powerful information technologies. As Landauer pointed out \cite{Landauer61}, \emph{information is physical}, so it should not look strange to try to bring together quantum mechanics and information theory. Indeed, it was soon realized that it is possible to use the laws of quantum physics to perform tasks which are unconceivable within the framework of classical physics. For instance, the discovery of quantum teleportation \cite{BeBa93}, superdense coding \cite{BW92}, quantum cryptography \cite{BB84,Eke91}, Shor's factorization algorithm \cite{Shor94} or Grover's searching algorithm \cite{Grover96}, are some of the remarkable achievements that have attracted the attention of many people, both scientists and non-scientists. This settles down quantum information as a genuine interdisciplinary field, bringing together researchers from different branches of physics, mathematics  and engineering.

While until recently it was mostly quantum information science that benefited from other fields, today the tools developed within its framework can be used to study  problems of different areas, like quantum many-body physics or quantum field theory. The basic reason behind that is the fact that quantum information develops a detailed study of quantum correlations, or quantum \emph{entanglement}. Any physical system described by the laws of quantum mechanics can then be considered from the perspective of quantum information by means of entanglement theory. 

It is the purpose of this introduction to give some elementary background about basic concepts of quantum information and computation, together with its possible relation to other fields of physics, like quantum many-body physics. We begin by considering the definition of a \emph{qubit}, and move then towards the definition of \emph{entanglement} and the convertibility properties of pure states by introducing  \emph{majorization} and the  \emph{von Neumann entropy}. Then, we consider the notions of \emph{quantum circuit} and \emph{quantum adiabatic algorithm}, and move towards what is typically understood by a \emph{quantum phase transition}, briefly sketching how this relates to \emph{renormalization} and \emph{conformal field theory}. We also comment briefly on some possible \emph{experimental implementations} of quantum computers.

\subsection*{What is a ``qubit''?}

A \emph{qubit} is a quantum two-level system, that is, a physical system described in terms of a Hilbert space ${\mathbb C}^2$. You can think of it as a spin-$\frac{1}{2}$ particle, an atom in which we only consider two energy levels, a photon with two possible orthogonal polarizations, or a ``dead or alive'' Schr$\ddot{{\rm o}}$dinger's cat. Mathematically, a possible orthonormal basis for this Hilbert space is denoted by the two orthonormal vectors $|0\rangle$ and $|1\rangle$. This notation is analogous to the one used for a classical bit, which can be in the  two ``states" $0$ or $1$. Notice, however, that the laws of quantum mechanics allow a qubit to physically exist  in \emph{any} linear combination of the states $|0\rangle$ and $|1\rangle$. That is, the generic state $|\psi\rangle$ of a qubit is given by
\begin{equation}
|\psi \rangle = \alpha |0\rangle + \beta |1\rangle \ , 
\end{equation}
where $\alpha$ and $\beta$ are complex numbers such that $|\alpha|^2 + |\beta|^2 = 1$. Given this normalization condition, the above state can always be written as
\begin{equation}
|\psi \rangle = e^{i \gamma} \left( \cos \left(\frac{\theta}{2}\right) |0\rangle + e^{i \phi} \sin \left(\frac{\theta}{2} \right)|1\rangle \right) \ , 
\end{equation}
where $\gamma$, $\theta$ and $\phi$ are some real parameters. Since the global phase $e^{i\gamma}$ has no observable effects, the physical state of a qubit is always parameterized in terms of two real numbers $\theta$ and $\phi$, that is, 
\begin{equation}
|\psi \rangle = \cos \left(\frac{\theta}{2} \right)|0\rangle + e^{i \phi} \sin \left(\frac{\theta}{2}\right) |1\rangle \ . 
\end{equation}
The angles $\theta$ and $\phi$ define a point on a sphere that is usually referred to as the \emph{Bloch sphere}. Generally speaking, it is possible to extend the definition of qubits and define the so-called \emph{qudits}, by means of quantum $d$-level systems. 

\subsection*{What is ``entanglement''?}

The definition of entanglement  varies depending on whether we consider only pure states or the general set of mixed states. Only for pure states, we say that a given state $|\psi\rangle$ of $n$ parties is \emph{entangled} if it is not a tensor product of individual states for each one of the parties, that is, 
\begin{equation}
|\psi \rangle \ne |v_1\rangle_1 \otimes |v_2 \rangle_2 \otimes \cdots \otimes |v_n\rangle_n \ . 
\end{equation}
For instance, in the case of $2$ qubits $A$ and $B$ (sometimes called ``Alice'' and ``Bob") the quantum state
\begin{equation}
|\psi^+\rangle = \frac{1}{\sqrt{2}} \left( |0\rangle_A \otimes  |0\rangle_B + |1\rangle_A \otimes |1\rangle_B \right) 
\label{maxen}
\end{equation}
is entangled since $|\psi^+\rangle \ne |v_A\rangle_A \otimes |v_B\rangle_B$. On the contrary, the state 
\begin{equation}
|\phi\rangle = \frac{1}{2} \left( |0\rangle_A \otimes |0\rangle_B + |1\rangle_A \otimes |0\rangle_B  + |0\rangle_A \otimes |1\rangle_B     + |1\rangle_A \otimes |1\rangle_B  \right)
\end{equation}
is not entangled, since 
\begin{equation}
|\phi\rangle = \left(\frac{1}{\sqrt{2}}\left( |0\rangle_A + |1\rangle_A \right) \right) \otimes \left(\frac{1}{\sqrt{2}}\left( |0\rangle_B + |1\rangle_B \right) \right) \ .
\label{separab}
\end{equation}
A pure state like the one from Eq.\ref{maxen} is called a \emph{maximally entangled state of two qubits}, or a \emph{Bell pair}, whereas a pure state like the one from Eq.\ref{separab} is called \emph{separable}. 

In the general case of mixed states, we say that a given state $\rho$ of $n$ parties is \emph{entangled} if it is not  a probabilistic sum of tensor products of individual states for each one of the parties, that is, 
\begin{equation}
\rho \ne \sum_k p_k  \ \rho^k_1 \otimes \rho^k_2 \otimes \cdots \otimes \rho^k_n \ , 
\end{equation}
with $\{ p_k \}$ being some probability distribution. Otherwise, the mixed state is called \emph{separable}. 

The essence of the above definition of entanglement relies on the fact that  entangled states of $n$ parties cannot be prepared by acting locally on each one of the parties, together with classical communication (telephone calls, e-mails, postcards...) among them. This set of operations is often referred to as ``local operations and classical communication'', or LOCC. If the actions performed on each party are probabilistic, as is for instance the case in which one of the parties  draws a random variable according to some probability distribution, the set of operations is called ``stochastic local operations and classical communication", or SLOCC. Entanglement is, therefore, a genuine quantum-mechanical feature which does not exist in the classical world. It carries non-local correlations between the different parties in such a way that they cannot be described classically, hence, these correlations are \emph{quantum correlations}. 

The study of the structure and properties of entangled states constitutes what is known as \emph{entanglement theory}. In this thesis, we shall always restrict ourselves to the entanglement that appears in pure states. We also wish to remark that the notation for the tensor product of pure states can be different depending on the textbook, in such a way that  $|v_A \rangle_A \otimes |v_B \rangle_B = |v_A \rangle_A |v_B \rangle_B = |v_A, v_B \rangle $. An introduction to entanglement theory, both for pure and mixed states, can be found for instance in \cite{Lew00}. 

\subsection*{Majorization and the von Neumann entropy}

\emph{Majorization theory} is a part of statistics that studies the notion of order in probability distributions\cite{Muirhead1903, Hardy78, Marshall79, Bathia97}. Namely, majorization states that given two probability vectors $\vec{x}$ and $\vec{y}$, the probability distribution $\vec{y}$ majorizes $\vec{x}$, written as $\vec{x} \prec \vec{y}$, if and only if 
\begin{equation}
\vec{x} = \sum_k p_k P_k \vec{y} \ , 
\end{equation}
where $\{ p_k \}$ is a set of probabilities and $\{ P_k \}$ is a set of permutation matrices. The above definition implies that the probability distribution $\vec{x}$ is more disordered than the probability distribution $\vec{y}$, since it can be obtained by a probabilistic sum of permutations of $\vec{y}$. More details on majorization theory, which is often used in this thesis, are given in Appendix A. 

Majorization theory has important applications in quantum information science. One of them is that it provides a criteria  for the interconvertibility of bipartite pure states under LOCC. More concretely, given 
two bipartite states $|\psi_{AB} \rangle$ and $|\phi_{AB} \rangle$ for parties $A$ and $B$, and given the spectrums $\vec{\rho}_{\psi}$ and $\vec{\rho}_{\phi}$ of their respective reduced density matrices describing any of the two parties, the state $|\psi_{AB} \rangle$ may be transformed to $|\phi_{AB} \rangle$ by LOCC if and only if \cite{Nielsen99}
\begin{equation}
\vec{\rho}_{\psi} \prec \vec{\rho}_{\phi} \ . 
\label{majjj}
\end{equation}

An important theorem from classical information theory that plays a role in the study of entanglement  is the so-called \emph{theorem of typical sequences}. In order to introduce it, let us previously sketch some definitions. Consider a source of letters $x$ which are produced with some probability $p(x)$. The \emph{Shannon entropy} associated to this source is defined as $H = -\sum_x p(x) \log_2 p(x)$. Given a set of $n$ independent sources, we say that a string of symbols $(x_1, x_2, \ldots , x_n)$ is $\epsilon$-typical if 
\begin{equation}
2^{-n (H - \epsilon)} \le p(x_1, x_2, \ldots , x_n) \le 2^{-n(H + \epsilon)} \ , 
\end{equation}
where $p(x_1, x_2, \ldots , x_n) \equiv p(x_1) p(x_2) \cdots p(x_n)$ is the probability of the string. The set of the $\epsilon$-typical sequences of length $n$ is denoted as $T(n,\epsilon)$. We are now in position of considering the theorem of typical sequences, which is composed of three parts:

\bigskip

{\bf Theorem 0.1 (of typical sequences):} {\it 
\begin{itemize}
\item{Given $\epsilon > 0$, for any $\delta > 0$ and sufficiently large $n$, the probability that a sequence is $\epsilon$-typical is at least $1-\delta$.}
\item{For any fixed $\epsilon > 0$ and $\delta > 0$, and sufficiently large $n$, the number $|T(n,\epsilon)|$ of $\epsilon$-typical sequences satisfies
\begin{equation}
(1-\delta)2^{n (H - \epsilon)} \le |T(n,\epsilon)| \le 2^{n(H + \epsilon)} \ . 
\end{equation}
}
\item{Let $S(n)$ be a collection of size at most $2^{nR}$, of length $n$ sequences from the source, where $R < H$ is fixed. Then, for any $\delta > 0$ and for sufficiently large $n$, 
\begin{equation}
\sum_{(x_1, x_2, \ldots , x_n) \ \in \ S(n)} p(x_1, x_2, \ldots , x_n) \le \delta \ . 
\end{equation}
}
\end{itemize}
}

\bigskip

It is not our purpose here to provide a detailed proof of this theorem (the interested reader is addressed for instance to \cite{Nielsen-Chuang}). We shall, however, make use of it in what follows. 

Let us introduce at this point a quantity which is to play a major role all along this thesis. Given a bipartite pure quantum state $|\psi_{AB} \rangle$, with reduced density matrices $\rho_A = {\rm tr}_B ( |\psi_{AB} \rangle \langle \psi_{AB}| )$ and $\rho_B = {\rm tr}_A ( |\psi_{AB} \rangle \langle \psi_{AB}| )$, the \emph{von Neumann entropy} of this bipartition is defined as
\begin{equation}
S \equiv S(\rho_A) = -{\rm tr} (\rho_A \log_2 \rho_A) = S(\rho_B) =  -{\rm tr} (\rho_B \log_2 \rho_B) \ , 
\end{equation}
where the equality follows from the fact that $\rho_A$ and $\rho_B$ share the same spectrum. This entropy is also called \emph{entanglement entropy}, since it provides a measure of the bipartite entanglement present in pure states. To be precise, the entanglement entropy measures the optimal rate at which it is possible to distill Bell pairs by LOCC in the limit of having an infinite number of copies of the bipartite system. 

Let us explain how the above consideration works. Given the bipartite pure state $|\psi_{AB} \rangle$, we write it in terms of the so-called \emph{Schmidt decomposition}:
\begin{equation}
|\psi_{AB}\rangle = \sum_x \sqrt{p(x)} |x_A\rangle_A|x_B\rangle_B \ , 
\end{equation}
where the square $p(x)$ of the Schmidt coefficients define the probability distribution that appears as the spectrum of the reduced density matrices for the two parties. The $n$-fold tensor product $|\psi_{AB}\rangle ^{\otimes n}$ can be written as
\begin{equation}
|\psi_{AB}\rangle ^{\otimes n} = \sum_{(x_1, x_2, \ldots , x_n)} \sqrt{p(x_1) p(x_2) \cdots p(x_n)} |x_{1A}, x_{2A}, \ldots , x_{nA} \rangle_A |x_{1B}, x_{2B}, \ldots , x_{nB} \rangle_B \ . 
\label{bippp}
\end{equation}
Let us now define a quantum state $|\phi_n\rangle$ obtained by omitting in Eq.\ref{bippp}
those strings $(x_1, x_2, \ldots , x_n)$ which are not $\epsilon$-typical: 
\begin{equation}
|\phi_n\rangle = \sum_{(x_1, x_2, \ldots , x_n) \ \in \ T(n,\epsilon)} \sqrt{p(x_1) p(x_2) \cdots p(x_n)} |x_{1A}, x_{2A}, \ldots , x_{nA} \rangle_A |x_{1B}, x_{2B}, \ldots , x_{nB} \rangle_B \ .  
\end{equation}
Since the previous state is not properly-normalized, we define the state $|\phi^{\prime}_n \rangle \equiv |\phi_n\rangle / \sqrt{\langle \phi_n | \phi_n \rangle }$. Because of the first part of the theorem of typical sequences, the overlap between $|\psi_{AB}\rangle^{\otimes n}$ and $|\phi^{\prime}_n\rangle$ tends to $1$ as $n \rightarrow \infty$. Furthermore, by the second part of the theorem we have that  $|T(n,\epsilon)| \le 2^{n(H + \epsilon)} = 2^{n(S + \epsilon)}$. Given these properties, a possible protocol  to transform copies of the state $|\psi_{AB}\rangle$ into Bell pairs by means of LOCC reads as follows: party $A$ may convert the state $|\psi_{AB}\rangle^{\otimes n}$ into the state $|\phi^{\prime}_n \rangle$ with high probability by performing a local measurement into its $\epsilon$-typical subspace. The largest Schmidt coefficient of $|\phi_n \rangle$ is $2^{-n(S - \epsilon)/2}$ by definition of typical sequence, and since the theorem of typical sequences also tells us that $1-\delta$ is a lower bound on the probability for a sequence to be $\epsilon$-typical, the largest Schmidt coefficient of $|\phi^{\prime}_n \rangle$ is at most $2^{-n(S - \epsilon)/2} / \sqrt{1-\delta}$. Let us now choose an $m$ such that 
\begin{equation}
\frac{2^{-n (S - \epsilon)}}{1-\delta} \le 2^{-m} \ . 
\end{equation}
Then, the spectrum of the reduced density matrices for $A$ and $B$ are majorized by the probability vector $(2^{-m}, 2^{-m}, \ldots , 2^{-m})^T$, and therefore the state $|\phi^{\prime}_n\rangle $ can be transformed into $m$ copies of a Bell state by means of local operations and classical communication. More specifically, in the limit $n \rightarrow \infty$ the ratio $m/n$ between the number of distilled Bell pairs and the original number of states exactly coincides with the entanglement entropy $S$. 

It is possible to see that the above distillation protocol is optimal, that is, it is not possible to distill more than $n S$ Bell pairs from a total of $n$ copies of a bipartite pure state in the limit $n \rightarrow \infty$. Because of this property, the von Neumann entropy is also called the \emph{distillable entanglement} of a pure bipartite system. Furthermore, it is possible to see that the entropy $S$ coincides as well with the \emph{entanglement of formation} of bipartite pure states, which is the optimal ratio $m/n$ describing the number $m$ of Bell pairs that are required to create $n$ copies of a given bipartite pure state by means of LOCC, in the limit $n \rightarrow \infty$. The von Neumann entropy constitutes then a genuine measure of the bipartite entanglement that is present in a given pure quantum  state. 

\subsection*{Quantum circuits and adiabatic quantum algorithms}

Much in analogy to the situation in classical computation, where it is possible to define a computation by means of logic gates applied to bits, a quantum computation may be defined in terms  of a set of  \emph{unitary gates} applied to qubits. These unitary gates may either be local, acting on a single qubit, or non-local, acting on several qubits at a time. An important example of a local gate is given by the so-called Hadamard gate:
\begin{equation}
U_H = \frac{1}{\sqrt{2}}
\begin{pmatrix}
1 & 1 \\
1 & -1 
\end{pmatrix}
\ , 
\end{equation}
which acts on the two-dimensional Hilbert space of a single qubit such that
\begin{eqnarray}
U_H |0\rangle & = & \frac{1}{\sqrt{2}} (|0\rangle + |1\rangle ) \nonumber \\
U_H |1\rangle & = & \frac{1}{\sqrt{2}} (|0\rangle - |1\rangle ) \ . 
\end{eqnarray}
Also, an important example of a non-local gate is the controlled-not gate $U_{CNOT}$:
\begin{equation}
U_{CNOT} = 
\begin{pmatrix}
1 & 0 & 0 & 0 \\
0 & 1 & 0 & 0 \\
0 & 0 & 0 & 1 \\
0 & 0 & 1 & 0 
\end{pmatrix}
\ , 
\end{equation}
acting on the four-dimensional Hilbert space of two qubits such that 
\begin{eqnarray}
U_{CNOT} |0,0\rangle &=& |0,0\rangle \nonumber \\
U_{CNOT} |0,1\rangle &=& |0,1\rangle \nonumber \\
U_{CNOT} |1,0\rangle &=& |1,1\rangle \nonumber \\
U_{CNOT} |1,1\rangle &=& |1,0\rangle  \ . 
\end{eqnarray}
In the example of the controlled-not gate, the first and second qubits are respectively called the \emph{controlling qubit} and the \emph{controlled qubit}, since the action of the gate on the second qubit depends on the value of the first one. It is possible to define more general \emph{controlled gates} similarly to the controlled-not gate, namely, if the controlling qubit is in the state $|0\rangle$ nothing is done on the second one, whereas if the controlling qubit is in the state $|1\rangle$ then some local unitary gate acts on the second qubit. 
The application of the different unitary gates that define a quantum computation on a system of qubits  can be represented in terms of \emph{quantum circuits}, such as the ones from Fig.\ref{hamon} and Fig.\ref{qcircu}. In a quantum circuit each wire represents a qubit, and the time flows from left to right. 
\begin{figure}
\begin{equation}
\Qcircuit @C=1.5em @R=1em {
   & \gate{U_H} & \qw } 
\ \ \ \ \ \ \ \ \ \ \ \ \ \ \
   \Qcircuit @C=1.5em @R=1em {
    & \ctrl{+1} & \qw \\
    & \targ      & \qw 
}      
\nonumber
\end{equation}
\caption{Quantum circuits representing the action of a Hadamard gate on a single qubit and a controlled-not gate on two qubits. The controlling qubit is denoted by a black dot, and the controlled qubit is denoted by the symbol $\oplus$.}  \label{hamon}
\end{figure}
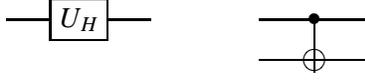
\begin{figure}
\begin{equation}
\Qcircuit @C=1.5em @R=1em {
 & \gate{U_H} & \ctrl{+3} & \qw & \targ & \qw & \qw & \qw & \qw & \qw & \qw & \qw & \meter \\
 & \qw               & \qw         & \qw & \qw &  \ctrl{3} & \qw & \targ & \gate{U_H} & \qw & \qw & \qw & \meter \\
  & \qw & \qw & \qw & \ctrl{-2} & \qw & \gate{U_H} & \ctrl{-1} & \qw & \qw & \qw & \qw & \meter \\
  & \qw & \targ & \gate{U_H} & \qw & \qw & \qw & \qw & \qw & \ctrl{1} & \qw & \qw & \meter \\
  & \qw & \qw & \qw & \qw & \targ & \qw & \qw & \qw & \targ & \gate{U_H} & \qw & \meter
 } 
 \nonumber
\end{equation}
\caption{A possible quantum circuit of $5$ qubits composed of Hadamard and controlled-not gates. Some measurements are performed on the qubits at the end of the quantum computation.}  \label{qcircu}
\end{figure}
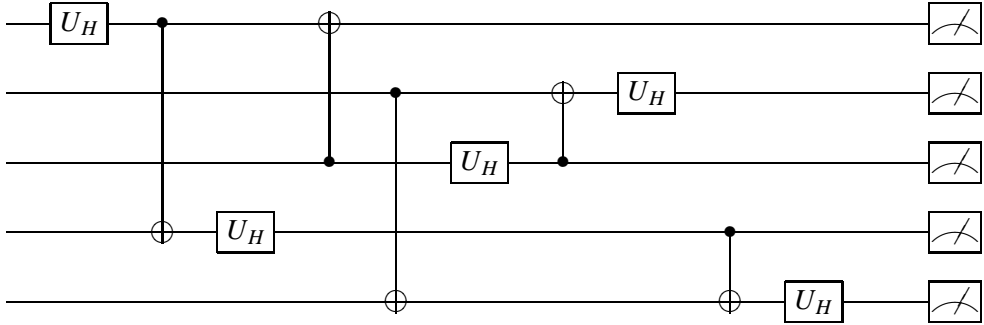

Independently of quantum circuits, it is possible to define alternative models to perform quantum computations, such as the adiabatic model of quantum optimization \cite{Farhi00_1}. The adiabatic quantum algorithm 
deals with the problem
of finding the ground state of a physical system represented by its Hamiltonian $H_P$. The basic idea is to perform an interpolation in time between some easy-to-build Hamiltonian $H_0$ and $H_P$, such that if the initial state of our system is a ground state of $H_0$, we may end up in a ground state of $H_P$ with high probability after evolving for a certain amount of time, as long as some adiabaticity conditions are fulfilled. 
For example, we could consider the time-dependent Hamiltonian 
\begin{equation}
H(t) = \left(1-\frac{t}{T}\right) H_0 + \frac{t}{T} H_P \ , \label{jamon}
\end{equation}
where $t \in [0,T]$ is the time parameter, $T$ being some computational interpolation time.
If $g_{min}$ represents the global minimum along the evolution 
of the energy gap between the ground state and the first excited state of the system, 
the adiabatic theorem implies that, if at $t=0$ the system is at ground state of $H_0$, in order to be at the ground state of $H_P$ at time $T$ with high probability it is required that $T \sim 1/  g_{min}^2$. The scaling properties with the size of the system of the minimum energy gap controls then the computational time of the quantum algorithm. Actually, the fact that the system evolves through a point of minimum gap implies that it approaches a quantum critical point, to be defined in what follows. A more detailed explanation of adiabatic quantum algorithms is given in Chapter 4. 

\subsection*{Quantum criticality in quantum many-body systems}

A \emph{quantum phase transition} is a phase transition between different phases of matter at zero temperature. Contrary to classical (also called ``thermal'') phase transitions, quantum phase transitions are driven by the variation of some physical parameter, like a magnetic field. The transition describes an abrupt change in the properties of the ground state of the quantum system due to the effect of quantum fluctuations. The point in the space of parameters at which a quantum phase transition takes place is called the \emph{critical point}, and separates quantum phases of different symmetry.  

Some properties of the system may display a characteristic behavior at a quantum critical point. For instance,  the correlators in a quantum many-body system may decay to zero as a power-law at criticality, which implies a divergent correlation length and therefore scale-invariance, while decaying exponentially at off-critical regimes. Since quantum correlations are typically maximum at the critical point, some entanglement measures may have a divergence. The ground-state energy may display non-analyticities when approaching criticality, and the energy gap between the ground state and the first excited state of the system may close to zero. Our definition of quantum phase transition is very generic and  does not necessarily involve all of the above behaviors. In fact, it is indeed possible to find quantum systems in which there is an abrupt  change of the inner structure of the ground state that can be detected by some properties but not by others \cite{Del04}. 

\begin{figure}[h]
\centering
\includegraphics[width=0.6\textwidth]{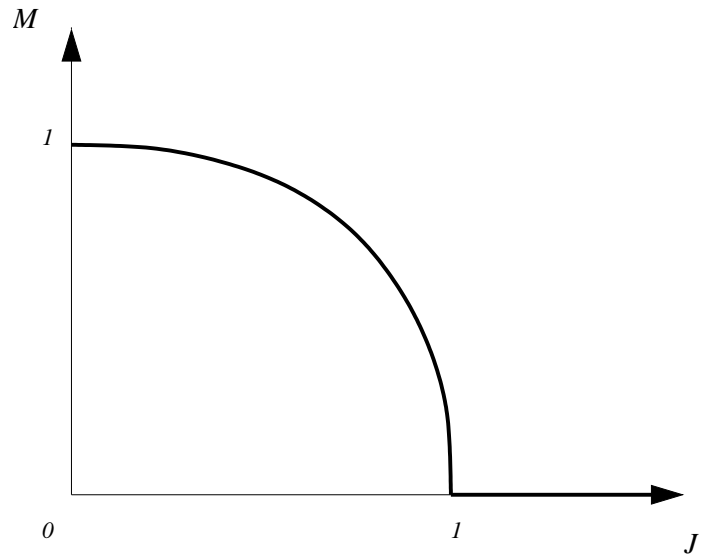}
\caption{Magnetization per particle in the ferromagnetic quantum Ising spin chain as a function of the coupling parameter, in the thermodynamic limit. The point $J=J^* = 1$ corresponds to a second-order quantum phase transition point.} 
\label{magising}
\end{figure}

Let us give a simple example of a quantum critical point: consider the $(1+1)$-dimensional\footnote{We use the field-theoretical notation $(1+1)$ to denote one spatial and one temporal dimension. Time is always to be kept fixed.} ferromagnetic quantum Ising spin chain, as defined by the  Hamiltonian
\begin{equation}
H =  -J\sum_{i = 1}^{N} \sigma_i^x \sigma_{i+1}^x - \sum_{i=1}^{N} \sigma_i^z \ , 
\label{Isinggg}
\end{equation}
where $\sigma_i^{\alpha}$ is the Pauli matrix $\alpha$ at site $i$ of the chain, $J \ge 0$ is a coupling parameter, and $N$ is the number of spins. At $J=\infty$ the ground state of the system is two-fold degenerate and consists of all the spins aligned ferromagnetically in the $x$-direction,  being its subspace spanned by the two vectors $|+,+,\ldots, + \rangle$ and $|-,-,\ldots ,-\rangle$, where $|+\rangle$ and $|-\rangle$ denote the two possible eigenstates of the pauli matrix $\sigma^x$. On the other hand, at $J = 0$ the ground state of the system consists of all the spins aligned along the $z$-direction, $|0,0,\ldots , 0 \rangle $, where $|0\rangle = \frac{1}{\sqrt{2}} (|+\rangle + |-\rangle)$. We now consider the behavior of the magnetization per particle of the ground state in the $z$-direction, as defined by the expected value $M \equiv \frac{<\sum_{i=1}^N \sigma_i^z >}{N}$. In the thermodynamic limit $N \rightarrow \infty$ this quantity tends to one when $J \rightarrow 0$, and tends to zero when $J \rightarrow \infty$. A detailed analysis of this model in this limit shows that there is a specific point  at which the magnetization per particle has a sudden change, as is represented in Fig.\ref{magising}. This behavior implies that the model undergoes a second-order quantum phase transition at the critical point $J = J^* = 1$ in the thermodynamic limit. 
 
One may wonder what is the symmetry that we are breaking in this simple example of a quantum phase transition: it is the symmetry $\mathbb{Z}_2$ that the Hamiltonian from Eq.\ref{Isinggg} has at high  values of the coupling parameter. In fact, this symmetry could even be further broken when $J \rightarrow \infty$ if some extremely small magnetic field in the $x$-direction were present in our system, selecting one of the two possible ground states within this phase. In such a case, it is said that the symmetry of the Hamiltonian is \emph{spontaneously broken}. 

A useful tool in the study of quantum critical systems is the \emph{renormalization group} \cite{Wilson71_1, Wilson71_2}, which describes the way in which a theory gets modified under scale transformations. Given some Hamiltonian depending on a set of parameters, the transformations of the renormalization group define a flow in the parameter space, and in particular the fixed points of those transformations correspond to theories which are invariant under changes of scale. Indeed, the essence of the renormalization procedure is the elimination of degrees of freedom in the description of a system. This point of view is one of the basis for the development of different numerical techniques that allow to compute basic properties of quantum many-body systems, as is the case of the so-called 
\emph{density matrix renormalization group} algorithm \cite{White92}. 

The behavior of many quantum critical models can also be explained by using tools from  \emph{conformal field theory} \cite{Ginsparg}. There are quantum many-body systems which can be understood as a regularization on a lattice of a quantum field theory, as is the case of the previously-discussed Ising quantum spin chain, which can be represented by the quantum field of a $(1+1)$-dimensional spinless fermion \cite{Rico2005}. When those quantum many-body systems become critical, their description in terms of a quantum field theory allows to see that the symmetry group is not composed of only scale transformations, but of the full group of \emph{conformal transformations}. In fact, conformal symmetry is  particularly powerful when applied to $(1+1)$-dimensional quantum systems, allowing to determine almost all the basic properties of the model in consideration just by means of symmetry arguments. We perform some conformal field theory calculations in this thesis, and some basic technical background is given in Appendix B. 

\subsection*{Experimental quantum computers}

There will exist some day a quantum computer? This apparently simple question is by no means easy to answer. Actually, it is the opinion of some scientists that it is eventually impossible to build a quantum computer because of the unavoidable problem of the \emph{decoherence} that any quantum system undergoes when it interacts with its environment. Nevertheless, other physicists think that these experimental drawbacks can be eventually in part ameliorated if the appropriate conditions are given.  The main requirements that any experimental proposal must match if its purpose is to faithfully represent a quantum computer are known as the \emph{DiVincenzo criteria} \cite{DIV95}, and so far there have been many different ideas to perform experimental quantum computation that try to fulfill as much as possible these conditions. Important proposals are those based on quantum optical devices, such as the \emph{optical photon quantum computer}, \emph{cavity quantum electrodynamics devices}, \emph{optical lattices}, or \emph{ion traps} \cite{CZ95}.  The idea of performing quantum computation by means of  \emph{nuclear magnetic resonance} (NMR) has been considered as well \cite{div95b, chf97, gc97}. Furthermore, proposals based on \emph{superconductor devices}, \emph{quantum dots} \cite{LD98}, and \emph{doped semiconductors} \cite{Kan98, VYW99} have also been considered by different people. The future development of these and other experimental techniques, and to what extent they can implement a many-qubit quantum computer, remains yet uncertain. A detailed discussion about experimental quantum computation can be found for instance in \cite{Nielsen-Chuang}.

\subsection*{What is this thesis about?}

We focus here on the fields of quantum information science, condensed-matter physics, and quantum field theory. While these three branches of physics can be regarded as independent by themselves, there are clear overlaps among them, such that knowledge from one field benefits the others. As we said, conformal field theory \cite{Ginsparg} has helped to understand the universality classes of many critical $(1+1)$-dimensional quantum many-body systems.  Also, the study of the entanglement present in the ground state of quantum Hamiltonians at a quantum phase transition shows direct analogies with those coming from the study of entropies in quantum field theory \cite{Sred93, Callan94, Preskill94, Kabat94, Kabat95, Holzhey94, Latorre03, Latorre04_1, Korepin04_1, Korepin04_2, Calabrese04, Calabrese05, Casini05_1, Casini05_2}.  These results in turn connect with the performance of numerical techniques like the density matrix renormalization group \cite{White92}, that allow to compute basic properties of some quantum many-body systems \cite{AKLT87, AKLT88,Ostlund95, Fannes92, GVidal03_1,GVidal04, MG04, Scholl05_2, VPC04, VGRC04, VW05, VCprb06, Scholl05, VC04,GVidal05,Anders06}. Indeed, quantum phase transitions are very much related to the model of adiabatic quantum computation \cite{Farhi00_1, Farhi00_2, Childs00, Farhi01, Childs02_1, Farhi02_1, Farhi02_2, Farhi05, Jordan05,  Roland02, Dam01, Aharonov04}, which poses today challenges within the field of computational complexity \cite{Cook71}. 

The work that we present in this thesis tries to be at the crossover of quantum information science, quantum many-body physics, and quantum field theory. We use tools from these three fields to analyze problems that arise in the interdisciplinary intersection. More concretely, in Chapter 1 we consider the irreversibility of renormalization group flows from a quantum information perspective by using majorization theory and conformal field theory. In Chapter 2 we compute the entanglement of a single copy of a bipartite quantum system for a variety of models by using techniques from conformal field theory and Toeplitz matrices. The entanglement entropy of the so-called Lipkin-Meshkov-Glick model is computed in Chapter 3, showing analogies with that of $(1+1)$-dimensional quantum systems. In Chapter 4 we apply the ideas of scaling of quantum correlations in quantum phase transitions to the study of quantum algorithms, focusing on Shor's factorization algorithm and quantum algorithms by adiabatic evolution solving an NP-complete and the searching problems. Also, in Chapter 5 we use techniques originally inspired by condensed-matter physics to develop classical simulations,  using the so-called matrix product states, of an adiabatic quantum algorithm. Finally, in Chapter 6 we consider the behavior of some families of quantum algorithms from the perspective of majorization theory. 

The structure within each Chapter is such that the last section always summarizes the basic results. Some general conclusions and possible future directions are briefly discussed in Chapter 7.  Appendix A, Appendix B and Appendix C respectively deal with some basic notions on majorization theory,  conformal field theory, and classical complexity theory.

    \chapter{Majorization along parameter and renormalization group flows}

Is it possible to somehow relate physical theories that describe Nature at different scales? Say, given a theory describing Nature at high energies, we should demand that the effective low-energy behavior should be obtained by integrating out the high-energy degrees of freedom, thus getting a new theory correctly describing the low-energy sector of the original theory. This should be much in the same way as Maxwell's electromagnetism correctly describes the low-energy behavior of quantum electrodynamics. 

This non-perturbative approach to the fundamental theories governing Nature was essentially developed by Wilson and is the key ingredient of the so-called renormalization group \cite{Wilson71_1, Wilson71_2, Wilson74}: effective low-energy theories can be obtained from high-energy theories by conveniently eliminating the high-energy degrees of freedom. To be more precise, the renormalization group is the mechanism that controls the modification of a physical theory through a change of scale. Renormalization group transformations then define a flow in the space of theories from high energies (ultraviolet theories) to low energies (infrared theories). Actually, it is possible to extend this idea, and the renormalization procedure can be more generically understood as the \emph{elimination of some given degrees of freedom} which we are not interested in because of some reason. The name ``renormalization group'' is used due to historical reasons, since the set of transformations does not constitute a formal group from a mathematical point of view. 

Since the single process of integrating out modes seems to apparently be an irreversible operation by itself, one is naturally led to ask whether  renormalization group flows are themselves irreversible. This question is in fact equivalent to asking whether there is a fundamental obstruction to recover microscopic physics from macroscopic physics, or more generally, whether there is a net information loss along renormalization group trajectories. While some theories may exhibit limit cycles in these flows, the question is under which conditions irreversibility remains. Efforts in this direction were originally carried by Wallace and Zia \cite{Wallace75}, while a key theorem was later proven by Zamolodchikov \cite{Zam86} in the context of (1+1)-dimensional quantum field theories: for every unitary, renormalizable, Poincar\'e invariant quantum field theory, there exists a universal $c$-function which decreases along renormalization group flows, while it is only stationary at (conformal) fixed points, where it reduces to the central charge $c$ of the conformal theory. This result sets an arrow on renormalization group flows, since it implies that a given theory can be the infrared (IR) realization of another ultraviolet (UV) theory only if their central charges satisfy the inequality $c_{IR} < c_{UV}$. 

The following question then arises: ``under which conditions irreversibility of renormalization group flows holds in higher dimensions?''. This has been addressed from different perspectives 
\cite{Capelli92, Zumbach94_1, Zumbach94_2, Generowicz97, Haagensen94, Bastianelli96, Anselmi98_1, Anselmi98_2, Cardy88, Osborn89, Osborn90, Capelli91, Latorre98, Latorre98_2, Osborn2000, Capelli2000, Capelli2000_2, Barnes04, Anselmi02}.  It is our purpose here to provide a new point of view about this problem based on the accumulated knowledge from the field of quantum information science, by focusing first on the case of $(1+1)$ dimensions. 

An important application of quantum
information to quantum many-body physics has been the use of majorization theory \cite{Muirhead1903,Bathia97, Hardy78, Marshall79} in order to analyze the structure present in the ground state -- also called vacuum -- of some models along
renormalization group flows. Following this idea,
in \cite{Lutken04} it was originally proposed that irreversibility along the flows may be rooted in properties concerning only the vacuum, without necessity of
accessing the whole Hamiltonian of the system and its full tower of
eigenstates. Such an irreversibility was casted into the idea of an \emph{
  entanglement loss} along renormalization group flows, which proceeded in three constructive steps for
(1+1)-dimensional quantum systems: first, due to the fact that the central
charge of a (1+1)-dimensional conformal field theory is in fact a genuine
measure of the bipartite entanglement present in the ground state of the
system \cite{Holzhey94, Latorre03, Latorre04_1, Korepin04_1, Korepin04_2, Calabrese04, Calabrese05, Casini05_1, Casini05_2}, there is a global loss of
entanglement due to the $c$-theorem of Zamolodchikov \cite{Zam86}; second, given a splitting of the system into two contiguous pieces, there is a 
monotonic loss of entanglement due to the
numerically observed monotonicity for the entanglement entropy between the
two subsystems along the
flow, decreasing when going away from the critical fixed -- ultraviolet -- point;
third, this loss of entanglement is seen to be fine-grained, since it follows
from a strict set of majorization ordering relations, numerically obeyed by the
eigenvalues of the reduced density matrix of the subsystems. This last step motivated the authors of \cite{Lutken04} to conjecture
that there was a \emph{fine-grained entanglement loss} along renormalization group flows rooted
\emph{only in properties of  the vacuum}, at least for (1+1)-dimensional quantum systems.  
In fact, a similar fine-grained entanglement loss had already been numerically
observed in \cite{Latorre03, Latorre04_1}, for changes in 
the size of the bipartition described by the corresponding ground-state
density operators, at conformally-invariant critical points.  

The aim of this Chapter is to analytically prove relations between conformal field theory, 
renormalization group and entanglement.
We develop, in the bipartite scenario, a detailed and analytical study 
of the majorization properties of the eigenvalue spectrum obtained 
from the reduced density matrices of the ground state for a variety of 
(1+1)-dimensional quantum models. 
Our approach is based on infinitesimal variations of the parameters defining
the model -- magnetic fields, anisotropies -- or deformations in the size of the block 
$L$ for one of the subsystems. We prove in these situations that there 
are strict majorization relations
underlying the structure of the eigenvalues of the considered reduced density 
matrices or, in other words, that there is a fine-grained entanglement loss. 
The result of our study is presented in terms of two
theorems. On the one hand,  we are able to prove continuous 
majorization relations as a function of the parameters defining the model under study. Some of these flows in parameter space may indeed be understood as renormalization group flows for a particular class of integrable theories, like the Ising quantum spin chain.  On the other hand, 
using the machinery of conformal field theory in the bulk we are able to 
prove exact continuous majorization
relations in terms of deformations of the size of the block $L$ that is
considered. We also provide explicit analytical examples for models with a boundary  based on 
previous work of Peschel, Kaulke and Legeza \cite{Peschel99, Peschel04_1, Peschel04_2}.

\section{Global, monotonous and fine-grained entanglement loss}

\begin{figure}[h]
\centering
\includegraphics[width=.35\textwidth]{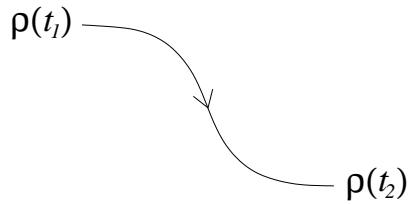}
\caption{A flow in the space of density matrices, driven by parameter $t$.}
\label{matrixflow}
\end{figure}

Consider the pure ground state $|\Omega \rangle$ of a given 
regularized physical system which depends on a particular set of parameters, and let us
perform a bipartition of the system into two pieces $A$ and $B$. The density
matrix for $A$, describing all the physical observables
accessible to $A$, is given by $\rho_A = {\rm tr}_B (|\Omega \rangle
\langle \Omega|)$ -- and analogously for $B$ --. Here we will focus our
discussion on the density matrix for the subsystem $A$, so we will drop the
subindex $A$ from our notation. Let us consider a change in
one of the parameters on which the resultant 
density matrix depends, say, parameter ``$t$'', which can be  
an original parameter of the system, or be related to the size of the region $A$. To be precise, we perform a change in the parameter space from $t_1$ to $t_2$, with $t_2 > t_1$. This involves a flow in the space of reduced density matrices from $\rho(t_1) $ to $\rho(t_2)$, as represented in Fig.\ref{matrixflow}. 

We wish to understand how this
variation of the parameter alters the inner structure of the ground state and,
in particular, how it modifies the entanglement between the two partys, $A$
and $B$. Because we are considering entanglement at two different points $t_2$
and $t_1$, let us assume that the entanglement between $A$ and $B$ is
larger at the point $t_1$ than at the point $t_2$, so we have an entanglement
loss when going from $t_1$ to $t_2$. 

Our characterization of this entanglement loss will progress through three
stages, refining at every step the underlying 
ordering of quantum correlations. These
three stages will be respectively called \emph{global}, \emph{monotonous}
and \emph{fine-grained} entanglement loss. 

\bigskip

\paragraph{Global entanglement loss.-} A possible way to quantify the loss
of entanglement between $A$ and $B$ when going from $t_1$ to $t_2$ is by means
of the entanglement entropy $S(\rho (t)) = -{\rm tr}(\rho (t) \log_2{\rho (t)})$.
Since at $t_2$ the two partys are less entangled than at $t_1$, we have that 

\begin{equation}
S(\rho(t_1)) > S(\rho(t_2)) \ ,
\label{global}
\end{equation}
which is a global assessment between points $t_2$ and $t_1$. This is what we
shall call \emph{global} entanglement loss.

\bigskip

\paragraph{Monotonous entanglement loss.-} A more refined condition
of entanglement loss  can be obtained by imposing the
monotonicity of the derivative of the entanglement entropy when varying
the parameter ``$t$''. That is, the infinitesimal condition

\begin{equation}
S(\rho(t)) > S(\rho(t+dt))
\label{monotomic}
\end{equation}
implies a stronger condition on the structure of the ground state under
deformations of the parameter along the flow in $t$. This monotonic behavior of the
entanglement entropy is what we shall call \emph{monotonous} entanglement loss. 

\bigskip

\paragraph{Fine-grained entanglement loss.-} When monotonous entanglement loss
holds, we can wonder whether the spectrum of $\rho(t)$ becomes more and more ordered as we
change the value of the parameter. It is then plausible to ask if it is
possible to make stronger claims than the inequalities given by Eq.\ref{global} and Eq.\ref{monotomic} 
and unveil some richer structure. The finest notion of reordering when
changing the parameter is then given by the monotonic majorization of the 
eigenvalue distribution along the flow. If we call
$\vec{\rho}(t)$ the vector corresponding to the probability distribution of
the spectrum arising from the density operator $\rho(t)$, then the infinitesimal condition

\begin{equation}
\vec{\rho}(t) \prec \vec{\rho}(t+dt) 
\label{finegrained}
\end{equation}
along the flow in $t$ reflects a strong ordering of the
ground state along the flow. This is what we call \emph{fine-grained} entanglement
loss, because this condition involves a whole tower of
inequalities to be simultaneously satisfied. This Chapter is
devoted to this precise majorization condition
in different circumstances when considering $(1+1)$-dimensional quantum systems. 
For background on majorization, see Appendix A.

\section{Majorization along parameter flows in
  $(1+1)$-dimensional quantum systems}
 
 Our aim in this section is to study strict continuous majorization relations along 
parameter flows, under the conditions of monotonicity of the eigenvalues of the reduced density matrix of the vacuum  
in parameter space. Some of these flows indeed coincide with renormalization group flows for some integrable theories, 
as is the case of the Ising quantum spin chain.

Before entering into the main theorem of this section, let us perform a small calculation which will turn to be very useful: we want to compute the
 reduced density matrix for an interval of length $L$ of the vacuum of a conformal field theory in $(1+1)$ dimensions -- see Appendix B for background on conformal field theory --. With this purpose,
 let $Z_L(q)=q^{-c/12}  {\rm tr}\left(q^{(L_0 + \bar L_0)}\right)$ denote the
partition function of a subsystem of size $L$ \cite{Holzhey94,Ginsparg}, 
where $q  = e^{2 \pi i \tau}$,  
$\tau =(i \pi)/(\ln{(L/\eta)})$, $\eta$ being an ultraviolet cut-off, and $L_0$ and $\bar L_0$ the 0th Virasoro operators.  Let $b \equiv c/12$
be a parameter that depends on the central 
charge and therefore on the universality class of the model. The unnormalized 
density matrix can then be written as $q^{-b} q^{(L_0 +
  \bar{L}_0)}$, since it can be understood as a propagator and $(L_0 +
\bar{L}_0)$ is proportional to the generator of translations in time -- which
corresponds to dilatations in the 
conformal plane -- \cite{Ginsparg}. Furthermore, we have that 
\be
{\rm tr}(q^{(L_0 + \bar{L}_0)}) = 1 + n_1q^{\alpha_1} + n_2q^{\alpha_2}+  \cdots \ ,  
\label{trace}
\ee
due to the fact that the operator $(L_0 + \bar{L}_0)$ is diagonalized in terms of highest-weight 
states $|h, \, \bar{h}\rangle$: $(L_0 + \bar{L}_0)  |h, \, \bar{h}\rangle = 
(h+\bar{h}) |h, \, \bar{h}\rangle$, with $h\ge 0$ and $\bar{h} \ge 0$; the 
coefficients $\alpha_1, \alpha_2, \ldots > 0$, $\alpha_i \ne \alpha_j \ 
\forall i \ne j$ are related to the eigenvalues of $(L_0 + \bar{L}_0)$, and $n_1,n_2, \ldots$ correspond to degeneracies. The normalized 
distinct eigenvalues of $\rho_L = \frac{1}{Z_L(q)}q^{-b} q^{(L_0 +
  \bar{L}_0)}$ are then given by
\be
\begin{split}
\lambda_1 &= \frac{1}{(1 + n_1q^{\alpha_1} + n_2q^{\alpha_2} +  \cdots)}  \\
\lambda_2 &= \frac{q^{\alpha_1}}{(1 + n_1 q^{\alpha_1} + n_2 q^{\alpha_2} + \cdots)}  \\
\vdots  \\
\lambda_l &= \frac{q^{\alpha_{(l-1)}}}{(1 + n_1q^{\alpha_1} + n_2q^{\alpha_2} + \cdots)}.
\label{eiid}
\end{split}
\ee

We are now in conditions of introducing the main result of this section, which can be casted into the following theorem: 

\bigskip

{\bf Theorem 1.1:}
{\it Consider a $(1+1)$-dimensional physical theory which depends on a set of real 
parameters $\vec{g} = (g_1,g_2,\ldots)$, such that
\begin{itemize}
\item there is a non-trivial conformal point $\vec{g}^*$, for which the model 
is conformally invariant in the bulk,
\item the deformations from $\vec{g}^*$ in parameter space in the positive 
direction of a given unity vector $\hat{e}$ preserve part of the conformal 
structure of the model, that is, the eigenvalues of the generic reduced density  
matrices of the vacuum $\rho(\vec{g})$ are still of the form given by 
Eq.\ref{eiid} with some parameter-dependent factors $q(\vec{g})$,
for values of the parameters $\vec{g} = \vec{g}^* + a \hat{e}$, and
\item the factor $q(\vec{g})$ is a monotonic decreasing function along the direction of  $\hat{e}$, that is, we demand that 
\be
\hat{e} \cdot \left( \vec{\nabla}_{\vec{g}} q(\vec{g})\right) =\frac{{\rm d}q(\vec{g})}{{\rm d}a}\le 0
\ee
along the flow. 
\end{itemize}

\begin{figure}[h]
\centering
\includegraphics[width=.24\textwidth,angle=-90]{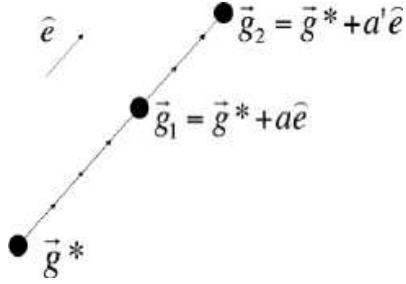}
\caption{A possible flow in the space of parameters in the direction of $\hat{e}$.}
\label{muevo}
\end{figure}
Then, away from the conformal point there is continuous majorization of the eigenvalues 
of the reduced density matrices of the ground state along the flow in 
the parameters $\vec{g}$ in the positive direction of $\hat{e}$ (see Fig.\ref{muevo}), that is,}
\be
\begin{split}
&\rho(\vec{g}_1) \prec \rho(\vec{g}_2) \ , \\
\vec{g}_1= \vec{g}^* + a \hat{e}&, \, \vec{g}_2 = \vec{g}^* + a' \hat{e}, \, a' \ge a \ .
\end{split}
\ee

\bigskip

\emph{Proof:}
Let us define the quantity $\tilde{Z}(q) \equiv   
(1 + n_1q^{\alpha_1} + n_2q^{\alpha_2} + \cdots )$, where it is assumed that $q=q(\vec{g})$, for values of $\vec{g}$ along the flow in $a$. Notice that at conformal points $\tilde{Z}(q(\vec{g}^*))$ is \emph{not} invariant under modular transformations, as opposed to the partition function  $ Z(q(\vec{g}^*))$. 
The behavior of the eigenvalues in terms of deformations with respect to the parameter $a$ follows from
\be
\frac{{\rm d}\tilde{Z}(q)}{{\rm d}a} =\frac{\tilde{Z}(q)-1}{q} \frac{ {\rm
    d}q}{{\rm d} a}
\le 0 \ ,
\label{zet}
\ee
and therefore
\be
\frac{ {\rm d} \lambda_1}{{\rm d}a} = \frac{{\rm d}}{{\rm d}a} \left(
\frac{1}{\tilde{Z}(q)} \right) \ge 0 \ .
\label{first}
\ee
Because $\lambda_1$ is always the largest eigenvalue $\forall a$, the first cumulant
automatically satisfies continuous majorization along the considered flow.
 The variation of the other eigenvalues $\lambda_l$ ($l > 1$) with respect to
$a$ reads as follows:
\begin{eqnarray}
\frac{ {\rm d} \lambda_l}{{\rm d}a} &=& \frac{{\rm d}}{{\rm d}a}
\left( \frac{q^{\alpha_{(l-1)}}}{\tilde{Z}(q)} \right)  \nonumber \\ 
&=& \frac{q^{\alpha_{(l-1)}-1}}{\tilde{Z}(q)} \left(\alpha_{(l-1)}
- \frac{\tilde{Z}(q)-1}{\tilde{Z}(q)} \right) \frac{{\rm d}q}{{\rm d}a} \ .
\label{bigone}
\end{eqnarray}
Let us concentrate on the behavior of the second eigenvalue $\lambda_2$. We observe that two different
situations can happen: 
\begin{itemize} 
\item{if 
\be
\left(\alpha_{1} - \frac{\tilde{Z}(q)-1}{\tilde{Z}(q)} \right) \ge
  0 \ ,
\ee
 then since $\alpha_{(l-1)} > \alpha_1 \ \forall l > 2$, we have that 
 \be
 \left(\alpha_{(l-1)} - \frac{\tilde{Z}(q)-1}{\tilde{Z}(q)} \right) > 0 \
  \forall l > 2 \ ,
  \ee 
  which in turn implies that 
  \be 
  \frac{ {\rm d} \lambda_l}{{\rm
  d}a} \le 0 \ \forall l \ge 2 \ . 
  \ee
  From this we have that the second cumulant
  satisfies 
\be
\frac{ {\rm d} (\lambda_1+\lambda_2)}{{\rm d}a} = -\frac{{\rm d}}{{\rm d}a} \left(
 \sum_{l > 2} \lambda_l \right) \ge 0 \ ,
\label{second}
\ee
thus fulfilling majorization. The same conclusion extends easily in this case to all
the remaining cumulants, and therefore majorization is satisfied by the whole
probability distribution.}
\item{if 
\be
\left(\alpha_{1} - \frac{\tilde{Z}(q)-1}{\tilde{Z}(q)} \right) <
  0 \ ,
\ee
then 
\be
\frac{ {\rm d} \lambda_2}{{\rm d}a} > 0 \ ,
\ee
and therefore  
\be
\frac{ {\rm d} (\lambda_1+\lambda_2)}{{\rm d}a} > 0 \ ,
\ee

 so the second cumulant
  satisfies majorization, but nothing can be said from the previous three equations about the
  remaining cumulants.}
\end{itemize}
Proceeding with this analysis for each one of the eigenvalues we see that, if these are
monotonically decreasing functions of $a$ then majorization is fulfilled for
the particular cumulant under study, but since $\alpha_{i+1} > \alpha_i \
\forall i$ we notice that once the first monotonically increasing eigenvalue is found,
majorization is directly satisfied by the whole distribution of eigenvalues,
therefore $\rho(\vec{g}_1) \prec \rho(\vec{g}_2)$ if $ 
\vec{g}_1= \vec{g}^* + a \hat{e}$,  $\vec{g}_2 = \vec{g}^* + a' \hat{e}$, and $a' \ge a$, as claimed. $\square$

\bigskip

An interesting application of Theorem 1.1 comes whenever $a$ can be related to the scale of a renormalization group transformation. Then it can be understood as a proof of fine-grained entanglement loss along a renormalization group flow
for a particular set of integrable theories, namely, those theories which fulfill the hypothesis of our theorem. We stress that, while it would probably be possible to obtain results based on perturbation theory in the neighborhood of the conformal point for non-integrable theories, 
our theorem is based on the alternative approach of  completely non-perturbative results under the assumption of integrability of the theory along the flow. 
This assumption is naturally fulfilled 
by many interesting models: 
we wish to illustrate this point with 
the analytical examples of similar situations for the Heisenberg and  \emph{XY} quantum spin 
chains with a boundary. At this point we wish to remark as well that, for those theories depending only on one parameter $g$, the monotonicity in the change of the parameter along a renormalization group flow between two fixed points is trivial, since between two zeros the $\beta$-function $\beta = -\frac{{\rm d}g}{{\rm d}\ln l}$, $l$ being the scale of the renormalization group transformation, can only be either  positive or negative, thus implying the monotonicity of the parameter when flowing from one fixed point to the other. Notice that our claim, which is majorization of the reduced density matrices of the vacuum, is  stronger.  

\subsubsection{A majorization lemma}

As a previous step in our derivations, let us state a useful lemma about majorization theory  which 
we shall constantly use in the forthcoming sections. We refer the reader to Appendix A for   
mathematical definitions and more background on majorization theory. The lemma reads as follows:

\bigskip

{\bf Lemma 1.1 \cite{Lutken04}:} {\it
If $\vec{p}_1 \prec \vec{p}_2$ and $\vec{q}_1 \prec \vec{q}_2$, then 
$(\vec{p}_1 \otimes \vec{q}_1) \prec (\vec{p}_2 \otimes \vec{q}_2)$.  
This means that majorization is preserved under the direct product 
operation. }

\bigskip

\emph{Proof:}
If $\vec{p}_1 \prec \vec{p}_2$ and $\vec{q}_1 \prec \vec{q}_2$ then 
$\vec{p}_1 = D_{p} \vec{p}_2$ and $\vec{q}_1 = D_{q} \vec{q}_2$ where 
$D_{p},  D_{q}$ are both doubly stochastic matrices. Therefore 
$(\vec{p}_1 \otimes \vec{q}_1) = (D_{p} \otimes D_{q}) (\vec{p}_2 \otimes
\vec{q}_2)$, where $(D_{p} \otimes D_{q})$ is a doubly stochastic matrix 
in the direct product space, and so $(\vec{p}_1 \otimes \vec{q}_1) \prec 
(\vec{p}_2 \otimes \vec{q}_2)$. $\square$

\subsection{Quantum Heisenberg spin chain with a boundary}

Consider the Hamiltonian of the Heisenberg quantum spin chain with a boundary
\begin{equation}
H = \sum_{i=1}^{\infty} \left(\sigma_i^x \sigma_{i+1}^x +  \sigma_i^y \sigma_{i+1}^y
+ \Delta \sigma_i^z \sigma_{i+1}^z \right) \ , 
\end{equation}
where $\Delta \ge 1 $ is the anisotropy parameter. This model is non-critical 
in the region defined by $\Delta > 1$ and critical at $\Delta = 1$. Notice that, since this is a uniparametric theory which can be mapped to a Gaussian free theory, any renormalization group transformation must be reflected in a change of the only existing parameter. Thus, the flow in $\Delta$ must necessarily coincide with a renormalization group flow. 

From the pure ground state of 
the system, we trace out the $N/2$ contiguous spins $i = 1, 2, \ldots , N/2$, getting an infinite-dimensional density matrix $\rho_{\Delta}$ in the limit $N \rightarrow \infty$ which describes half of the system, and such that it can be written as a thermal density matrix of 
free fermions \cite{Peschel99, Peschel04_1, Peschel04_2}. Its eigenvalues are given by 
\begin{eqnarray}
\rho_{\Delta}(n_0, n_1, \ldots, n_{\infty}) &=& \frac{1}{Z_{\Delta}} e^{-
    \sum_{k = 0}^{\infty} n_{k} \epsilon_{k}} \nonumber \\
    &=& \rho_{\Delta}(n_0) \rho_{\Delta}(n_1) \cdots \rho_{\Delta}(n_{\infty}) \ ,
\end{eqnarray}
with $\rho_{\Delta}(n_k) = \frac{1}{Z_{\Delta}^{k}}e^{-n_{k}
  \epsilon_{k}}$, where $Z_{\Delta}^{k} = (1 + e^{-\epsilon_{k}})$ is 
the partition function for the mode $k$, $n_{k} = 0, 1$, for $k = 0, 1, \ldots, \infty$ and
 with dispersion relation
\be
\epsilon_k = 2 k \ {\rm arcosh} (\Delta) \ .
\ee
 The physical branch of the function ${\rm
  arcosh}(\Delta)$ is defined for $\Delta \ge 1$ and is a monotonic increasing 
function of $\Delta$. On top, the whole partition function $Z_{\Delta}$ can 
be decomposed as an infinite direct product of the different free fermionic modes:  
\begin{equation}
Z_{\Delta} = \prod_{k = 0}^{\infty} \left( 1 + e^{-\epsilon_k} \right) \ .
\end{equation}

From the last equations, it is not difficult to see that $\rho_{\Delta} 
\prec \rho_{\Delta'}$ if $\Delta \le \Delta'$. Fixing the attention on a 
particular mode $k$, we evaluate the derivative of the largest
probability for this mode, $P^{k}_{\Delta} = (1 + e^{-
  \epsilon_{k}})^{-1}$. This derivative is seen to be
\be
\frac{{\rm d} P^{k}_{\Delta}}{{\rm d}\Delta} = \frac{2 k}{(1 +
  e^{-\epsilon_{k}})^2 \sqrt{\Delta^2 - 1}}> 0 \ ,
\ee
for $k = 1, 2, \ldots \infty$ and $0$ for $k = 0$. It follows 
from this fact that all the modes independently majorize their respective 
probability distributions as $\Delta$ increases, with the peculiarity that 
the $0$th mode remains unchanged along the flow, since its probability 
distribution is always $(\frac{1}{2},\frac{1}{2})^T$. The particular behavior 
of this mode is responsible for the appearance of the ``cat" state 
that is the ground state for large values of $\Delta$ -- notice that in that limit 
the model corresponds to the quantum Ising model without magnetic field --. These results, 
together with the Lemma 1.1, make this example obey majorization along the flow in the parameter, which can indeed be understood as a renormalization group flow because of the reasons mentioned at the beginning of the example. 

\subsection{Quantum $XY$ spin chain with a boundary}

\begin{figure}
\centering
\includegraphics[width=.6\textwidth]{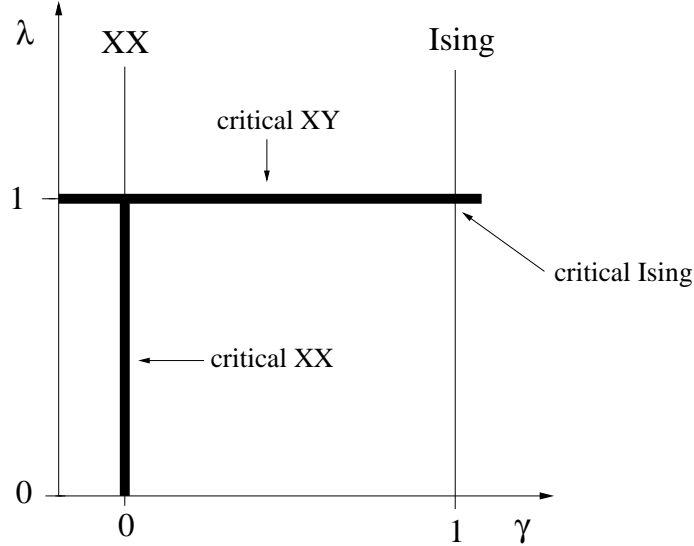}
\caption{Phase diagram of the quantum $XY$-model.}
\label{phaseXY}
\end{figure}

Similar results to the one obtained for the Heisenberg model can be obtained
 for a different model.  Let us consider the quantum 
$XY$-model with a boundary, as described by the Hamiltonian
\begin{equation}
H = -\sum_{i = 1}^{\infty} \left( \frac{(1+\gamma)}{2} \sigma_i^x \sigma_{i+1}^x + 
\frac{(1-\gamma)}{2} \sigma_i^y \sigma_{i+1}^y
+ \lambda \sigma_i^z \right) \ ,
\label{gege}
\end{equation}
where $\gamma$ can be regarded as the anisotropy parameter and $\lambda$ as 
the magnetic field. The phase diagram of this model is shown in
Fig.\ref{phaseXY}, where one can see that there exist different critical
regions depending on the values of the parameters, corresponding to different
universality classes \cite{BH71, Latorre03, Latorre04_1, Korepin04_1, Korepin04_2}. Similarly to the previous example, this model is integrable and can be mapped to a Gaussian free theory with a  mass  parameter depending on a particular combination of both $\lambda$ and $\gamma$ once the kinetic term has been properly normalized (see \cite{Rico2005}). A renormalization group flow can then be understood as a set of flows in the plane of $\lambda$ and $\gamma$. 

Consider the ground state 
of Eq.\ref{gege}, and trace out the contiguous spins $i = 1, 2, \ldots, N/2$ in the limit 
$N \rightarrow \infty$. The resulting density matrix $\rho_{(\lambda,\gamma)}$ 
can be written as a thermal state of free fermions, and its eigenvalues are 
given by \cite{Peschel99, Peschel04_1, Peschel04_2}:
\be
\rho_{(\lambda,\gamma)}(n_0, n_1, \ldots, n_{\infty}) =
\frac{1}{Z_{(\lambda,\gamma)}} 
e^{- \sum_{k = 0}^{\infty} n_{k} \epsilon_{k}} \ ,
\ee
where $n_{k} = 0,1$, and the single-mode energies $\epsilon_{k}$ are given by 
\be
\epsilon_{k} = 
\begin{cases}
2 k  \epsilon \ , & {\rm if} \ \lambda < 1 \\
(2 k+1)  \epsilon \ , & {\rm if} \ \lambda > 1 \ , 
\end{cases}
\label{ddispersion}
\ee
with $k = 0, 1, \ldots , \infty$. The parameter $\epsilon$ is defined by the relation
\be
\epsilon = \pi \frac{I(\sqrt{1-x^2})}{I(x)} \ ,
\label{epsss}
\ee
$I(x)$ being the complete elliptic integral of the first kind
\be
I(x) = \int_0^{\pi/2} \frac{d \theta}{\sqrt{1 - x^2 \sin^2 (\theta)}}
\label{Ixxx}
\ee
and $x$ being given by 
\be
x = 
\begin{cases}
(\sqrt{\lambda^2 + \gamma^2 - 1})/\gamma \ , & {\rm if} \ \lambda < 1 \\
\gamma / (\sqrt{\lambda^2 + \gamma^2 - 1})  \ , & {\rm if} \ \lambda > 1 \ , 
\end{cases}
\label{eks}
\ee
where the condition $\lambda^2 + \gamma^2 > 1$ is assumed for a correct behavior of the above expressions (external region of the
Baruoch-McCoy circle \cite{BH71}). 

We observe that the probability distribution defined by the eigenvalues of 
$\rho_{(\lambda, \gamma)}$ is again the direct product of distributions for each 
one of the separate modes. Therefore, in order to study majorization we can 
focus separately on each one of these modes, in the same way as we already did
in the previous 
example. We wish now to consider our analysis in terms of the flows 
with respect to the magnetic field $\lambda$ and with respect to the
anisotropy $\gamma$ in a separate way. Other trajectories in the parameter space may induce different behaviors, and a trajectory-dependent analysis should then be considered for each particular case.

\subsubsection{Flow along the magnetic field $\lambda$}

We consider in this subsection a fixed value of $\gamma$ while the value of 
$\lambda$ changes, always fulfilling the condition $\lambda^2+ \gamma^2 >
1$. Therefore, at this point we can drop $\gamma$ from our notation.
We separate the analysis of majorization for the regions $1 < \lambda <
\infty$ and $+\sqrt{1-\gamma^2} < \lambda < 1$ for reasons that will become 
clearer during the study example but that already can be realized just by looking at 
the phase space structure in Fig.\ref{phaseXY}.

\bigskip 

\paragraph{{\bf Region $1 < \lambda < \infty$.-}} We show that $\rho_{\lambda} \prec
\rho_{\lambda'}$ if $\lambda \le \lambda'$. In this region of parameter 
space, the largest probability for the mode $k$ is $P^{k}_{\lambda} 
= (1+e^{- \epsilon_{k}})^{-1}$.  The variation of $P^{k}_{\lambda}$
with respect to $\lambda$ is
\be
\frac{{\rm d}P^{k}_{\lambda}}{{\rm d} \lambda} =  \frac{ (2k + 1) e^{-(2
k+1) \epsilon}}{\left( 1 + e^{- (2 k + 1) \epsilon} \right)^2} 
\frac{{\rm d} \epsilon}{{\rm d}\lambda} \ .
\ee
A direct computation using Eq.\ref{epsss}, Eq.\ref{Ixxx} and Eq.\ref{eks} shows that $\frac{{\rm d}\epsilon}{{\rm d}\lambda} > 0$. Therefore, $\frac{{\rm d}P^{k}_{\lambda}}{{\rm d} \lambda} > 0$ for $k = 0,1,
\ldots, \infty$. This derivation shows mode-by-mode majorization when
$\lambda$ increases. Combining this result with the Lemma 1.1, we see that this example obeys majorization.

\bigskip

\paragraph{{\bf Region $+\sqrt{1-\gamma^2} < \lambda < 1$.-}} For this case, we show 
that $\rho_{\lambda} \prec \rho_{\lambda'}$ if $\lambda \ge \lambda'$. 
In particular, the probability distribution for the $0$th fermionic mode 
remains constant and equal to $(\frac{1}{2},\frac{1}{2})^T$, which brings again 
a ``cat'' state for low values of $\lambda$. Similarly  to the latter case, 
the largest probability for mode $k$ 
is $P^{k}_{\lambda} = (1+e^{-\epsilon_{k}})^{-1}$, with
\be
\epsilon_{k} = 2 k \pi \frac{I(\sqrt{1-x^{2}})}{I(x)} = 2k \epsilon \ ,
\ee
and $x = (\sqrt{\lambda^2 + \gamma^2 - 1})/\gamma$. 
Its derivative with respect to $\lambda$ is
\be
\frac{{\rm d}P^{k}_{\lambda}}{{\rm d} \lambda} = \frac{ 2k  e^{-2 k
\epsilon}}{\left( 1 + e^{- 2 k \epsilon} \right)^2} \frac{{\rm
    d}\epsilon}{{\rm d}\lambda} \ .
\label{deri}
\ee

It is easy to see that this time $\frac{{\rm d}\epsilon}{{\rm d}\lambda} <
0$, and therefore $\frac{{\rm d}P^{k}_{\lambda}}{{\rm d} \lambda} < 0$ for $k =
1, 2, \ldots, \infty$, which brings majorization individually for each one of 
these modes when $\lambda$ decreases. The mode $k = 0$ calls for special 
attention. From Eq.\ref{deri} it is seen that
$\frac{{\rm d}P^{k=0}_{\lambda}}{{\rm d} \lambda} = 0$, therefore the probability 
distribution for this mode remains equal to $(\frac{1}{2},
\frac{1}{2})^T$ all along the flow. This is a marginal mode that brings the 
system to a ``cat" state that appears as ground state of the system 
for low values of $\lambda$. Notice that this peculiarity is rooted on the 
particular form of the dispersion relation given in Eq.\ref{ddispersion}, which is proportional to $2 k$ instead of $2k + 1$ 
for this region in parameter space. These results, together with the 
Lemma 1.1, prove that this example also fulfills majorization. 

\subsubsection{Flow along the anisotropy $\gamma$}

In this subsection, the magnetic field $\lambda$ is fixed and the 
anisotropy $\gamma$ is the only free parameter of the model, still 
fulfilling $\lambda^2 + \gamma^2 > 1$. Thus, at this point we can drop
$\lambda$ from our notation. We will see that $\rho_{\gamma} 
\prec \rho_{\gamma'}$ if $\gamma \ge \gamma'$, in the two regions $1 < 
\lambda < \infty$ and $+\sqrt{1-\gamma^2} < \lambda < 1$. In particular, 
in the region  $+\sqrt{1-\gamma^2} < \lambda < 1$, the probability
distribution for the $0$th fermionic mode remains constant and equal 
to $(\frac{1}{2},\frac{1}{2})^T$. Let us consider the biggest probability 
for the mode $k$,
$P^{k}_{\gamma} =  (1+e^{- \epsilon_{k}})^{-1}$, with 
$\epsilon_{k} = \omega \epsilon$, where 
\be
\omega = 
\begin{cases}
2 k \ , & {\rm if} \ \lambda < 1 \\
(2 k + 1)  \ , & {\rm if} \ \lambda > 1 \ , 
\end{cases}
\label{omega}
\ee
and $\epsilon$ as defined in the preceding sections. It is easy to verify that 
\be
\frac{{\rm d} P^{k}_{\gamma}}{{\rm d} \gamma} =  \frac{\omega e^{- \omega
    \epsilon_{k}}}{(1+e^{-\omega \epsilon_{k}})^2} \frac{{\rm d}
  \epsilon}{{\rm d} x} \frac{{\rm d} x}{{\rm d} \gamma}  < 0  
\label{proba}
\ee
for $k = 0, 1, \ldots, \infty$ if $\lambda > 1$ and for $k = 1, 2,
\ldots, \infty$ if $\lambda < 1$. The mode $k = 0$ for $\lambda < 1$
needs of special attention, since $\frac{{\rm
    d}P^{k=0}_{\lambda}}{{\rm d}
  \lambda} = 0$, and therefore the probability distribution for this mode remains 
constant and equal to $(\frac{1}{2},\frac{1}{2})^T$ all along the flow. 
These results, together with the Lemma 1.1,
show that this case obeys again majorization along the flow in the parameter.

\section{Majorization with $L$ in $(1+1)$-dimensional conformal field theories}

A similar study to the one presented in the previous section about majorization along flows in parameter space can be now performed exclusively at the conformal point for flows in the size of the block under consideration. Here we present an analytical derivation of majorization relations for any
$(1+1)$-dimensional conformal field theory without boundaries -- or in the bulk\footnote{The case in which boundaries are present in the system must be properly 
considered from the point of view of the so-called \emph{boundary} conformal field theory. This has been done by H.Q. Zhou et \emph{al.} in 
\cite{Huan05}. For technical background on conformal field theory without boundaries, see Appendix B.} --  in the bipartite 
scenario when the size of the considered subsystems changes, that is to say, 
under deformations in the interval of the accessible 
region for one of the two partys. This size will be represented 
by the length $L$ of the space interval for which we consider the reduced
density matrix $\rho_L$ after tracing out all the degrees of freedom
corresponding to the rest of the universe. Our main result in this section can
be casted into the following theorem:

\bigskip

{\bf Theorem 1.2:} {\it $\rho_L \prec \rho_{L'}$ if $L \ge L'$ for all possible 
$(1+1)$-dimensional conformal field theories in the bulk. }

\bigskip

\emph{Proof:}
Since the factors $q$ are now monotonic functions of the size of the interval $L$, the proof of this theorem is analogous to the proof of Theorem 1.1, with the only exception that now the cumulants are monotonically decreasing (instead of increasing) functions along the flow in $L$. Taking this into account, it immediately follows that  $\rho_L \prec \rho_{L'}$ if $L \ge L' $. This proof is 
valid for all possible $(1+1)$-dimensional conformal field theories in the bulk, since 
it only relies on completely general assumptions. $\square$

\subsection{Critical quantum $XX$ spin chain with a boundary}

Let us give an example of a similar situation to the one presented in Theorem 1.2  for the particular case of the
quantum $XX$-model with a boundary, for which the exact spectrum of $\rho_L$ can be explicitly
computed. The Hamiltonian of the model without magnetic field is given 
by
\be
H = \sum_{i=1}^{\infty} (\sigma_i^x \sigma_{i+1}^x + \sigma_i^y
\sigma_{i+1}^y ). 
\ee
The system as described by this model is critical since it is gapless.  Notice 
that the ultraviolet cut-off coincides with the
  lattice spacing and the theory is naturally regularized, hence $\eta =
  1$. Taking the ground state and tracing 
out all but a block of $1,2,\ldots ,L$ contiguous spins, the density matrix $\rho_L$
describing this block can be written, in the large-$L$ limit, as a thermal 
state of free fermions \cite{Peschel99, Peschel04_1, Peschel04_2}:
\be
\rho_L = \frac{e^{-H'}}{Z_L} ,
\end{equation}
$Z_L$ being the partition function for a given $L$, $H' = \sum_{k =0}^{L-1} 
\epsilon_k d^{\dag}_k d_k$, with fermionic creation and annihilation operators $d^{\dag}_k$, $d_k$ and 
dispersion relation
\be
\epsilon_k = \frac{\pi^2}{2 \ {\rm ln}L} (2k+1) \ \ \ \ k = 0, 1, \ldots, L-1 \ .
\ee
The eigenvalues of the density matrix $\rho_L$ can then be written in terms of 
non-interactive fermionic modes
\be
\begin{split}
\rho_L(n_0, n_1, \ldots, n_{L-1}) &= \frac{1}{Z_L} e^{-\sum_{k = 0}^{L-1} n_k \epsilon_k} \\
&= \rho_L(n_0)  \cdots \rho_L(n_{L-1}) \ ,
\end{split}
\ee
with $\rho(n_k) = \frac{1}{Z_L^{k}}e^{-n_{k}
  \epsilon_{k}}$, where $Z_L^{k} = (1 + e^{-\epsilon_{k}})$ is 
the partition function for the mode $k$, and $n_{k} = 0, 1$, $\forall k$. It is worth 
noticing that the partition function of the whole block $Z_L$ factorizes as a 
product over the $L$ modes:
\begin{equation}
Z_L = \prod_{k = 0}^{L-1} \left( 1 + e^{-\epsilon_k} \right) \ .
\end{equation}

Once the density matrix of the subsystem is well characterized with
respect to its size $L$, it is not difficult to prove that $\rho_L \prec \rho_{L'}$ 
if $L \ge L'$. In order to see this, we will fix our attention on the
majorization within each mode and then we will apply Lemma 1.1
for the whole subsystem. We initially have to 
observe the behavior in $L$ of the largest probability defined by each individual 
distribution for each one of the modes, that is, $P^{k}_L =
1/Z^{k}_L = (1 + e^{-\epsilon_{k}})^{-1}$, for $k= 0, 1, 
\ldots, L-1$. It is straightforward to see that 
\begin{equation}
\frac{{\rm d} P^{k}_L}{{\rm d}L} = \frac{e^{-\epsilon_{k}}}{ \left(1 + 
e^{-\epsilon_{k}}\right)^2} \frac{{\rm d} \epsilon_{k}}{{\rm d} L} < 0 \ ,
\end{equation}
which implies that $P^{k}_L$ decreases if $L$ increases $\forall
k$. This involves majorization within each mode $k = 0, 1, 
\ldots, L-2$ when decreasing $L$ by one unit. In addition, we need to 
see what happens with the last mode $k = L-1$ when 
the size of the system is reduced from $L$ to $L-1$. Because this mode 
disappears for the system of size $L-1$, its probability distribution 
turns out to be represented by the probability vector $(1,0)^T$, which 
majorizes any probability distribution of two components. Combining 
these results with Lemma 1.1, we see 
that this example for the quantum \emph{XX}-model provides a similar situation 
for a model with a boundary to the one presented in Theorem 1.2.

\section{Conclusions of Chapter 1}

In this Chapter we have analyzed majorization relations along parameter and renormalization group flows for a variety of models in $(1+1)$ dimensions. We have also provided in a rigorous way
explicit and detailed proofs for all the majorization 
conjectures raised in some papers on quantum spin chains \cite{Lutken04,
  Latorre03, Latorre04_1}. In order to be more specific:
\begin{itemize} 
\item{We have proven the existence of a fine-grained entanglement loss for $(1+1)$-dimensional 
quantum systems along uniparametric flows, when 
perturbations in parameter space preserve part of the
conformal structure of the partition function, and some monotonicity
conditions hold as well. These flows may coincide with renormalization group flows in some cases. We also considered similar situations which can be treated analytically, arising 
in the Heisenberg and \emph{XY} models with a boundary.} 
 \item{We have also developed 
a completely general proof of
majorization relations underlying the structure of the vacuum with respect 
to the size of the block $L$ for all possible
$(1+1)$-dimensional conformal field theories in the bulk. An example of a similar situation has been considered for the 
particular case of the \emph{XX}-model with a boundary.} 

\end{itemize}

These results provide solid mathematical grounds for the
existence of majorization relations along renormalization group flows underlying the structure of
the vacuum of $(1+1)$-dimensional quantum spin chains. 
It would be interesting to relate the results of this Chapter to possible
extensions of the $c$-theorem \cite{Zam86} to systems with more than
$(1+1)$ dimensions. While other approaches are also possible \cite{Capelli92, Zumbach94_1, Zumbach94_2, Generowicz97, Haagensen94, Bastianelli96, Anselmi98_1, Anselmi98_2, Cardy88, Osborn89, Osborn90, Capelli91, Latorre98}, 
majorization may be a unique tool in
order to assess irreversibility of renormalization group flows in terms of 
properties of the vacuum only, and
some numerical results in this direction have already been observed in
systems of different dimensionality for flows in the parameter space
\cite{Wellard04, LMG05}. The analytical derivation and the consideration of the consequences for higher-dimensional systems 
of the properties presented here for $(1+1)$ dimensions remains an open problem.

    \chapter{Single-copy entanglement in $(1+1)$-dimensional quantum systems}


How much entanglement is contained in a given quantum many-body system? This simple but fundamental question has been considered for systems close to and at quantum phase transitions by means of analyzing very different entanglement measures  \cite{Osterloh02, Osborne02, Au02, Botero04, Hein04, Keating05, Wolf06, Gio06, Bart06, Orus05, VCprb06, EC05, Popp05, Del04, Holzhey94, Latorre03, Latorre04_1, Jin03, Jin04, Korepin04_1, Korepin04_2, Fan04, Calabrese04, Calabrese05, Casini05_1, Casini05_2, Sred93, Eisert06}. All these different ways of measuring entanglement lead to results which complement each other and which help us to understand the precise way in which the ground state of critical models is organized. While the concurrence measures the pairwise entanglement that is present in the system between two of its specific  constituents \cite{Wootters98}, the entanglement entropy measures the entanglement that appears between two different blocks in a bipartition, in turn showing very interesting connections to the entropic area law found for systems such as black holes \cite{Sred93, Callan94, Preskill94, Kabat94, Kabat95}. A detailed analysis of the entanglement entropy in critical quantum spin chains unveils a universal logarithmic scaling law with the size of the block under consideration, which admits an explanation in terms of the underlying conformal field theory in $(1+1)$ dimensions \cite{Holzhey94, Latorre03, Latorre04_1, Korepin04_1, Korepin04_2, Calabrese04, Calabrese05, Casini05_1, Casini05_2}. Furthermore, it is now well understood that the good performance of density matrix renormalization group algorithms in $(1+1)$ dimensions relies very much on this property\footnote{The relation between scaling of entanglement and the performance of classical numerical simulations for different quantum systems will be addressed in detail in Chapters 4 and 5.} \cite{VCprb06}. 

Our aim in this Chapter is to study an entanglement measure which, very much like the entanglement entropy, is proven to have intriguing scaling properties for $(1+1)$-dimensional quantum systems. We call this measure \emph{single-copy entanglement} \cite{EC05, Pes05}, and its operational definition comes naturally motivated by a practical reason: while the entanglement entropy measures the average amount of entanglement possible to be distilled from a bipartite system in the limit of having an infinite number of copies of the system \cite{Bennett96}, the single-copy entanglement measures the amount of entanglement present in the more realistic case of having just \emph{one} copy of the system, in a way to be precisely defined later. As we shall see, we are able to \emph{analytically} compute the asymptotic leading scaling behavior of the single-copy entanglement for all $(1+1)$-dimensional conformal field theories in the bulk, together with its first-order correction. At that point in our derivations a surprise will appear: the entanglement contained in a single specimen of a critical $(1+1)$-dimensional system is seen to be, asymptotically, \emph{half} the entanglement that is available in the ideal case of having an infinite number of copies. This result is reinforced by an analysis from the point of view of quasi-free fermionic systems in $(1+1)$ dimensions which leads again to similar conclusions: whenever the entanglement entropy scales logarithmically in the size of the system, the single-copy entanglement scales asymptotically as half of the entanglement entropy. 
Furthermore, and in order to make our study more complete, we also analyze the behavior of single-copy entanglement away from criticality for the specific example of the $XY$ quantum spin chain. Let us then begin our study by formally defining what the single-copy entanglement is.

\section{Operational definition of the single-copy entanglement}

\begin{figure}[h]
\centering
\includegraphics[width=.8\textwidth]{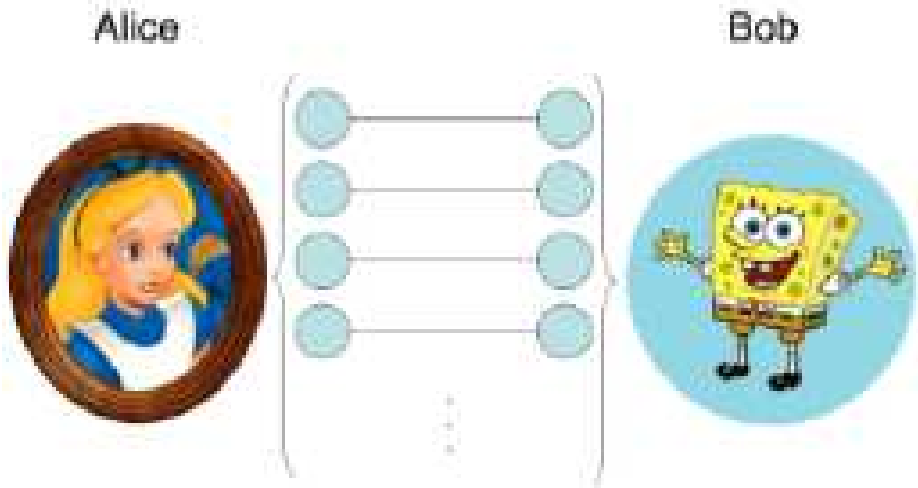}
\caption{Scenario defining the entanglement entropy. Alice and Bob share an infinite number of copies of the bipartite system, and wish to distill EPR-pairs by performing LOCC.}
\label{entrab}
\end{figure}

Let us ask ourselves the following question: how much entanglement is contained in an infinite number of copies of a pure bipartite system $|\psi_{AB}\rangle$? Let us be more specific with the term ``how much", by posing the question differently: what is the maximal rate at which EPR-pairs $ \frac{1}{\sqrt{2}}\left(|0\rangle_A |0\rangle_B + |1\rangle_A |1\rangle_B \right)$ can be distilled from an infinite number of copies of a pure bipartite system $|\psi_{AB}\rangle$, just by invoking local operations and classical communication (LOCC) between the two parties? The answer to this question was originally found by Bennett et \emph{al.} in \cite{Bennett96}: if we are able to distill $M$ EPR-pairs  from $N$ copies of a pure bipartite system $|\psi_{AB}\rangle$, the rate $M/N$ coincides, in the infinite-copy limit, with the entanglement entropy between the two partys, namely
\be
\lim_{N\rightarrow \infty} \frac{M}{N} = S(\rho_A) = -{\rm tr}(\rho_A \log_2 \rho_A) =  S(\rho_B) = -{\rm tr}(\rho_B \log_2 \rho_B) \ ,
\label{assimp}
\ee
$\rho_A$ and $\rho_B$ respectively being the reduced density matrices of the two partys $A$ (Alice) and $B$ (Bob). This situation corresponds to the one represented in Fig.\ref{entrab}. 

\begin{figure}[h]
\centering
\includegraphics[width=.8\textwidth]{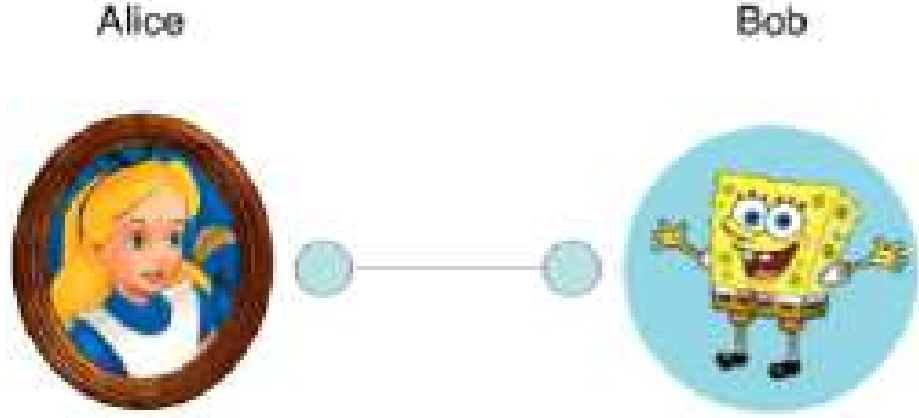}
\caption{Scenario defining the single-copy entanglement. Alice and Bob share only one copy of the bipartite system, and wish to distill a maximally entangled state of the largest possible dimension by performing LOCC.}
\label{singab}
\end{figure}

While the above definition of entanglement entropy obviously makes sense, having an infinite number of copies of the system at hand is an unrealistic situation from the experimental point of view. Thus, let us now ask ourselves this  variant of the above original question: how much entanglement is contained in a single specimen of a pure bipartite system $|\psi_{AB}\rangle$?
Or, equivalently, what is the largest entanglement content that any
apparatus could potentially distill by LOCC from just one bipartite entangled system at hand?
This scenario is represented in Fig.\ref{singab}.
 
 The maximum entanglement that it is possible to obtain by distillation 
with LOCC in the single-copy case can be measured by the largest dimension of a maximally 
entangled state that can be distilled with certainty from the single specimen. 
That is, for a pure bipartite state $|\psi_{AB}\rangle$ 
with reduced density matrices  $\rho_A$ and $\rho_B$, 
we write for the single-copy
entanglement  
\be
E_1(\rho_A) = E_1(\rho_B) = \log_2 (M) 
\ee
if 
\be 
|\psi_{AB}\rangle \longmapsto |\psi_M\rangle \ {\rm under} \  {\rm LOCC} \ ,
\ee
where 
\be
|\psi_M\rangle \equiv \frac{1}{\sqrt{M}} \sum_{q=1}^M 
|q\rangle_A|q\rangle_B
\ee
is a maximally entangled state of dimension $M$. 
Now, we recall the result that the interconversion of bipartite pure states
under LOCC in the single-copy case is governed by the following
majorization relation for the reduced density matrices \cite{Nielsen99}: 
\be
|\psi_{AB}\rangle \longrightarrow |\widetilde{\psi}_{AB}\rangle \ \ {\rm under \ LOCC} \ \iff \rho_A \prec \widetilde{\rho}_A \ ,
\ee
where $\widetilde{\rho}_A$ is the reduced density matrix of the converted state $|\widetilde{\psi}_{AB}\rangle$ for the party $A$\footnote{Of course the same relation holds as well for the party $B$.}. Replacing in the above condition $|\widetilde{\psi}_{AB}\rangle = |\psi_M\rangle$ and  $\widetilde{\rho}_A=\frac{1}{M}{\mathbb I}_M$, ${\mathbb I}_M$ being the $M \times M$ identity matrix, and considering the definition of majorization between probability distributions in terms of a set of inequalities to be satisfied by partial sums of its components -- see Appendix A --, we find the inequality 
\be
\lambda_1 \le \frac{1}{M} \Rightarrow M \le \frac{1}{\lambda_1} \ ,
\ee
$\lambda_1$ being the largest eigenvalue of $\rho_A$. Given the above upper bound for $M$, one finds that 
\be
E_1(\rho_A) = -\log_2 \lambda_1  = E_1(\rho_B)  \ .
\label{single}
\ee
Therefore, the single-copy entanglement can be directly computed by looking only at the \emph{largest eigenvalue} of the reduced density matrix of the system under consideration. This situation is very different from that of the entanglement entropy, where all the eigenvalues of the reduced density matrix contribute to the final quantity. 

\section{Exact conformal field theoretical computation}

Now we wish to show the exact and analytical computation of the single-copy entanglement in the case of $(1+1)$-dimensional conformal field theories in the bulk. We remind that the systems described by these theories correspond to the continuum limit of a variety of regularized quantum critical theories defined on a chain. For technical background, see Appendix B. 

As we saw in the previous Chapter, the reduced 
density matrix for a block of size $L$ describing the vacuum of a $(1+1)$-dimensional conformal field theory can be written as \cite{Orus05,Holzhey94, Ginsparg}
\be 
	\label{rho}
	\rho_L=\frac{1}{Z_L(q)} q^{-c/12} 
	 q^{(L_0+\bar L_0)} \ ,
\ee
where $c$ is the central charge of the theory, $L_0$ and $\bar{L}_0$ are the $0$th 
holomorphic and antiholomorphic
Virasoro operators, $Z_L(q)$ is the partition function, $q=e^{2 \pi i \tau}$,
 and $\tau= (i \pi)/ (\ln (L/\eta))$,
$\eta$ being a regularization ultraviolet cut-off. For
critical quantum chains we have that  $\eta = 1$, 
which corresponds to the lattice
spacing, and which is to be understood in our forthcoming 
calculations. 

The largest eigenvalue of the density matrix $\rho_L$ corresponds to the
zero mode of $(L_0 + \bar L_0)$, that is, 
\be
	\label{largest}
	\lambda_1= \frac{1}{Z_L(q)} q^{-c/12} \ ,
\ee
since for this mode $|0\rangle$ we have that $(L_0+\bar L_0)\vert 0\rangle=0$. We then get a first
expression for the single-copy entanglement:
\be
\label{singlecopydef}
	E_1(\rho_L)= - \log_2{\lambda_1} =  \log_2{\left( Z_L(q) q^{c/12} \right)} \ .
\ee
The leading behavior for the partition
function can be computed 
when  $L$ is large by taking advantage of its invariance
under modular transformations.
The needed transformation corresponds to $\tau \rightarrow -1/\tau$,
which amounts to 
$Z_L(q)=Z_L(\tilde q)$, $q=e^{-2 \pi^2/\ln L}$,
$\tilde q=e^{-2\ln L} = 2^{-2\log_2 L}$. It is now possible to expand the partition function in powers
of $\tilde q$, since all the eigenvalues of the operator $(L_0 + \bar{L}_0)$ 
are positive, and find that the leading contribution originates 
from the central charge:
\be
 \log_2 Z_L(\tilde q) = -\frac{c}{12} \log_2 \tilde
	q+O\left(\frac{1}{L}\right) = \frac{c}{6}\log_2 L+O\left(\frac{1}{L}\right) \ .
\ee
This result translates into an explicit expression for the
single-copy entanglement
\be
\label{singlecopyL}
	E_1(\rho_L)=\frac{c}{6}\log_2 L - \frac{c}{6} \frac{(\pi \log_2 e)^2}{\log_2 L}+
	O  \left(\frac{1}{L}\right) \ .
\ee
We wish to point out that  the above result is exact up to polynomial corrections
in $1/L$ since no further powers of $1/\log_2 L$  appear in
the expansion when $L$ is large. 

Similar conformal field theory manipulations were used to prove
that the von Neumann entropy for the same reduced density matrix $\rho_L$ 
is given by \cite{Holzhey94}
\be
S(\rho_L)=-\frac{c}{6}\log_2 \tilde q+ O\left(\frac{1}{L}\right) \ , 
\ee
which implies the following  direct relation between
entropy and single-copy entanglement:
\be
\label{relation}
E_1(\rho_L)=\frac{1}{2} S(\rho_L)- \frac{c}{6} \frac{(\pi \log_2 e)^2}{\log_2 L}+
O\left(\frac{\log_2{L}}{L}\right) \ , 
\ee
where the last subleading
correction is easily calculated by comparing the results from
\cite{Holzhey94} and our expression given in Eq.\ref{singlecopyL}. 
It should be noted here that the above result completely fixes
 the leading eigenvalue of the reduced density 
matrix of the block of size $L$ to be dictated by its
entropy within the large-$L$ limit, that is, 
\be
\lim_{L\to \infty}\left(\frac{\lambda_1}{2^{S(\rho_L)/2}}\right)=1 \ .
\ee
Corrections to this limit can be obtained from Eq.\ref{relation}.
Quite remarkably, we also notice that all the eigenvalues will inherit the same leading behavior
and differ by their subleading corrections controlled by the 
conformal weights corresponding to the universality class of the particular model in consideration. 

\section{Exact computation in quasi-free fermionic quantum spin chains}

We aim now to reinforce the previously 
achieved result by investigating the same question from an alternative point of view, namely, 
 we investigate all translationally invariant 
quantum spin models which can, under 
a Jordan-Wigner transformation,
be written as an isotropic 
quadratic Hamiltonian in fermionic operators. 

The Jordan-Wigner transformation relates the Pauli operators
in the quantum spin system to spinless fermionic operators $\{c_j\}$ 
obeying the fermionic anticommutation relations
\begin{eqnarray} 
\{ c_j,   c_k\}&=&0 \nonumber \\
\{c_j^\dagger, c_k^\dagger\} &=& 0 \nonumber \\
\{  c_j^\dagger,   c_k\}&=&\delta_{jk} \ , 
\end{eqnarray}
according to
\begin{eqnarray}
	  \sigma_l^x &=& 
	\frac{1}{2} \prod_{n=1}^{l-1} (1- 2   c_n^\dagger   c_n)
	(  c_l^\dagger +   c_l) \nonumber \\
	  \sigma_l^y &=& \frac{1}{2i} \prod_{n=1}^{l-1} 
	(  c_l^\dagger -   c_l)
	(1- 2   c_n^\dagger   c_n) \nonumber \\
	  \sigma_l^z &=&   c_l^\dagger   c_l -\frac{1}{2} \ .
\end{eqnarray}

Consider now an infinite quantum spin system in $(1+1)$ dimensions that 
corresponds to a general translationally invariant 
isotropic quasi-free fermionic model. These 
correspond to chain systems whose Hamiltonian can be
cast into the form
\begin{equation}
	H= \sum_{l,k} c_l^\dagger A_{l-k} c_k
\end{equation}	
with $A_l=A_{-l}\in\rr$. The ground state of $H$ is a quasi-free 
fermionic state, that is,  a state that is
completely characterized by the second moments 
of the fermionic operators. Notice that, while some of the spin chains described by this setting can be considered as well within the framework of conformal field theory in $(1+1)$ dimensions, there may also be models that do not correspond to any such conformal field theory.

Our claim is the following: 
if the entropy of entanglement satisfies
\begin{equation}\label{assumption}
	S(\rho_L) = \xi \log_2(L) + O(1) \ ,	
\end{equation}
for some $\xi>0$, 
then the single-copy entanglement satisfies
\begin{equation}\label{result}
	E_1(\rho_L)  = \frac{1}{2} S(\rho_L) + O(1) \ .
\end{equation}
That is, if we find that the entropy of entanglement
scales asymptotically as the logarithm of $L$ -- as typically observed for 
 this class of systems at criticality --
then we can infer that the leading behavior of the single-copy entanglement will
 asymptotically be exactly one half of it. 
Notice that this does not fix such a relationship
in the case that, for example, the system is gapped and 
the entropy of entanglement saturates (we shall consider an example of non-critical behavior
 within the next section). Let us now show how we arrive to the previous statement.

\begin{figure}[h]
\psfrag{a}{\small $i \delta$}
\psfrag{b}{\small $-i \delta$}
\psfrag{c}{\small $\epsilon/2$}
\psfrag{d}{\small $1+\epsilon$}
\psfrag{e}{\small $-1-\epsilon$}
\centering
\includegraphics[width=.6\textwidth]{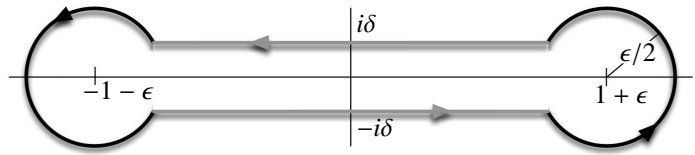}
\caption{Contour of integration to be taken in case of both
the entropy of entanglement and the single-copy 
entanglement.}\label{fig1}
\end{figure}

The reduced state of a block of length $L$
is entirely specified by the eigenvalues of the
real symmetric 
$L\times L$  Toeplitz matrix $T_L$, with 
$l$-th row being given by 
	$(t_{-l+1},t_{-l+2}, ..., t_0 ,...,t_{L-l})$.
The latter numbers are for an infinite quasi-free fermionic quantum chain found to be 
\be
t_l = \frac{1}{2\pi} \int_0^{2\pi} g(k) e^{-ilk} dk \ ,
\ee
where $g:\cc\rightarrow \cc$ is the so-called symbol
\cite{Lieb61,BH71,Ehrh97}, which essentially
characterizes the fermionic model. The fact that $T_L$ is a 
Toeplitz matrix reflects the translational invariance of the model. The real eigenvalues of $T_L$
will be labeled as $\mu_1,...\mu_L \in[-1,1]$.
They can be found from the zeroes of the
characteristic polynomial
$F:\cc\rightarrow \cc$,
\begin{equation}
	F(z )= \det\left( z {\mathbb I}_L - T_L \right) \ . 
\end{equation}
The entropy of entanglement can then be obtained as 
\cite{Jin03,Korepin04_1,Korepin04_2,Keating05}
\be  
S(\rho_L) = \sum_{l=1}^{L} f_S(1,\mu_l) \ ,
\ee 
where $f_S:\rr^+\times \cc\rightarrow \cc$ 
is defined as 
\be 
f_S(x,y) = -\left(\frac{x+y}{2}\right)\log_2{\left(\frac{x+y}{2}\right)} - \left(\frac{x-y}{2}\right)\log_2{
\left(\frac{x-y}{2}\right)} \ .
\ee
In fact, we can write \cite{Jin03,Korepin04_1,Korepin04_2,Keating05}
\begin{equation}
	S(\rho_L)= \lim_{\varepsilon \rightarrow0 }
	\lim_{\delta\rightarrow 0 }
	\frac{1}{2\pi i}
	\int f_S(1+\varepsilon, z) \frac{F'(z)}{F(z)} dz \ .
\end{equation}
The contour of the integration in the complex plane is shown in Fig.\ref{fig1}. 
In turn, we may also write for the single-copy entanglement \cite{EC05}
\be 
E_1 (\rho_L)= \sum_{l=1}^{L} f_1(0,\mu_l) \ , 
\ee
in terms of the above $\mu_1,...,\mu_L$, where now 
$f_1:\rr^+ \times \cc \rightarrow\cc$ is to be defined as 
\be
 f_1 (\varepsilon,z) = -\log_2 \left( \frac{1+(z^2+\varepsilon^2)^{1/2}}{2}\right) \ .
\ee
Respecting the
cuts of the logarithm (see \cite{EC05}), we may now cast
$E_1(\rho_L)$ into the form
\begin{equation}
	E_1(\rho_L)=\lim_{\varepsilon \rightarrow 0 } 
	\lim_{\delta \rightarrow 0 }
	\frac{1}{2\pi i}
	\int f_1(\varepsilon, z) \frac{F'(z)}{F(z)} dz \ .
\end{equation}
Now we take advantage of the fact that $T_L$ is a real symmetric Toeplitz matrix,
which means that we can assess the asymptotic behavior
of their determinants using proven instances of the Fisher-Hartwig
conjecture \cite{Jin03,Korepin04_1,Korepin04_2,Keating05,Lieb61,BH71,Ehrh97}.
We wish to remark at this point that the observation that we only refer to proven instances
of the Fisher-Hartwig conjecture derives from the fact that we are only 
considering isotropic models \cite{Keating05}. 
Concerning the function $F:\cc\rightarrow \cc$, 
the Fisher-Hartwig conjecture allows us to write 
\begin{equation}
\frac{F'(z)}{F(z)}
	 =a (z)  L -  b(z)  \log_2 L   + O(1) 
\end{equation}
in the large-$L$ limit, where 
\be
b(z)= -2 \sum_{r=1}^R  \beta(z)\beta'(z) \ ,
\ee
with $\beta:\cc\rightarrow\cc$ such that \cite{Keating05}
\be 
z \rightarrow \frac{1}{2 \pi i}\log_2\left(\frac{z+1}{z-1}\right) \ .
\ee  
The number $R$,  in turn corresponds to half
the number of discontinuities of  the above symbol $g(k)$ in 
the interval $[0,2\pi)$.
Now, if we assume the validity of the logarithmic scaling of the entropy given in the expression 
of Eq.\ref{assumption}, we know that, necessarily,  
\begin{equation}
	\lim_{\varepsilon \rightarrow 0 }
	\lim_{\delta\rightarrow 0 }
	\int f_S(1+\varepsilon, z) a(z) dz=0 \ , 
\end{equation}
since no linear dependence in $L$ must appear.  
Moreover, we know that $S(\rho_L)\geq E_1(\rho_L)$, which can easily be proven from their respective mathematical definitions -- apart from the intuition that many copies of a system may help in entanglement distillation --. Therefore, in the large-$L$ limit we must also necessarily have  
\begin{equation}
	\lim_{\varepsilon \rightarrow 0 } 
	\lim_{\delta \rightarrow 0 }
	\int f_1(\varepsilon, z) a(z) dz=0 \ . 
\end{equation}
Consequently, we only have to consider the 
logarithmically divergent term.
For the entropy of entanglement 
the only relevant contour integral reads
\begin{equation}
	I_S = \lim_{\varepsilon \rightarrow 0 }
	\lim_{\delta\rightarrow 0 }
	\frac{1}{2\pi i}
	\int f_S(1+\varepsilon, z) b(z) dz \ .
\end{equation}
In turn, for the single-copy entanglement the relevant
contour integral becomes
\begin{equation}
	I_1 = \lim_{\varepsilon \rightarrow 0 }
	\lim_{\delta\rightarrow 0 }
	\frac{1}{2\pi i}
	\int f_1(\varepsilon, z) b(z) dz \ .
\end{equation}
Taking into account that $b(z)$ is analytic outside the interval $[-1,1]$,
 the contributions of the circle pieces vanish in the two 
cases. Hence, we finally arrive at 
\begin{eqnarray}
      S(\rho_L) &=& \frac{R}{\pi^2} 
      \int_{-1}^1 dx
	\frac{f_S(1,x)}{1-x^2} \log_2(L) + O(1) \nonumber \\
	E_1(\rho_L) &=&  \frac{R}{\pi^2}  \int_{-1}^1 dx
	\frac{f_1(0,x)}{1-x^2} \log_2(L) + O(1) \ .
\end{eqnarray}
Since $f_1(0,x)=-\log_2((1+|x|)/2)$ for $x\in[-1,1]$, we have that within the large-$L$ limit,  
\begin{eqnarray}
	S(\rho_L)&=& \frac{R}{3}\log_2 L + O(1) \nonumber \\
	 E_1(\rho_L)&=& \frac{R}{6}\log_2 L + O(1) \ ,
\label{erre}
\end{eqnarray}
which in turn implies the validity of the expression that we anticipated in Eq.\ref{result}. 
We have therefore proven that, in this class of models, whenever the system has a logarithmic asymptotical scaling of the entanglement entropy, the single-copy entanglement is exactly half the asymptotically available in the infinite-copy case in its leading contribution. We wish to remark as well that, from Eq.\ref{erre}, the number $R$ 
 precisely  corresponds to the central charge $c$ for those models that are governed by an underlying conformal symmetry.  For instance, for the quantum $XX$ spin chain, we have that $R=c=1$, corresponding to the universality class of a free boson.


\section{Single-copy entanglement away from criticality}

In this section we exhibit an explicit example for which the relation 
between single-copy entanglement
and entanglement entropy can be demonstrated 
near but off the critical region. We consider
the $XY$ quantum spin chain with a boundary, with Hamiltonian 
\begin{equation}
H = -\sum_{i = 1}^{\infty} \left( \frac{(1+\gamma)}{2} \sigma_i^x \sigma_{i+1}^x + 
\frac{(1-\gamma)}{2} \sigma_i^y \sigma_{i+1}^y
+ \lambda \sigma_i^z \right) \ ,
\end{equation}
studied in Chapter 1. Again, we
consider the chain of semi-infinite length with a boundary, 
where the spins $i = 1, 2, \ldots , N/2$ with $N \rightarrow \infty$ have been traced out from the ground state of the system. The resultant density matrix $\rho_{(\lambda , \gamma)}$ can be written as a thermal density operator of a system of spinless fermions with creation and annihilation operators $d^\dagger_k$ and $d_k$ in the following way \cite{Peschel04_2}:
\be
	\rho_{(\lambda , \gamma)} = \frac{e^{-H}}{{\text{tr}}\left(e^{-H}\right)} \ \ \ \ \ \  
	H=\sum_{k} \epsilon_k d^\dagger_k d_k \ , 
\ee
where 
\be
\epsilon_{k} = 
\begin{cases}
2 k  \epsilon \ , & {\rm if} \ \lambda < 1 \\
(2 k+1)  \epsilon \ , & {\rm if} \ \lambda > 1 \ , 
\end{cases}
\label{dispersion}
\ee
$k \in \nn$, and $\lambda\in \rr$ 
is the parameter controlling the external magnetic field, 
$\lambda^*=1$ corresponding to the quantum phase transition point. We also have that  
\be
 \epsilon = \pi \frac{I(\sqrt{1-x^2})}{I(x)} \ , 
 \ee
$I(x)$ being the complete 
elliptic integral of the first kind, 
\be
I(x) = \int_0^{\pi/2}\frac{d \theta}{\left(1 - x^2 \sin^2 (\theta)\right)^{1/2}} \ .
\ee 
Furthermore, $x$ is related to the parameters $\lambda$ and $\gamma$ defining the model as follows: 
\be
x = 
\begin{cases}
(\sqrt{\lambda^2 + \gamma^2 - 1})/\gamma \ , 
	& {\rm if} \ \lambda < 1, \\
\gamma / (\sqrt{\lambda^2 + \gamma^2 - 1})  \ , 
	& {\rm if} \ \lambda > 1 \ , 
\end{cases}
\label{eks}
\ee
with the condition $\lambda^2 + \gamma^2 > 1$ (external region of the Baruoch-McCoy circle
\cite{BH71}). A computation of the single-copy entanglement  
with respect to this partitioning 
can be performed in terms of $\epsilon$, transforming sums into 
integrals by means of the Euler-McLaurin expansion, and finding
\be 
\label{singlecopynoncritical1}
E_1(\rho_{L\rightarrow \infty,\epsilon})= 
\frac{\pi^2 \log_2 e}{24 \epsilon}- \frac{\epsilon \log_2 e}{24}
+ O(e^{-\epsilon})
\ee
if $\lambda <1$ and
\be 
\label{singlecopynoncritical2}
E_1(\rho_{L\rightarrow \infty,\epsilon})
= \frac{\pi^2 \log_2 e}{24 \epsilon}+\frac{1}{2}+ \frac{\epsilon \log_2 e}{12}
+O(e^{-\epsilon})
\ee
if $\lambda >1$. No subleading corrections in powers of $\epsilon$ do appear
in the expansion. On the other hand it is easy to see by explicit evaluation that the entropy of entanglement can be related to
the single copy-entanglement by 
\be
S(\rho_{L\rightarrow \infty,\epsilon})
=\left(1-\epsilon\frac{\partial}{\partial \epsilon}\right)
E_1(\rho_{L\rightarrow \infty,\epsilon}) \ ,
\ee
which shows that
\be
\label{relationnoncritical}
	\lim_{\epsilon\rightarrow0}\left(
	E(\rho_{L\rightarrow \infty,\epsilon})-\frac{1}{2}S(\rho_{L\rightarrow \infty,\epsilon})\right)=0 \ .
\ee
We notice that the limit $\epsilon \rightarrow 0$ is precisely the limit where
the theory becomes critical, that is when $\lambda\to \lambda^*=1$. The above expression for finite $\epsilon$ gives us corrections away from criticality to the $1/2$ factor between the entanglement entropy and the single copy entanglement that has been discussed in the preceding sections. These corrections vanish as the system approaches criticality, as we have explicitly seen in this example.  

\section{Conclusions of Chapter 2}

In this Chapter we have analyzed the single-copy entanglement, that is, the entanglement that it is possible to deterministically distill by using local operations and classical communication when only one copy of a bipartite system is at hand, in quantum systems in $(1+1)$ dimensions. We have carried our analysis mainly from the point of view of conformal field theory in $(1+1)$ dimensions in the bulk and quasi-free fermionic models  in order to analyze critical systems, and also studied the behavior close to but away from criticality for the integrable example of the $XY$ quantum spin chain.  To be more precise:  
\begin{itemize}
\item{For $(1+1)$-dimensional conformal field theories we have proven that the leading scaling behavior of the single-copy entanglement is exactly \emph{half} the asymptotic behavior of the entanglement entropy. The first-order correction to the leading term has also been explicitly computed.}
\item{For quasi-free fermionic quantum systems we have proven that if the asymptotic scaling of the entanglement entropy is logarithmic, then the asymptotic scaling of the single-copy entanglement is also logarithmic, with a prefactor that is exactly \emph{half} the one of the entanglement entropy.}
\item{For the example of the semi-infinite $XY$ quantum spin chain, we have computed the single-copy entanglement away from criticality and have observed that the factor $1/2$ between the entropy and the single-copy entanglement is \emph{only} recovered when the system approaches the quantum phase transition point.}
\end{itemize}

The main conclusion is, therefore, that for $(1+1)$-dimensional quantum systems at criticality the single-copy entanglement and the entanglement entropy for a system described by a reduced density matrix $\rho_L$ typically obey the law  
\begin{equation}
	\lim_{L\rightarrow\infty} \left(\frac{S(\rho_L)}{E_1(\rho_L)}=2\right) \ .
\end{equation}
For systems obeying the above relation we can say that in a {\it single run}, with a single invocation of a physical device
acting on only one physical system, it is possible to obtain half the
entanglement per specimen that is asymptotically available in the infinite-copy limit. 
Furthermore, all these results also show relationships 
between the largest eigenvalue of the reduced vacuum $\rho_L$
and its full spectrum for a very large class of quantum systems.

    \chapter{Entanglement entropy in the Lipkin-Meshkov-Glick model}

Most of the analytical studies of the entanglement properties of quantum many-body systems close to criticality have been focused on the particular case of  $(1+1)$-dimensional systems, like the ones that we considered in the previous Chapters. Few models have been discussed so far in higher dimensions
\cite{Syljuasen03_2,Hamma04,Hamma05_1, Hamma05_2, Hamma05_3, Orus04_1,Orus04_2,Wellard04,Sred93,Cramer05,Wolf06,Gio06,Bart06, Kabat94, Kabat95, VC04, GVidal05, Anders06} either due to the
absence of an exact diagonalization of the system or to a difficult numerical treatment. 
We can in part understand this difficulty because of the existing link between
the connectivity of a system and its entanglement entropy: one should
naively think that, the bigger the connectivity of the system is, the bigger
the amount of quantum correlations present in the ground state of
the model should be, especially when the system is close to a quantum critical point. A classical 
numerical treatment of the model can become then 
very inefficient, as we shall in detail explain in the forthcoming Chapters 4 and 5.  
The idea in favor of this is rather simple: the more connected a system is, the more interactions it has,
therefore the more entangled its ground state should be and the more difficult it should be to get its fundamental properties -- like the ground-state energy or the correlation functions -- 
by means of a classical numerical treatment.

Actually, with some insight it is possible to
make a non-accurate quantitative statement about the previous idea: given a system of $N$ particles in
$(d+1)$ dimensions, $d$ being the number of spatial dimensions of the underlying lattice, if we believe that at criticality the entropy of 
entanglement $S$ is to scale proportionally to the area of the boundary of the region that
separates the two subsystems under consideration, as is the case of bosonic systems \cite{Sred93, Au02, Cramer05}, then it is not difficult to
check that the entropy of a bipartition of the system between $N/2$ contiguous particles and the rest has to roughly scale like 
\be
S \sim N^{\frac{d-1}{d}} \ .
\label{escala}
\ee
Critical fermionic systems may differ from the above law by means of an $O(\log_2 N)$ multiplicative factor \cite{Wolf06, Gio06, Bart06}. From the above reasoning we can see that the bigger the dimensionality $d$ is -- which is
directly related to the connectivity of the system --, the stronger the
scaling of the entanglement entropy should be. The case of a conformally-invariant critical system with $d=1$ 
has to be treated separately since the entropy has a \emph{logarithmic} divergence, as we already 
remarked in previous Chapters. This intuitive 
relation between entanglement and connectivity will be considered again in Chapter 4, when
studying the scaling of entanglement in quantum algorithms. 

In this context, the Lipkin-Meshkov-Glick model \cite{Lipkin65,Meshkov65,Glick65} 
has drawn much attention since it allows for a very efficient numerical treatment as
well as analytical calculations. Furthermore, it provides a
useful counter-example of the previous intuitive relation between entanglement and
connectivity: in a system defined on a simplex -- totally connected network --, and
contrary to the intuition that we have specified before, the entanglement in the
system behaves \emph{as if} the system were $(1+1)$-dimensional. This is a
consequence of the role played by the symmetries within the description of the
model, as we shall see. Entanglement can be increased by the connectivity, but
 can also be ``killed'' by the symmetries in some cases.

First introduced by Lipkin, Meshkov
and Glick in nuclear physics, this model has been the
subject of intensive studies during the last two decades.  It
is of interest in order to describe in particular the Josephson
effect in two-mode Bose-Einstein condensates \cite{Milburn97,Cirac98}.
Its entanglement properties have been already discussed through the concurrence, which
exhibits a cusp-like behavior at the critical point \cite{JVidal04_1,JVidal04_2,Dusuel04_4} as well as interesting dynamical properties \cite{JVidal04_3}. Similar results have also been
obtained in the Dicke model \cite{Dicke54,Lambert04_1,Lambert04_2} which can be mapped
onto the Lipkin-Meshkov-Glick model in some cases \cite{Reslen05}, or in the reduced BCS model
\cite{Dusuel05}. Let us mention as well that the entanglement entropy has also been 
calculated for the anti-ferromagnetic 
Lipkin-Meshkov-Glick model \cite{Unanyan04} for which the ground state is known 
exactly \cite{Unanyan03,JVidal04_2}. Here we analyze the von Neumann entropy computed from the ground state of the Lipkin-Meshkov-Glick
model.  We show that, at criticality, it behaves logarithmically with the size of the blocks $L$ used
in the bipartite decomposition of the density matrix with a prefactor that depends on the anisotropy
parameter tuning the underlying universality class. We also discuss the dependence of the entropy with the magnetic field and stress the close analogy of the found results with those of $(1+1)$-dimensional quantum systems.

%
%

\section{The Lipkin-Meshkov-Glick model}

\begin{figure}
\centering
\includegraphics[scale=.5]{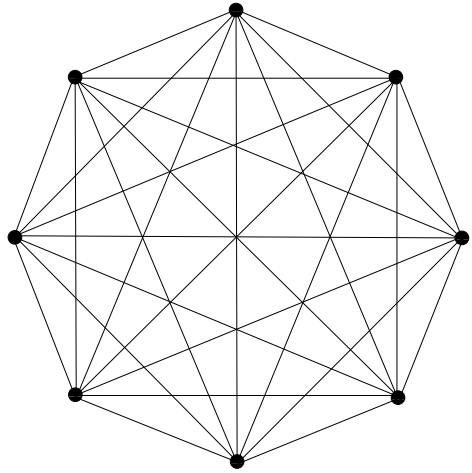}
\caption{Complete graph -- or simplex -- of $8$ vertices.} 
\label{simplex}
\end{figure}

The Lipkin-Meshkov-Glick model is defined by the Hamiltonian
%
%
\begin{equation}
H=-\frac{\lambda}{N} \sum_{i<j} \left( \sigma_i^x \sigma_j^x + \gamma
\sigma_i^y \sigma_j^y
\right) - h \sum_{i=1}^N \sigma_i^z \ ,
\label{hamil0}
\end{equation}
where $\sigma_k^{\alpha}$ is the Pauli matrix at position
$k$ in the direction $\alpha$, and $N$ the total number of spins. This
Hamiltonian describes a set of spins one-half located at the vertices of a
$N$-dimensional simplex -- complete graph, as shown in Fig.\ref{simplex} -- interacting via a
ferromagnetic coupling $\lambda>0$ in the $xy$-spin plane, $\gamma$
being an anisotropy parameter and $h$ an external magnetic field
applied along the $z$ direction.  

Given that the model is defined on a simplex, the symmetry under permutations of particles
allows us to rewrite the Hamiltonian from Eq.\ref{hamil0} in terms of the total spin operators $J^{\alpha}= \sum_{i=1}^N
\sigma_i^{\alpha} / 2$. The previous Hamiltonian
can then be expressed as
%
%
\begin{eqnarray}
H&=& - {\lambda \over N} (1+\gamma) \left({\bf J}^2-J^z J^z -N/2 \right)
-2h J^z  \nonumber \\
&&- {\lambda \over 2 N} (1-\gamma)\left(J^+ J^+ + J^- J^- \right) \ , \label{hamil}
\end{eqnarray}
%
%
where ${\bf J}^2$ is the representation of spin $N/2$ of the Casimir operator and $J^{\pm} \equiv J^x \pm i J^y$. In
the following, we set for simplicity $\lambda=1$ and since the spectrum of $H$ is even
under the transformation $h \leftrightarrow -h$ \cite{JVidal04_3}, we restrict our
analysis to the region $h\geq 0$. Furthermore, we only consider the maximum spin sector $J=N/2$ to
which the full spectrum of the Hamiltonian from Eq.\ref{hamil} belongs.
A convenient basis of this subspace is spanned by the so-called Dicke states 
$|N/2,M \rangle$ which are invariant under the permutation of spins and are eigenstates of
${\bf J}^2$ and $J^z$ with eigenvalues $N(N+2)/4$ and $M=-N/2, -N/2+1, \ldots,
N/2-1, N/2$, respectively.

\section{Entanglement within different regimes}

We consider the von Neumann entropy
associated with the ground state reduced density matrix $\rho_{L,N}$ of
a block of size $L$ out of the total $N$ spins, $S_{L,N} \equiv S(\rho_{L,N}) = -{\rm
tr}\: (\rho_{L,N} \log_2 \rho_{L,N})$ and analyze its behavior as $L$ is
changed, both keeping $N$ finite or sending it to infinity.  Notice that since the ground
state reduced density matrix is spanned by the set of $(L+1)$
Dicke states, the entropy of entanglement obeys the constraint $S_{L,N} \le
{\rm log}_2 (L+1)$ for all $L$ and $N$, where the upper bound corresponds to
the entropy of the maximally mixed state $\rho_{L,N} = {\mathbb I}/(L+1)$ in
the Dicke basis. This
argument implies that entanglement, as measured by the von Neumann
entropy, cannot grow faster than the typical logarithmic scaling law
observed in $(1+1)$-dimensional quantum spin chains at conformally-invariant critical points
 \cite{Holzhey94,Latorre03,Latorre04_1}. Entanglement has thus been drastically reduced by the symmetry under permutations of the model, 
 as we hinted at the beginning of the Chapter\footnote{One should take care with this statement, since there are other models which are symmetric under permutations of particles and such that the entropy of entanglement is very large, as are for instance those systems described by the Laughlin wavefunction \cite{Laughlin83}.}.  

%
%

\subsection{The $\gamma - h$ plane}

\begin{figure}
\centering
\includegraphics[width=0.7\textwidth]{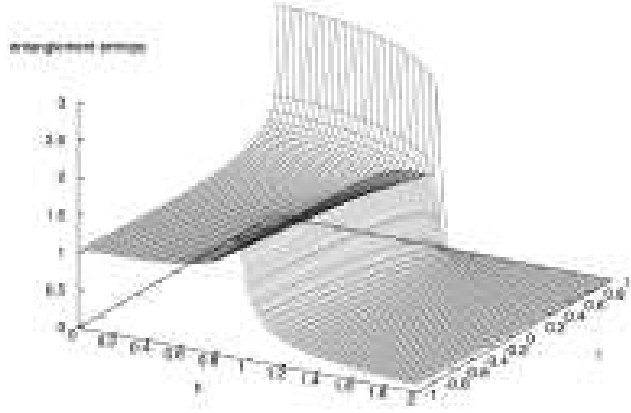}
\caption{Entanglement entropy for $N=500$ and $L=125$ as a function of $h$ and $\gamma$.} 
\label{fig:Shg500}
\end{figure}

In order to study the different entanglement regimes, we compute the entropy in the
plane spanned by $\gamma$ and $h$.  The numerical computation can be done by taking advantage of the Hamiltonian symmetries to reduce
the complexity of the task to a polynomial growth in $N$. Results are displayed in
Fig.\ref{fig:Shg500} for $N=500$ and $L=125$. For $\gamma \ne 1$, one clearly
observes a peak at the critical point $h=1$ whereas the entropy goes to
zero at large $h$ since the ground state is then a fully polarized state in
the field direction. In the zero field
limit, the entropy saturates when the
size of the system increases and goes to $S_{L,N}=1$ for $\gamma = 0$ where the ground
state approaches a GHZ-like ``cat'' state as in the
Ising quantum spin chain \cite{Latorre03,Latorre04_1,Lutken04,Orus05}. By contrast, for $\gamma = 1$,
the entropy increases with the size of the system in the region $0 \le h < 1$
and jumps directly to zero at $h=1$ as we shall now discuss.

%
%

\subsection{Analytical study of the isotropic case}

In the isotropic case ($\gamma=1$), it is possible to
analytically compute the entropy of entanglement since, at this point, the Hamiltonian is
diagonal in the Dicke basis.  The
ground-state energy is given by $E_0 (h,\gamma=1) = 
-{N \over 2} + \frac{2}{N} M^2  -2 h M $, with 
%
%
\begin{equation}
M=\Bigg\{ 
\begin{array}{ccc}
I(h N /2) \ , & {\rm if}&  0\leq h < 1 \\ 
N/2 \ ,& {\rm if}&  h \geq 1 
\end{array}
\ ,
\label{MGS}
\end{equation}
%
%
and the corresponding eigenvector is simply $|N/2,M\rangle$. Here, $I(x)$ denotes the round value of $x$. 

To calculate the entropy, it is convenient to introduce the number $n$ of
spins ``up" so that $M=n-N/2$, and to write this state in a bipartite
form. Indeed, since Dicke states are completely symmetric under any
permutation of sites, it is straightforward to see that the ground
state can be written as a sum of byproducts of Dicke states
%
%
\begin{eqnarray}
\label{dicke}
|N/2,n-N/2\rangle &=&\sum_{l=0}^L p_l^{1/2} |L/2,l-L/2\rangle \otimes  \\
&&  |(N-L)/2 ,n-l-(N-L)/2\rangle \nonumber \ ,
\label{decoo}
\end{eqnarray}
%
%
where the partition is made between two blocks of
size $L$ and $(N-L)$ and 
%
%
\begin{equation}
p_l = 
{\left(
\begin{array}{c}
L
\\
l
\end{array}
\right)
\left(
\begin{array}{c}
N-L
\\
n-l
\end{array}
\right)
\over
\left(
\begin{array}{c}
N
\\
n
\end{array}
\right)
} \ ,
\end{equation}
%
%
defining an hypergeometric probability distribution. The expression given in
Eq.\ref{decoo} corresponds to the Schmidt decomposition of the ground state
of the system. The entropy of this state for this bipartition is then simply 
given by $S_{L,N} (h,\gamma=1)=-\sum_{l=0}^L p_l \log_2{p_l}$. 
In the limit  $N,  L \gg 1$, the hypergeometric distribution of the $p_l$ can be
recast into a Gaussian distribution 
\begin{equation}
p_l \simeq p_l^g=\frac{1}{\sqrt{2 \pi} \sigma} e^{-\left(\frac{(l-\bar
l)^2}{2\sigma^2}\right)} \ ,
\label{ga}
\end{equation}
of mean value $\bar{l}=n {L \over N}$ and variance 
%
%
\begin{eqnarray}
\label{variance}
\sigma^2 = n(N-n)\frac{(N-L)L}{N^3} \ ,
\end{eqnarray}
%
%
where we have retained the sub-leading term in $(N-L)$ to explicitly
preserve the symmetry $S_{L,N}=S_{N-L,N}$. The entropy then reads
%
%
\begin{equation}
\label{gaussian}
- \int_{-\infty}^\infty\ {\rm d}l\  p_l^g  \log_2{p_l^g}
= \frac{1}{2} \left(\log_2{e}+ \log_2{2\pi} + \log_2{\sigma^2} \right) \ , 
\end{equation}
%
%
and only depends on its variance as expected for a Gaussian
distribution\footnote{This result has also been obtained in the context 
of the ferromagnetic Heisenberg chain \cite{Popkov05}.}.  Of course, for $h \ge 1$, the
entanglement entropy is exactly zero since the ground state is, in this case,
fully polarized in the magnetic field direction ($n=N$).  For $h \in [0,1)$ and in 
the limit $N, L \gg 1$, Eq.\ref{MGS}, Eq.\ref{variance} and Eq.\ref{gaussian} lead to
%
%
\begin{equation}
S_{L,N} (h,\gamma=1) \sim \frac{1}{2} \log_2\left(\frac{L(N-L)}{N}\right) \ .
\label{eq:entropy_iso}
\end{equation}
%
%

Moreover, the  dependence of the entropy on the magnetic field is given by
%
%
\begin{equation}
S_{L,N} (h,\gamma=1) -S_{L,N} (h=0,\gamma=1) \sim \frac{1}{2} \log_2 {\left(1-h^2 \right)} \ ,
\label{eq:Shiso}
\end{equation}
%
and thus diverges, at fixed $L$ and $N$, in the limit $h\rightarrow 1^-$.

%
 %
%

\subsection{Numerical study of the anisotropic case}

Let us now discuss the more general situation $\gamma \ne 1$ for which no
simple analytical solution exists. In this case, the ground state is a
superposition of Dicke states with coefficients that can be easily
determined by exact numerical diagonalizations. Upon tracing out $(N-L)$
spins, each Dicke state decomposes as in Eq.\ref{dicke}. 
It is then easy to build the $(L+1)\times(L+1)$ ground state reduced density
matrix and to compute its associated entropy.  

%
%
\begin{figure}
\centering
\includegraphics[angle=-90, width=0.7\textwidth]{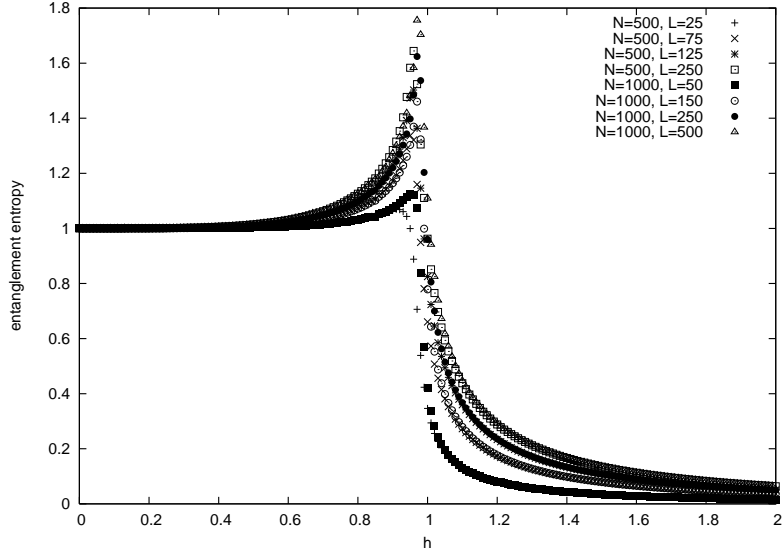}
\caption{Entanglement entropy at $\gamma = 0$ as a function of $h$ for
  different values of $N$ and $L$. Outside of the critical region, the entropy only depends on the 
 ratio $L/N$.}  
\label{fig:Shg0N}
\end{figure}
%
%

\begin{figure}
\centering
\includegraphics[angle=-90, width=0.7\textwidth]{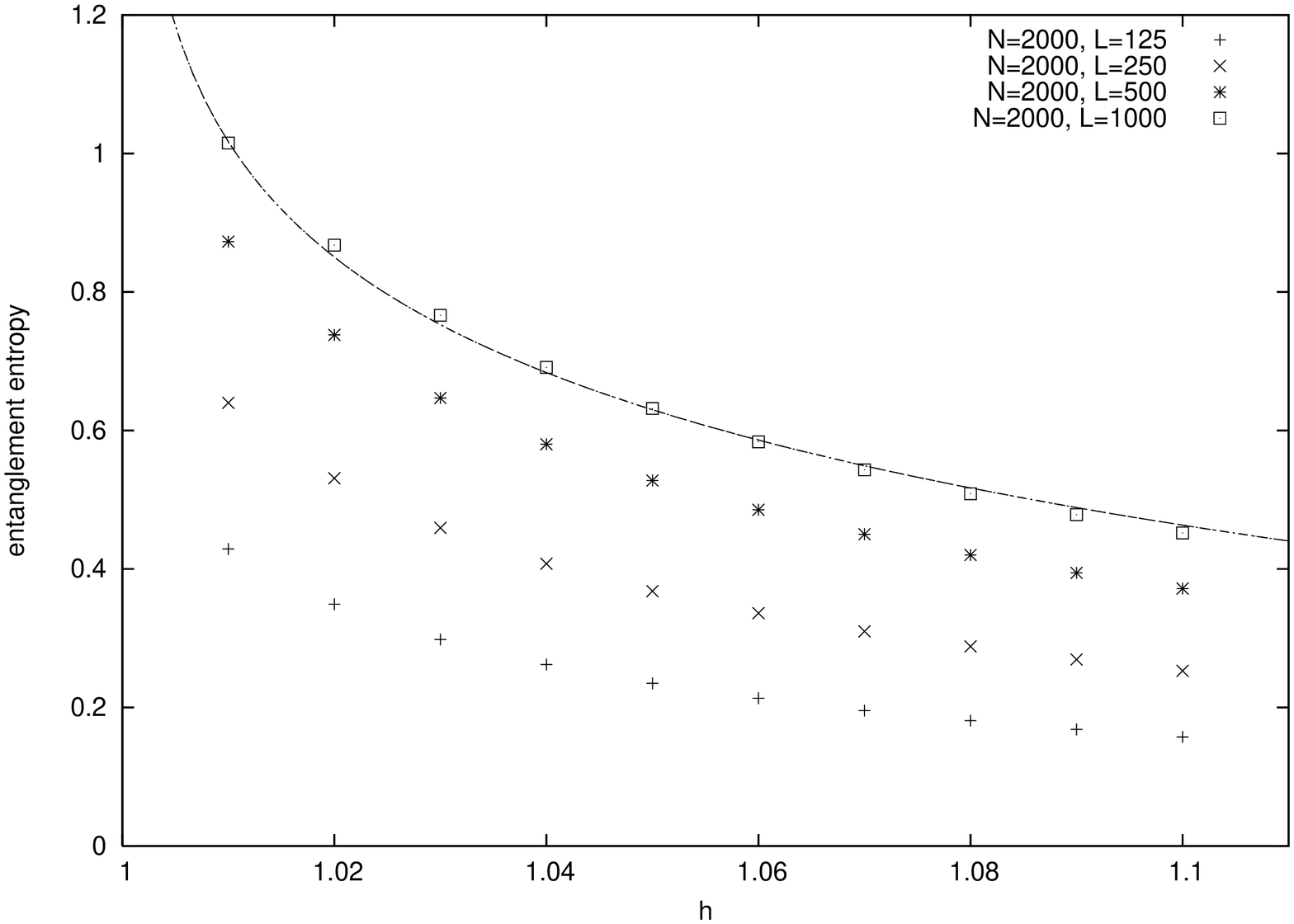}
\caption{Entanglement entropy as a function of $h$ near the critical point for $\gamma=0$. The full line corresponds to the fitting law from Eq.\ref{eq:Shani} with $a=1/6$.}
\label{fig:Shg02000}
\end{figure}

We have displayed in Fig.\ref{fig:Shg0N}, the behavior of the entropy as a
function of $h$, for different values of the ratio $L/N$ and for $\gamma=0$. 
For $h\neq 1$, the entropy only depends on the ratio $L/N$. For any $\gamma$, 
at fixed $L/N$ and in the limit $h\rightarrow \infty$, the entropy goes to
zero since the ground state becomes then fully polarized in the field
direction. Notice that the entropy also vanishes, at $h>1$, in the limit 
$L/N \rightarrow 0$ where the entanglement properties become trivial. 
In the zero field limit, the entropy goes to a constant which depends on
$\gamma$ and equals 1 at $\gamma=0$ since the ground state is then a GHZ-like 
state made up of spins pointing in $\pm x$ directions. 
Close to criticality, the entropy displays a logarithmic divergence,  
which we numerically find  to obey the law 
\begin{equation}
S_{L,N} (h,\gamma) \sim -a \log_2{|1-h|} \ ,
\label{eq:Shani}
\end{equation}
where $a$ is close to 1/6 for $N,L \gg 1$ as can be seen in Fig.\ref{fig:Shg02000}.
%
%
%
%

\begin{figure}
\centering
\includegraphics[angle=-90, width=0.7\textwidth]{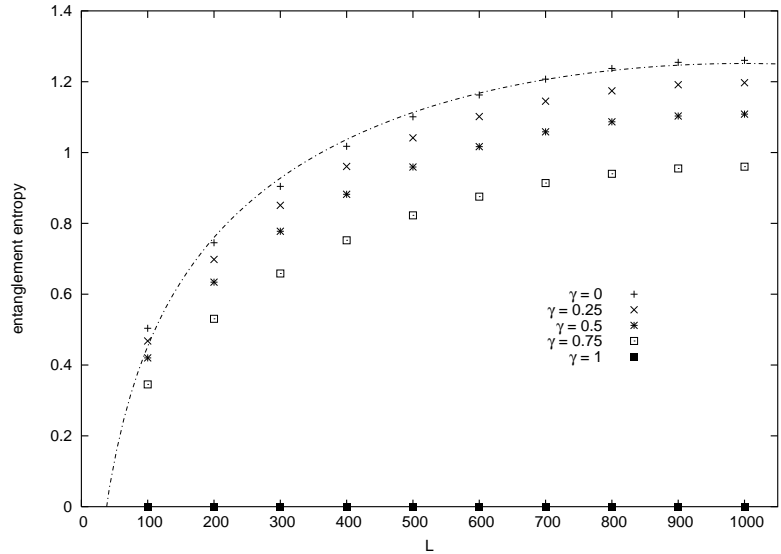}
\caption{Entanglement entropy as a function of $L$ at the critical point for 
different  $\gamma$ and $N=2000$. The full line corresponds to the fitting law from Eq.\ref{eq:entropy_ani} with $b=1/3$.}
\label{fig:Sh1L2000}
\end{figure}

At the critical point, the entropy has a 
nontrivial behavior that we have studied focusing on 
the point $\gamma=0$ which is representative of the class $\gamma \ne
1$. There, the entropy also scales logarithmically  with $L$ as in the 
isotropic case, but with a different prefactor. 
More precisely, we find
%
%
\begin{equation}
S_{L,N} (h=1,\gamma \neq 1) \sim b \log_2\left(\frac{L(N-L)}{N}\right) \ .
\label{eq:entropy_ani}
\end{equation}
%
%
For the finite-size systems investigated here, the prefactor varies when 
either the ratio $L/N$ or $\gamma$ is changed, as  can be seen in  
Fig.\ref{fig:Sh1L2000}. However,  in the thermodynamic limit  $N, L \gg 1$ 
(and finite $L/N$), $b=1/3$ fits well our numerical results.

%
%
\begin{figure}
\centering
\includegraphics[angle=-90, width=0.7\textwidth]{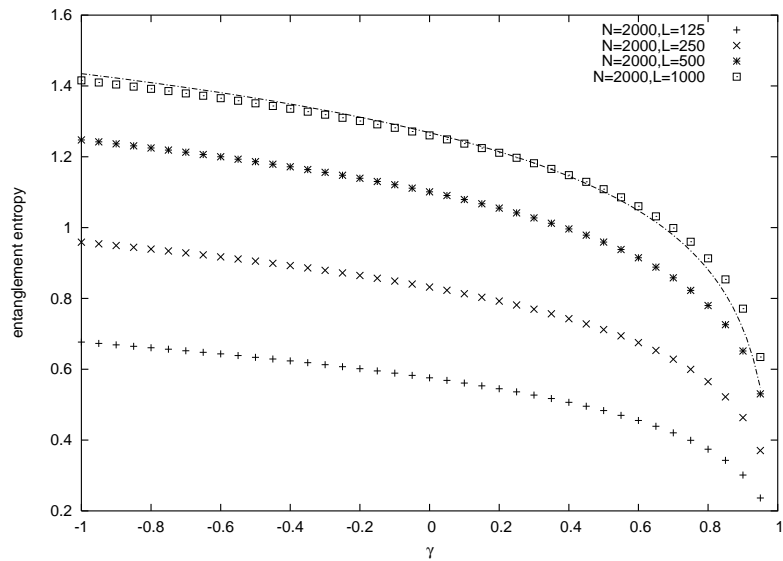}
\caption{Entanglement entropy at the critical point $h=1$ as a 
function of $\gamma$. The full line corresponds to the fitting law from Eq.\ref{eq:Sgamma} with $f=1/6$.}
\label{fig:Sh1g2000}
\end{figure}

%
%
In addition, at fixed $L$ and $N$, the entropy also depends on the anisotropy 
parameter logarithmically as
%
%
\begin{equation}
S_{L,N}(h=1,\gamma) - S_{L,N}(h=1,\gamma=0)\sim f  \log_2(1-\gamma) \ ,
\label{eq:Sgamma}
\end{equation}
%
%
for all $-1 \le \gamma < 1$ as can be seen in Fig.\ref{fig:Sh1g2000}.
Here again, it is likely that, in the thermodynamic limit, $f$  
has a simple (rational) value which, from our data, seems to be $1/6$.  
%
%
%
%
It is important to keep in mind that the limit $\gamma \rightarrow 1$ and 
the thermodynamic limit do not commute so that Eq.\ref{eq:Sgamma} 
is only valid for $\gamma \neq 1$.

Actually, the logarithmic behavior of the laws given in Eq.\ref{eq:Shani}, Eq.\ref{eq:entropy_ani} and Eq.\ref{eq:Sgamma} has been very recently confirmed by yet unpublished analytical computations \cite{Julien06}, but with values of $a$ and $b$ that differ from those obtained in simulations. More precisely, it has been proven that the exact coefficients $a$ and $b$ governing the logarithmic behaviors of Eq.\ref{eq:Shani} and Eq.\ref{entropy_ani} are $1/4$ and $1/2$ respectively, instead of the values $1/6$ and $1/3$ obtained from the numerical computations. The same analytical study confirms the value of $1/6$ for coefficient $f$ in Eq.\ref{eq:Sgamma}.

%

\section{Comparison to quantum spin chains}

Let us now compare the previous results with those found in the $(1+1)$-dimensional quantum $XY$ model. 
As for the Lipkin-Meshkov-Glick model, the $XY$ quantum spin chain has two different universality classes 
depending on the anisotropy parameter. At the critical point, the entropy 
has been found to behave as \cite{Latorre03,Latorre04_1,Jin03}
%
%
\begin{equation}
S_{L,N} \sim {c \over 3} \log_2\left(\frac{L(N-L)}{N}\right),
\label{scaling}
\end{equation}
%
%
where $c$ is the central charge of the corresponding $(1+1)$-dimensional conformal field theory
\cite{Holzhey94} (see Appendix B). For the isotropic case, the critical model is indeed
described by a free boson theory  with $c=1$ whereas the anisotropic 
case corresponds to a free fermion theory with $c=1/2$.
It is striking to see that the entropy in the Lipkin-Meshkov-Glick model has the same 
logarithmic dependence with some prefactor which, as in the $(1+1)$-dimensional case, only 
seems to depend on the universality class -- see Eq.\ref{eq:entropy_iso} and 
Eq.\ref{eq:entropy_ani} --.
Concerning the dependence with the magnetic field and with the anisotropy
parameter, it is also worth noting that logarithmic behaviors of  
Eq.\ref{eq:Shiso}, Eq.\ref{eq:Shani}, and  Eq.\ref{eq:Sgamma} are similar 
to those found in the $XY$ quantum spin chain \cite{Latorre03, Latorre04_1} except that 
the prefactors in the Lipkin-Meshkov-Glick model are different. A list of analogies between the
results of the Lipkin-Meshkov-Glick model and the $XY$ quantum spin chain in the limit $N \gg L \gg 1$ is given in Table \ref{xyLipkin-Meshkov-Glick}. 
\begin{table}
\begin{center}
{\small
\begin{tabular}{|c||c|}
    \hline & \\
     $XY$ quantum spin chain & Lipkin-Meshkov-Glick model \\
     & \\
$H = -\sum_{i=1}^{N} \left( \frac{(1+\gamma)}{2} \sigma_i^x \sigma_{i+1}^x + 
\frac{(1-\gamma)}{2} \sigma_i^y \sigma_{i+1}^y
+ \lambda \sigma_i^z \right)$ & $H=-\frac{1}{N} \sum_{i<j} \left( \sigma_i^x \sigma_j^x + \gamma
\sigma_i^y \sigma_j^y\right) - h \sum_{i=1}^{N} \sigma_i^z $ \\
     & \\ \hline \hline
     & \\
     $S_L(\lambda,\gamma = 0) \sim \frac{1}{3} \log_2(L)$& $S_L(h,\gamma = 1) \sim \frac{1}{2} \log_2(L)$ \\
     & \\
     $S_L(\lambda,\gamma=0) - S_L(\lambda=0,\gamma=0) \sim \frac{1}{6} \log_2\left(1-\lambda^2
     \right)$& $S_L(h,\gamma=1) - S_L(h=0,\gamma=1) \sim \frac{1}{2} \log_2\left(1-h^2
     \right)$ \\
     & \\
     $S_L(\lambda=1,\gamma=1) \sim \frac{1}{6} \log_2(L)$ & $S_L(h=1,\gamma=0) \sim \frac{1}{3} \log_2(L)$ \\
     & \\
     $S_L(\lambda, \gamma=1) \sim - \frac{1}{6} \log_2(m) $& $S_L(h,\gamma=0) \sim - \frac{1}{4} \log_2 |1-h|$ \\
     & \\
     $S_L(\lambda=1,\gamma) - S_L(\lambda=1,\gamma=1) \sim \frac{1}{6} \log_2(\gamma)$ 
     & $S_L(h=1,\gamma) - S_L(h=1,\gamma=0) \sim \frac{1}{6} \log_2(1-\gamma)$ \\
     & \\

    \hline    
\end{tabular}
}
  \caption{Comparison of results between the $XY$ quantum spin chain and the Lipkin-Meshkov-Glick
  model, when $N \gg L \gg 1$.}
  \label{xyLipkin-Meshkov-Glick}
\end{center}
\end{table}
Also, and just as a remark, it is possible to numerically check that the  behavior of this model with respect to
majorization (see Appendix A) for $\gamma \ne 1$ and as $h$ departs from its critical value is completely analogue 
to the case of the quantum $XY$ model \cite{Lutken04,Orus05}, which was analytically
studied in Chapter 1. Namely, the whole set of eigenvalues of the reduced density matrices of the ground state obey strict majorization relations as $h$ grows, while for decreasing $h$ one of the eigenvalues of the reduced density matrix in consideration drives the system towards a GHZ-like state in such a way that
majorization is only strictly obeyed in the thermodynamic limit. This behavior implies a 
very strong sense of order of the correlations present in the ground state, in complete analogy to the
behavior of the $XY$ quantum spin chain.  

\section{Conclusions of Chapter 3}

In this Chapter we have studied the entanglement properties of a quantum spin
model defined on a simplex. We have seen that: 
\begin{itemize}
\item{Contrary to the intuitive idea that the quantum
correlations present in the system increase together with the connectivity of
the model, here the symmetries force the entropy
to scale \emph{as if} the system were defined on a chain.}
\item{Also, the Lipkin-Meshkov-Glick model presents striking similarities with the $XY$ quantum spin chain:
not only their phase diagrams are almost identical, but the scaling properties of
the entanglement of the ground state seem to obey the same laws but with appropriate proportionality coefficients.} 
\end{itemize}
The observed similarity in the behavior of this model to $(1+1)$-dimensional quantum systems is indeed very pleasant, since quantum spin chains 
have been heavily studied and their properties are very well-known.
Some of their properties seem to be directly translated into systems
which, a priori, are not defined in $(1+1)$ dimension, like the Lipkin-Meshkov-Glick model. 
Nevertheless, most of the situations that one finds when considering 
models which are not defined on a chain turn out to be much more intrincated, as we
will see in the next two Chapters. Perhaps, a perturbative analysis around the Lipkin-Meshkov-Glick model -- for instance removing a few number
of links in the simplex and thus slightly breaking the symmetry present in the problem -- could allow to analytically study non-trivial properties of quantum many-body systems of high dimensionality.

    \chapter{Entanglement entropy in quantum algorithms}

The previous Chapters were focused on the properties of quantum many-body systems, basically from a condensed matter and field theoretical point of view. In particular, we saw that it is possible to apply   tools from quantum information science -- such as majorization and entanglement theory -- to obtain a better understanding of the properties of these systems. We will now see that these tools can also be used to understand better problems arising in the area of quantum information and quantum computation. 

In this and the forthcoming Chapters our aim is to study a physical system which is very close to the spirit of quantum many-body physics: we wish to understand the properties and behavior of \emph{quantum computers and quantum algorithms}. Indeed, a quantum computer is nothing but a physical system which is governed by the laws of quantum mechanics and on which we can perform physical actions -- algorithms -- such that the device is able deliver solutions to specific problems. Of course, the kind of problems that we can solve by using a quantum computer is necessarily limited by quantum physics itself, being this properly formalized by the area of quantum complexity theory \cite{Aharonov02}. Furthermore, it is plausible to think of a quantum computer as a device made of qubits which interact among themselves in some way. Therefore, \emph{a quantum computer can be understood as an interacting quantum many-body system}.  The full machinery from quantum many-body physics can then in principle be applied to analyze the performance of quantum algorithms. In particular, there is a very strong connection between quantum algorithms and quantum phase transitions, as we shall see.  

From the point of view of quantum computation, the design of new quantum algorithms is a great theoretical challenge. The most relevant property in order to understand these algorithms is clearly the role entanglement plays in quantum computational speedup, while some other properties seem to play a role as well, as we shall see in Chapter 6 with majorization \cite{LM02, OLM03, OLM04}. Regarding entanglement, several results have been found \cite{Ahn00, Knight00, Jozsa02, Parker02, Kendon04, GVidal03_1, GVidal04} which suggest that entanglement is at the heart
of the power of quantum computers. An important and remarkable 
result was obtained by Vidal \cite{GVidal03_1},
who proved that large entanglement between the qubits of a quantum register is a
\emph{necessary} condition for exponential speed-up in  quantum
computation. To be precise, a quantum register such that the maximum Schmidt
number of any bipartition is bounded at most by a polynomial in
the size of the system can be simulated efficiently  by
classical means. The classical simulation scheme proposed in \cite{GVidal03_1} was, indeed, a time-dependent version of the density matrix renormalization group algorithm, based on the efficient updates in time of the quantum register defined in terms of a matrix product state \cite{AKLT87, AKLT88}. Those methods are, indeed, tools for the classical simulation of the dynamics of a quantum many-body system which are also useful in the simulation of a quantum computation, since any quantum algorithm can be understood as the time evolution of a quantum many-body system \cite{Aharonov04}. Here we just sketch the basic idea of Vidal's algorithm, and leave all the specific details of this and other classical simulation protocols for the next Chapter. 

The figure of merit $\chi$ proposed in \cite{GVidal03_1}
is the maximum Schmidt number of any bipartitioning
of the quantum state or, in other words, the maximum rank of the
reduced density matrices for any possible splitting. It can be
proven that $\chi \ge 2^{S(\rho)}$, where the von Neumann entropy
$S(\rho)$ refers to the reduced density matrix of any of the two
partitions. From now on, in this and also in all the forthcoming Chapters we shall use the following computer-science notation: the number of qubits in the quantum register will be denoted by $n$, and $N=2^n$ denotes the dimensionality of the computational Hilbert space, as opposed to the condensed matter notation of the previous Chapters, were $N$ was the number of particles present in the system. Using this notation, Vidal proved that
if $\chi = O({\rm poly}(n))$ at every step of the
computation in a quantum algorithm, then it can be efficiently
classically simulated. Exponential speed-up over
classical computation is only possible if
at some step along the computation  $\chi \sim {\rm exp}(n^a)$, or
$S(\rho) \sim n^b$,  $a$ and $b$ being positive constants. In order to
exponentially accelerate the performance of classical computers
any quantum algorithm must necessarily create an exponentially
large amount of $\chi$ at some point.

As we saw in the previous Chapters, a topic of intense research concerns the behavior of
entanglement in systems undergoing a quantum phase transition
\cite{Sachdev}. More generally,
when a splitting of a $(d+1)$-dimensional
spin system is made, the von Neumann entropy of the ground state for the reduced
density matrix of one of the subsystems $S(\rho) = -{\rm tr}(\rho
\log _2 \rho)$ at the critical point should typically display a universal
leading scaling behavior determined by the \emph{area} of the
region partitioning the whole system \cite{Sred93, Au02, Cramer05}, with at most logarithmic corrections if the system is fermionic \cite{Wolf06, Gio06, Bart06}. As hinted in the previous Chapter, 
this result depends on the connectivity of the Hamiltonian.
Using a naive reasoning, we saw there that the leading universal scaling 
behavior for the entropy of an exact bipartition of the system should typically be written in terms of the number of particles $n$ as

\begin{equation}
S(\rho) \sim n^{\frac{d-1}{d}}
\end{equation}
for a $(d+1)$-dimensional critical non-fermionic system with sufficiently local interactions,
 which reduces to a logarithmic law for $d = 1$. This explicit
dependence of entanglement on dimensionality turns out to shed new light
into some well established results from quantum computation.

A similar situation is present in quantum adiabatic algorithms,
originally introduced by Farhi et \emph{al.} in \cite{Farhi00_1}, where the
Hamiltonian of the system depends on a control parameter $s$ which
in turn has a given time dependence. The Hamiltonians related to
adiabatic quantum computation for solving some NP-complete
problems (such as 3-SAT or Exact Cover) can be directly mapped to
interacting non-local spin systems, and therefore we can extend the
study of entanglement to include this kind of Hamiltonians. This
point of view has the additional interest of being directly
connected to the possibility of efficient classical simulations of
the quantum algorithm, by means of the protocol proposed in
\cite{GVidal03_1}.

Here we analyze the scaling of the entropy of
entanglement in several quantum algorithms. More concretely, we
focus on Shor's quantum factoring algorithm \cite{Shor94}
and on a quantum algorithm by adiabatic evolution solving the
Exact Cover NP-complete problem \cite{Farhi00_1, Farhi00_2, Childs00, Farhi01, Childs02_1, Farhi02_1, Farhi02_2, Farhi05, Jordan05}, finding for both of
them evidence (either analytical or numerical) of a quantum exponential speedup with linear scaling
of quantum correlations -- as measured by the entropy --, which seems to prohibit the possibility of
an efficient classical simulation. We furthermore make an analytical study of the
adiabatic implementation of Grover's quantum search algorithm
\cite{Grover96, Roland02, Dam01}, in which entanglement is a
bounded quantity between calls to the quantum oracle 
even at the critical point, regardless of the size
of the system. Let us begin, then, by considering the behavior of the factoring quantum algorithm. 

\section{Entanglement in Shor's factoring quantum algorithm}
 It is  believed that the reason why
Shor's quantum algorithm for factorization \cite{Shor94}
beats so clearly its classical rivals is rooted in
the clever use it makes of quantum entanglement. Several attempts
have been made in order to understand the behavior of the quantum
correlations present along the computation \cite{Jozsa02, Parker02, Kendon04}. In our
case, we will concentrate in the study of the scaling
behavior for the entanglement entropy of the system. We shall first
remember both Shor's original \cite{Shor94} and phase-estimation
\cite{Cleve98} proposals of the factoring algorithm and afterwards
we shall move to the analytical study of their quantum correlations.

\subsection{The factoring quantum algorithm}
The interested reader is addressed to \cite{Shor94, Cleve98, Nielsen-Chuang, Galindo02} 
 for precise details. Given an odd integer $N$ to
factorize, we pick up a random number $a \in [1,N]$. We make the
assumption that $a$ and $N$ are co-primes -- otherwise the greatest
common divisor of $a$ and $N$ would already be a non-trivial
factor of $N$ --. There exists a smaller integer $r \in [1,N]$,
called the \emph{order} of the modular exponentiation $a^x \ {\rm
mod} \ N$, such that $a^r \ {\rm mod} \ N = 1$. Let us assume that
the $a$ we have chosen is such that $r$ is even and $a^{r/2} \
{\rm mod} \ N \ne -1$, which happens with very high probability,
 bigger than or equal to $1/(2 \log_2N)$. This is the case of
interest because then the greatest common divisor of $N$ and
$a^{r/2} \pm 1$ is a non-trivial factor of $N$. Therefore, the original
factorization problem has been reduced to the order-finding problem of
the modular exponentiation function $a^x \ {\rm mod} \ N$, and it
is at this point where quantum mechanics comes at work. The
procedure can be casted in two different (but equivalent) ways:

\subsubsection{Shor's proposal for order-finding}

\begin{figure}[h]
\begin{equation}
\Qcircuit @C=1.5em @R=1em {
 \lstick{|0\rangle^{\otimes k}}  & \ustick{(k)} \qw & \gate{U_H^{\otimes k}} &  \multigate{1}{U_f} & \gate{QFT} & \meter  \\
  \lstick{|0\rangle^{\otimes n}}  & \ustick{(n)} \qw &     \qw   & \ghost{U_f} & \qw   & \meter 
}  \nonumber
\end{equation}
\caption{Quantum circuit for the order-finding algorithm for the
modular exponentiation function. The source and target registers have $k$ and $n$ qubits respectively. } \label{shor}
\end{figure}
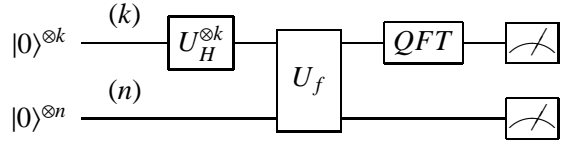

We make use of two quantum registers: a source register of $k$
qubits such that $2^k \in [N^2, 2  N^2]$, and a target register
of $n = \lceil \log_2N \rceil$ qubits. The quantum circuit of the
quantum algorithm is shown in Fig.\ref{shor}, where we are making
use of the Hadamard gate initially acting over the $k$ qubits of
the source, the unitary implementation of the modular
exponentiation function
\begin{equation}
U_f |q\rangle |x\rangle = |q\rangle |(x + a^q) \ {\rm mod} \ N\rangle \ , 
\label{modular} 
\end{equation}
where $|q\rangle$ and $|x\rangle$ respectively belong to the
source and target registers, and the quantum Fourier transform
operator
\begin{equation}
QFT |q\rangle = \frac{1}{2^{k/2}}\sum_{m=0}^{2^k-1}e^{2 \pi i q m
/ 2^k }|m\rangle \ . \label{quan}
\end{equation}
All these operations can be efficiently implemented by means of
one and two-qubit gates. Finally, a suitable classical treatment
of the final measurement of this quantum algorithm provides us
with $r$ in few steps, and therefore the prime factorization of
$N$ in a time $O((\log_2N)^3)$.

\bigskip

\subsubsection{Phase-estimation proposal for order-finding}
\begin{figure}[h]
\begin{equation}
\Qcircuit @C=1.5em @R=1em {
 \lstick{|0\rangle^{\otimes k}}  & \ustick{(k)} \qw & \gate{U_H^{\otimes k}} &  \ctrl{1} & \gate{QFT} & \meter  \\
  \lstick{|1\rangle^{\otimes n}}  & \ustick{(n)} \qw &     \qw   & \gate{V_f} & \qw   & \meter 
}  \nonumber
\end{equation}
\caption{Phase-estimation version of the quantum circuit for the
order-finding algorithm. The controlled operation is $\Lambda
(V_f)$. The source and target registers have $k$ and $n$ qubits respectively.} \label{shor2}
\end{figure}
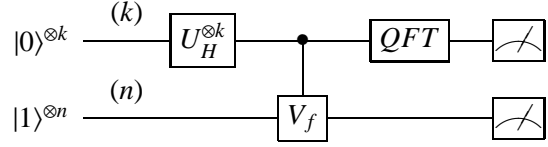

We shall address the specific details of the generic quantum phase-estimation 
algorithm in Chapter 6 and refer the interested reader to \cite{Cleve98} for more information.
For order-finding purposes, the quantum circuit is similar to the one shown in the previous
section but slightly modified, as is shown in Fig.\ref{shor2}. The
unitary operator $V_f$ to which the phase-estimation procedure is
applied is defined as

\begin{equation}
V_f |x\rangle = |(a \ x) \ {\rm mod} \ N \rangle \label{modular2}
\end{equation}
(notice the difference between Eq.\ref{modular2} and Eq.\ref{modular}), being diagonalized by eigenvectors

\begin{equation}
|v_s\rangle = \frac{1}{r^{1/2}} \sum_{p = 0}^{r-1} e^{- 2 \pi i s
p / r}|a^p \ {\rm mod} \ N\rangle \label{vv}
\end{equation}
such that

\begin{equation}
V_f |v_s\rangle = e^{2 \pi i s / r}|v_s\rangle \ , \label{veigen}
\end{equation}
and satisfying the relation $\frac{1}{r^{1/2}} \sum_{s = 0}^{r-1}
|v_s\rangle = |1\rangle$. The operator is applied over the target
register being controlled on the qubits of the source in such a
way that

\begin{equation}
\Lambda (V_f) |j\rangle |x\rangle = |j\rangle V_f^j |x\rangle \ ,
\label{lambda}
\end{equation}
where by $\Lambda (V_f)$ we understand the full controlled
operation acting over the whole system, which can be efficiently
implemented in terms of one and two-qubit gates. As in the
previous case, the information provided by a final measurement of
the quantum computer enables us to get the factors of $N$ in a
time $O((\log_2 N)^3)$.

\bigskip

\subsection{Analytical results}
We choose to study the amount of entanglement between the source
and the target register in the two proposed quantum circuits,
right after the modular exponentiation operation $U_f$
from Fig.\ref{shor} or the controlled $V_f$ operation
from Fig.\ref{shor2}, and before the quantum Fourier transform in
both cases. At this step of the computation, the pure quantum
state of the quantum computer is easily seen to be exactly the
same for both quantum circuits, and is given by

\begin{equation}
|\psi \rangle = \frac{1}{2^{k/2}} \sum_{q = 0}^{2^k-1} |q\rangle
|a^q \ {\rm mod} \ N\rangle \ , \label{status}
\end{equation}
and therefore the density matrix of the whole system is

\begin{equation}
|\psi\rangle \langle \psi| = \frac{1}{2^k} \sum_{q, q' =
0}^{2^k-1} \left(|q\rangle \langle q'|\right) \ \left(|a^q \ {\rm
mod} \ N\rangle \langle a^{q'} \ {\rm mod} \ N|\right) \ .
\label{mat}
\end{equation}
Tracing out the quantum bits corresponding to the source, we get
the density matrix of the target register, which reads

\begin{equation}
\rho_{{\rm target}} = {\rm tr_{source}}(|\psi\rangle \langle
\psi|) = \frac{1}{2^k} \sum_{p, q, q' = 0}^{2^k-1} \left(\langle
p|q\rangle \langle q'|p\rangle\right) \ \left(|a^q \ {\rm mod} N
\rangle \langle a^{q'} \ {\rm mod} N|\right) \ ,\label{target}
\end{equation}
that is,

\begin{equation}
\rho_{{\rm target}} = \frac{1}{2^k} \sum_{p = 0}^{2^k-1} |a^p \
{\rm mod} \ N \rangle \langle a^p \ {\rm mod} \ N| \sim
\frac{1}{r} \sum_{p=0}^{r-1} |a^p \ {\rm mod} N \rangle \langle
a^p \ {\rm mod} N| \ . \label{rho}
\end{equation}
The last step comes from the fact that $a^r \ {\rm mod} \ N = 1$,
where $r \in [1, N]$ denotes the order of the modular exponentiation. If
$2^k$ were a multiple of $r$ there would not be any approximation
and the last equation would be exact. This is not necessarily the
case, but the corrections to this expression are $O(1/2^k)$,
thus being exponentially small in the size of the system.

It follows from Eq.\ref{rho} that the rank of the
reduced density matrix of the target register at this point of the
computation is
\begin{equation}
{\rm rank}(\rho_{{\rm target}}) \sim r \ . \label{ranking}
\end{equation}
Because $r \in [1, N]$, this rank is usually $O(N)$. If this were
not the case, for example if $r$ were $O(\log_2N)$, then the
order-finding problem could be efficiently solved by a classical
naive algorithm and it would not be considered as classically
hard. Because $N$ is exponentially big in the number of qubits,
we have found a particular bipartition of the system (namely, the
bipartition between the source register and the target register)
and a step in the quantum algorithm in which the entanglement, as
measured by the rank of the reduced density matrix of one of the
subsystems, is exponentially big. This implies in turn that
Shor's quantum factoring algorithm can not be efficiently
classically simulated by any protocol in \cite{GVidal03_1} owing
to the fact that at this step $\chi = O(N)$, therefore
constituting an inherent exponential quantum speed-up based on an
exponentially big amount of entanglement. It is worth noticing
that the purpose of the entanglement between the two registers
consists on leaving the source in the right periodic state to be processed
by the quantum Fourier transform. Measuring the register right
after the entangling gate disentangles the two registers while
leaving the source in a periodic state, and this effect can only
be accomplished by previously entangling source and target. These
conclusions apply both to Shor's original proposal (circuit of
Fig.\ref{shor}) and to the phase-estimation version (circuit of
Fig.\ref{shor2}).

The behavior of the rank of the system involves that the entropy
of entanglement of the reduced density matrix at this point will
essentially scale linearly with the number of qubits, $S(\rho_{\rm target}) = \log_2 r
\sim \log_2 N \sim n$, which is the hardest of all the possible scaling
laws. We will find again this strong behavior for the entropy in the following section, when considering an adiabatic quantum algorithm solving an optimization NP-complete problem. 

\section{Entanglement in an adiabatic NP-complete optimization algorithm}

We now turn to analyze how entanglement scales for a quantum algorithm
based on adiabatic evolution  \cite{Farhi00_1}, designed to solve
 the Exact Cover NP-complete problem \cite{Farhi01}. Basic background on NP-completeness 
 and classical complexity theory can be found in Appendix C.  We first briefly
 review the proposal and, then,  we
consider the study of the properties of the system, in particular
the behavior of the entanglement entropy for a given bipartition
of the ground state.

\subsection{The adiabatic quantum algorithm}

The adiabatic model of quantum optimization algorithm deals with the problem
of finding the ground state of a given system represented by its
Hamiltonian. Many relevant computational problems, such as 3-SAT \cite{Cook71},
can be mapped to this situation. The method is briefly summarized
as follows: we start from a time dependent Hamiltonian of the form

\begin{equation}
H(s(t)) = (1-s(t)) H_0 + s(t) H_P \ , \label{ham}
\end{equation}
where $H_0$ and $H_P$ are the initial and problem Hamiltonian
respectively, and $s(t)$ is a time-dependent function satisfying
the boundary conditions $s(0) = 0$ and $s(T) = 1$ for a given $T$.
The desired solution to a certain problem is encoded in the
ground state of $H_P$. The gap between the ground and the first
excited state of the instantaneous Hamiltonian at time $t$ will be
called $g(t)$. Let us define $g_{min}$ as the global minimum of
$g(t)$ for $t$ in the interval $[0, T]$. If at time $T$ the ground
state is given by the state $|E_0; T\rangle$, the adiabatic
theorem states that if we prepare the system in its ground state
at $t=0$, which is assumed to be easy to prepare, and let it
evolve under this Hamiltonian, then

\begin{equation}
|\langle E_0; T|\psi(T)\rangle|^2 \geq 1 - \epsilon^2
\label{probab}
\end{equation}
provided that

\begin{equation}
\frac{{\rm max} |\frac{dH_{1,0}}{dt}|}{g^2_{min}} \leq \epsilon
\label{cond}
\end{equation}
where $H_{1,0}$ is the Hamiltonian matrix element between the
ground and first excited state, $\epsilon << 1$, and the
maximization is taken over the whole time interval $[0,T]$.
Because the problem Hamiltonian encodes the solution of the
problem in its ground state, we get the desired solution with high
probability after a time $T$. A closer look at the adiabatic
theorem tells us that $T$ dramatically depends on the scaling of
the inverse of $g_{min}^2$ with the size of the system. More
concretely, if the gap is only polynomially small in the number of
qubits (that is to say, it scales as $O (1/{\rm poly}(n))$, the
computational time is $O({\rm poly}(n))$, whereas if the gap is
exponentially small ($O(2^{-n})$) the algorithm makes use of an
exponentially big time to reach the solution.

The explicit functional dependence of the parameter $s(t)$ on time
can be very diverse. The point of view we adopt in this Chapter 
is such that this time dependence is not taken into account,
as we study the properties of the system as a function of $s$,
which will be understood as the Hamiltonian parameter. We will in
particular analyze the entanglement properties of the ground state
of $H(s)$, as adiabatic quantum computation assumes that the
quantum state remains always close to the instantaneous ground
state of the Hamiltonian all along the computation. Notice that we
are dealing with a system which is suitable to undergo a quantum
phase transition at some critical value of the Hamiltonian
parameter in the thermodynamic limit, and therefore we expect to achieve the largest quantum
correlations when evolving close to this point. The question is how these large quantum
correlations scale with the size of the system when dealing with
interesting problems. This is the starting point for the next two
sections.

\subsection{Exact Cover}

The Exact Cover NP-complete problem is a particular case of the
3-SAT problem, and is defined as follows: given the $n$ boolean
variables $\{x_i\}_{i=1,\ldots n}$, $x_i = 0,1 \ \forall \ i$,
where $i$ is regarded as the bit index, we define a \emph{clause}
of Exact Cover involving the three qubits $i$, $j$ and $k$  (say,
clause ``$C$") by the equation $x_i + x_j + x_k = 1$. There are
only three assignments of the set of variables $\{x_i, x_j, x_k
\}$ that satisfy this equation, namely, $\{1,0,0\}$, $\{0,1,0\}$
and $\{0,0,1\}$. The clause can be more specifically expressed in
terms of a boolean function in Conjunctive Normal Form (CNF) as
\begin{eqnarray}
\phi_{C}(x_i,x_j,x_k) &&= (x_i \lor x_j \lor x_k)\land(\neg x_i
\lor \neg x_j \lor \neg x_k)\land(\neg x_i \lor \neg x_j \lor
x_k) \nonumber \\
&&\land(\neg x_i \lor x_j \lor \neg x_k)\land(x_i \lor \neg x_j
\lor \neg x_k) \ , \label{CNF}
\end{eqnarray}
so $\phi_{C}(x_i,x_j,x_k) = 1$ as long as the clause is properly
satisfied. An \emph{instance} of Exact Cover is a collection of
clauses which involves different groups of three bits. The
problem is to find a string of bits $\{x_1, x_2 \ldots , x_n \}$
which satisfies all the clauses.

This problem can be mapped into finding the ground state of the
Hamiltonian $H_P$ of a spin-$1/2$ system in the following way: given a clause $C$ define
the Hamiltonian associated to this clause as

\begin{eqnarray}
H_C &=&
\frac{1}{2}(1+\sigma_i^z)\frac{1}{2}(1+\sigma_j^z)\frac{1}{2}(1+\sigma_k^z)
 \nonumber \\ &+&
\frac{1}{2}(1-\sigma_i^z)\frac{1}{2}(1-\sigma_j^z)\frac{1}{2}(1-\sigma_k^z)
 \nonumber \\ &+&
\frac{1}{2}(1-\sigma_i^z)\frac{1}{2}(1-\sigma_j^z)\frac{1}{2}(1+\sigma_k^z)
 \nonumber \\ &+&
\frac{1}{2}(1-\sigma_i^z)\frac{1}{2}(1+\sigma_j^z)\frac{1}{2}(1-\sigma_k^z)
 \nonumber \\ &+&
\frac{1}{2}(1+\sigma_i^z)\frac{1}{2}(1-\sigma_j^z)\frac{1}{2}(1-\sigma_k^z)
 \ , \label{hamit}
\end{eqnarray}
where we have defined $\sigma^z |0\rangle = |0\rangle$, $\sigma^z
|1\rangle = -|1\rangle$. Note the analogy between Eq.\ref{CNF} and Eq.\ref{hamit}.
 The quantum states of the
computational basis that are eigenstates of $H_C$ with zero
eigenvalue (ground states) are the ones that correspond to the bit
string which satisfies $C$, whereas the rest of the computational
states are penalized with an energy equal to one\footnote{In the next Chapter we shall consider a different implementation of $H_C$.}. Now, we
construct the problem Hamiltonian as the sum of all the
Hamiltonians corresponding to all the clauses in our particular
instance, that is to say,

\begin{equation}
H_P = \sum_{C \ \in \ {\rm instance}} H_C \ , \label{hamil}
\end{equation}
so the ground state of this Hamiltonian corresponds to the quantum
state whose bit string satisfies \emph{the maximum number} of clauses (all of them if the clauses are mutually compatible). We have
reduced the original problem stated in terms of boolean logic to
the hard task of finding the ground state of a two and three-body
interactive spin Hamiltonian with local magnetic fields. Observe that
the couplings depend on the particular instance we are dealing
with, and that the spin system has not an a priori well defined
dimensionality neither a well defined lattice topology, in
contrast with some usual simple spin models.

We now define our s-dependent Hamiltonian $H(s)$ as a linear
interpolation between an initial Hamiltonian $H_0$ and $H_P$:

\begin{equation}
H(s) = (1-s)H_0 + s H_P \label{finalham}
\end{equation}
where we take the initial Hamiltonian $H_0$ to be that resulting from the interaction with a
magnetic field in the $x$ direction:

\begin{equation}
H_0 = \sum_{i = 1}^n \frac{d_i}{2}(1-\sigma_i^x) \ , \label{h0}
\end{equation}
where $d_i$ is the number of clauses in which qubit $i$ appears,
and $\sigma^x |+\rangle = |+\rangle$, with $|+\rangle =
\frac{1}{\sqrt{2}}(|0\rangle + |1\rangle)$, so the ground state of
$H_0$ is an equal superposition of all the possible computational
states. Observe that $H(s)$ is, apart from a constant factor, a
sum of terms involving local magnetic fields in the $x$ and $z$
direction, together with two and three-body interaction coupling
terms in the $z$ component. We can thus expect this system to undergo a
quantum phase transition (in the limit of infinite $n$) as $s$ is
shifted from $0$ to $1$. The numerical study of this phenomena is the aim of
the next section.

\subsection{Numerical results up to 20 qubits}

We have randomly generated instances for Exact Cover with only one
possible satisfying assignment and have constructed the
corresponding problem Hamiltonians. Instances are produced by
adding clauses randomly until there is exactly one satisfying
assignment, starting over if we end up with no satisfying
assignments. According to \cite{Farhi01}, these are believed to be
the most difficult instances for the adiabatic algorithm. Our
analysis proceeds as follows:

\subsubsection{Appearance of a quantum phase transition}

\begin{figure}
\centering
\includegraphics[angle=-90, width=0.7\textwidth]{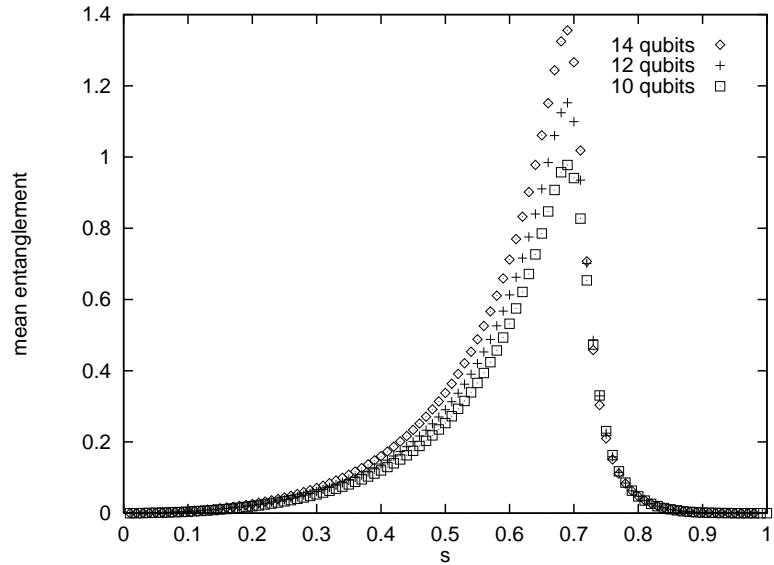}
\caption{Evolution of the entanglement entropy between the two
blocks of size $n/2$ when a bipartition of the system is made, on
average over 300 different instances with one satisfying
assignment. A peak in the correlations appears for $s_c \sim 0.7$
in the three cases.} \label{mix-300}
\end{figure}

We have generated 300 Exact Cover instances -- 300 random
Hamiltonians with a non-degenerated ground state -- and have
calculated the ground state for 10, 12 and 14 qubits for different
values of the parameter $s$ in steps of $0.01$. We then consider a
particular bipartition of the system into two blocks of $n/2$
qubits, namely, the first $n/2$ qubits versus the rest, and have
calculated the entanglement entropy between the two blocks. For
each of the randomly generated Hamiltonians we observe a peak in
the entanglement entropy around a critical value of the parameter
$s_c \sim 0.7$. We have averaged the obtained curves over the 300
instances and have obtained the plot from Fig.\ref{mix-300}.

\begin{figure}
\centering
\includegraphics[angle=-90, width=0.7\textwidth]{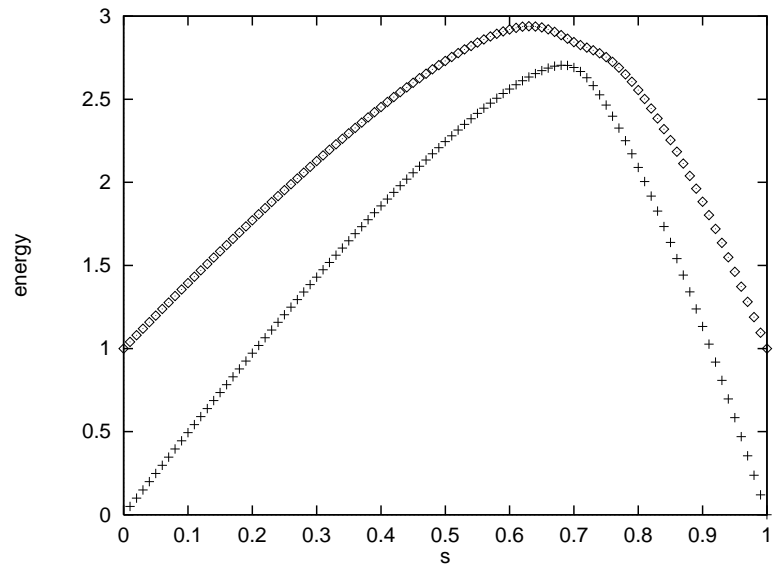}
\caption{Energies of the ground state and first excited state for
a typical instance with one satisfying assignment of Exact Cover
in the case of 10 qubits (in dimensionless units). The energy gap approaches its minimum at $s_c
\sim 0.7$.} \label{gap-10}
\end{figure}

The point at which the entropy of entanglement reaches its maximum
value is identified as the one corresponding to the critical point
of a quantum phase transition in the system (in the limit of
infinite size). This interpretation is reinforced by the
observation of the typical energy eigenvalues of the system. For a
typical instance of 10 qubits we observe that the energy gap
between the ground state and the first excited state reaches a
minimum precisely for a value of the parameter $s_c \sim 0.7$ (see
Fig.\ref{gap-10}).


We observe from Fig.\ref{mix-300} that the peak in the entropy is
highly asymmetric with respect to the parameter $s$. A
study of the way this peak seems to diverge near the critical
region seems to indicate that the growth of entanglement is slower
at the beginning of the evolution and fits remarkably well a curve
of the type $S \sim \log_2{| \log_2{(s-s_c)}|}$, whereas the falling
down of the peak is better parameterized by a power law $S \sim
|s-s_c|^{-\alpha}$ with $\alpha \sim 2.3$, $\alpha$ being a
certain critical exponent. These laws governing the critical
region fit better and better the data as the number of qubits is
increased.

\subsubsection{Analysis of different bipartitions of the system}

\begin{figure}
\centering
\includegraphics[angle=-90, width=0.7\textwidth]{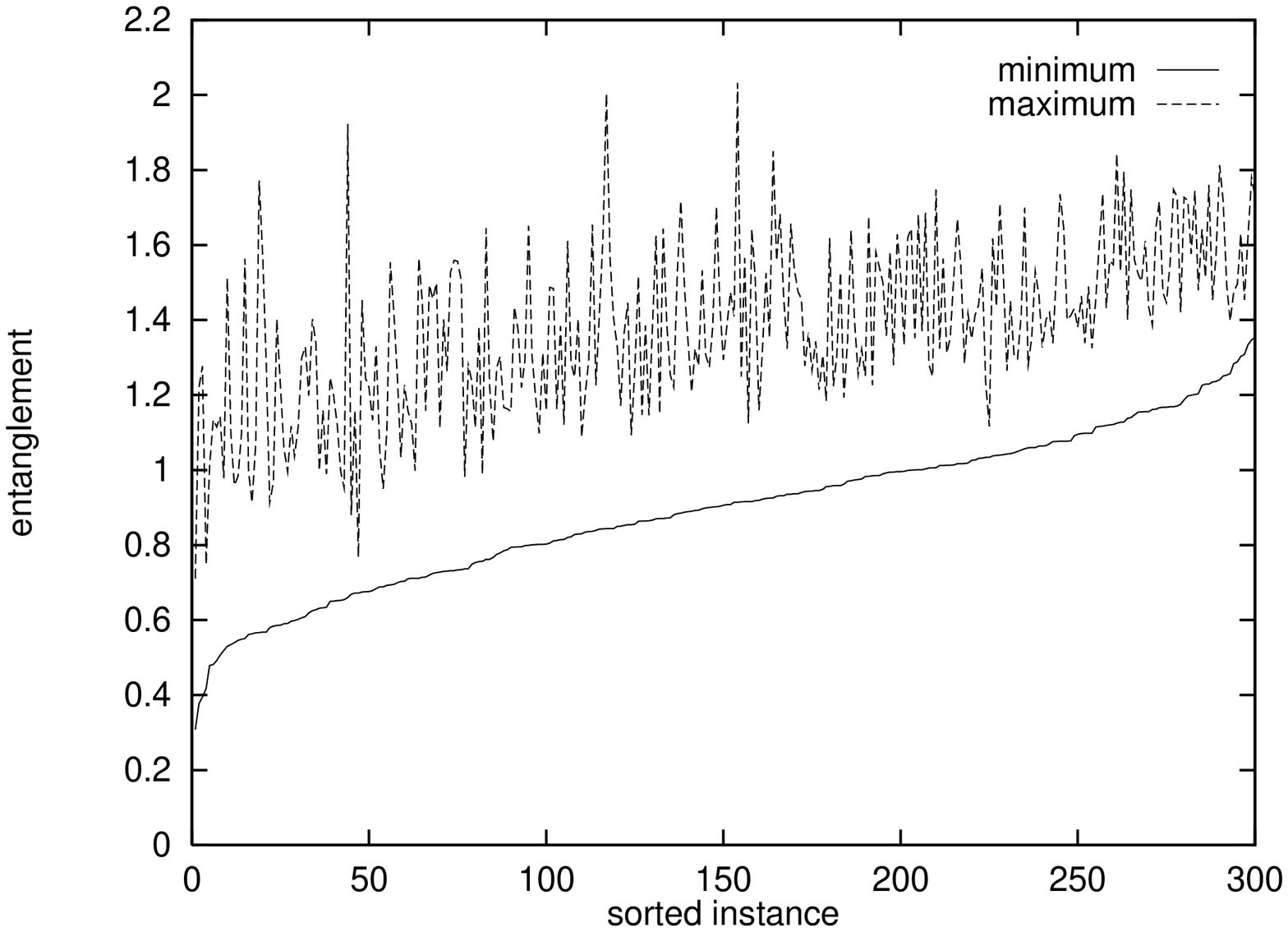}
\caption{Minimum and maximum entropy over all possible
bipartitions of a $10$-qubit system for each of the $300$
randomly generated instances of Exact Cover. Instances are sorted
such that the minimum entanglement monotonically increases.}
\label{part1}
\end{figure}
\begin{figure}
\centering
\includegraphics[angle=-90, width=0.7\textwidth]{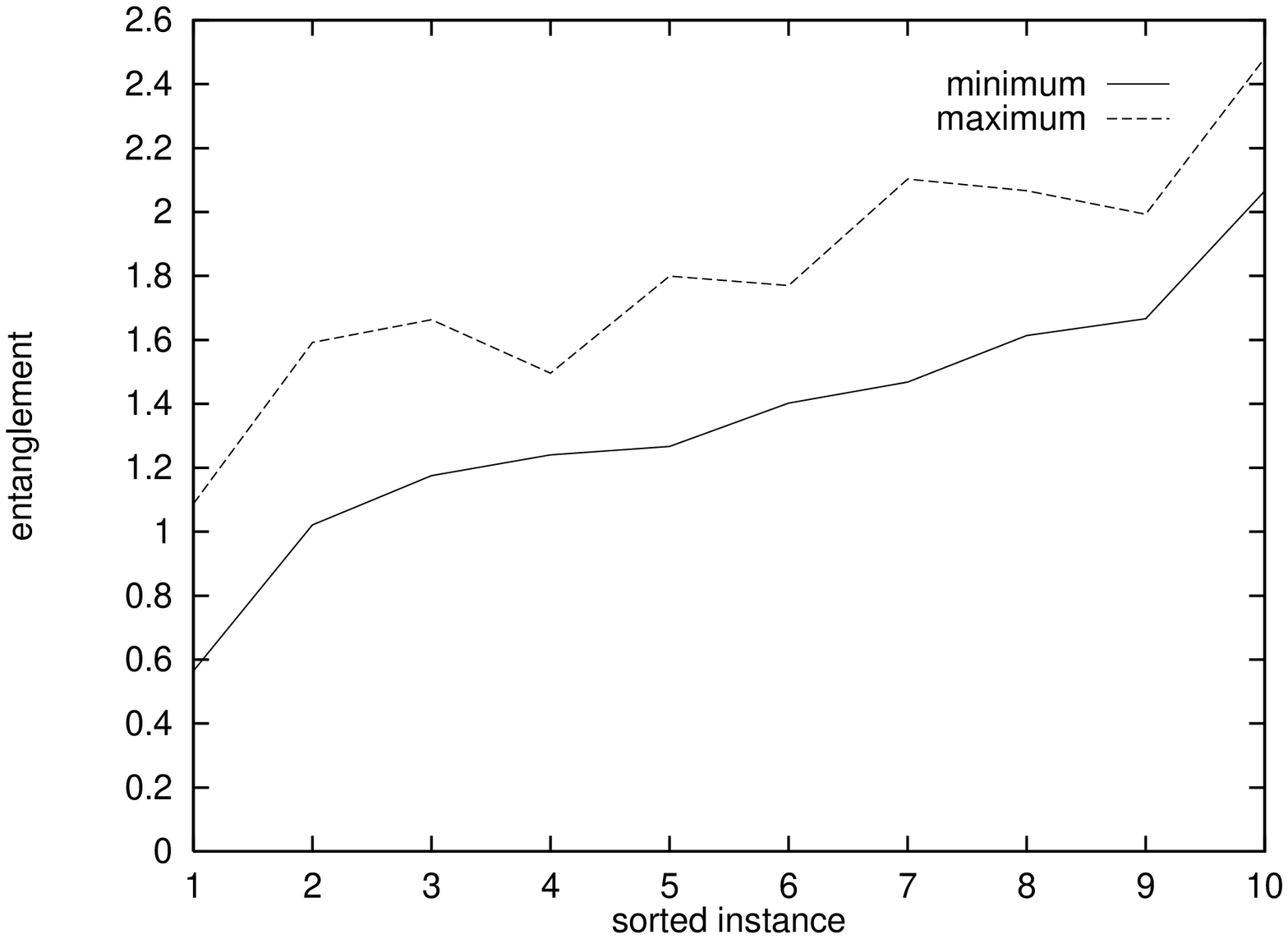}
\caption{Minimum and maximum entropy over $64$ bipartitions of a
$16$-qubit system for $10$ randomly generated instances of Exact
Cover. Instances are sorted such that the minimum entanglement
monotonically increases.} \label{part2}
\end{figure}

An explicit numerical analysis for $10$ qubits tells us that all
possible bipartitions for each one of the instances produce
entropies at the critical point of the same order of magnitude -- as
expected from the non-locality of the interactions --. This is
represented in Fig.\ref{part1}, where we plot the minimum and
maximum entanglement obtained from all the possible bipartitions of
the system for each one of the generated instances (points are
sorted such that the minimum entropy monotonically increases).

Similar conclusions follow from the data plotted in
Fig.\ref{part2}, where we have considered again the same
quantities but looking at $64$ randomly-chosen bipartitions of the ground state for
$10$ different instances of $16$ qubits. According to these
results we restrict ourselves in what follows to the analysis of a
particular bipartition of the system, namely the first $n/2$
qubits versus the rest.

\begin{figure}
\centering
\includegraphics[angle=-90, width=0.7\textwidth]{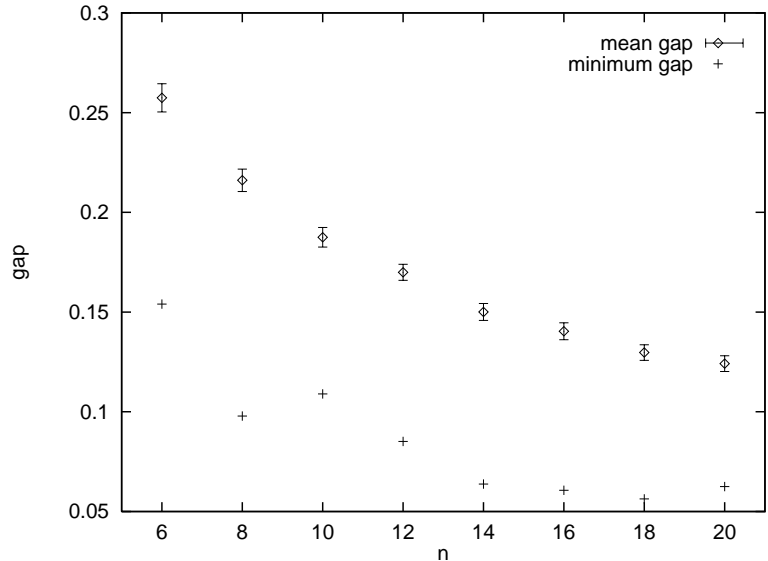}
\caption{Scaling of the minimum energy gap (in dimensionless units) with the size of the system,
both in the worst case and in the mean case over all the randomly
generated instances. Error bars give $95$ per cent of confidence
level for the mean.} \label{gap}
\end{figure}
\begin{figure}
\centering
\includegraphics[angle=-90, width=0.7\textwidth]{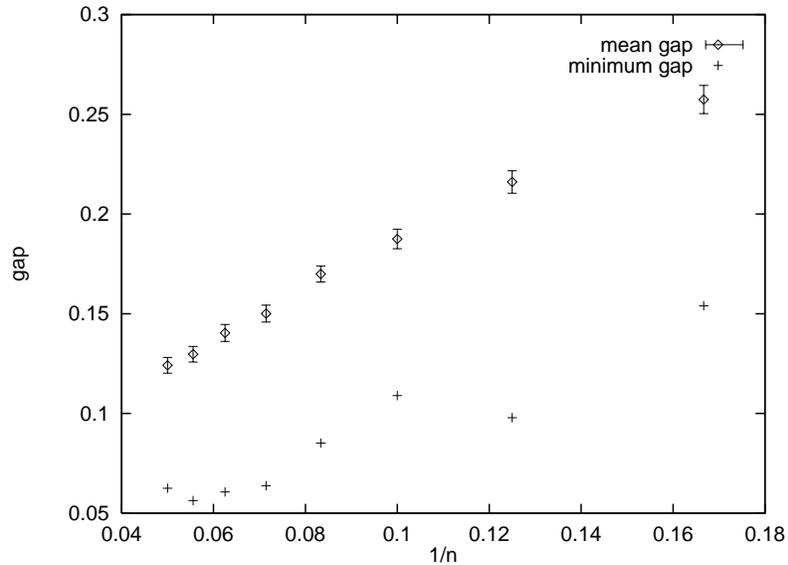}
\caption{Minimum energy gap (in dimensionless units) versus the inverse size of the system, both
in the worst case and in the mean case over all the randomly
generated instances. Error bars give $95$ per cent of confidence
level for the mean. The behavior of the mean is apparently linear. }
\label{gap2}
\end{figure}

\begin{figure}
\centering
\includegraphics[angle=-90, width=0.7\textwidth]{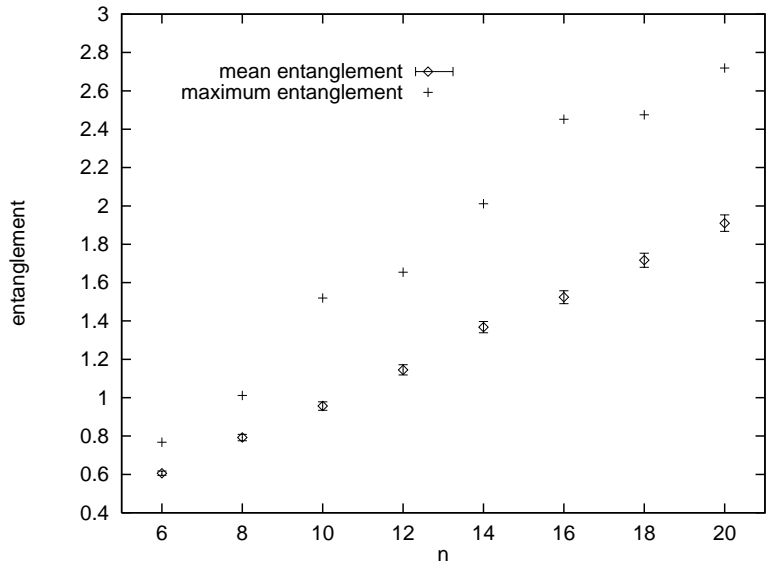}
\caption{Scaling of the entanglement entropy for an equally sized
bipartition of the system, both in the worst case and in the mean
case over all the randomly generated instances. Error bars give
$95$ per cent of confidence level for the mean. The data are
consistent with a linear scaling. } \label{ent}
\end{figure}

\subsubsection{Scaling laws for the minimum energy gap and the entanglement
entropy}

To characterize the finite-size
behavior of the quantum phase transition, we have generated 300
random instances of Exact Cover with only one satisfying
assignment from 6 to 20 qubits, and studied the maximum von
Neumann entropy for a bipartition of the system as well as the
minimum gap, both in the worst case and in the mean case over all
the randomly generated instances. We must point out that the scaling
laws found in this section are limited to the small systems we can handle with
our computers in an exact way. Increasing the number of qubits may lead to corrections in
the numerical results, which should be of particular importance for a more precise
time-complexity analysis of the adiabatic algorithm.   
 Fig.\ref{gap} represents the
behavior of the gap in the worst and mean cases. From
Fig.\ref{gap2} we observe that the gap seems to obey a scaling
law of the type $O(1/n)$, $n$ denoting the number of qubits, which
would guarantee a polynomial-time quantum computation. This law is in
agreement with the results in \cite{Farhi01}, and are in concordance with 
the idea that the energy gap typically vanishes as the inverse of the volume
in condensed matter systems (here the volume is the number of qubits). 
Error bars in the two plots give $95$ per cent of confidence level in the numerically
calculated mean.

We have also considered the scaling behavior of the
entanglement entropy for an equally sized bipartition of the
system, again both in the worst and in the mean case. The obtained data
from our simulations are plotted in Fig.\ref{ent} -- where error
bars give $95$ per cent of confidence level in the mean -- and seem
to be in agreement with a linear scaling of entanglement
as a function of the size of the number of qubits. More
concretely, a numerical linear fit for the mean entanglement
entropy gives us the law $S \sim 0.1  n$. Observe that the entropy
of entanglement does not  saturate at its maximum allowed
value (which would be $S_{{\rm max}} = n/2$ for $n$ qubits), so we can say
that only twenty percent of all the possible potential available
entanglement appears in the quantum algorithm. Linearity in the
scaling law would imply that this quantum computation by adiabatic
evolution, after a suitable discretization of the continuous time
dependence, could not be classically simulated by the protocol of
\cite{GVidal03_1}. Given that the scaling of the gap seems to
indicate that the quantum computation runs in a polynomial time in
the size of the system, our conclusion is that apparently we are
in front of an exponentially fast quantum computation that seems
extremely difficult (if not impossible) to be efficiently
simulated by classical means. This could be an inherent quantum
mechanical exponential speedup that can be understood in terms of
the linear scaling of the entropy of entanglement. Note also the
parallelism with the behavior of the entanglement found in Shor's
algorithm in the previous section. As a remark, our numerical analysis shows that the quantum
algorithm is difficult to simulate classically in an efficient way, which does not
necessarily imply that the quantum computer runs exponentially faster than the
classical one, as our time-complexity analysis is limited to 20 qubits.

The linear behavior for the entropy with respect to the size of
the system could in principle be expected according to the
following qualitative reasoning. Naively, the entropy was expected
to scale roughly as the area of the boundary of the splitting. This area-law is in some 
sense natural: since the entropy 
value is the same for both density matrices arising from the two subsystems, it
can only be a function of their shared properties, and these are geometrically
encoded in the area of the common boundary. For a system of $n$
qubits, we observe again that this implies a scaling law for the entropy of
an exact bipartition like $S \sim
n^{\frac{d-1}{d}}$ (which reduces to a logarithm for $d = 1$). Our
system does not have a well defined dimensionality, but owing to
the fact that there are many random two and three-body
interactions, the effective dimensionality of the system
should be very large. Therefore, we expect a linear (or almost
linear) scaling, which is what we numerically obtained. While this reasoning is not valid for critical fermionic systems, it differs only by at most a logarithmic multiplicative correction which we did not see in our computations. The
data seems to indicate that such an effective dimensionality is
around $d \sim n$, thus diverging as $n$ goes to infinity.

It is possible to compare our seemingly linear scaling of
the mean entropy of entanglement with the known results obtained by averaging
this quantity over the entire manifold of $n$-qubit pure states, with respect
to the natural Fubini-Study measure. According to the results conjectured in
\cite{Page93} and later proved in \cite{Sen96}, the average entropy for an
equally-sized bipartition of a random $n$-qubit pure state in the large $n$
limit can be approximated by $S \sim (n/2) - 1/(2 \ \ln{2})$ (in our
notation), therefore displaying as well a linear scaling law (but different
from ours). In fact, this is an indicator that most of the $n$-qubit pure
states are highly entangled, and that adiabatic quantum computation naturally
brings the system close to these highly entangled regions of the pure state
manifold.

\begin{figure}
\centering
\includegraphics[angle=-90, width=0.7\textwidth]{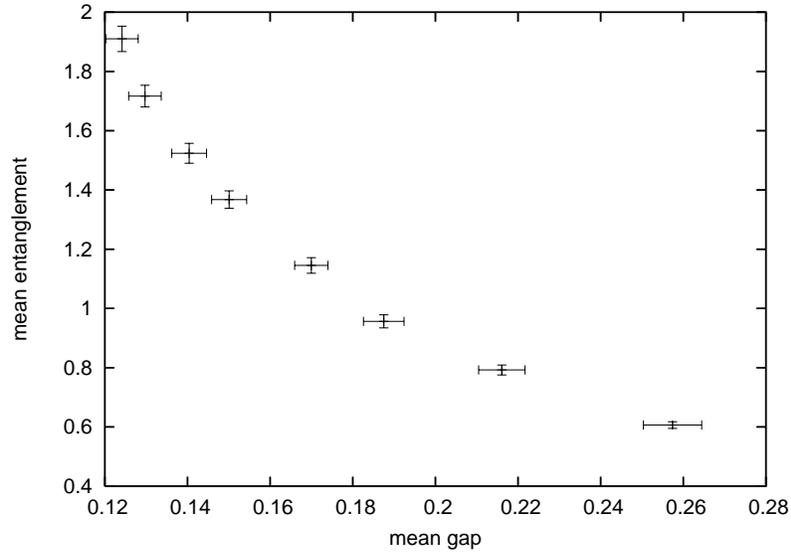}
\caption{Mean entropy of entanglement versus mean size of the energy gap (in
  dimensionless units).
Error bars give $95$ per cent of confidence level for the means.
Each point corresponds to a fixed number of qubits. }
\label{entgap1}
\end{figure}
\begin{figure}
\centering
\includegraphics[angle=-90, width=0.7\textwidth]{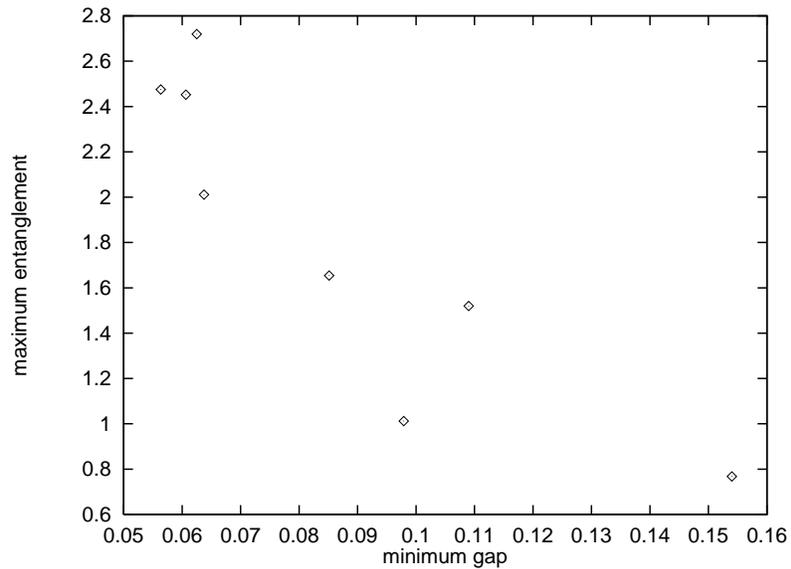}
\caption{Maximum entropy of entanglement versus minimum size of
the energy gap (in dimensionless units). Each point corresponds to a fixed number of qubits.}
\label{entgap2}
\end{figure}

\subsubsection{The entanglement-gap plane}

The plots in
Fig.\ref{entgap1} and Fig.\ref{entgap2} show the behavior of the peak
in the entanglement versus the gap, both again in the average and
the worst case for all the generated instances. Clearly, as the
gap becomes smaller the production of entanglement in the
algorithm increases. A compression of the energy levels correlates with
high quantum correlations in the system.

\subsubsection{Convergence of the critical points}

\begin{figure}
\centering
\includegraphics[angle=-90, width=0.7\textwidth]{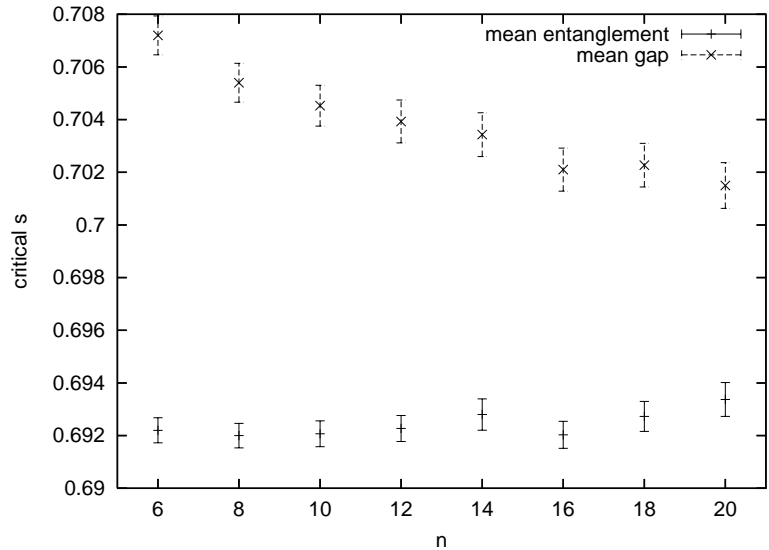}
\caption{Mean critical point for the energy gap and for the entropy.
Error bars give $95$ per cent of confidence level for the means.
Note that they tend to approach as the size of the system is
increased.} \label{move}
\end{figure}

The critical point $s_c$ seems to be bounded by the values of $s$
associated with the minimum gap
and the maximum entropy.
Actually, the  value of the critical point
corresponding to the minimum size of the energy gap is
systematically slightly bigger than the value of the critical point
corresponding to the peak in the entropy. By increasing the size
of the system these two points  converge towards the same
value, which would correspond to the true critical point of a
system of infinite size. This effect is neatly observed in
Fig.\ref{move}, which displays the values of $s$ associated
with the mean critical
points both for the gap and for the entropy as a function of $n$.

\subsubsection{Universality}

The above results suggest that the system comes
close to  a quantum phase transition.
The characterization we have presented based on the study of
averages over instances reconstructs its universal behavior. Results do not
depend on particular microscopic details of the Hamiltonian, such
as the interactions shared by the spins or the strength of local
magnetic fields. Any adiabatic algorithm solving a $k$-sat problem
and built in the same way we have done for Exact Cover should
display on average exactly the same properties we have found
\emph{regardless of the value of $k$}, which follows from
universality ($k=1$ is a particular case, as its
Hamiltonian is non-interacting). Linear scaling of entanglement should therefore be a
universal law for this kind of quantum algorithms. The
specific coefficients of the scaling law for the entropy should be a function
only of the connectivity of the system, that is on the type
of clauses defining the instances.

\begin{figure}
\centering
\includegraphics[angle=-90, width=0.7\textwidth]{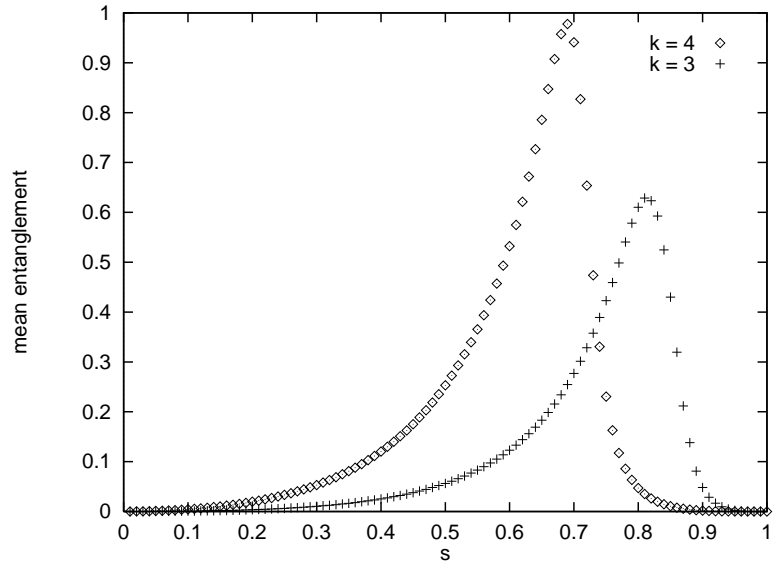}
\caption{Entanglement as a function of the Hamiltonian parameter
for clauses of Exact Cover involving $3$ ($k = 3$) and $4$ ($k=4$)
qubits, for a $10$-qubit system, averaged over all the randomly generated
instances.} 
\label{univ1}
\end{figure}

We have explicitly checked this assertion by numerical simulations
for clauses of Exact Cover but involving $4$ qubits ($x_i + x_j +
x_k + x_l = 1$), which is a particular case of $4$-SAT. In
Fig.\ref{univ1} we plot the behavior of the entropy of
entanglement for a $10$-qubit system for these type of clauses and
compare it to the same quantity calculated previously for the
clauses involving $3$ qubits (the common Exact Cover Hamiltonian).
We observe again the appearance of a peak in the entropy, which
means that the system is evolving close to a quantum phase
transition.

\begin{figure}
\centering
\includegraphics[angle=-90, width=0.7\textwidth]{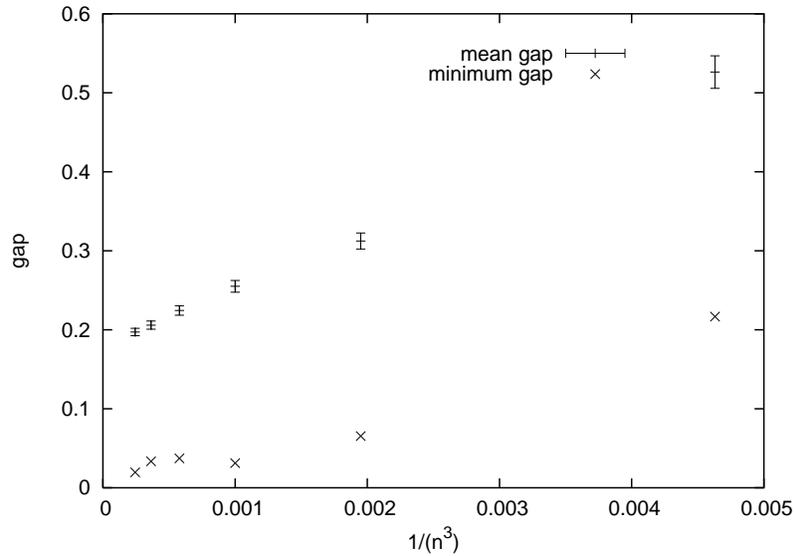}
\caption{Minimum energy gap (in dimensionless units) versus
 $1/(n^3)$,
both in the worst and in the mean cases over all the randomly
generated instances of clauses involving $4$ qubits, up to
$n = 16$. Error bars give $95$ per cent of confidence
level for the mean. The behavior seems to be linear.}
\label{univ3}
\end{figure}
\begin{figure}
\centering
\includegraphics[angle=-90, width=0.7\textwidth]{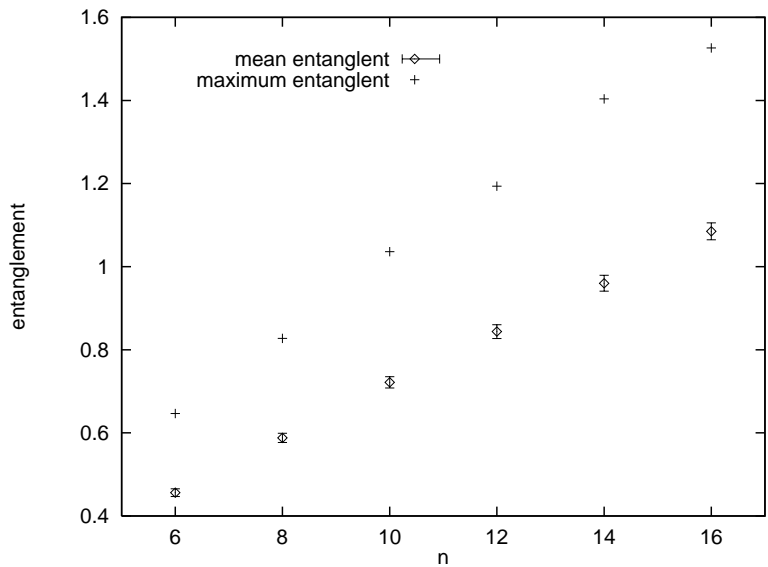}
\caption{Scaling of the entanglement entropy for an equally sized
bipartition of the system, both in the worst and in the mean
cases over all the randomly generated instances of clauses
involving $4$ qubits, up to $n = 16$. Error bars give
$95$ per cent of confidence level for the mean. The data are
consistent with a linear scaling. } \label{univ2}
\end{figure}

Fig.\ref{univ3} and Fig.\ref{univ2} respectively show the scaling
of the energy gap in the mean and worst case and the scaling of
the peak in the entropy in the mean and worst case as well, up to
$16$ qubits. Error bars give again $95$ per cent of confidence
level for the means. The behavior is similar to the one already
found for the instances of Exact Cover involving $3$ qubits
(Fig.\ref{gap2} and Fig.\ref{ent}), which supports the idea of the
universality of the results. The minimum energy gap seems to scale
in this case as $\sim \frac{1}{n^3}$ ($n$ being the number of
qubits), which would guarantee again a polynomial-time quantum
adiabatic evolution.

\section{Entanglement in adiabatic quantum searching algorithms}

Grover's quantum algorithm solves the problem of finding a ``needle in a haystack", which is mathematically defined as finding a specific element of an unsorted database by means of calls to an oracular function. If the database is composed of $2^n$ elements, $n$ being the number of bits, then the best classical algorithm for solving this problem takes $O(2^n)$ time as measured in calls to the oracle, whereas Grover's quantum algorithm takes only $O(2^{n/2})$ calls to the quantum implementation of the oracular function \cite{Grover96}. Optimality of Grover's quantum algorithm has been proven as well \cite{BBBV97}.  

Let us now consider the adiabatic implementation of Grover's quantum
searching algorithm in terms of a Hamiltonian evolution
\cite{Grover96, Roland02, Dam01} and study its properties as a function
of the number of qubits and the parameter $s$. For this problem, it is
possible to
compute all the results analytically, so we shall get a closed
expression for the scaling of entanglement. As a side remark,
 it is worth noting
that the treatment made in
\cite{GVidal03_1} is not valid for the oracular model of quantum computation,
 as it is assumed that all quantum gates are known in advanced. Independently of
this issue, we shall see that the system remains weakly entangled
between calls to the oracle.

\subsection{Adiabatic quantum search}

Grover's searching algorithm \cite{Grover96} can be implemented in
adiabatic quantum computation by means of the $s$-dependent
Hamiltonian

\begin{equation}
H(s) = (1-s)(I - |\psi \rangle \langle \psi |) + s (I-|x_0\rangle \langle
x_0|) \ , \label{grahm}
\end{equation}
where $|\psi \rangle \equiv \frac{1}{2^{n/2}} \sum_{x = 0}^{2^n-1}
|x\rangle$, $n$ is the number of qubits, and $|x_0\rangle$ is the
marked state. The computation takes the quantum state from an
equal superposition of all  computational states
directly to the state $|x_0\rangle$, as long as the evolution
remains adiabatic. The time the algorithm takes to succeed depends
on how we choose the parameterization of $s$ in terms
of time. Our aim here is to compute the amount of entanglement 
present in the register and need not deal with the explicit
dependence of the parameter $s$ on 
time and its consequences (see \cite{Roland02, Dam01}
for further information about this topic).

It is straightforward to check that the Hamiltonian from Eq.\ref{grahm}
has its minimum gap between the ground and first excited states at
$s = 0.5$, which goes to zero exponentially fast as the number of
qubits in the system is increased. Therefore, this Hamiltonian
apparently seems to undergo a quantum phase transition in the
limit of infinite size at $s=0.5$. 
Quantum correlations approach their  maximum
for this value of $s$.

\subsection{Analytical results}

It can be seen (see for example \cite{Das03}) that the ground state
energy of the Hamiltonian given in Eq.\ref{grahm}
corresponds to the expression

\begin{equation}
E_-(s) = \frac{1}{2} \left(1-\sqrt{(1-2s)^2
+\frac{4}{2^n}s(1-s)}\right) \ , \label{ground}
\end{equation}
$s$ denoting the Hamiltonian parameter. The corresponding
normalized ground state eigenvector is given by

\begin{equation}
|E_-(s)\rangle = a |x_0\rangle + b \sum_{x \neq x_0} |x\rangle \ ,
\label{groundstate}
\end{equation}
where we have defined the quantities

\begin{eqnarray}
a &\equiv& \alpha \ b \nonumber \\
&& \nonumber \\
b^2 &\equiv& \frac{1}{2^n - 1 + \alpha^2} \nonumber \\
&& \nonumber \\
\alpha &\equiv&
\frac{2^n-1}{2^n-1-\left(\frac{2^n}{1-s}\right)E_-(s)} \ .
\label{expressions}
\end{eqnarray}

In all the forthcoming analysis we will assume that the marked
state corresponds to $|x_0\rangle = |0\rangle$, which will not
alter our results. The corresponding density matrix for the ground
state of the whole system of $n$ qubits is then given by

\begin{equation}
\rho_n = b^2(\alpha^2-2\alpha +1)|0\rangle\langle 0| + b^2
|\phi\rangle \langle \phi| + b^2(\alpha-1)(|\phi\rangle\langle 0|
+ |0\rangle \langle \phi|) \ , \label{density}
\end{equation}
where we have defined $|\phi\rangle$ as the the unnormalized sum
of all the computational quantum states (including the marked
one), $|\phi \rangle \equiv \sum_{x = 0}^{2^n-1} |x\rangle$.
Taking the partial trace over half of the qubits, regardless of
what $n/2$ qubits we choose, we find the reduced density matrix

\begin{equation}
\rho_{n/2} = b^2(\alpha^2 - 2 \alpha + 1)|0' \rangle \langle 0'| +
2^{n/2} b^2 |\phi'\rangle \langle \phi'| + b^2 (\alpha - 1)
(|\phi'\rangle \langle 0'| + |0'\rangle \langle \phi'|) \ ,
\label{densityn2}
\end{equation}
where we understand that $|0'\rangle$ is the remaining marked
state for the subsystem of $n/2$ qubits and $|\phi'\rangle \equiv
\sum_{x=0}^{2^{n/2}-1} |x\rangle$ is the remaining unnormalized
equally superposition of all the possible computational states for
the subsystem. Defining the quantities

\begin{eqnarray}
A &\equiv& \frac{\alpha^2 + 2^{n/2} - 1}{\alpha^2 + 2^n - 1}
\nonumber \\
&& \nonumber \\
B &\equiv& \frac{\alpha + 2^{n/2} -1}{\alpha^2 + 2^n - 1}
\nonumber \\
&& \nonumber \\
C &\equiv& \frac{2^{n/2}}{\alpha^2 + 2^n - 1} \label{definitions}
\end{eqnarray}
(note that $A + (2^{n/2} - 1) C = 1$), the density operator for
the reduced system of $n/2$ qubits can be expressed in matrix
notation as

\begin{equation}
\rho_{n/2} =
\begin{pmatrix}
  A & B & \cdots & B \\
  B & C & \cdots & C \\
  \vdots & \vdots & \ddots & \vdots \\
  B & C & \cdots & C
\end{pmatrix} 
\label{matrix}
\end{equation}
in the computational basis, where its dimensions are $2^{n/2} \times 2^{n/2}$. We clearly see
that the density matrix has a rank equal to $2$. Therefore, because
${\rm rank}(\rho) \ge 2^{S(\rho)} \ \forall \rho$ (where $S(\rho)$
is the von Neumann entropy of the density matrix $\rho$) we
conclude that $S(\rho_{n/2})$, which corresponds to our
entanglement measure between the two blocks of qubits, is always
$\le 1$. This holds true even for non symmetric bipartitions of
the complete system. Regardless of the number of qubits,
entanglement in Grover's adiabatic algorithm is always a
\emph{bounded} quantity for any $s$, in contrast with the results
obtained in the previous sections for Shor's factoring algorithm
and for the Exact Cover problem. Grover's adiabatic quantum
algorithm essentially makes use of very little entanglement between calls to the quantum oracle,
 but even this bounded quantity of quantum correlations is enough to give a
square-root speedup.

We have explicitly calculated the von Neumann entropy for
$\rho_{n/2}$. Because the rank of the reduced density matrix is
two, there are only two non-vanishing eigenvalues that contribute
in the calculation which are

\begin{equation}
\lambda_{\pm} = \frac{1}{2}\left(1 \pm \sqrt{1 - 4 (2^{n/2}-1) (A
C - B^2)}\right) \ . \label{ei}
\end{equation}
We analyze the limit $n \rightarrow \infty$ for $s \ne 0.5$ and $s
= 0.5$ separately. 

\subsubsection{Entropy at $s \ne 0.5$}

In the limit of very high $n$ we can approximate the ground state
energy given in Eq.\ref{ground} by

\begin{equation}
E_-(s) \sim \frac{1}{2}\left(1-\sqrt{1-4 s (1-s)}\right) \ .
\label{approx}
\end{equation}
Therefore, the quantity

\begin{equation}
\alpha \sim \frac{1}{1 - \left(\frac{E_-(s)}{1-s}\right)}
\label{apal}
\end{equation}
diverges at $s=0.5$, which implies that this limit can not be
correct for that value of the parameter. 
 The closer we are to $s = 0.5$, the
bigger is $\alpha$. In this limit we find that

\begin{eqnarray}
A &\sim& \frac{\alpha^2 + 2^{n/2}}{\alpha^2 + 2^n}
\nonumber \\
&& \\
B &\sim& \frac{\alpha + 2^{n/2}}{\alpha^2 + 2^n}
\nonumber \\
&& \\
C &\sim& \frac{2^{n/2}}{\alpha^2 + 2^n} \ , \label{limit}
\end{eqnarray}
where all these quantities tend to zero as $n \rightarrow \infty$.
It is important to note that the convergence of the limit depends
on the value of $\alpha$ or, in other words, how close to $s=0.5$
we are. The closer we are to $s=0.5$, the slower is the
convergence, and therefore any quantity depending on these
parameters (such as the entropy) will converge slower to its
asymptotical value. For the eigenvalues of the reduced density
matrix we then find that when $n \rightarrow \infty$

\begin{equation}
\lambda_{\pm} \rightarrow \frac{1}{2} (1 \pm 1) \ , \label{eigenv}
\end{equation}
so $\lambda_+ \sim 1$ and $\lambda_- \sim 0$, and therefore the
asymptotical entropy is

\begin{equation}
S(s\ne 0.5, n\rightarrow \infty) = -\lambda_+ \log_2\lambda_+ -
\lambda_- \log_2\lambda_- = 0 \ . \label{entropia}
\end{equation}
The convergence of this quantity is slower as we move towards $s =
0.5$.

\subsubsection{Entropy at $s=0.5$}

We begin our analysis by evaluating the quantities at $s = 0.5$
and then taking the limit of big size of the system. We have that
$\alpha (s = 0.5) = \frac{2^n - 1}{2^{n/2}-1} \sim 2^{n/2}$. From
here it is easy to get the approximations

\begin{eqnarray}
A &\sim& \frac{1}{2} \nonumber \\
&& \nonumber \\
B &\sim& \frac{1}{2^{n/2}} \nonumber \\
&& \nonumber \\
C &\sim& \frac{1}{2^{n/2 + 1}} \ , \label{abc}
\end{eqnarray}
and therefore

\begin{equation}
\lambda_{\pm} \sim \frac{1}{2}\left(1 \pm \sqrt{1 - 4 \ 2^{n/2}
\left(\frac{1}{4} \frac{1}{2^{n/2}} - \frac{1}{2^n}\right)}\right)
= \frac{1}{2} \pm \frac{1}{2^{n/4}} \ , \label{eigensdos}
\end{equation}
so $\lambda_{\pm} \rightarrow \frac{1}{2}$ and $S(s=0.5, n
\rightarrow \infty) = 1$. According to Eq.\ref{eigensdos} we can
evaluate the finite size corrections to this behavior and find the
scaling of the entropy with the size of the system for very large
$n$. The final result for the entropy at the critical point reads

\begin{equation}
S(s=0.5, n \gg 1 ) \sim 1 - \frac{4}{\ln{2}} 2^{-n/2} \ .
\label{scaling}
\end{equation}
Note that the entropy remains bounded and 
tends to $1$ for $s = 0.5$ as a square root in the
exponential of the size of the system, which is the typical factor
in Grover's quantum algorithm.

\begin{figure}
\centering
\includegraphics[angle=-90, width=0.7\textwidth]{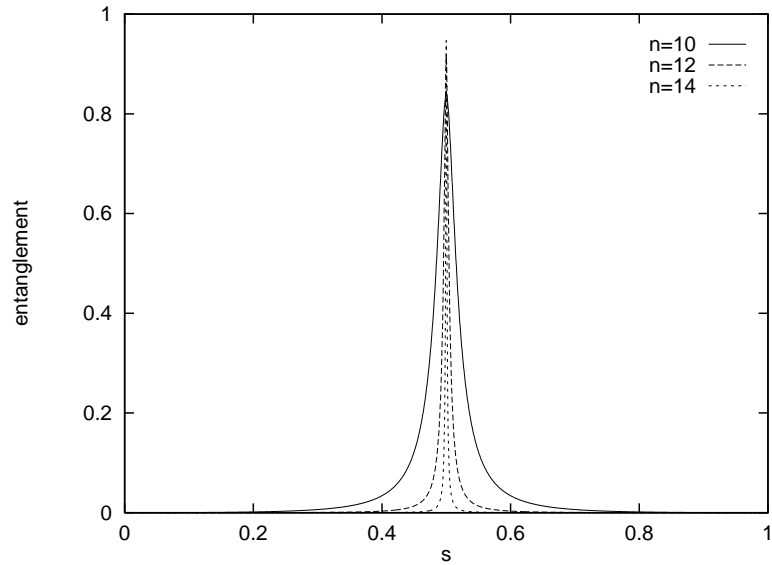}
\caption{Von Neumann entropy for the reduced system as a function
of $s$ for $10$, $12$ and $14$ qubits. As the size of the system
increases the entropy tends to zero at all points, except at $s =
0.5$ in which tends to $1$.} \label{pic}
\end{figure}
\begin{figure}
\centering
\includegraphics[angle=-90, width=0.7\textwidth]{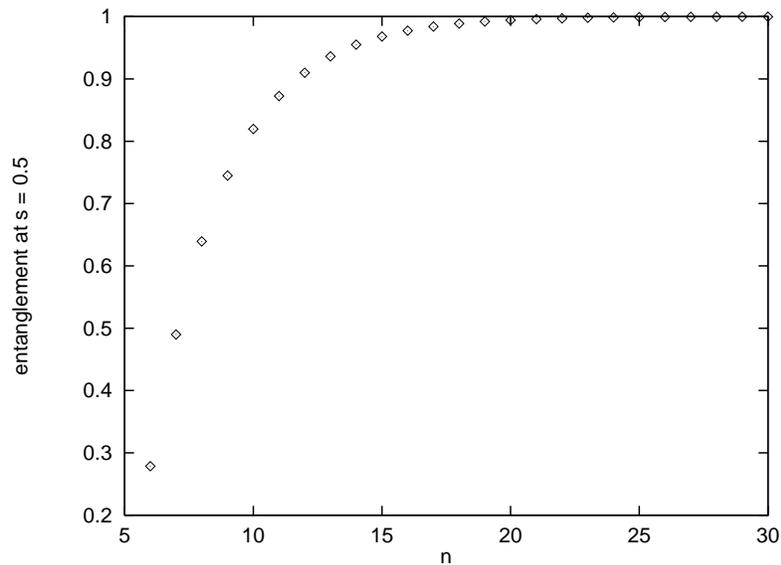}
\caption{Von Neumann entropy for the reduced system at $s = 0.5$
as a function of $n$. For infinite size of the system there is a
saturation at $1$.} \label{scale}
\end{figure}

We have represented the evolution of the entanglement entropy as a
function of $s$ for different sizes of the system in Fig.\ref{pic}
and have plotted in Fig.\ref{scale} the maximum value of the
entropy along the computation as a function of the size of the
system according to the expression given in Eq.\ref{scaling}. We can now compare the two plots with Fig.\ref{mix-300} and
Fig.\ref{ent} in the previous section.  The behavior for 
the entropy in
Grover's adiabatic algorithm is dramatically different to the one
observed in the NP-complete problem. 
Entanglement gets saturated in Grover's adiabatic
algorithm \emph{even at the point at which the gap vanishes}, 
which reminds us of short ranged quantum correlations
in non-critical quantum spin chains\footnote{A somehow
similar situation is present in $(1+1)$-dimensional quantum spin
chains outside of the critical region, where the entanglement
entropy also reaches a saturation when increasing the size of the
system \cite{Latorre03, Latorre04_1, Rico2005}. 
Saturation does not appear in higher dimensional systems.}. 

Let us note that, in the limit of
infinite size, the quantum state in Grover's algorithm is separable
with respect to any bipartition of the system (and therefore not
entangled, as it is a pure state) for any $s$ except for $s =
0.5$. All the entanglement along the algorithm is concentrated at this
point, but this entanglement is still a bounded quantity and
actually equal to $1$. Consequently, a small amount of
entanglement appears essentially only at one point when the size
of the system is big, whereas the rest of the algorithm
needs to handle just separable states. We point out that these results
apply as well to the traditional discrete-time implementation of Grover's searching
algorithm, as the states between iterations are the same as in the adiabatic
version for discrete $s$ values. 

\section{Conclusions of Chapter 4}

In this Chapter we have studied the scaling of the entanglement
entropy in several quantum algorithms. In order to be precise:

\begin{itemize}
\item{We have analytically proven that Shor's factoring quantum algorithm makes use of an
exponentially large amount of entanglement in the size of the system between the target
register and the source register after the modular exponentiation
operation, which in turn implies the impossibility of an efficient
classical simulation by means of the protocol of Vidal in \cite{GVidal03_1}.}
\item{We have provided numerical
evidence for a universal linear scaling of the entropy with the
size of the system together with a polynomially small gap in a
quantum algorithm by adiabatic evolution devised to solve the
NP-complete Exact Cover problem, therefore obtaining a
polynomial-time quantum algorithm which would involve exponential
resources if simulated classically, in analogy to Shor's algorithm.
Universality of this result follows from the fact that the quantum
adiabatic algorithm evolves close to a quantum phase transition and
the properties at the critical region do not depend on
particular details of the microscopic Hamiltonian (instance) such as
interactions among the spins or local magnetic fields.}
\item{We have also proven that the von Neumann entropy remains bounded by $1$ between calls to the quantum oracle in Grover's adiabatic algorithm regardless of the size of
the system and even at the critical point. More concretely, the
maximum entropy approaches one as a square root in the size of
the system, which is the typical Grover's scaling factor.}
\end{itemize}

Our results show that studying the scaling of the entropy is a
useful way of analyzing entanglement production in quantum
computers. Results from the study of quantum many-body systems 
 can be directly applied to bring further
insight into the analysis of the quantum correlations present in a quantum computer. Different
entanglement scaling laws follow from different situations
according to the amount of correlations involved, as can be seen
in Table \ref{scalinglaws}. A quantum algorithm can be understood
as the simulation of a system evolving close to a 
quantum phase transition. The amount of entanglement
involved depends on the effective dimensionality of the system,
which in turn governs the possibilities of certain efficient
classical simulation protocols.
\begin{table}[h]
\begin{center}
\begin{turn}{90}
$\longleftarrow$ less entanglement
\end{turn}
\begin{tabular}{|c||c|}
    \hline
     & \\
   Problem & Scaling of the entanglement entropy  \\
     & \\ \hline \hline
     & \\ 
   Adiabatic Exact Cover's quantum algorithm  & $S = O(n)$  \\
     & \\
   Shor's quantum factoring algorithm & $S = O(\log_2  r) \sim O(n)$  \\
     & \\
   Critical $(d+1)$-dimensional fermionic lattices & $S=O(n^{\frac{d-1}{d}} \log_2 n)$ \\
     & \\
   Critical $(d+1)$-dimensional bosonic lattices & $S = O(n^{\frac{d-1}{d}})$  \\
     & \\
   Critical $(1+1)$-dimensional spin chains & $S = O(\log_2 n)$  \\
     & \\
   Non-critical $(1+1)$-dimensional spin chains & $S = O(1)$ \\
     & \\
   Adiabatic Grover's quantum algorithm & $S = O(1)$ \\
     & \\
    \hline
\end{tabular}
\end{center}
  \caption{Entanglement scaling laws in different problems, 
in decreasing complexity order.}
  \label{scalinglaws}
\end{table}

These scaling laws provide also a new way of understanding some
 aspects from one-way quantum computation. It is known that
the so-called cluster state of the one-way quantum computer can be
generated by using Ising-like interactions on a planar
$(2+1)$-dimensional lattice \cite{Raus01, Brieg01, Raus02}.
This fact can be related to the at least linear (in the size of a box)
behavior of the entropy for
spin systems in $(2+1)$ dimensions.
$(1+1)$-dimensional models seem not to be able to efficiently create
the highly-entangled cluster state. Again,
this fact can be traced to the logarithmic scaling law 
of the entropy in spin chains which is insufficient
to handle the large amount of entanglement to carry out for instance 
Shor's algorithm. Note also that $(d+1)$-dimensional
systems with $d \ge 3$ bring unnecessarily large entanglement.

Quantum phase transitions stand as demanding systems
in terms of entanglement. They are very hard to simulate classically.
It is then reasonable to try to bring NP-complete problems to
a quantum phase transition setup, which quantum mechanics
handles naturally.  

    \chapter{Classical simulation of quantum algorithms using matrix product
  states}

In Chapter 4 we saw that understanding the detailed behavior and properties of quantum many-body systems plays a role in different areas of physics. Those systems whose properties can be analytically found are typically called \emph{integrable} systems and  
offer a way to study, for instance, the low-energy sector of different models. It is a pity, though, that many of the models that we know
are not integrable, in the sense that it is not even known whether it is possible or not to study in an exact way their fundamental properties. The realistic alternative is, then, to use different techniques based on 
 numerical simulations by means of computer programs, so that we can get a detailed understanding of the system.   

While it is possible to numerically study the low-energy properties of any model by means of an exact diagonalization of the quantum Hamiltonian or related techniques, such a possibility is always limited to a relatively small number of particles due to the exponential growth in the size of the Hilbert space. Indeed, this is at the heart of the motivation to build a quantum computer, as originally proposed by Feynman \cite{Feynman82}. Using standard present technology, a faithful numerical study of the ground-state properties of a general quantum Hamiltonian can be achieved for systems up to the order of 20 spins, as we did in the previous Chapter. Luckily enough, other numerical techniques are possible. For instance, quantum Montecarlo algorithms have provided good results for some systems while they fail for some others due to the presence of the so-called sign problem \cite{Hirsch85}. Another example of successful numerical technique has been the density matrix renormalization group (DMRG) algorithm, as introduced by White in \cite{White92}. While it was soon realized that DMRG produced extremely accurate results when  computing the ground-state energy of quantum systems in one spatial dimension, it was also realized that the method did not work so well when applied to higher dimensional systems \cite{Noack94, Liang94}. Even in the $(1+1)$-dimensional case, there was a difference in the performance of the algorithm between open and periodic boundary conditions, and between non-critical and critical systems, the former being the more successful in both cases. Nevertheless, DMRG has been the algorithm of reference for computing the low-energy properties of quantum models with one spatial dimension during the last decade. 

After the appearance of DMRG, a notorious result was found by Ostlund and Rommer in \cite{Ostlund95}, where they showed that the original DMRG algorithm can be completely understood in terms 
of the so-called matrix product states. Originally introduced in the valence-bond model of Affleck, Kennedy, Lieb and Tasaki \cite{AKLT87, AKLT88}, generalized by Fannes, Nachtergaele and Werner  \cite{Fannes92}, and rediscovered in the field of quantum information science by Vidal  \cite{GVidal03_1}, matrix product states have been proved to be an extremely useful tool in order to develop numerical techniques for computing the low-energy properties together with the dynamics of sufficiently local Hamiltonians in one spatial dimension \cite{GVidal04, MG04, Scholl05_2, VPC04, VGRC04, VW05, VCprb06, Scholl05}, and have inspired as well several numerical techniques to study higher-dimensional systems \cite{VC04,GVidal05,Anders06}. 

The natural question arises of whether matrix product states can
be applied to simulate the dynamics of a quantum computer. The content of
this Chapter is aimed to show that this is indeed possible and
that we can handle relatively large simulations with controlled 
accuracy. We call the parameter controlling the size of the matrices $\chi$, which was already introduced in Chapter 4, and which can in turn be related to the entanglement entropy $S$ of a considered bipartition of the system like $\chi \ge 2^{S(\rho)}$. As we shall see, the total time cost of the simulation scales polynomially in parameter $\chi$. Thus, 
we expect this approximation scheme
to fail whenever the inherently needed $\chi$ is $O(2^n)$, $n$ being the number of qubits of the quantum register. 
Nevertheless, it may be possible in some of these cases that by keeping only $\chi =
O({\rm poly}(n))$ in the simulation we already get a reasonable
approximation to the exact computation. This is indeed the case of the quantum algorithm that we consider here. We study the numerical performance of the classical simulation scheme for quantum computations originally proposed by Vidal in \cite{GVidal03_1} based on matrix product states, when applied to the simulation of an adiabatic quantum algorithm solving the Exact Cover NP-complete problem. The performance of this quantum algorithm was already addressed in Chapter 4, where we saw that the typical entanglement entropy of the system for a given bipartition tends to scale roughly as $S \sim 0.1 n$, which makes the parameter $\chi$ exponentially big in the number of qubits and thus forbids the possibility of an \emph{exact} classical simulation. Nevertheless, the fact that the coefficient in front of the scaling law of the entropy is small inspires us to think that, perhaps, it should still be possible to perform a relatively good \emph{approximated} classical simulation of the quantum algorithm by keeping a small amount of $\chi$ along the evolution. Notice that this is a necessary, while not sufficient condition to have a good approximation of the evolution of the quantum algorithm. Let us then proceed in what follows with an explanation of what matrix product states are and how do they inspire numerical simulation algorithms for time evolution, moving then to our explicit results on the numerical simulation of a quantum computer. 

\section{The matrix product state ansatz}

A matrix product state is a parameterization of a pure quantum state of $n$ local systems (like, for instance, qubits) in terms of the amount of bipartite entanglement present in the state. Here we derive this ansatz from two different perspectives: on the one hand, we show how matrix product states appear from the point of view inspired by Affleck, Kennedy, Lieb and Tasaki in \cite{AKLT87,AKLT88} based on projectors on some ancillary unphysical particles; on the other hand, we show how it is possible to obtain a matrix product state by means of a series of Schmidt decompositions of the quantum state at hand, in the way  done by Vidal in \cite{GVidal03_1}. These two perspectives complement each other, and give different insights about the significance of the different parameters and quantities that appear in the ansatz. 

\subsubsection{Derivation by means of projectors}

\begin{figure}[h]
\begin{center} 
  \includegraphics[width=1.0\linewidth]{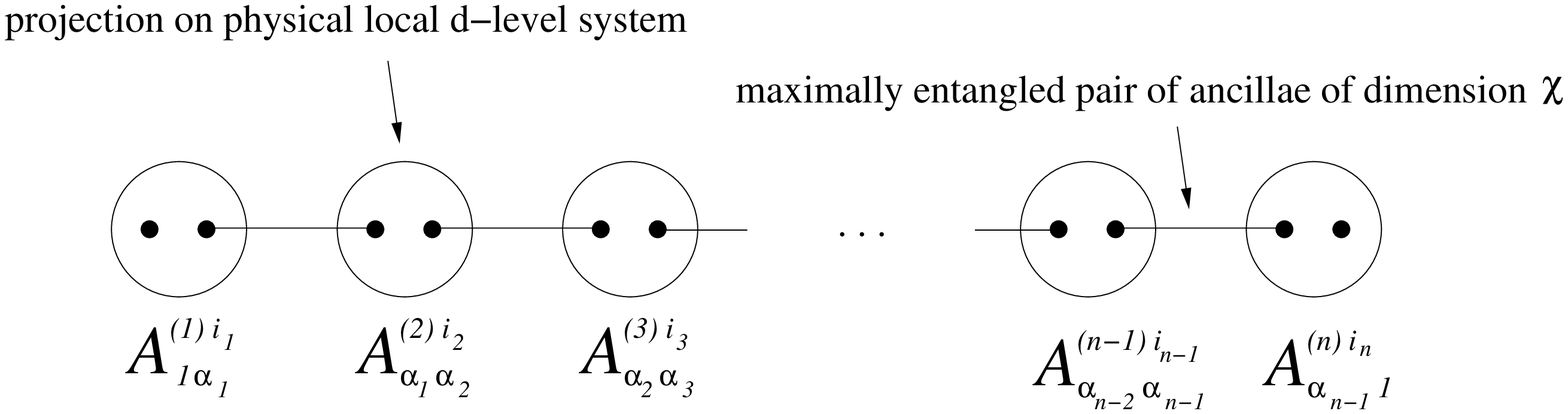}
  \caption{Graphical representation of a matrix product state in terms of projections. The projectors act on a dimerized state of maximally entangled pairs of dimension $\chi$.}
\end{center}
\label{projectors}
\end{figure}

Let us consider a set of $n$ physical local $d$-level systems, described by (pure) quantum state given by 

\begin{equation}
|\psi \rangle = \sum_{i_1=1}^d \sum_{i_2=1}^d \cdots \sum_{i_n=1}^d c_{i_1,i_2,\ldots,i_n}|i_1,i_2,\ldots,i_n\rangle \ ,
\label{estado}
\end{equation}
where the states $|i_l\rangle$, $l = 1, 2, \ldots, n$ denote a local $d$-level basis, and $c_{i_1,i_2,\ldots,i_n}$ are the corresponding $d^n$ coefficients specifying the state. We now consider the following picture. First, imagine that the local systems are placed on a linear chain. Second, let us represent the physical local $d$-level systems by means of two ancillary unphysical particles, each one of them being described by a Hilbert space of dimension $\chi$, together with a projector from the joint ancillary Hilbert space of dimension $\chi^2$ to the physical Hilbert space of dimension $d$. We also assume that the state of the ancillary particles (without the projectors) is in a dimerized state of maximally entangled pairs of dimension $\chi$. The projector on the local Hilbert space at site $l$ is represented by the three-index tensor 

\begin{equation}
A^{(l) i_l}_{\alpha_{l-1} \alpha_{l}} \ ,
\label{tensorrr}
\end{equation}

where the index $i_l = 1, 2, \ldots, d$ corresponds to the physical local Hilbert space, while the indexes $\alpha_{l-1} = 1, 2, \ldots, \chi $ and $\alpha_l = 1, 2, \ldots, \chi$ correspond to the two ancillary Hilbert spaces. This is represented in Fig.\ref{projectors}. 

\bigskip

At every site, and for each value of the physical index, we have then a matrix. Because the ancillary particles are in a dimerized state of maximally entangled pairs, the coefficients $c_{i_1,i_2,\ldots,i_n}$ of the system are decomposed as products of matrices, hence the name of matrix product state. The explicit form of the state is

 \begin{equation}
\label{statemps}
 \vert \psi\rangle =
\sum_{\{i\}}\sum_{\{\alpha \}}A^{(1)i_1}_{1\alpha_1}
A^{(2)i_2}_{\alpha_1\alpha_2}
\dots A^{(n)i_n}_{\alpha_{n-1}1}\vert i_1, i_2,\dots,i_n\rangle \ ,
 \end{equation}
where the sums are to be understood from now on over the complete range of the set of physical indices $\{i\}$ and ancillary indices $\{\alpha\}$. 

\subsubsection{Derivation by means of Schmidt decompositions}

Consider again the same set of $n$ physical local $d$-level systems described by the pure state of Eq.\ref{estado}, where we assume as before that the local systems are sorted from $1$ to $n$ in such a way that they can be thought as placed on a linear chain. Following Vidal \cite{GVidal03_1}, if we perform the Schmidt decomposition between the local system $1$ and the remaining $n-1$ we can write the state as

\begin{equation}
|\psi \rangle = \sum_{\alpha_1} \lambda^{(1)}_{\alpha_1} |\phi^{(1)}_{\alpha_1}\rangle |\phi^{(2 \cdots n)}_{\alpha_1}\rangle \ , 
\label{smdec}
\end{equation}
where $\lambda^{(1)}_{\alpha_1}$ are the Schmidt coefficients, $|\phi^{(1)}_{\alpha_1}\rangle$ and $|\phi^{(2 \cdots n)}_{\alpha_1}\rangle $ are the corresponding left and right Schmidt vectors, and $\alpha_1 = 1, 2, \ldots, d$. By expressing the left Schmidt vector in terms of the original local basis for system $1$ the state can then be written as

\begin{equation}
|\psi \rangle = \sum_{i_1, \alpha_1} \Gamma^{(1) i_1}_{1 \alpha_1} \lambda^{(1)}_{\alpha_1} |i_1\rangle |\phi^{(2 \cdots n)}_{\alpha_1}\rangle \ ,
\label{gamaequation}
\end{equation}
$\Gamma^{(1) i_1}_{1\alpha_1}$ being the appropriate coefficients of the change of basis, that is, $ |\phi^{(1)}_{\alpha_1}\rangle = \sum_{i_1}  \Gamma^{(1) i_1}_{1 \alpha_1} |i_1 \rangle $. At this point, we expand each Schmidt vector $ |\phi^{(2 \cdots n)}_{\alpha_1}\rangle$ in the original local basis for system $2$, that is, 

\begin{equation}
 |\phi^{(2 \cdots n)}_{\alpha_1}\rangle = \sum_{i_2} |i_2\rangle |\omega^{(3 \cdots n)}_{\alpha_1 i_2} \rangle \ . 
 \label{cosita}
 \end{equation}
 We now write the unnormalized quantum state  $|\omega^{(3 \cdots n)}_{\alpha_1 i_2} \rangle$  in terms of the at most $d^2$ eigenvectors of the joint reduced density matrix for systems $(3, 4, \ldots, n)$, that is, in terms of the right Schmidt vectors $|\phi^{(3 \cdots n)}_{\alpha_2} \rangle $ of the particular bipartition between the first two local systems and the rest, together with the corresponding Schmidt coefficients $\lambda^{(2)}_{\alpha_2}$: 

\begin{equation}
|\omega^{(3 \cdots n)}_{\alpha_1 i_2} \rangle = \sum_{\alpha_2} \Gamma^{(2) i_2}_{\alpha_1 \alpha_2} \lambda^{(2)}_{\alpha_2} |\phi^{(3 \cdots n)}_{\alpha_2} \rangle \ . 
\label{ecc}
\end{equation}
Replacing the last two expressions into Eq.\ref{gamaequation} we obtain 

\begin{equation}
|\psi \rangle = \sum_{i_1, \alpha_1, i_2, \alpha_2} \Gamma^{(1)i_1}_{1 \alpha_1} \lambda^{(1)}_{\alpha_1} \Gamma^{(2) i_2}_{\alpha_1 \alpha_2} \lambda^{(2)}_{\alpha_2} |i_1 i_2 \rangle |\phi^{(3 \cdots n)}_{\alpha_2} \rangle \ . 
\label{iuid}
\end{equation}
 
 By iterating the above procedure we finally get a representation of the quantum state in terms of some tensors $\Gamma$ and some vectors $\lambda$: 
 
  \begin{equation}
\label{gammalambda}
 |\psi\rangle =
\sum_{\{i\}}\sum_{\{\alpha \}}\Gamma^{(1)i_1}_{1\alpha_1} \lambda^{(1)}_{\alpha_1}
\Gamma^{(2)i_2}_{\alpha_1\alpha_2} \lambda^{(2)}_{\alpha_2}
\dots \lambda^{(n-1)}_{\alpha_{n-1}}\Gamma^{(n)i_n}_{\alpha_{n-1}1}\vert i_1, i_2,\dots,i_n\rangle \ .
 \end{equation} 
 
Several remarks are to be considered at this point. First, notice that the above decomposition immediately provides the Schmidt vectors $\lambda$ of all the possible contiguous bipartitions of the system. Second, the state from Eq.\ref{gammalambda} is indeed a reparametrization of a matrix product state of the form given in Eq.\ref{statemps} if we define the matrices at site $l$ in the following way: 

\begin{equation}
\label{agama}
A^{(l)i_l}_{\alpha_{l-1} \alpha_{l}} \equiv \Gamma^{(l) i_l}_{\alpha_{l-1} \alpha_l} \lambda^{(l)}_{\alpha_l} \ .
\end{equation}
Third, we see that the maximum allowed rank of the different indices $\alpha_l$, $l = 1, 2, \ldots, n-1$, is site-dependent, since the size of the Hilbert spaces considered when performing the consecutive  Schmidt decompositions depends on the site. In particular, we have that, at most,  $\alpha_l = 1, 2, \ldots, d^l$ for $l = 1, 2, \ldots, \lfloor n/2 \rfloor$, and $\alpha_l = 1, 2, \ldots, d^{(n-l)}$ for $l = \lfloor n/2 \rfloor + 1, \lfloor n/2 \rfloor + 2, \ldots, n-1$.  Actually, the fact that the maximum allowed range of the matrix indices is site-dependent can also be 
seen from Eq.\ref{statemps} by performing an appropriate set of concatenated singular value decompositions of the matrices defining the state. In practice, however, many of the Schmidt coefficients for the different contiguous bipartitions of the system shall be equal to (or almost equal to) zero depending on the particular state being considered. Let us then call $\chi(l,\mathcal{P})$ the local Schmidt rank for the bipartition between the $l $ and the $ l+1$ sides for a given permutation $\mathcal{P}$ of the particles. We shall now define $\chi$ as the maximum Schmidt rank over all the possible bipartitions of the system, that is 

\begin{equation}
\chi \equiv \underset{l,\mathcal{P}}{{\rm max}} \  \chi(l,\mathcal{P}) \ .
\label{chichichi}
\end{equation}
We immediately see from this definition that the parameter $\chi$ controlling the maximum possible size of the matrices in a matrix product state of $n$ particles is, indeed, a measure of the maximum bipartite entanglement that is present in the system. This representation is very appealing, since quantum states with low (bipartite) entanglement can then be represented by small matrices, while highly-entangled states must necessarily be described by matrices of large size, corresponding to the idea that the more entangled a system is, the harder it is to perform an exact classical description of it.

The above picture can be made specific by noticing that $\chi \ge d^{S}$, where $S$ is the entanglement entropy (measured in e-dits) corresponding to any possible bipartition of the system. The study of the scaling of the entanglement entropy can thus be translated into the study of the possibility or not of an efficient representation of the quantum state in terms of a matrix product state. To be more precise, matrix product states allow a representation of the state in terms of $O(nd\chi^2)$ parameters instead of the original $d^n$ coefficients. Therefore, those quantum states with $\chi = O({\rm poly}(n))$ can be efficiently classically represented by a matrix product state, while those where $\chi = O(2^n)$ cannot. In fact, the computation of the expected values of local observables can be done in $O(\chi^3)$ time, thus being efficient for systems with small enough $\chi$. This is an important property, since it means that the matrix product state representation is not only nice, but useful as well, in the sense that it allows to compute important physical quantities, like correlators, in an efficient way. Any possible  parameterization of a quantum state which does not allow to efficiently compute physical properties is not a useful parameterization for computational purposes. How to efficiently compute correlators with matrix product states can be found for instance in \cite{Scholl05}. 

The matrix product state parameterization has been very useful in computing low-energy properties of some sufficiently local Hamiltonians, and also the dynamics of quantum states. We shall not explain here the details of some optimization algorithms like DMRG, and the interested reader is addressed to the huge amount of existing literature about this (see for example \cite{Scholl05, Chi06}). We do sketch, however, the basic ideas on how to proceed for computing dynamical evolutions with matrix product states. In fact, some optimization algorithms, like euclidean time evolution, can also be understood in terms of the dynamical procedures that we explain in what follows. 

\subsection{Computing dynamics}

In this section we explain how to compute the dynamics of a matrix product state. Our model for dynamical evolution is based on the application of a set of unitary gates acting either on one or two local $d$-level systems, which could perfectly correspond to a discretization of the continuous time evolution driven by a generic one and two-body Hamiltonian. 

Let us begin this explanation by considering the effect of a unitary gate $U^{(l)}$ acting over a single $d$-level system $l$. The consequence of this operation involves an updating of the matrices  $A^{(l)}$ at site $l$ that goes as follows:

\begin{equation}
\label{one-qubit-update}
{A'}^{(l)i'_l}_{\alpha_{l-1}\alpha_l}=\sum_{i_l} U^{(l)}_{i_li'_l}A^{(l)i_l}_{\alpha_{l-1}\alpha_l} \ .
\end{equation}

Notice that this type of local gates does not affect the ancillary indices. Entanglement
is thus unaffected, which is a necessary condition since we are just performing a local
operation.

The effect of non-local unitary gates acting on different local systems is less obvious. 
We initially consider the case of a non-local gate $U^{(l,l+1)}$ involving contiguous local systems $l$ and $l+1$.
Let us define 
\begin{equation}
\label{two-qubit}
\sum_{i_l, i_{l+1}} U^{(l,l+1)}_{i'_l i'_{l+1},i_l i_{l+1}}
A^{(l)i_l}_{\alpha_{l-1}\alpha_{l}}
A^{(l+1)i_{l+1}}_{\alpha_l\alpha_{l+1}}\equiv
\Theta^{i'_l i'_{l+1}}_{\alpha_{l-1}\alpha_{l+1}} \ . 
\end{equation}

Unlike with local gates, the action of
an interacting gate does not preserve the product
form of the tensors $A$. To reestablish
the matrix product state structure we need to rewrite $\Theta$
using a Schmidt decomposition. The procedure to follow
is to compute the reduced density matrix from the bipartition of the system
between the $l$ and $l+1$ sides, which for the $l+1$ side reads 

\begin{equation}
\rho^{ij}_{\alpha_{l+1}\beta_{l+1}}=\sum_{k, \alpha_{l-1}}|\lambda^{(l-1)}_{\alpha_{l-1}}|^2 {\Theta}^{k
  i}_{\alpha_{l-1}\alpha_{l+1}}{\Theta^*}^{k j}_{\alpha_{l-1}\beta_{l+1}} \ ,
\end{equation}  
where we have made use of the at most
$\chi$ known Schmidt coefficients $\lambda^{(l-1)}_{\alpha_{l-1}}$ for the
cut between the $l-1$ and the $l$ sides. After diagonalizing $\rho$ using
$(i,\alpha_{l+1})$ and $(j,\beta_{l+1})$ as composed indices, we directly read from the
eigenvalues the at most $d\chi$ updated Schmidt coefficients ${\lambda '}^{(l)}_{\alpha_l}$ for this
bipartition, together with the updated matrices ${A'}^{(l+1)i_{l+1}}_{\alpha_l \alpha_{l+1}}$ 
from the coefficients of the eigenvectors. Finally, the new tensors for system
$l$ are easily calculated as ${A'}^{(l)i_l}_{\alpha_{l-1}\alpha_l} = \sum_{i_l, \alpha_{l+1}}
{A'}^{(l+1)i_{l+1}}_{\alpha_l \alpha_{l+1}} \Theta^{i_l i_{l+1}}_{\alpha_{l-1} \alpha_{l+1}}$. Non-local gates between  non-contiguous systems
can be reduced to the previous case by  using SWAP gates, producing 
a typical overhead of $O(n)$ operations. Notice that all our manipulations can be done in a time that grows like $O(\chi^3)$. 

As we have seen, non-local gates entangle the system by increasing the size of the matrices that must be kept in the classical simulation scheme. Each time an entangling gate is operated on two neighboring systems, the index of the connected ancillae is multiplied by $d$. 
To keep the numerical simulation under control, a (non-unique) truncation scheme is needed to stop the exponential growth of ancillary dimensions. The ability in this truncation is the key element for the success of the time-evolution algorithm. Here we explain two possible truncation schemes, the first one based on the original proposal of Vidal \cite{GVidal03_1} of an optimal local truncation, and the second one inspired on the methods of Verstraete and Cirac \cite{VPC04, VGRC04, VW05, VC04} based on an optimal non-local truncation procedure. 

Before entering into the details of the possible truncation schemes, let us introduce a graphical representation of the quantum state that shall be useful in what follows. We represent the tensor $A^{(l) i_l}_{\alpha_{l-1} \alpha_l}$ at site $l$ by the following diagram:

\bigskip

\begin{figure}[h]
\begin{center} 
  \includegraphics[width=0.3\linewidth]{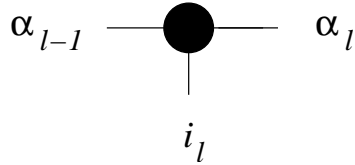}
  \caption{Diagrammatic representation of the tensor  $A^{(l) i_l}_{\alpha_{l-1} \alpha_l}$ at site $l$.}
\end{center}
\label{matricilla}
\end{figure}

With this notation, a matrix product state like the one from Eq.\ref{statemps} is represented by means of the following tensor network:

\bigskip

\begin{figure}[h]
\begin{center} 
  \includegraphics[width=1.0\linewidth]{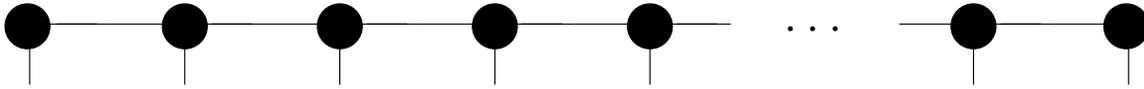}
  \caption{Diagrammatic representation of  a matrix product state in terms of a tensor network.}
\end{center}
\label{mpsnetwork}
\end{figure}
In the above figure we have decided to drop off the name of the indices of the matrices since they do not bring any extra information. Each one of the dots represents a specific particle. Vertical lines correspond to the indices of the physical Hilbert spaces and run up to $d$, while horizontal links between the dots correspond to the ancillary indices and run at most up to $\chi$. Now, we are in a position to discuss the different truncation procedures. 

\subsubsection{Local truncation scheme}

\begin{figure}[h]
\begin{center} 
\includegraphics[width=1.0\linewidth]{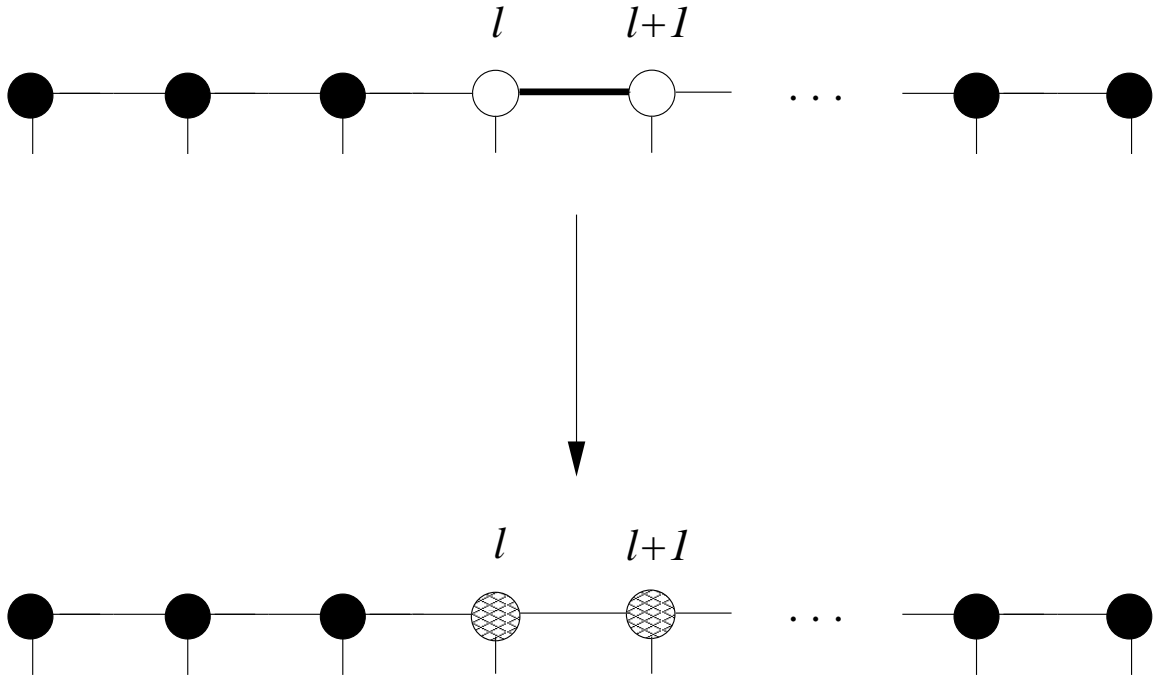}
\caption{Local truncation scheme. Black dots correspond to old matrices, white dots correspond to updated matrices after the unitary evolution, and the thick link line has a rank at most $d\chi$.  Only  matrices at sites $l$ and $l+1$ are truncated (indicated by dashed dots), and this is done by keeping only the most relevant $\chi$ terms of the corresponding Schmidt decomposition.}  
\label{lalala}
\end{center}
\end{figure}

After the application of a non-local gate on the adjacent systems $l$ and $l+1$, the obtained matrix product state is identical to the original one with the only exception that matrices for sites $l$ and $l+1$ have been updated, and the rank of the link connecting these two matrices has been multiplied by $d$. A possible truncation procedure is to \emph{only} change the matrices at sites $l$ and $l+1$, computing two new matrices with ancillary indices up to $\chi$, in such a way that the difference with the original state is minimum (or analogously, the overlap with the original state is maximum). This is a local scheme, since it only affects the two very specific matrices of the whole matrix product state that were touched by the action of the unitary gate. It is easy to see that optimality in this truncation is achieved by keeping the $\chi$ terms in the range of the common index that correspond to the largest eigenvalues $|{\lambda '}^{(l)}_{\alpha_l}|^2$ of the reduced density matrices of the bipartition of the system between the sites $l$ and $l+1$. The diagrammatic representation of this truncation is shown in Fig.\ref{lalala}. 

Notice that given the locality of the procedure, this scheme seems to be a good way to implement a truncation in order to eventually parallelize the code of the classical simulation algorithm. More precisely, one could think of different nodes of a computer network, each one of them storing one matrix (or a finite set of them). This truncation scheme would only involve communication between the two nodes on which the non-local gate operates, leaving the rest $n-2$ nodes untouched, and therefore involving a small amount of information to be sent between different nodes.

\subsubsection{Non-local truncation scheme}

\begin{figure}[h]
\begin{center} 
\includegraphics[width=1.0\linewidth]{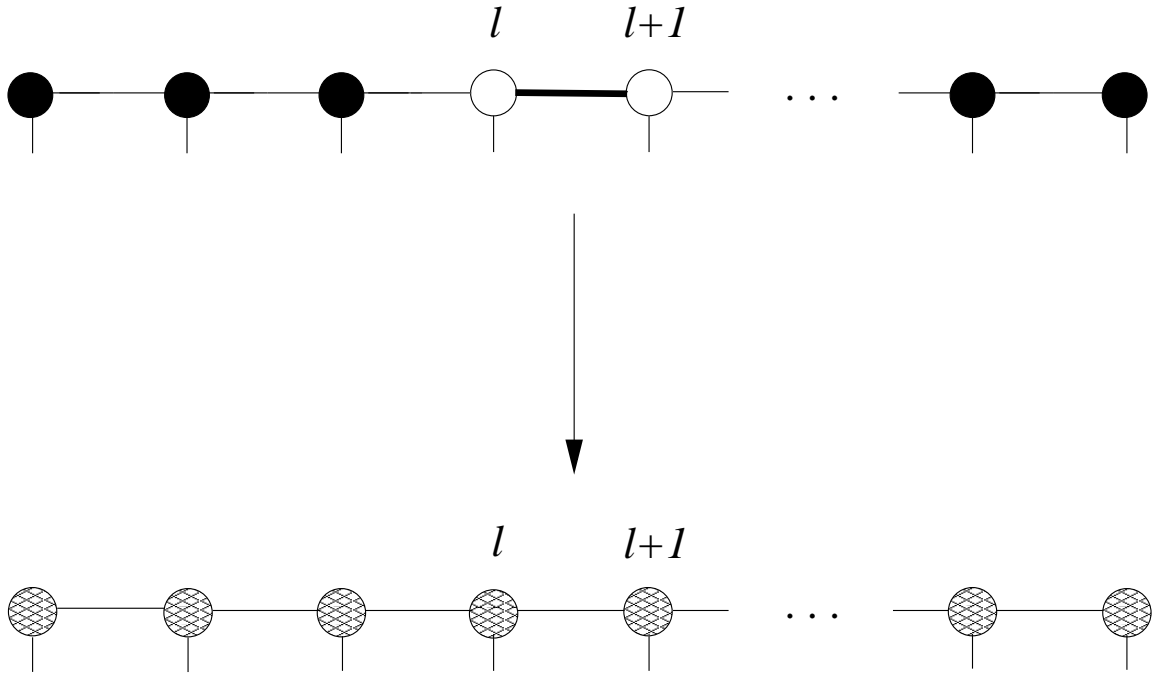}
\caption{Non-local truncation scheme. Black dots correspond to old matrices, white dots correspond to updated matrices after the unitary evolution, and the thick link line has a range $d\chi$.  We find new matrices at every site (indicated by dashed dots) with ancillary indices up to $\chi$ such that the distance to the original state is minimized.}  
\label{nonlocaltronco}
\end{center}
\end{figure}

Given the above local procedure, we can see that there exists the possibility to improve the precision in the truncation by means of a non-local updating of the matrices that define the matrix product state. The main idea is as follows: instead of performing an optimal truncation only in matrices at sites $l$ and $l+1$, perform an optimal truncation in \emph{all} the matrices defining the matrix product state, that is, find a new state with new matrices for all the sites with ancillary indices up to $\chi$ such that the distance to the original state is minimum. This is represented in Fig.\ref{nonlocaltronco}. 

In order to find the new optimal matrices it is possible to proceed in the following way. Let us call $|\psi'\rangle$ the exact state after the non-local unitary evolution, and $|\tilde{\psi}\rangle$ the new matrix product state that we use to approximate $|\psi'\rangle$. We wish to maximize the quantity $\left| \langle \psi' | \tilde{\psi}\rangle \right|^2$ over all possible matrix product states $|\tilde{\psi}\rangle$ with ancillary indices up to $\chi$ with the normalization constraint $|\langle \tilde{\psi} | \tilde{\psi} \rangle |^2 = 1$. In order to perform this minimization, we fix all the matrices of  $|\tilde{\psi}\rangle$  to a fixed value except the first one, and maximize the overlap with respect to the first matrix with the appropriate normalization constraint, which can be done in $O(\chi^3)$ time\footnote{This is valid for the case of open boundary conditions that we analyze here. Periodic boundary conditions may involve a larger computational time than our case.}.  Once the values of the first matrix are found, we repeat the procedure maximizing with respect to the second matrix and finding a better approximation to the original exact state. The complete maximization is then performed by repeating this procedure sequentially for every site, and sweeping back and forth along the system until some desired convergence is achieved. 

\begin{figure}[h]
\begin{center} 
  \includegraphics[width=1.0\linewidth]{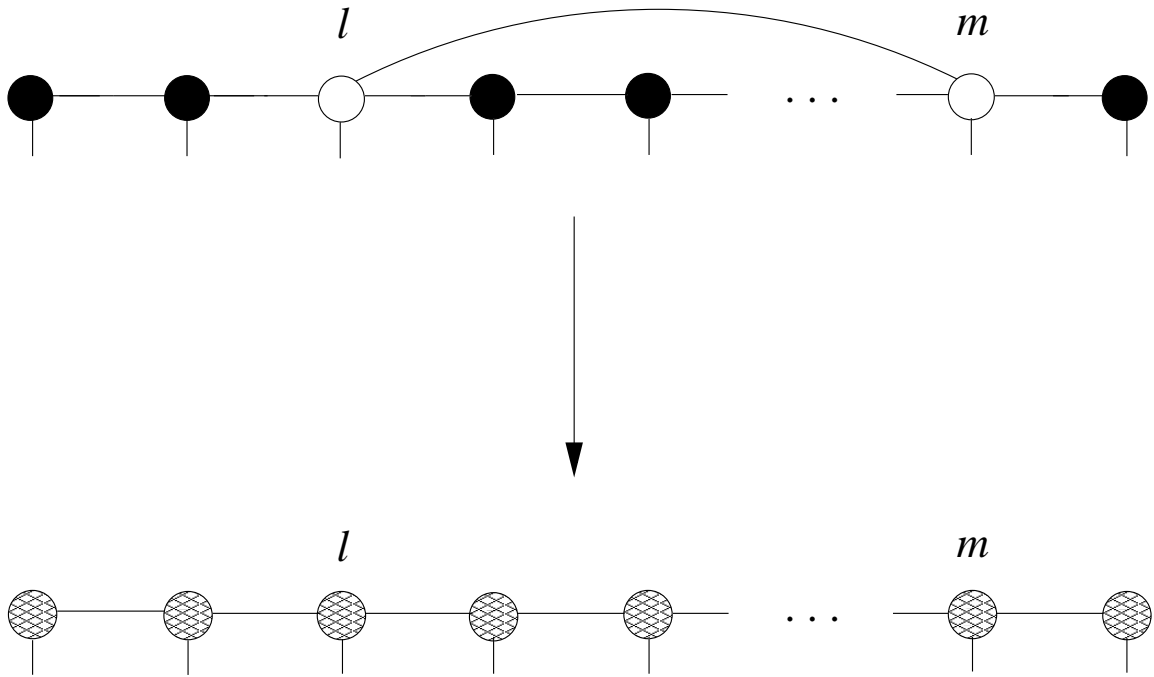}
  \caption{Non-local truncation scheme. Black dots correspond to old matrices, and white dots correspond to updated tensors after the unitary evolution. The action of a non-local gate has created a new link of rank $d^2$ between sites $l$ and $m$. We find new matrices at every site (indicated by dashed dots) with ancillary indices up to $\chi$ such that the distance to the original state is minimized.}  
\label{nonlodue}
\end{center}
\end{figure}

Indeed, this truncation scheme does not require the non-local gate to be necessarily applied on adjacent systems. Imagine that we wish to apply a non-local gate $U^{(l,m)}$ between distant systems $l$ and $m$. It is possible to see that any such unitary matrix $U^{(l,m)} \in {\rm U}(d^2)$ can always be written as $U^{(l,m)} = \sum_{a,b} C_{ab} O^{(l)}_a \otimes O^{(m)}_b$, where $O^{(l)}_a$ and $O^{(m)}_b$ are $2d^2$ local operators acting respectively on sites $l$ and $m$ ($d^2$ operators per site), and $C_{ab}$ are $d^4$ coefficients\footnote{It is possible to see this property by expressing the unitary operator as the exponential of a local basis for the algebra u$(d^2)$ and performing a Taylor expansion.}. Performing a singular value decomposition of the coefficient $C_{ab}$, this can be written as $C_{ab} = \sum_{\mu} U_{a \mu} D_{\mu} V_{\mu b}$, and therefore the original unitary matrix can be expressed as $U^{(l,m)} = \sum_{\mu} \tilde{O}^{(l)}_{\mu} \otimes \tilde{O}^{(m)}_{\mu}$, where we have defined the operators $\tilde{O}^{(l)}_{\mu} \equiv \sum_a U_{a \mu} O^{(l)}_a D_{\mu}^{1/2}$ and  $\tilde{O}^{(m)}_{\mu} \equiv \sum_b V_{\mu b} O^{(m)}_b D_{\mu}^{1/2}$. Applying these operators on the original matrix product state is equivalent to redefine the tensors at sites $l$ and $m$ in such a way that we add a new index $\mu$ of rank $d^2$: 

\begin{eqnarray}
{A'}^{(l) i'_l}_{\mu \ ; \ \alpha_{l-1} \alpha_l} &\equiv& \sum_{i_l} A^{(l) i_l}_{\alpha_{l-1} \alpha_l} \tilde{O}^{(l)}_{\mu \ ; \ i'_l i_l} \\ \nonumber
{A'}^{(m) i'_m}_{\mu \ ; \ \alpha_{m-1} \alpha_m} &\equiv& \sum_{i_m} A^{(m) i_m}_{\alpha_{m-1} \alpha_m} \tilde{O}^{(m)}_{\mu \ ; \ i'_m i_m}  \ . 
\label{potito}
\end{eqnarray}
Given the above equation, we see that after the application of the unitary gate, the sites $l$ and $m$ get linked by a common index $\mu$. This is another way of understanding how non-local gates entangle the system, namely, by creating new bonds between the sites on which they act. Now, it is possible to perform again a non-local truncation much in the same way as before, by finding new matrices for all the sites with only two ancillary indices up to $\chi$ and also in $O(\chi^3)$ time as well. This is represented in Fig.\ref{nonlodue}. 

We shall expect better accuracies for this non-local truncation scheme than for the local truncation procedure, basically because we optimize over a larger set of parameters, and because we do not have to necessarily implement SWAP operations in order to perform non-local gates between distant systems, thus reducing the number of truncations to be applied in the simulation. Nevertheless, this scheme has the drawback that the number of operations to be done at each truncation step is bigger than in the local case. Also, the fact that the truncation is non-local makes it a bad candidate for a possible parallelization of the numerical code, since all the nodes of the computer network should communicate among themselves at each truncation step in order to perform the approximation of the exact state by a new matrix product state.  

\section{Classical simulation of an adiabatic quantum algorithm solving Exact Cover}

In this section we show the results of a simulation of a quantum algorithm using matrix product states. More precisely, we have implemented the local truncation scheme explained in the previous section to the simulation of a quantum adiabatic algorithm solving hard instances of the Exact Cover NP-complete problem. The performance of this algorithm was already analyzed in detail in the previous Chapter by means of an exact numerical computation of its properties up to $20$ qubits. There we saw that the entanglement entropy of a typical bipartition of the system seems to scale as $S \sim 0.1 n$, $n$ being the number of qubits. We also found that the linear scaling of the entanglement entropy forbids the possibility of an efficient numerical simulation with the methods of \cite{GVidal03_1}. The reason becomes clear now, since a linear scaling of the entanglement entropy involves an exponentially big $\chi$ in the number of qubits, and therefore any algorithm based on matrix product states must necessarily handle matrices of exponential size in order to get a result sufficiently close to the exact one. In any case, the possibility of a numerical simulation of this quantum algorithm by using matrix product states is motivated in part by the fact that the coefficient of the scaling law for the entanglement entropy seems to be rather small (only $0.1$). Thus, even though we should need an exponentially big $\chi$ to perform a very accurate simulation of the adiabatic quantum algorithm, it could be possible that already good simulations can be performed by keeping a relatively small $\chi$. Furthermore, the performance of a classical simulation of a quantum algorithm by using the matrix product state ansatz may bring further insight on the way entanglement is used along the quantum evolution. As we shall see, the basic features of the quantum algorithm can still be observed even in the case of a highly-truncated simulation with very small $\chi$. 

Let us sketch the basic features of our simulation. 
First, let us remind that classically hard instances of Exact Cover seem to appear
at the so-called easy-hard-easy transition around 
$m\sim 0.8 n$ \cite{Crawford96}, $m$ being the number of clauses and $n$ being the number of qubits. We have generated such hard instances, with the additional 
property of having only a unique satisfying assignment. The generation of 
hard instances is in itself a hard problem for which we have developed specific
algorithms, essentially based on the iterative addition of random and 
non-redundant clauses until the number of
solutions of the instance is one.  
The quantum algorithm for a given Exact Cover instance
follows the adiabatic evolution
of the ground state of a Hamiltonian defined by $H(s)=(1-s) H_0 + s H_P$,
where the adiabatic parameter is $s=t/T$ and $t$ runs up to
a total predetermined time $T$. We take the
initial Hamiltonian to be $H_0=\sum_{i=1}^n \frac{d_i}{2} (1-\sigma_i^x)$
where $d_i$ stands for the number of clauses where
qubit $i$ enters. The non-local 
problem Hamiltonian corresponds to
the sum of clauses defined as 

\begin{equation}
H_P=\sum_{C \ \in \ {\rm instance}} (z_i+z_j+z_k-1)^2 \ ,
\label{hache}
\end{equation}
where $z_i=(1-\sigma^z_i)/2$ has eigenvalues 0 and 1, and $C$ stands
for a clause involving bits $i$, $j$ and $k$ in the specific instance. Notice the difference between the  problem Hamiltonians from Eq.\ref{hache} and from Eq.\ref{hamil}. Both  Hamiltonians describe correctly the solution to an Exact Cover instance in its ground state. The essential difference between them is that while the Hamiltonian of Eq.\ref{hamil} has three-body interactions, the Hamiltonian of Eq.\ref{hache} has not. The problem Hamiltonian that we use in this Chapter is built only from one and two-body terms, together with local magnetic fields, and its evolution can therefore be classically simulated by the algorithms based on matrix product states that we have already discussed, based on the efficient updatings of the register after performing one and two-body unitary gates.  At the level of eigenvalues, notice that the only difference between the two Hamiltonians comes on the eigenvalues of the excited states, thus keeping the properties of the low-energy sector untouched. In fact, it is easy to see by means of direct simulations that an adiabatic quantum algorithm based on this problem Hamiltonian shows  the same important features as the ones already described in Chapter 4, in particular the appearance of a quantum phase transition at $s_c \sim 0.69$ in the thermodynamic limit, together with a linear scaling of the entanglement entropy with the number of qubits with a small scaling coefficient of the order of $0.1$. 

Exact simulations of quantum algorithms by adiabatic evolution solving hard
instances of satisfiability problems were carried
so far up to 30 qubits \cite{Hogg03}. Here we present the possibility of performing \emph{approximated} simulations of this quantum algorithm beyond that number. 

\subsection{Discretization of the continuous time evolution in unitary gates}

Let us now turn to discuss the detailed way matrix product states can handle 
the simulation of the adiabatic evolution of Exact
Cover. The simulation
needs to follow a time evolution controlled by 
the $s$-dependent Hamiltonian. This continuous
unitary time evolution from time $0$ to time $T$ can be discretized as follows:

\begin{equation}
U(T,0)=U(T,T-\Delta)\dots U(2\Delta,\Delta) U(\Delta,0) \ ,
\label{ovo}
\end{equation}
where the increment 
$\Delta\equiv \frac{T}{M}$ defines the discretization, $M$ being a
positive integer. Our simulations indicate that we can take the value $\Delta=0.125$
while keeping sufficient accuracy -- as compared to smaller $\Delta$ -- 
in all of them. After $l$ steps $s=\frac{t}{T}=\frac{l \Delta}{T}=\frac{l}{M}$,
that is $l=0,\dots M$. 

At any point $l$ along the evolution the unitary operator $U((l+1)\Delta,l \Delta)$
needs further subdivision into elementary one and two-qubit gates. This
requires the use of Trotter's formula to second order \cite{Suzuki90, Suzuki91, Sorn98}: 

\begin{equation}
U((l+1)\Delta,l \Delta)=e^{i \Delta H(s)} \sim \left(
e^{i \frac{\delta}{2} (1-s) H_0} e^{i \delta 
s H_P} e^{i\frac{\delta}{2} (1-s) H_0}\right)^{\Delta\over \delta} \ ,
\label{tro}
\end{equation}

where the partition in $H_0 : H_P : H_0$ minimizes the number of two-qubit gates as compared to the alternative partition $H_P : H_0 : H_P$. 
We have verified as well that we can maintain a faithful classical simulation
by choosing  $\delta=\Delta$.
Notice that the split of exponentials in the Trotter's expansion of Eq.\ref{tro} 
 is chosen so that $H_0$
is explicitly separated from $H_P$, so that this brings the advantage
that  both pieces of 
the Hamiltonian  can be decomposed in 
mutually commuting one and two-qubit gates: 

\begin{equation}
\label{h0diag}
e^{i \frac{\delta}{2}(1-s) H_0}=\prod_{i=1}^n e^{i \frac{\delta}{4}
(1-s) d_i(1-\sigma_i^x)}\ ,
\end{equation}
and
\begin{eqnarray}
\label{hpdiag}
\nonumber
e^{i \delta s H_P}=&\prod_{C \ \in \ {\rm instance}} e^{i \delta s (z_i+z_j+z_k-1)^2}\\
\nonumber
=&\prod_{C \ \in \ {\rm instance}} e^{i \delta s (z_i^2-2 z_i)} e^{i \delta s (z_j^2-2 z_j)}
 e^{i \delta s (z_k^2-2 z_k)}e^{i \delta s}\\
&e^{i 2\delta s z_i z_j}
e^{i 2\delta s z_i z_k}e^{i 2\delta s z_j z_k }\ .
\end{eqnarray}
The complete adiabatic evolution
is thus finally discretized in terms of the sequential action of the above one and two-qubit gates.

\subsection{Numerical results of a simulation with matrix product states}

The exact simulation of a quantum computer using matrix product states is then completely
defined. As we said before, we have chosen the local truncation scheme in order to implement our algorithm. It is possible to see that the total running time of the simulation algorithm scales as
$O(T n m \chi^3)$. This 
reasonable truncation 
carries, though, an inherent -- but always under control -- loss of norm of the quantum state, since
the sum of the retained squared eigenvalues will not
reach $1$. As we shall see, larger $\chi$'s allow
for more faithful simulations, as expected. 

We have implemented a number of optimizations 
upon the above basic scheme which are worth mentioning. 
For any non-local gate there is
an overhead of SWAP operations that damage the 
precision of the computation. To minimize this effect,
every three-qubit clause is operated as follows: we 
bring together the three qubits with SWAPs of the 
left and right qubits keeping the central one 
fixed and, then, we operate  
the two-qubit gates. Before returning the qubits
to their original positions we check if any
of them is needed in the next gate. If so,
we save whatever SWAP may be compensated between the
two gates. Ordering of gates is also used to
produce a saving of $\sim 2/3$ of the naive SWAPs. 
Diagonalization of the relevant reduced density matrices in 
the allowed Hilbert space of minimum dimension  
is used as well. A further improvement is to keep
a both dynamical and local $\chi$, so that ancillary indices
at the different partitions are allowed to take independent values and
grow up to site-dependent and time-dependent limits. This last procedure, though, 
has shown essentially no big improvement upon a naive fixed $\chi$ strategy. Let us now explain in what follows the different results of our simulations.

\subsubsection{Instantaneous expected energy}

\begin{figure}[h]
 \begin{center}
  \includegraphics[angle=270,width=0.7\linewidth]{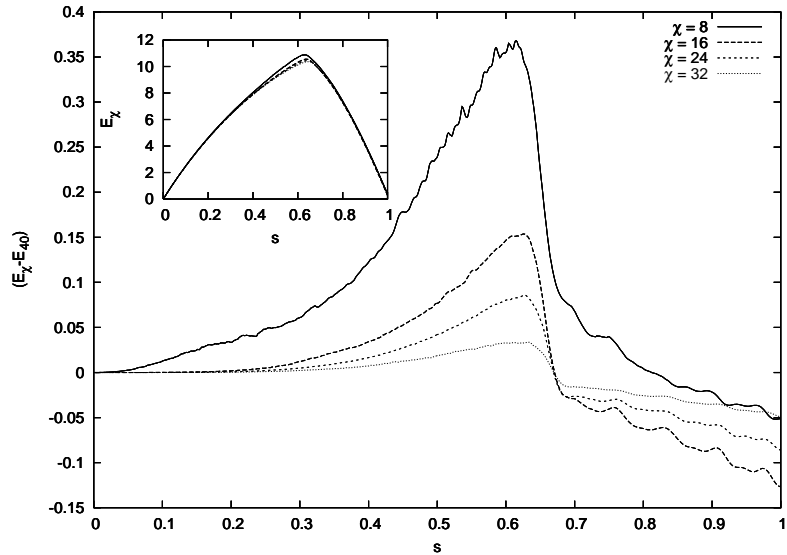}
  \caption{Computation of the absolute error, as compared to the $\chi = 40$ case, of
 the expected value of the Hamiltonian (in dimensionless units)
along the adiabatic evolution for a typical instance with
$30$ qubits and $24$ clauses for $T=100$ as $\chi$ increases. 
Note the increasing precision with larger $\chi$ as $s$ approaches the 
phase transition from the left-hand-side. In the inset, the instantaneous expected energy is
plotted (in dimensionless units). A similar behavior is also obtained for other instances, getting
 \emph{perfect} solution at the end of the computation (zero energy).}
 \label{FigEnergy}
 \end{center}
\end{figure}

We first simulate the adiabatic algorithm with the requirement that the right solution is found 
for a typical instance of $n=30$ qubits with $m=24$ clauses and 
$T=100$. Along the evolution we compute the expected value of the Hamiltonian
of the system, which can be calculated in $O(\chi^3)$ time. 
Our numerical data are shown in Fig.\ref{FigEnergy}. The system remains remarkably 
close to the instantaneous ground-state all along the approximated evolution and, as we can see, the maximum absolute error with respect to our best classical simulation ($\chi = 40$) comes when evolving close to the quantum phase transition point. We also see convergence in the error while the system approaches the critical point. This minimum absolute error in the ground-state energy is, when close to criticality, of the order of $10^{-2} -10^{-3}$, smaller than the typical value of the energy gap for $30$ qubits -- as hinted by extrapolating the data from Fig.\ref{gap} in Chapter 4 --. A bigger $\chi$ may bring a better precision by using a larger, but eventually affordable, time cost in the simulation. 
 
The error in the expected energy is minimized as $\chi$ increases.
It is noteworthy to observe how the error in the simulation of the adiabatic algorithm
increases at the phase transition point. We have also numerically checked in our simulations 
that it is precisely at this
point where each qubit makes a decision towards its final
value in the solution. Physically, the algorithm builds entanglement
up to the critical point where the solution is singled out
and, thereon, the evolution seems to drop the superposition of wrong states 
in the register.

\subsubsection{Loss of norm}

\begin{figure}
  \centering
  \includegraphics[angle=270,width=0.7\linewidth]{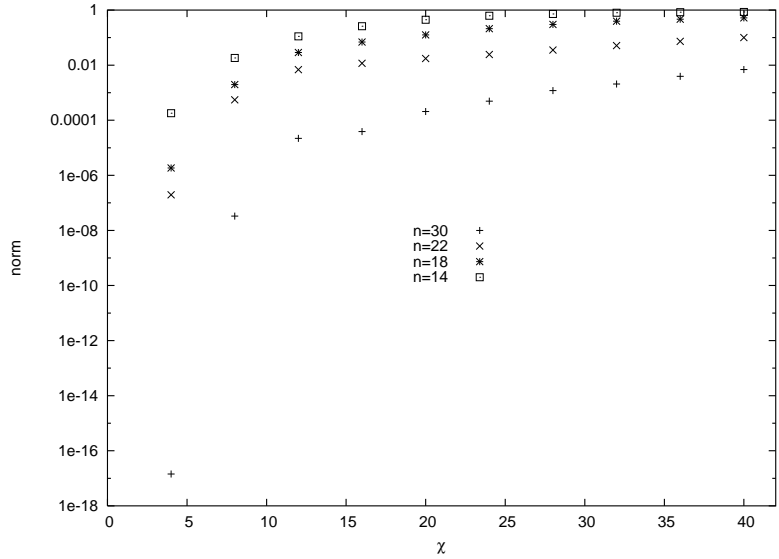}
  \caption{Final norm in the register as a function of $\chi$ in logarithmic
 scale, for instances of $14,18,22$ and $30$ qubits.}
  \label{FigUnitarity}
\end{figure}

We plot in Fig.\ref{FigUnitarity} the norm of the quantum state at the end of the simulation as
a function of $\chi$ in logarithmic scale, for typical instances of $14,18,22$ and
$30$ qubits. The remarkable fact is that some observables, like
the energy, appear to be
very robust against this inaccuracy, while the behavior of this norm was already expected 
 not to be good, since this is precisely the parameter in which we are truncating with respect to the exact evolution, and furthermore its accumulation is multiplicative as time evolves. 

\begin{figure}
  \centering
  \includegraphics[angle=270,width=0.7\linewidth]{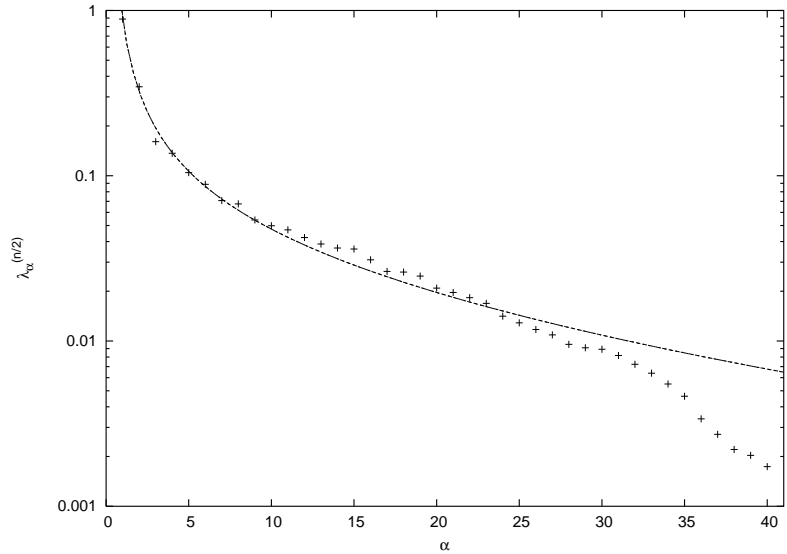}
  \caption{Decay of the Schmidt coefficients for a typical instance of $30$ qubits in logarithmic scale, with $\chi=40$. The behavior seems to be approximately described by a law of the kind $\log_2 (\lambda^{(n/2)}_{\alpha}) = a + \frac{b}{\sqrt{\alpha}} + 
c\sqrt{\alpha}$, for appropriate coefficients $a,b$ and $c$ (solid line).}
  \label{FigDecay}
\end{figure}

\subsubsection{Decay of the Schmidt coefficients}

Our simulations also allow to compute the
decay of the $\chi$ Schmidt coefficients $\lambda^{(l)}_{\alpha}$, $\alpha = 1, 2, \ldots , \chi$, 
at any site $l$ and at any step of the computation. At the closest point to criticality, and for the central
bipartition of the system, these can be approximately fitted by the law 
$\log_2 (\lambda^{(n/2)}_{\alpha}) = a + \frac{b}{\sqrt{\alpha}} + 
c\sqrt{\alpha}$, with appropriate instance-dependent coefficients $a,b$ and $c$. The behavior for a typical instance of $30$ qubits is shown in Fig.\ref{FigDecay}.

\subsubsection{100-qubit instance}

\begin{figure}
  \centering
  \includegraphics[angle=270,width=0.7\linewidth]{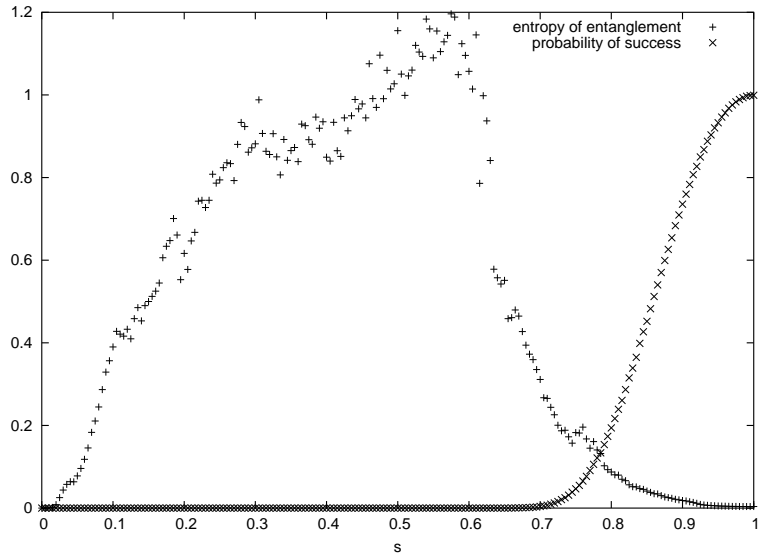}
  \caption{ Entanglement entropy of a bipartition and probability of being at the correct solution as a function of $s$ for a simulation with $\chi=14$ of adiabatic evolution solving a hard instance of $n=100$ bits and $m=84$ clauses.}
  \label{Fign100}
\end{figure}

The ultimate goal of finding the correct solution 
appears also to be very robust in the simulations we have performed. The exact
probability of success can be calculated in $O(\chi^2)$ time as
well. As a symbolic example, our program has solved an instance with
$n=100$ bits, that is, the adiabatic evolution algorithm
has found the correct product state out of $2^{100} \sim 10^{30}$ possibilities for
a hard instance with $m=84$ clauses and $T=2000$. The simulation
was done with a remarkably small value of $\chi=14 \ll 2^{50} = \chi_{max}$ and
is presented in Fig.\ref{Fign100}. Notice that while the entanglement entropy shows fluctuations because it is directly related to the truncation parameter of the simulation, the probability of success follows a smooth behavior, being almost zero at the beginning of the evolution, and jumping directly to one precisely when close to the quantum critical point. 

\begin{figure}
  \centering
  \includegraphics[angle=270,width=0.7\linewidth]{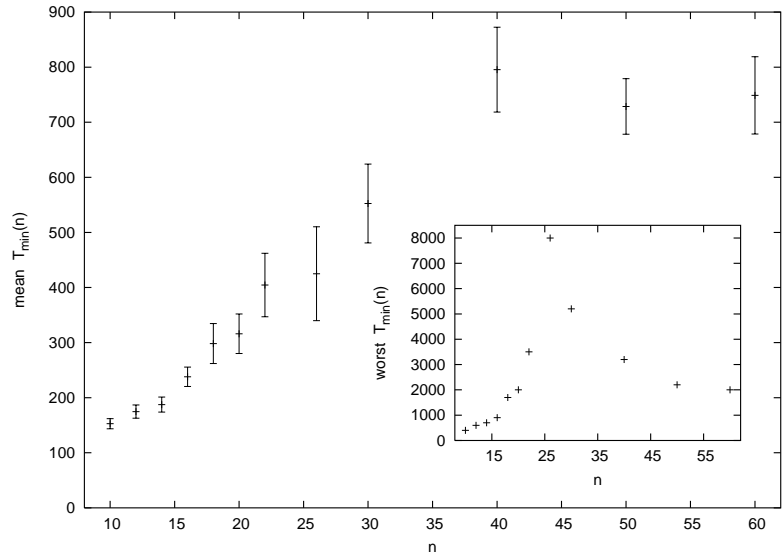}
  \caption{Mean and worst cases of the accumulated statistics up to $n=60$ for $T_{min}(n)$ (in dimensionless units) such that
an instance is solved. Averages are performed over $200$ instances
for each $n$, except for $n=50,60$ with respectively $199,117$ instances. 
Error bars give $95$ per cent of confidence level in the mean.}
  \label{FigStat}
\end{figure}

\subsubsection{Time statistics}

The robustness of evolving towards the correct solution is 
found for any number of qubits
and small $\chi$. To analyze further the performance of this classical simulation, we 
have launched a search for the minimal $T_{min}(n)$ that solves samples
of $n$-qubit hard instances in the following way: for a set of small values of
$\chi$, we try a random instance with an initial $T$, for instance $T=100$. 
If the solution is
found, we proceed to a new instance, and if not, we restart
with a slower adiabatic evolution with, for instance, $T=200$. This
slowing down of the algorithm
is performed until a correct solution is found and
the minimum successful $T_{min}$ is stored. 
Our results are shown in Fig.\ref{FigStat}. 
The average over $n$-qubit instances of $T_{min}(n)$ appears
to grow very slowly with $n$, though the extreme cases need increasingly 
larger times up to $n=25$. The slowing-down in the plots for a large number of qubits is a side-effect of the inherent difficulty to generate hard instances of Exact Cover for large $n$.  
We want to remind as well  that finding an instance that needs
a very large $T_{min}$ is no counterproof for the validity of the adiabatic
algorithm, as alternative interpolating paths may solve
the instance efficiently \cite{Farhi02_2}. 

\begin{figure}
  \centering
  \includegraphics[angle=270,width=0.7\linewidth]{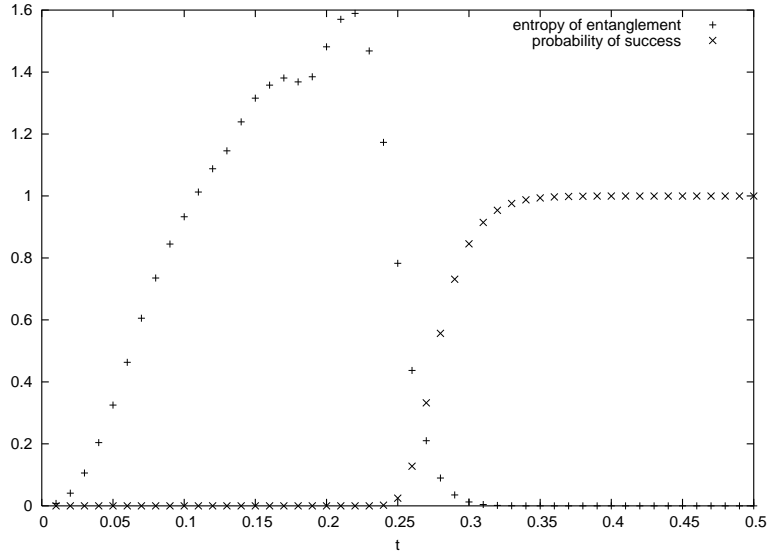}
  \caption{Euclidean time evolution solving a typical instance of $26$ qubits with $\chi=6$. The algorithm finds the correct solution much faster than the simulations of adiabatic quantum computation. The sudden jump in the probability of success comes again at the maximal point for the entanglement entropy.}
  \label{FigEu}
\end{figure}

\subsubsection{Solving hard classical instances by euclidean time evolution}

Independently of the fact that our simulation describes in an approximate way the behavior of an adiabatic quantum algorithm, we can think of it as a plausible classical algorithm for solving hard instances of an NP-complete problem. In fact, if our aim is to solve instances of Exact Cover, all that is required is a classical algorithm to find the ground-state of the problem Hamiltonian $H_P$ from Eq.\ref{hache}. A possibility is to perform an evolution in euclidean time, that is, to simulate the evolution driven by the non-unitary operator

\begin{equation}
e^{-H_P t} \ . 
\label{unuu}
\end{equation}
The above evolution is not physical, since it is not unitary and therefore does not correctly preserve the probabilities as the parameter $t$ (the euclidean time) flows. In any case, it is easy to see that if we have a (possibly not normalized) quantum state such that it has a non-zero overlap with the ground state of $H_P$, the action of the operator from Eq.\ref{unuu} over the state will eventually drive the original state towards the only fixed point of the map at $t \rightarrow \infty$, which is the ground state of $H_P$. In practice, the action of the above operator over an equally-weighted superposition of all possible computational states will drive the original state towards the ground state of $H_P$ with very high probability at times bigger than the inverse of the first gap of the system.
This optimization algorithm can be easily implemented by using the same time-evolution procedures described before in terms of matrix product states. Evolution in euclidean time shall not be unitary, though, but this particularity does not affect any of the essential features of the updating and truncation schemes previously explained. 

The performance of the evolution in euclidean time for solving hard instances of Exact Cover is remarkably good, as compared to the performance of the simulation of the adiabatic quantum algorithm. This new classical algorithm finds the correct solution to the instances much faster than our previous simulations of adiabatic evolution. As an example, we show in Fig.\ref{FigEu} the result of a simulation for a typical instance of $26$ qubits with $\chi = 6$. The behavior of the euclidean time evolution algorithm resembles very much the one of the adiabatic evolution, in the sense that the probability of success remains very close to zero, until some specific point in the evolution is reached, where it jumps to one very quickly. It is also interesting to notice that this point corresponds, once more, to the point of maximum entanglement in the evolution, as measured by the entanglement entropy. Since the ground state of $H_P$ is non-degenerate and separable, and since we begin with an equal superposition of all the possible states of the computational basis, the entropy must begin at zero and eventually die in zero, so it must necessarily reach a maximum at some point along the evolution. Remarkably, the point of maximum entropy coincides again with the jump in the probability of success. Note that even though the system is not evolving close to any quantum phase transition (like the one of the adiabatic quantum algorithm), the behavior along the evolution is very analogous to the one observed in those cases (compare Fig.\ref{FigEu} and Fig.\ref{Fign100}). Again, maximum entanglement brings the correct solution to the problem, although our algorithm is entirely classical. 

\section{Conclusions of Chapter 5}

In this Chapter we have shown that it is possible to implement approximated classical simulations of quantum algorithms by the use of matrix product states with controlled accuracy. More specifically: 

\begin{itemize}
\item{We have implemented a simulation based on matrix product states of an adiabatic quantum algorithm solving the NP-complete Exact Cover problem. This simulation is made precise by means of an  optimal local truncation scheme, and provides robust results for quantities like the expected energy or the probability of success, with a relatively small size of the involved matrices.}
\item{We have solved a hard $100$-qubit instance of Exact Cover by means of a highly-truncated simulation of the adiabatic evolution algorithm. This classical simulation finds the correct product state out of $2^{100} \sim 10^{30}$ possibilities by using matrices whose indices range up to $\chi = 14$,  much smaller than the necessary $2^{50}$ for an exact simulation.}
\item{We have seen that the mean time that our approximated classical simulations take to succeed increases slowly with the number of qubits, though not a definite scaling law can be inferred given the inherent difficulty to generate very hard instances of Exact Cover for a large number of qubits.} 
\item{Matrix product states algorithms for dynamical evolution can also be applied for simulating the non-unitary evolution in euclidean time, which we have shown to be a classical optimization algorithm that solves hard instances of Exact Cover much more efficiently than the classical simulations of the adiabatic algorithm.} 
\end{itemize}
 
The results presented here could be extended in several directions. For instance, it should be possible to study the performance of the optimal non-local truncation scheme and to compare it with the one we have considered here. Also, the performance of a parallelization of the numerical code that we have considered here could be analyzed. More generically, it should also be plausible to extend the rigid structure of a matrix product state to other tensor networks specifically adapted to the particular problem or instance in consideration, much in the same way as PEPS do in $(2+1)$-dimensional systems \cite{VC04}. Finally, the study of the performance of all the ideas exposed here but with other quantum algorithms is a direction to be considered as well. For instance, it should be possible to see the behavior of a classical simulation of Shor's factoring algorithm by using matrix product states or related techniques. As we saw in Chapter 4, Shor's algorithm is yet another quantum algorithm which inherently makes use of an exponentially big amount of $\chi$ in the number of qubits. The effect of truncations in that algorithm are, though, not evident. Perhaps, a classical simulation of Shor's quantum algorithm using the ideas of this Chapter could be a good candidate for a new classical factorization algorithm.

    \chapter{Majorization arrow in quantum algorithm design}

Finding underlying mathematical structures
in efficient quantum
algorithms is one of the problems that quantum computation deals with. 
The fact that there  is only a short list of ideas behind quantum algorithm design hints how difficult it is to come up with new quantum techniques and strategies to efficiently solve important problems.  
Grover's quantum searching algorithm \cite{Grover96} exploits
calls to an oracle by enhancing a particular state, actually implementing a rotation in
the relevant Hilbert space associated to the problem. Shor's factoring quantum algorithm  \cite{Shor94} exploits the 
periodicity of an initial quantum state using a minimum of
Hadamard and controlled-phase gates at the core of the quantum Fourier transform.
Based on more general quantum mechanical principles, the idea
of using adiabatic evolution to carry quantum computation \cite{Farhi00_1}  
has proven suitable for performing Grover's algorithm
and has been numerically studied as a candidate for
attacking NP-complete problems, as we saw in Chapters 4 and 5. Also, 
the so-called quantum walks in continuous time have
proven to efficiently solve a classically hard problem \cite{Childs02_3}, 
whereas quantum random walks in discrete-time have proven to bring also 
Grover's square-root speed-up in a problem of quantum search \cite{Shenvi03}. 
Many other quantum algorithms can be mapped to the above families, 
being then based on the same basic principles.

Some attempts to uncover the properties of
quantum algorithms  have already been explored. 
One relevant instance is undoubtedly  the
role of entanglement  \cite{Ahn00, Knight00, Jozsa02, Parker02, Kendon04, GVidal03_1, GVidal04}, 
which was already considered in detail in the preceding two Chapters.
In fact, although entanglement is a natural resource to
 be exploited in quantum algorithm
design, there are known examples of faster-than-classical oracle-based quantum algorithms
where the quantum register remains in a product state between calls to the quantum oracle all
along the computation, though the speed-up is only by a factor of two \cite{Bernstein97, Mosca99,Cleve98}. In this Chapter we will concentrate on quite
a different proposal. The basic idea is that there is an underlying 
 strong majorization behavior in some quantum algorithms that seems to play a role as well.
 
More concretely,
we study the evolution in different  quantum algorithms, with respect to majorization, of the probability distribution 
arising in the evolving quantum state from the probabilities of the final outcomes, as introduced in \cite{LM02}. We consider
several families of quantum algorithms based on distinct properties. As a first
step, we analyze the majorization behavior of the family of quantum
phase-estimation algorithms, comparing their performance with respect to majorization to that of Grover's algorithm \cite{LM02}, and
 giving also the explicit example of a slightly different quantum algorithm solving a hidden affine function problem by means of calls to an oracle \cite{Bernstein97,Mosca99,Cleve98}.  
 We also consider here
 the class of adiabatic algorithms \cite{Farhi00_1} by studying the behavior of the adiabatic algorithm implementing 
 a quantum search \cite{Grover96,Roland02,Dam01}.
 Efficiency is seen to depend
on the interpolating time path taken along the evolution \cite{Farhi02_2, Roland02, Dam01}, and we observe that 
optimality in adiabatic quantum searching appears when
step-by-step majorization is present. Finally, 
quantum walks provide exponential speed-up
over classical oracle-based random walks \cite{Childs02_3}, and again a  
manifest strong majorization behavior is detected. 
Let us begin, then, by considering the way in which we understand majorization theory as applied to the study of
 quantum algorithms. 

\section{Applying majorization theory to quantum
algorithms}

The way we relate majorization theory -- as defined in Appendix A --  
to quantum algorithms is 
as follows: let $|\psi^{(m)} \rangle$ be the pure state
representing the register of a quantum computer at an operating
stage labeled by $m = 1 \ldots M$, where $M$ is the total number
of steps in the algorithm, and let $N$ be the dimension of the
Hilbert space. If we denote as $\{|i\rangle\}_{i=1}^N$ the basis
in which the final measurement is to be performed, we
can naturally associate a set of sorted probabilities $p_i$, $i =
1 \ldots N$, to this quantum state in the following way: decompose
the register state in the measurement basis such that

\begin{equation}
|\psi^{(m)} \rangle = \sum_{i = 1}^{N}a_i^{(m)}|i \rangle \ .
\label{decomp}
\end{equation}
The probability distribution associated to this state is

\begin{equation}
\vec p^{(m)}=\{p_i^{(m)}\}\qquad
 p_{i}^{(m)} \equiv |a_{i}^{(m)}|^2 = |\langle i | \psi^{(m)} \rangle
|^2  \ ,
\label{probabs}
\end{equation}
where $i = 1 \ldots N$. This corresponds to the probabilities of all the possible outcomes if the computation were to be stopped at stage $m$ and a measurement were performed. A quantum algorithm will be said to majorize this probability
distribution between steps $m$ and $m+1$ if and only if \cite{LM02, OLM03, OLM04}

\begin{equation}
\vec p^{(m)} \prec \vec p^{(m+1)} \ .
\label{majflow}
\end{equation}
Similarly, a quantum algorithm will be said to reversely majorize this
probability distribution between steps $m$ and $m+1$ if and only
if

\begin{equation}
\vec p^{(m+1)}  \prec \vec p^{(m)} \ .
\label{minflow}
\end{equation}

If Eq.\ref{majflow} is
step-by-step verified, then there is a net flow of probability
towards the value of highest weight, in such a way that the
probability distribution will be steeper and steeper as  time flows in the algorithm.
In physical terms, this can be stated as a very
particular constructive interference behavior, namely, a
constructive interference that has to step-by-step satisfy a set of $N-1$ 
constraints -- see Appendix A -- at each time step. The quantum
algorithm monotonically builds up the solution by means of
this very precise reordering of the probability distribution.

It is important to note that majorization is checked on
a particular basis. Step-by-step majorization is, then, 
a basis-dependent concept. Nevertheless there is 
a preferred basis, namely, the basis defined by the
final measurement of the quantum register. This typically 
(though not necessarily always)  corresponds to the 
 computational basis of the quantum computer. 
 The principle we analyze
is rooted in the physical and practical possibility to 
arbitrarily  stop the computation at any time and 
perform a measurement. Generically speaking, we analyze the majorization 
properties of the probability distribution
of the possible outcomes of our measurement apparatus along the time-flow
in the algorithm.

\subsubsection{Natural majorization}

Let us now define the concept of natural majorization for quantum
algorithms. Working with the probability amplitudes in the basis
$\{|i\rangle\}_{i=1}^N$ as defined in Eq.\ref{decomp}, the action of a generic
unitary gate at step $m$ makes the amplitudes evolve to
step $m+1$ in the following way:

\begin{equation}
a_i^{(m+1)} = \sum_{j=1}^N U_{ij} a_j^{(m)} \ ,
\label{amp}
\end{equation}
where $U_{ij}$ are the matrix elements in the chosen basis of the
unitary evolution operator. By inverting this evolution, we can write

\begin{equation}
a_i^{(m)} = \sum_{j=1}^N C_{ij} a_j^{(m+1)} \ ,
\label{amp2}
\end{equation}
where $C_{ij}$ are the matrix elements of the inverse unitary evolution, which is of course unitary as well. 
Taking the square-modulus we find

\begin{equation}
|a_i^{(m)}|^2 = \sum_{j=1}^N |C_{ij}|^2 |a_j^{(m+1)}|^2 + {\rm interference \ terms} \ .
\label{mquad}
\end{equation}
Should the interference terms disappear, majorization would be
verified in a ``natural'' way between steps $m$ and $m+1$  since
the initial probability distribution could be obtained from the
final one just by the action of a doubly stochastic matrix with
entries $|C_{ij}|^2$. We shall refer to this property as ``natural
majorization'': majorization which naturally emerges from the
unitary evolution due to the lack of interference terms when
making the square-modulus of the probability amplitudes.
Similarly, we can define the concept of ``natural reverse majorization'',
which follows in a straightforward way: there will be ``natural
reverse majorization'' between steps $m$ and $m+1$ if and only if there is
``natural majorization'' between steps $m+1$ and $m$. 
As we shall see, this very specific kind of majorization shall appear in some of our forthcoming calculations. 

\section{Majorization in quantum phase-estimation algorithms}

Quantum phase-estimation algorithms \cite{Kitaev95,Cleve98,Mosca99,Shor94,Bernstein97,Nielsen-Chuang} 
are a good example of a wide class of quantum algorithms 
to begin our study. Their key ingredients 
are the use of the
quantum Fourier transform operator and the promise of a
specific structure of the initial state.
In \cite{LM02}, it has been numerically checked that the
canonical form of the quantum Fourier transform majorizes
 step-by-step the probability distribution attached
to the computational basis.
Here we analytically address this problem and provide a proof of how the notion of
majorization formulated in \cite{LM02} explicitly operates in the
special case of quantum phase-estimation algorithms.
To be more specific, our purpose now is to present a detailed
proof of the following proposition: majorization works step-by-step in the
quantum Fourier transform of quantum phase-estimation algorithms. 
The whole property is based on the idea that Hadamard operators act by majorizing 
the probability distribution given 
the symmetry of the quantum state, and such a symmetry is partially 
preserved under the action of both Hadamard and controlled-phase gates \cite{LM02}.

\subsection{The quantum phase-estimation algorithm}

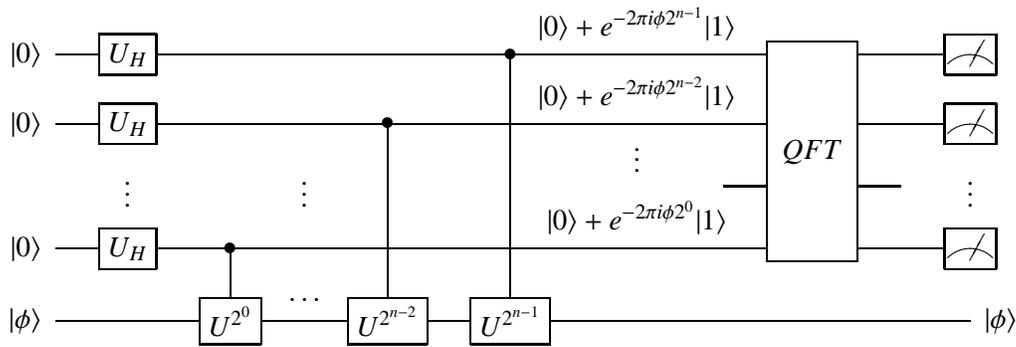
\begin{figure}[h]
\begin{equation}
\Qcircuit @C=1.5em @R=1em {
 \lstick{|0\rangle}  & \gate{U_H} &  \qw & \qw & \qw & \ctrl{4} & \qw & \ustick{|0\rangle + e^{-2 \pi i \phi 2^{n-1}} |1\rangle} \qw& \qw & \qw & \multigate{3}{QFT}&  \qw & \meter \\ 
 \lstick{|0\rangle} & \gate{U_H} &  \qw & \qw & \ctrl{3} & \qw & \qw &\ustick{|0\rangle + e^{-2 \pi i \phi 2^{n-2}} |1\rangle} \qw & \qw & \qw & \ghost{QFT}& \qw & \meter \\
                                &   \vdots       &            &  \vdots &          &    &  &                       \ustick{\vdots}                                     & & & \ghost{QFT}&   \qw       &    \vdots \\
 \lstick{|0\rangle} & \gate{U_H} &  \ctrl{1} & \qw & \qw   & \qw & \qw & \ustick{|0\rangle + e^{-2 \pi i \phi 2^{0}} |1\rangle} \qw & \qw & \qw & \ghost{QFT}& \qw & \meter \\
 \lstick{|\phi \rangle} &  \qw             &  \gate{U^{2^0}} & \ustick{\cdots} \qw &  \gate{U^{2^{n-2}}} & \gate{U^{2^{n-1}}} & \qw & \qw & \qw & \qw & \qw & \qw & \rstick{|\phi\rangle} \qw  } 
 \nonumber
\end{equation}
\caption{Quantum circuit for the quantum phase-estimation algorithm.} \label{QPE}
\end{figure}

Quantum phase-estimation algorithms were originally introduced by Kitaev in \cite{Kitaev95}, and the basic problem that they aim 
to solve can be stated as follows.
Given a unitary operator $U$ and one of its eigenvectors
$|\phi\rangle$, estimate the phase of the corresponding eigenvalue
$U|\phi\rangle = e^{-2\pi i \phi}|\phi\rangle$, $\phi \in [0,1)$ up to
$n$ bits of accuracy. An efficient solution was found in
\cite{Cleve98} and can be summarized in the following steps, as
represented by the quantum circuit of Fig.\ref{QPE}:

\bigskip 

(i) Prepare the pure state $|\psi^{(i)} \rangle = |0,0, \ldots , 0\rangle |\phi
\rangle$, where $|0,0, \ldots , 0\rangle$ is called the source register state
of $n$ qubits and $|\phi \rangle$ is the target state  where we have stored
the given eigenvector of the unitary operator $U$.

\bigskip

(ii) Apply Hadamard operators
\begin{equation}
U_H^{(i)} =\frac{1}{\surd2}\left( \sigma^x_i + \sigma^z_i \right)
\label{had}
\end{equation}
over all the qubits $i$ in the source state, $i = 1, 2, \ldots , n$.

\bigskip

(iii) Apply bit-wise controlled $U^j$ gates over the target state
as shown in the Fig.\ref{QPE}, where each $U^j$ gate corresponds to
the application of $j$ times the proposed $U$-gate with $j = 0, 1\ldots n-1$.

\bigskip

\bigskip

\begin{figure}[h]
\begin{equation}
\Qcircuit @C=1.5em @R=1em {
   & \gate{U_H} & \gate{U_2} & \ustick{\cdots} \qw & \gate{U_n} & \qw & \qw & \qw & \qw & \qw & \qw &          \qw & \qw & \qw & \qw \\
   & \qw & \ctrl{-1} & \qw & \qw & \gate{U_H} & \gate{U_2} & \ustick{\cdots} \qw & \gate{U_{n-1}} & \qw & \qw & \qw & \qw & \qw & \qw \\
  & \qw & \qw & \qw & \qw & \qw & \ctrl{-1} & \qw & \qw & \ustick{\cdots} \qw & \qw & \qw & \qw & \qw & \qw & \\
       &  \vdots &   & & & & \vdots & & & & & & & \vdots \\
  & \qw & \qw & \qw & \qw & \qw & \qw & \qw & \qw & \qw & \qw & \gate{U_H} & \gate{U_2} & \qw & \qw \\
  & \qw & \qw & \qw & \ctrl{-5} & \qw & \qw & \qw &  \ctrl{-4} & \qw & \qw & \qw & \ctrl{-1} & \gate{U_H} & \qw 
 } 
 \nonumber
\end{equation}
\caption{Canonical decomposition of the quantum Fourier transform
operator. By $U_j$ we denote the unitary gate
$|0\rangle \langle 0| + e^{2 \pi i /2^j}|1\rangle \langle 1 |$, to be controlled $j-1$ qubits below.}  \label{QFtrans}
\end{figure}
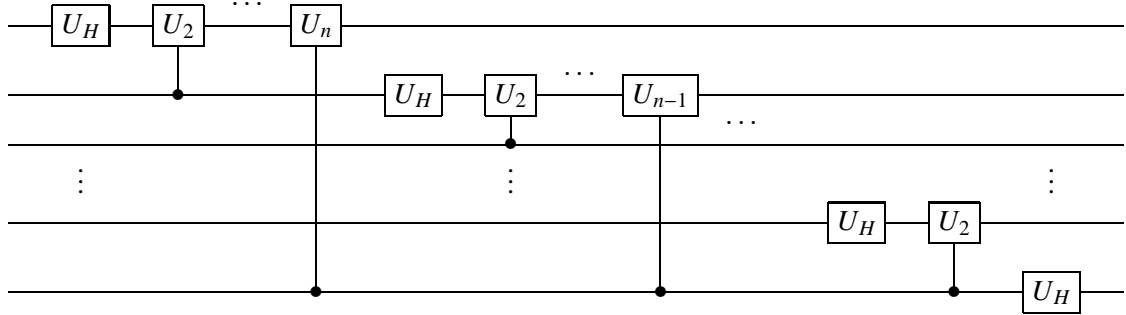

\bigskip

(iv) Apply the  quantum Fourier transform operator
\begin{equation}
 QFT|q\rangle =
\frac{1}{2^{n/2}}\sum_{q'=0}^{2^n-1}e^{2\pi i q q'/2^n}|q'\rangle
\label{fourier}
\end{equation}
over the source register state.

\bigskip

(v) Make a measurement of the source state of the system. This
provides with high probability the corresponding eigenvalue of $U$
with the required precision.

\subsection{Analytical results}

Let us now go through the steps of the algorithm focusing on how the
majorization of the considered set of probabilities of the computational states
evolve.
The application of the Hadamard gates in step (ii) to the initial state
 produces a lowest element of majorization by means of step-by-step reverse majorization,
\begin{equation}
|\psi^{(ii)} \rangle = 2^{-n/2}\sum_{x=0}^{2^n-1}|x \rangle |\phi
\rangle \ ,
\label{stepdos}
\end{equation}
yielding the probability distribution $p_x^{(ii)} = 2^{-n} \ \forall
x$. The outcome of the controlled $U^j$ gates in step (iii) is
the \emph{product} state
\begin{eqnarray}
\label{steptres}
& |\psi^{(iii)} \rangle = 2^{-n/2}\left(|0\rangle
+ e^{-2 \pi i 2^{n-1}
 \phi}|1\rangle \right)\cdots \left(|0\rangle + e^{-2 \pi i 2^0 \phi}|1\rangle \right)|\phi\rangle
   \nonumber \\
& = 2^{-n/2}\sum_{x=0}^{2^n-1}e^{-2 \pi i x \phi }|x\rangle
|\phi\rangle \ .
\end{eqnarray}
Since the action of these gates adds only local phases in
 the computational basis, the uniform distribution for the
 probabilities
is maintained ($p_x^{(iii)}=2^{-n} \ \forall x$).

Verifying majorization for the global action of the quantum Fourier transform is simple.
After step (iv) the quantum state becomes
\begin{equation}
 |\psi^{(iv)} \rangle = 2^{-n}\sum_{x,y=0}^{2^n-1}e^{-2 \pi i x (\phi - y/2^n) }|y\rangle |\phi\rangle \ .
\label{stepcuatro}
\end{equation}
We then have the probability distribution
\begin{equation}
  p_{y}^{(iv)}=\left|2^{-n} \sum_{x=0}^{2^n-1}e^{-2 \pi i x (\phi -
  y/2^n)}\right|^2 \qquad \forall y\ .
\label{finalprob}
\end{equation}
Global majorization between steps (ii) and (iv)
holds \cite{LM02}.
The remaining step (v) corresponds to a measurement whose
output is controlled with the probability distribution
$p_y^{(iv)}$.

While global majorization of the probability distribution is somehow straightforward to see, 
step-by-step majorization is less obvious. To this aim, the mathematical result that we shall prove reads as follows:
the quantum Fourier transform  majorizes step-by-step the
probability distribution calculated in the computational basis
as used in the quantum phase-estimation algorithm. 
This fact is seen to emerge from two important properties. 
It is, first, essential that the initial state entering the quantum Fourier transform has
a certain symmetry to be discussed. Second, the order of the
action of Hadamard and controlled-phase gates maintains as much of
this symmetry as to be used by the rest of the algorithm. To be precise,
Hadamard gates take the role of majorizing the
probability distribution as long as some relative phases are properly
protected. Controlled-phase transformations do preserve such a
symmetry, as we shall see. 

The above property arises in three steps: the first one consists on
a majorization lemma, the
second one is a lemma concerning the preservation of phases, and finally
the third one is the analysis of the controlled-phase
operators in the quantum Fourier transform.  As hinted above, we shall observe that
the only relevant operators for the majorization procedure are the
Hadamard gates acting over the different qubits, while
controlled-phase operators, though providing entanglement,
turn out to be immaterial for majorization purposes.

\subsubsection{A majorization lemma}

Let us first introduce the concept of  ``H($j$)-pair'', central to
this discussion. Consider a Hadamard gate  $U_{H}^{(j)}$ acting on qubit $j$ of
the quantum register. In general, the quantum register would
correspond to a superposition of states. This superposition
can be organized in pairs, each pair being characterized
by the fact that the Hadamard operation on qubit $j$ will mix
the two states in the pair. Let us illustrate
this concept with the example of a general quantum state of two qubits:
\begin{eqnarray}
|\psi \rangle &&  =  \alpha|00\rangle + \beta|01\rangle + \gamma|10\rangle + \delta|11\rangle  \nonumber \\
&& = \underbrace{ (\alpha|00\rangle + \gamma|10\rangle)}_{{\rm H(0)-pair}} + \underbrace{(\beta|01\rangle + \delta|11\rangle)}_{{\rm H(0)-pair}} \nonumber \\
&& = \underbrace{ (\alpha|00\rangle + \beta|01\rangle)}_{{\rm H(1)-pair}}+ \underbrace{ (\gamma|10\rangle + \delta|11\rangle )}_{{\rm H(1)-pair}} \ .
\label{pairs}
\end{eqnarray}
The second line corresponds to 
organizing the state as $H(0)$-pairs, because each pair
differs only on the $0$th qubit value. The third line, instead, 
organizes the state on $H(1)$-pairs, since each
pair differs only on the first qubit value.  We now formulate the following
lemma:

\bigskip

\textbf{Lemma 6.1:} {\it Let $|\psi \rangle$ denote a pure
quantum state of $n$ qubits,
 with the property that the probability amplitudes of the
computational H($j$)-pairs differ only by a phase for a
given qubit $j$. Then, the probability distribution resulting from
$U_{H}^{(j)}|\psi \rangle$ in the computational basis majorizes the one resulting from $|\psi \rangle$.}

\bigskip

\emph{Proof:}
The state $|\psi \rangle$ can
always be written as:
\begin{eqnarray}
& |\psi \rangle = a_1|0, 0, \ldots, 0^j, \ldots , 0 \rangle + a_1e^{i \delta_1}|0, 0,  \ldots , 1^j, \ldots , 0 \rangle \nonumber \\
& + \cdots + a_{2^{n-1}}|1, 1,  \ldots , 0^j , \ldots , 1 \rangle +
a_{2^{n-1}}e^{i \delta_{2^{n-1}}}|1, 1, \ldots , 1^j ,\ldots 1 \rangle \ .
\label{majlemma1}
\end{eqnarray}
The above expression makes it explicit that the amplitudes for every pair of states
that can be mixed by a Hadamard transformation on the qubit $j$ only
differ by a phase. The Hadamard gate $U_{H}^{(j)}$ will mix all these
pairs. The two states in every pair
are equal in all their qubits except for the $j$th one. After
the application of the $U_{H}^{(j)}$ we have
\begin{eqnarray}
& U_{H}^{(j)}|\psi \rangle = 2^{-1/2} \biggl( a_1\left(1+e^{i\delta_1} \right) |0, 0, \ldots , 0^j,
\ldots , 0 \rangle + a_1\left(1-e^{i \delta_1} \right) |0 , 0, \ldots , 1^j , \ldots ,0 \rangle \nonumber \\
& +  \cdots + a_{2^{n-1}} \left(1+e^{i\delta_{2^{n-1}}} \right) |1, 1, \ldots , 0^j , 
\ldots , 1 \rangle + a_{2^{n-1}} \left(1-e^{i \delta_{2^{n-1}}} \right) |1, 1, \ldots ,
1^j ,\ldots , 1 \rangle \biggr) \ . \nonumber \\
&  
\label{majlemma2}
\end{eqnarray}
We have to find a set
of probabilities $p_k$ and permutation matrices $P_k$  such that
\begin{equation}
\textstyle \begin{pmatrix}
  |a_1|^2 \\
  |a_1|^2 \\
   \vdots \\
  |a_{2^n-1}|^2 \\
  |a_{2^n-1}|^2 \\
\end{pmatrix}
=\sum_k p_k P_k
\begin{pmatrix}
 |a_1|^2(1+\cos{(\delta_1)}) \\
  |a_1|^2(1-\cos{(\delta_1)}) \\
   \vdots \\
  |a_{2^n-1}|^2(1+\cos{(\delta_{2^n-1})}) \\
  |a_{2^n-1}|^2 (1-\cos{(\delta_{2^n-1})})\\
\end{pmatrix} \ ,
\label{majlemma3}
\end{equation}
and the unique solution to this probabilistic mixture is
\begin{eqnarray}
&  p_1 = p_2 = \frac{1}{2} \nonumber \\
& \textstyle P_1 =\begin{pmatrix}

  1 &        &          &          &     \\
    & 1      &          &          &     \\
    &        &   \ddots &          &     \\
    &        &          &  1       &     \\
    &        &          &          & 1   \\

\end{pmatrix}
;  \quad P_2 =
\begin{pmatrix}
 0 & 1 &        &   &   \\
 1 & 0 &        &   &   \\
   &   & \ddots &   &   \\
   &   &        & 0 & 1 \\
   &   &        & 1 & 0 \\
\end{pmatrix} \ .
\label{majlemma4}
\end{eqnarray}
The permutation matrix
 $P_1$ is nothing but the identity
matrix and $P_2$ is a permutation of the probabilities of each
pair which has undergone Hadamard mixing. This completes the
proof of the lemma. $\square$

\bigskip

The lemma we have just presented states that Hadamard transformations do
order the probability distribution when the input state has a special
structure, namely, those amplitudes to be mixed only differ by 
phases. This is the key element pervading in the quantum phase-estimation algorithm: 
Hadamard transformations and controlled-phase transformations carefully
preserve such a structure when needed, as we shall now see.

\subsubsection{A phase-preservation lemma}

Let us now prove the following lemma:

\bigskip

\textbf{Lemma 6.2:} {\it Consider  the Hadamard gate $U_{H}^{(j)}$ acting on qubit $j$, 
and the quantum state $|\psi^{(iii)}\rangle$ from Eq.\ref{steptres}
with the property that the probability amplitudes of the
computational H($i$)-pairs differ only by a phase which only depends on $i$, 
$\forall i$. Then, the quantum state $U_{H}^{(j)}|\psi^{(iii)}\rangle$ is such that the
H($i$)-pairs differ only by a phase $\forall i \neq j$.}

\bigskip

This lemma implies that the quantum Fourier transform works in such a way that states
to be mixed by Hadamard transformations only differ by a phase all
along the computation, until the very moment when  the Hadamard
operator acts. In other words, the structure of gates 
respects the relative weights of the H($i$)-pairs.

Before proving the Lemma 6.2 let us build some intuition by considering first an example. We start by introducing a new notation for the phases appearing in
the source quantum state of Eq.\ref{steptres} to be operated by the
quantum Fourier transform operator by defining $\beta_x \equiv -2 \pi x \phi $.
Then
\begin{equation}
|\psi^{(iii)} \rangle = 2^{-n/2}\sum_{x=0}^{2^n-1}
e^{i\beta_x}|x\rangle \ .
\label{newsteptres}
\end{equation}
Notice that since $x = \sum_{i = 0}^{n-1} x_i 2^i$, we can write 
\begin{equation}
\beta_x = \sum_{i=0}^{n-1} -2 \pi x_i 2^i \phi \equiv
\sum_{i=0}^{n-1}x_i \alpha_i \ ,
\label{beta}
\end{equation}
where $\alpha_i \equiv -2 \pi 2^i \phi $.
As an example of this notation, let us write the state $|\psi^{(iii)}\rangle$ in the case
of three qubits:
\begin{eqnarray}
|\psi^{(iii)} \rangle =& &\frac{1}{2^{3/2}} \left(|000 \rangle + e^{i
 \alpha_2}|100
 \rangle + e^{i \alpha_1}|010 \rangle +  e^{i(\alpha_2+ \alpha_1)}|110 \rangle \right) \nonumber \\
&+&\frac{1}{2^{3/2}}\left(|001 \rangle + e^{i \alpha_2}|101
 \rangle
+ e^{i \alpha_1}|011 \rangle +  e^{i(\alpha_2 + \alpha_1)}|111
 \rangle \right)e^{i \alpha_0} \ .
\label{example1}
\end{eqnarray}
We have factorized the $\alpha_0$ phase in the second
line of the above equation. Alternatively, we can choose to factorize $\alpha_1$,
\begin{eqnarray}
|\psi^{(iii)} \rangle =& &\frac{1}{2^{3/2}} \left(|000 \rangle + e^{i \alpha_2}|100 \rangle + e^{i \alpha_0}|001 \rangle +  e^{i(\alpha_2 + \alpha_0)}|101 \rangle \right) \nonumber \\
&+& \frac{1}{2^{3/2}}\left(|010 \rangle + e^{i \alpha_2}|110
\rangle + e^{i \alpha_0}|011 \rangle +  e^{i(\alpha_2 +
\alpha_0)}|111 \rangle \right)e^{i \alpha_1} \ ,
\label{example2}
\end{eqnarray}
or $\alpha_2$,
\begin{eqnarray}
|\psi^{(iii)} \rangle = & &\frac{1}{2^{3/2}} \left(|000 \rangle + e^{i \alpha_1}|010 \rangle + e^{i \alpha_0}|001 \rangle +  e^{i(\alpha_1 + \alpha_0)}|011 \rangle \right) \nonumber \\
&+& \frac{1}{2^{3/2}} \left(|100 \rangle + e^{i \alpha_1}|110
\rangle + e^{i \alpha_0}|101 \rangle +  e^{i(\alpha_1 +
\alpha_0)}|111 \rangle \right)e^{i \alpha_2} \ .
\label{example3}
\end{eqnarray}
On the whole, the initial state for three qubits can be factorized in
 these three different ways.
This example shows that there are three different possibilities to 
  write the quantum state by focusing on a particular
qubit.
 The above property is easily extrapolated to the general case of
$n$ qubits: we can always write the quantum state $|\psi^{(iii)} \rangle$ in
$n$ different ways by factorizing a particular phase in the second
line.

\bigskip

\emph{Proof:}
In the general case we can factorize the
$\alpha_j$ phase so that the pure state is
written as
\begin{eqnarray}
|\psi^{(iii)}\rangle
 = & &\frac{1}{2^{n/2}} \left( |0, 0, \ldots , 0^{j}, \ldots , 0 \rangle + \cdots + e^{i\sum_{k \neq j} \alpha_k}|1, 1, \ldots , 0^{j} , \ldots , 1 \rangle \right) \nonumber \\
&+& \frac{1}{2^{n/2}} \left( |0, 0, \ldots , 1^{j}, \ldots , 0 \rangle
+ \cdots + e^{i\sum_{k \neq j} \alpha_k}|1, 1,\ldots , 1^{j} , \ldots , 1
\rangle \right)e^{i\alpha_j} \ .
\label{general}
\end{eqnarray}
Then, the action of $U_{H}^{(j)}$ transforms the state as
follows:
\begin{eqnarray}
 U_{H}^{(j)}|\psi^{(iii)} \rangle = & & \frac{(1+e^{i \alpha_{j}})}{2^{(n+1)/2}}
\left( |0 , 0, \ldots , 0^j , \ldots , 0 \rangle + \cdots +
e^{i\sum_{k\neq j} \alpha_k}|1, 1, \ldots ,  0^j , 
\ldots , 1 \rangle \right)
 \nonumber \\
&+&\frac{(1-e^{i
\alpha_{j}})}{2^{(n+1)/2}} \left( |0 , 0 , \ldots , 1^j , \ldots , 0 \rangle + \cdots + e^{i\sum_{k\neq j} \alpha_k} |1, 1, \ldots , 1^j , \ldots , 1
\rangle \right) \ . \nonumber \\
& & 
\end{eqnarray}
The resulting state still preserves the necessary
symmetry property to apply Lemma 6.2 to
the  rest of qubits $i\not= j$. The reason is that the effect of the operator has been
to split the quantum state in two pieces which individually retain the property that all
the H($i$)-pairs differ only by a phase for $i\not= j$. If we now
apply another Hadamard operator over a different qubit, for instance qubit $j-1$,  each of
these two quantum states splits in turn in two pieces
\begin{eqnarray}
& &U_{H}^{(j-1)} U_{H}^{(j)}|\psi^{(iii)} \rangle = \nonumber \\
& &\frac{(1+e^{i \alpha_{j}})(1+e^{i \alpha_{j-1}})}{2^{(n+2)/2}}
 \left( |0,0, \ldots, 0^{j-1}, 0^j ,\ldots,0 \rangle
 + \cdots + e^{i \beta_{\tilde{x}}}|1,1, \ldots , 0^{j-1} , 0^j ,
\ldots , 1\rangle \right)
 \nonumber \\
&+& \frac{(1+e^{i \alpha_{j}})(1-e^{i \alpha_{j-1}})}{2^{(n+2)/2}}
 \left( |0 , 0, \ldots ,1^{j-1},  0^j, \ldots, 0 \rangle
 + \cdots + e^{i \beta_{\tilde{x}}}|1, 1, \ldots , 1^{j-1}, 0^j ,
\ldots , 1 \rangle \right)
 \nonumber \\
&+& \frac{(1-e^{i \alpha_{j}})(1+e^{i \alpha_{j-1}})}{2^{(n+2)/2}}
 \left( |0, 0, \ldots , 0^{j-1},  1^j , \ldots, 0\rangle
 + \cdots + e^{i \beta_{\tilde{x}}}|1, 1, \ldots , 0^{j-1} ,1^j ,
\ldots ,1 \rangle \right)
 \nonumber \\
&+& \frac{(1-e^{i \alpha_{j}})(1-e^{i \alpha_{j-1}})}{2^{(n+2)/2}}
 \left( |0, 0, \ldots , 1^{j-1},  1^j ,\ldots , 0 \rangle
 + \cdots + e^{i \beta_{\tilde{x}}}|1, 1, \ldots , 1^{j-1} , 1^j ,
\ldots , 1 \rangle \right)  \ , \nonumber \\
& &
\label{sgon}
\end{eqnarray}
where $\beta_{\tilde{x}}$ is the phase defined in Eq.\ref{beta} for the $n$-bit string 
$\tilde{x} = (1,1, \ldots , 0^{j-1} , 0^j ,\ldots , 1)$. 
The register now consists of
a superposition of four quantum states, each one made of amplitudes
that only differ by a phase.
Further application of a Hadamard gate over yet a different qubit
would split each of the four states again in two pieces in a way that the
symmetry would again be preserved within each piece. This splitting takes place each
 time a particular Hadamard acts. Thus, all Hadamard gates operate
 in turn producing majorization while not spoiling the symmetry
property needed for the next step. This completes the proof of the
 phase-preserving Lemma 6.2. $\square$

\subsubsection{Analysis of the controlled-phase operators}

It is still necessary to verify that the action of
controlled-phase gates does not interfere with the majorization
action carried by the Hadamard gates. Let us concentrate on the
action of $U_{H}^{(n-1)}$,
 which is the first Hadamard operator applied in the
 canonical decomposition of the quantum Fourier transform. Originally we had
\begin{eqnarray}
|\psi^{(iii)} \rangle = & &\frac{1}{2^{n/2}} \left( |0, 0, \ldots, 0
 \rangle + \cdots + e^{i\sum_{k \neq n-1} \alpha_k}|0, 1, \ldots , 1  \rangle \right)  \nonumber \\
&+&  \frac{1}{2^{n/2}} \left( |1, 0, \ldots,0  \rangle + \cdots
+ e^{i\sum_{k \neq n-1} \alpha_k}|1, 1, \ldots , 1 \rangle
\right)e^{i\alpha_{n-1}}  \ ,
\label{first}
\end{eqnarray}
where we have taken the $\alpha_{n-1}$ phase-factor out.
After the action of  $U_{H}^{(n-1)}$ we get
\begin{eqnarray}
  U_{H}^{(n-1)}|\psi^{(iii)} \rangle  =& &  \frac{(1+e^{i \alpha_{n-1}})}{2^{(n+1)/2}}
\left(|0, 0,\ldots,0 \rangle + \cdots + e^{i\sum_{k\neq n-1} \alpha_k}|0, 1, \ldots,1\rangle \right) \nonumber \\
& +& \frac{ (1-e^{i \alpha_{n-1}})}{2^{(n+1)/2}} \left(|1, 0, \ldots , 0\rangle + \cdots + e^{i\sum_{k\neq n-1} \alpha_k} |1, 1 ,\ldots , 1 \rangle
\right) 
 \equiv  |a\rangle + |b\rangle \ . \nonumber \\
 & & 
\label{second}
\end{eqnarray}
We repeat our previous observation that the  state resulting
from the action of
 $U_{H}^{(n-1)}$
 can be considered as the sum of two states, which we
have called $|a \rangle$  and $|b\rangle$. For each of these two
states the amplitudes of the $H(i)$-pairs
$\forall i \neq n-1$ still differ only by a phase.

We can now analyze the effect of the controlled-phase operators.
Following the structure of the quantum Fourier transform operator (see Fig.\ref{QFtrans})
we  focus on what happens after applying a general controlled-phase operator on
the $(n-1)$th qubit of the  quantum state
$U_{H}^{(n-1)}|\psi^{(iii)}\rangle$ (the following procedure is easily
extrapolated to the controlled-phase operators acting over the
rest of the qubits). If the control qubit is the $l$th one, $l
\neq n-1$, then the operator will only add phases over those
computational states from Eq.\ref{second} such that both the $(n-1)$th
and the $l$th qubits are equal to $1$, so we see that it will
only act on the $|b\rangle$ state. Let us
write $|b\rangle$ by factorizing  the $l$th phase as follows:
\begin{eqnarray}
 |b \rangle = & &\frac{(1-e^{i\alpha_{n-1}})}{2^{(n+1)/2}} \left(
|1,0,\ldots, 0^l, \ldots, 0  \rangle + \cdots +
e^{i\sum_{k\neq l, n-1}\alpha_k}|1,1,\ldots, 0^l,
\ldots,1 \rangle \right) \nonumber \\
 & +& \frac{ (1-e^{i\alpha_{n-1}})}{2^{(n+1)/2}}
\left(|1,0,\ldots, 1^l, \ldots,0  \rangle + \cdots
+e^{i\sum_{k\neq l, n-1}\alpha_k}|1,1,\ldots ,
1^l, \ldots,1 \rangle \right)e^{i\alpha_l} \ . \nonumber \\ 
& & 
\label{de}
\end{eqnarray}
It is now clear  that the action of the controlled-phase gate only adds a global phase in the
second piece of $|b\rangle$, which can always be absorbed by means of a convenient 
redefinition of the phase $\alpha_l$. Hence we see that no
relevant change is made in the quantum state concerning
majorization, because the amplitudes of the computational
$H(i)$-pairs $\forall i \neq n-1$  still differ only by a
single phase which only depends on $i$. The action of controlled-phase operators only
amounts to a redefinition of  phases, which does not affect the
necessary property for the Lemma 6.1 to hold. We see
that the needed phase redefinition can be easily made each time one of
these operators acts over a particular qubit.

From all the above considerations and lemmas, it immediately follows that the quantum Fourier transform operator majorizes
step-by-step the probability distribution in phase-estimation
algorithms, as we wished to show. 
We wish to emphasize the fact that controlled-phase operators play no role on majorization,
 though they provide entanglement. On the contrary, local Hadamard operators act
exactly in the complementary way, providing majorization
without providing entanglement.
We also note  that the majorization
arrow in the quantum algorithm is based
on two ingredients. On the one hand we have the special
 properties of the quantum state, and on the other hand we have
 the structure of the quantum Fourier transform. A quantum Fourier transform acting on an arbitrary state
would fail to obey majorization. 

One may be tempted to say at this point that Shor's quantum factoring algorithm \cite{Shor94} obeys a majorization arrow, since it can be
 completely understood in terms of a certain quantum-phase estimation algorithm, as we already saw in Chapter 4 (see Fig.\ref{shor2}). 
 Notice, though, that there is a subtle but key difference between the quantum phase-estimation procedure explained here and the one being used
 in Shor's algorithm, namely, the target register in Shor's algorithm is not in a particular eigenstate of the unitary operator of Eq.\ref{modular2}, but 
 in a given superposition of all of them. This difference makes step-by-step majorization in Shor's 
 quantum factoring algorithm fail. To see how this actually happens, let us remind that in Shor's quantum factoring algorithm the source state to be
 processed by the $QFT$ operator is not the one from Eq.\ref{steptres}, but the state 
 \begin{equation}
 \sqrt{\frac{r}{2^n}} \sum_{i=0}^{2^n/r-1} |i r + l\rangle \ ,
 \label{shorequation}
 \end{equation}
 for a particular $l = 0, 1, \ldots, r-1$ (or a superposition of all of them according to Eq.\ref{status}), where $r$ is the period of the modular exponentiation function $f(x) = a^x \ {\rm mod} \ N$, 
 with a randomly chosen $a \in [1,N]$, $N$ being the number to be factorized. The number of qubits $n$ of the source register is chosen 
 such that $2^n \in [N^2, 2N^2]$. The non-trivial instances of Shor's algorithm come whenever $r$ is both even and $O(N)$, as we saw in Chapter 4. We notice
 that whenever $r$ is even, then $ir+l$ is either even if $l$ is even, or odd if $l$ is odd, $\forall i$. Therefore, the single bit that determines the parity of $ir+l$ 
 will always be either $0$ or $1$, which implies that the corresponding qubit will always be either $|0\rangle$ or $|1\rangle$ in all the states of the superposition
 from Eq.\ref{shorequation}. It is clear, then, that the action of a Hadamard gate on that specific qubit does not majorize the probability distribution of the final outcomes. Even in the 
 case of removing that qubit from the register, there typically are other qubits in the quantum state from Eq.\ref{shorequation} that have the same value in all the states of the superposition, as happens already in the simple case $r=4$, and about which we can not have any a priori information. The whole computation must be then carried without the possibility of removing these qubits, whose evolution breaks step-by-step majorization. Nevertheless, majorization seems to be working locally in the neighborhood of the final peaks of the distribution rather than globally on the whole set of probabilities. 
As a matter of fact, it is also true that our derivations rely very much on the specific
decomposition of the quantum Fourier transform in terms of individual gates.
The underlying quantum circuit is not unique and majorization may not
be present if alternative decompositions are considered. 
   
\subsection{Natural majorization and comparison with quantum searching}

We now turn to investigate further the way majorization has
emerged in the quantum phase-estimation algorithm 
as compared to majorization in other quantum algorithms, 
such as Grover's searching algorithm \cite{Grover96, LM02}. 

For a search in an unstructured database of a particular item, the
best known classical algorithm takes asymptotically
$O(2^n)$ steps in succeeding (where $2^n \equiv N$ is the
number of entries). However, and as we already said in Chapter 4, 
Grover was able to discover a
quantum mechanical algorithm that implements a quadratic speed-up
as compared to the best classical one, that is, Grover's quantum algorithm
makes use of $O(2^{n/2})$ steps. We do not enter here
into precise details about the construction of this quantum
algorithm, and will only make a few comments on the way it proceeds.
The interested reader is addressed to \cite{Grover96}. 

The analysis of Grover's algorithm can be reduced to a two-dimensional Hilbert
 space
 spanned by the state we are searching $|x_0\rangle$
 and some orthogonal state $|x_0^{\bot}\rangle$\cite{Nielsen-Chuang}.
 The unitary evolution of the
quantum state is given by the repeated application of a given 
kernel $K$ which amounts to a rotation
\begin{equation}
\textstyle K =
\begin{pmatrix}
 \cos{(\theta)} &  -\sin{(\theta)}   \\
 \sin{(\theta)} &  \cos{(\theta)}   \\
\end{pmatrix} \ , 
\label{kernel}
\end{equation}
where $\cos{(\theta)} = 1-2/2^n$.
Other choices of kernels are possible but
the one from the above equation is optimal \cite{Galindo01,Galindo02}. The
initial state of the computation is an equal superposition of all
the computational states, written as $|\psi\rangle =
2^{-n/2}|x_0\rangle + \left( 1-2^{-n} \right)^{1/2}|x_0^{\bot}\rangle$
in this two-dimensional notation. For a given intermediate computation
step the state  $(\alpha, \beta)^T$ will be transformed to
 $(\alpha', \beta')^T$. If we wish to express the
initial amplitudes in terms of the final ones, we have:
\begin{equation}
\textstyle
\begin{pmatrix}
 \alpha   \\
 \beta   \\
\end{pmatrix}
=
\begin{pmatrix}
\ \ \ \alpha'\cos{(\theta)} + \beta'\sin{(\theta)} \\
 -\alpha'\sin{(\theta)} + \beta'\cos{(\theta)} \\
\end{pmatrix} \ .
\label{evol}
\end{equation}
We now take the square-modulus of the amplitudes, obtaining:
\begin{eqnarray}
|\alpha|^2 &=& \cos^2(\theta) \ |\alpha'|^2 + \sin^2(\theta) \ |\beta'|^2
 + 2\cos{(\theta)} \sin{(\theta)} \ {\mathcal Re}(\alpha^{\prime *} \beta')  \nonumber \\
 |\beta|^2 &=& \sin^2(\theta) \ |\alpha'|^2 + \cos^2(\theta) \ |\beta'|^2 -
2\cos{(\theta)} \sin{(\theta)} \ {\mathcal Re}(\alpha^{\prime *} \beta') \ .
 \label{relations}
\end{eqnarray}
If the interference terms were to vanish then
 majorization would follow in a straightforward way from the above
relations. 
But it is not the case. Yet
it has been proven that step-by-step majorization in Grover's algorithm exists
\cite{LM02}, although the way it arises is not so directly
related to
the unitary evolution in the way suggested here.

Let us turn back to majorization in the quantum phase-estimation
algorithm and its relation to unitary evolution. We write a generic $n$-qubit state $|\psi\rangle$ to be operated
 by a Hadamard gate acting on the $j$th qubit as

\begin{eqnarray}
& |\psi \rangle = c_0|0, 0, \ldots , 0^j , \ldots ,0 \rangle + c_j|0, 0, \ldots , 1^j , \ldots , 0 \rangle \nonumber \\
& + \cdots + c_{2^{n}-1-j}|1, 1, \ldots , 0^j , \ldots , 1 \rangle +
c_{2^{n}-1}|1, 1, \ldots , 1^j , \ldots , 1 \rangle \ ,
\label{estinicial}
\end{eqnarray}
where we are focusing on the  coefficients of the different $H(j)$-pairs. Applying
the Hadamard gate over the $j$th qubit we get

\begin{eqnarray}
& U_{H}^{(j)} |\psi\rangle = 2^{-1/2}  (c_0 + c_j)|0 , 0, \ldots, 0^j ,\ldots ,0 \rangle + 2^{-1/2}(c_0 - c_j)|0 , 0,\ldots ,1^j ,\ldots,0  \rangle  \nonumber \\
& + \cdots + 2^{-1/2}(c_{2^{n}-1-j}+c_{2^n-1})|1, 1, \ldots , 0^j , \ldots , 1 \rangle \nonumber \\
& + 2^{-1/2}(c_{2^ n-1-j} - c_{2^{n}-1})|1, 1, \ldots , 1^j , \ldots , 1 \rangle  \ .
\label{estfinal}
\end{eqnarray}

For a given pair of original amplitudes $c_{m-j}$ and $c_m$ we now find final amplitudes $c'_{m-j}$ and $c'_m$ to be related
to the initial ones as follows: 

\begin{equation}
\textstyle
\begin{pmatrix}
 c_{m-j}   \\
 c_m  \\
\end{pmatrix}
= \frac{1}{2^{1/2}}
\begin{pmatrix}
c'_{m-j} + c'_m \\
c'_{m-j} - c'_m  \\
\end{pmatrix} \ .
\label{evol2}
\end{equation}
Taking the square-modulus of the amplitudes in the above expression we have
\begin{eqnarray}
 |c_{m-j}|^2 &=& \frac{1}{2} |c'_{m-j}|^2 + \frac{1}{2} |c'_m|^2
 +{\mathcal Re}(c^{\prime *}_{m-j} c'_m)  \nonumber \\
|c_{m}|^2 &=& \frac{1}{2} |c'_{m-j}|^2 + \frac{1}{2} |c'_m|^2
 -{\mathcal Re}(c^{\prime *}_{m-j} c'_m)  \ .
\label{relmodul}
\end{eqnarray}

As in the Grover's previous example, we observe that if
interference terms disappeared majorization would
arise from this set of relations. In such a case, we  would only have to choose the
set of probabilities and permutation matrices given in Eq.\ref{majlemma4}
to prove majorization.  For those
terms to vanish,  very specific properties
for the coefficients $c_{m-j}$ and $c_m$ must hold. It
can be checked that the interference terms
vanish if and only if
\begin{eqnarray}
 c_{m-j} &=& a_{m-j} \nonumber \\
 c_m &=& a_{m-j}e^{i \delta_{m-j}}  \ ,
\label{ifandonlyif}
\end{eqnarray}
where $a_{m-j}$ is real.

The above case is indeed the case of quantum phase-estimation
algorithms. Recalling our previous lemmas, it is possible to see that the
interference terms vanish also step-by-step, and therefore step-by-step majorization
arises as a natural consequence of the unitary evolution of the algorithm. Notice that the quantum state from 
Eq.\ref{steptres} has a very specific structure so that natural majorization
is verified step-by-step along the evolution through the quantum Fourier transform circuit. 
In a way we can say that previous steps in the algorithm prepare the source state 
in this particular and unique form, in order to be processed by the $QFT$ operator.

\subsection{The quantum hidden affine function determination algorithm}

We now wish to see how all the above properties work in a specific example of 
quantum algorithm, namely, we study majorization in a quantum algorithm
solving a particular hidden affine function problem \cite{Bernstein97} as a generalization
of Deutsch's problem \cite{Deutsch85}. Further studies have
provided a range of
fast quantum algorithms for solving different generalizations  
 \cite{Cleve98,Mosca99}. The case that we present here 
  is one of the multiple variations that appear in
\cite{Mosca99}, but our main results are also valid 
 for the whole set of quantum
algorithms that solve similar situations. As we shall see, this algorithm can indeed be understood
in terms of a slight variation of the general quantum phase-estimation algorithm previously discussed.

Let us consider the following problem \cite{Mosca99}: given an integer-$N$ 
function $f: \mathbb{Z}_N \rightarrow \mathbb{Z}_N$, $f(x) = mx + b$, where
$x, m, b \in \mathbb{Z}_N$, find out the value of $m$.
A classical analysis reveals that no information about $m$ can
be obtained with only one evaluation of the function $f$.
Conversely, given the unitary operator $U_f$ acting in a
reversible way such that

\begin{equation}
U_f |x\rangle |y\rangle = |x\rangle |y + f(x)\rangle \ ,
\label{f}
\end{equation}
-- where the sum is to be interpreted as modulus $N$ -- there is a
quantum algorithm solving this problem with only one single query to
$U_f$. The requested quantum algorithm proceeds as follows: let us take $N=2^n$, $n$ being the number
of qubits. Perform then the following steps:

\bigskip

(i) Prepare two $n$-qubit registers (source and target) in the state $|0, 0,  \ldots , 0 \rangle |\psi_1 \rangle$, where $|\psi_1 \rangle = QFT^{-1}|1, 1, \ldots, 1\rangle$, and 
$QFT^{-1}$ denotes the inverse quantum Fourier transform in a Hilbert space of dimension $N$.

\bigskip

(ii) Apply the operator $QFT$ over the source register.

\bigskip

(iii) Apply the operator $U_f$ over the whole quantum state (source and target registers).

\bigskip

(iv) Apply the operator $QFT^{-1}$ over the source register.

\bigskip

(v) Measure the source register and output the measured value.

\bigskip

\begin{figure}[h]
\begin{equation}
\Qcircuit @C=1.5em @R=1em {
 \lstick{|0\rangle^{\otimes n}}  & \ustick{(n)} \qw & \gate{QFT} &  \multigate{1}{U_f} & \gate{QFT^{-1}} & \meter  \\
  \lstick{|\psi_1 \rangle}  & \ustick{(n)} \qw &     \qw   & \ghost{U_f} & \qw   & \qw  
}  \nonumber
\end{equation}
\caption{Quantum circuit solving the hidden affine function problem. Both source and target registers are 
 assumed to be respectively composed of $n$ qubits.} 
\label{circ}
\end{figure}
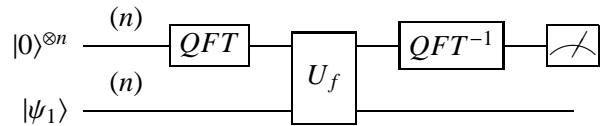

The different steps concerning this process are summarized in Fig.\ref{circ}.

\bigskip

We now show how the proposed quantum algorithm leads
to the solution of the problem. Our analysis raises 
observations concerning the way both entanglement and majorization
behave along the evolution.

In step (i) of the algorithm the quantum state is 
not entangled, since that the quantum Fourier transform -- and
its inverse -- applied 
on a well defined state in the
computational basis leads to a separable state (see, for
example, \cite{Nielsen-Chuang}). 
That is, the quantum state $|0,0,\ldots ,
0\rangle |\psi_1\rangle$ is completely separable.
In step (ii)  the algorithm evolves through a
quantum Fourier transform in the source register.
This action leads to a step-by-step reverse majorization of the probability
distribution of the possible outcomes while it does
not use neither create any entanglement. Moreover,  
natural reverse majorization is at work due to the absence of interference terms.

Next, it is easy to verify that the quantum state
\begin{equation}
|\psi_1 \rangle = \frac{1}{2^{n/2}}\sum_{y=0}^{2^n-1}e^{-2 \pi i y/2^n}|y\rangle
\label{secreg}
\end{equation}
is an eigenstate of the operation $|y\rangle \rightarrow |y +
f(x)\rangle $ with eigenvalue $e^{2 \pi i f(x)/2^n}$. Thus, after 
the third step, the quantum state reads
\begin{equation}
\frac{1}{2^{n/2}}\sum_{x = 0}^{2^n-1}e^{2 \pi i f(x)/2^n}|x\rangle |\psi_1 \rangle = \frac{e^{2 \pi i b/2^n}}{2^{n/2}} \left( \sum_{x=0}^{2^n-1}e^{2 \pi i mx/2^n}|x\rangle \right) |\psi_1 \rangle \ .
\label{equation}
\end{equation}
The probability distribution of possible outcomes has not been
modified, thus not affecting majorization. Furthermore, the pure
quantum state of the first register can be written as
$QFT|m\rangle$ (up to a phase factor), so this step has not eventually 
created any entanglement among the qubits of the system right after the application of the quantum oracle.

In step (iv)  of the algorithm, 
the action of the operator $QFT^{-1}$ over the
first register leads to the state $e^{2 \pi i
b/2^n}|m\rangle |\psi_1\rangle$. A subsequent  measurement in the
computational basis over the first register provides the desired
solution. Recalling our previous results, we see that
the inverse quantum Fourier transform naturally majorizes
step-by-step the probability distribution attached to the
different outputs. Notice also that the separability of the
quantum state still holds step-by-step. This observation completes
our analysis of this example. 

\section{Majorization in adiabatic quantum searching algorithms}

Our aim now is to study the majorization behavior of quantum adiabatic algorithms, which were already considered in the two previous Chapters. 
Here, we choose to analyze a very specific instance of the quantum adiabatic algorithm, namely, we consider the quantum adiabatic algorithm 
that solves the problem of searching in an unstructured database. As we shall see, the effects of a
change of path between the initial and the problem Hamiltonian imply also a change of behavior in the algorithm from 
the majorization's perspective. More concretely, those paths
leading to optimality in the quantum algorithm do lead as well to
step-by-step majorization, while the converse is not necessarily
true. We do not repeat here the details of how do adiabatic quantum algorithms work, 
since they were already explained in Chapter 4. We do, however, sketch a couple of its basic properties. 

The quantum adiabatic evolution method has been successfully
applied to the searching problem \cite{Roland02,Dam01,Das03}.
 Let the initial state be
$|\psi \rangle = \frac{1}{\sqrt{N}}\sum_{x = 1}^N |x\rangle$,
$N$ being  the number of entries of the database, and let the
initial and problem Hamiltonian respectively be $H_0 = I - |\psi\rangle
\langle \psi|$ and $H_P = I - |x_0\rangle \langle x_0|$,
$|x_0\rangle$ being the marked state. The interpolating Hamiltonian $H(s(t)) = (1-s(t)) H_0 + s(t) H_P$ depends on a 
time-dependent parameter $s(t)$ satisfying the boundary conditions $s(0) = 0$ and $s(T) = 1$, $T$ being the computational time of the adiabatic algorithm. 
This scheme leads to different
results depending on whether we apply the adiabatic condition
globally (that is, in the whole time interval $[0,T]$) or locally
(at each time $t$). In what follows, we consider these two
situations without entering into precise details of the involved
calculations.
For further information, we refer the reader to \cite{Roland02, Dam01}
and references therein.

\subsection{Numerical results}

We have performed a numerical analysis of the way in which majorization appears in the quantum adiabatic searching algorithm. 
Our study can be divided into two parts, regarding whether we demand the adiabatic condition to be fulfilled either 
globally or locally along the evolution. 

\subsubsection{Analysis of the fastest global adiabatic evolution}

Let us suppose that we demand the usual adiabatic condition given
in Eq.\ref{cond} of Chapter 4 to be satisfied globally in the whole
interval $[0,T]$. This does not involve any particular restriction
on the $t$-dependence of  $s(t)$, so we can choose $s(t) = t/T$, leading
to a linear evolution of the Hamiltonian. Under these
circumstances, it can be proven \cite{Roland02,Dam01} that the global
adiabatic condition is verified provided that
\begin{equation}
T \geq \frac{N}{\epsilon} \ ,
\label{timmme}
\end{equation}
$\epsilon$ being the probability amplitude of not being at the ground-state 
of $H_P$ at time $T$. 
Hence, this quantum algorithm needs a computational time of $O(N)$ to hit the right 
solution with high probability, so the global adiabatic searching does
not lead to an increasing efficiency with respect to a classical
searching.

In what follows we call $P_+(t)$ the probability of being at the marked
state at time $t$ 
and similarly $P_-(t)$ the probability of being at one of the remaining $N-1$ basis states different from the desired one at time $t$.
Notice
that, given the symmetry of the problem, $P_-(t)$ will 
exactly be the same for all those basis states different from the
marked one all along the evolution. In order to
analyze majorization, we recall the set of inequalities given in
Eq.\ref{deftwo} of Appendix A to be satisfied at each majorizing time
step. Let us make the observation that the maximum probability at all times
is indeed $P_+(t)$, while the other probabilities will remain
smaller than this quantity all along the computation and equal to
$P_-(t)$. It is possible to see that the whole set of $N$ cumulants
that arise from the probability distribution follows the same basic behavior as time flows. 
Because of that, we present here the 
behavior of the first two non-trivial cumulants $P_+(t)$ and 
$P_+(t) + P_-(t)$, as the rest of them do not lead to different 
conclusions.

\begin{figure}
\centering
\includegraphics[angle=-90, width=0.7\textwidth]{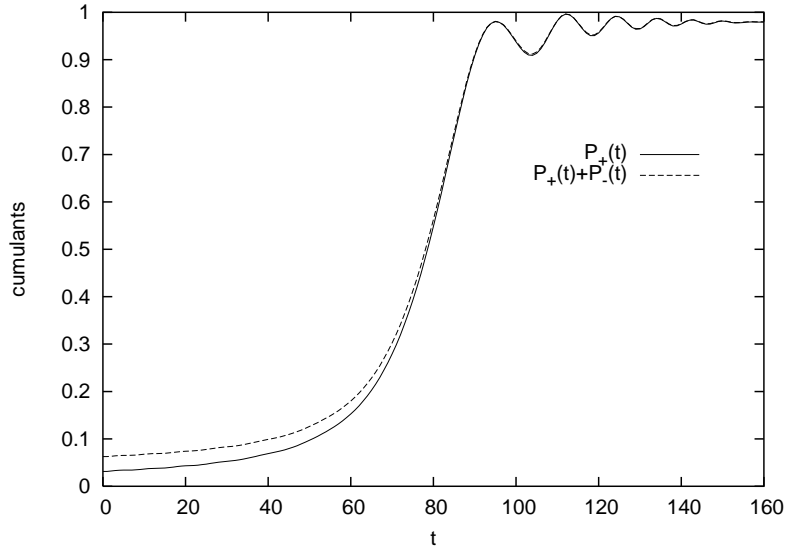}
\caption{Quantum searching using global adiabatic evolution with parameters
$\epsilon = 0.2$, $N = 32$ and $T = 160$.} 
\label{adi1}
\end{figure}

We have performed exact numerical simulations of the quantum algorithm in the fastest allowed
case saturating the bound from Eq.\ref{timmme} ($T = \frac{N}{\epsilon}$) and have found the time 
evolution for the two cumulants. The results for $\epsilon = 0.2$
and $N = 32$ are shown in Fig.\ref{adi1}.
From our numerical analysis we conclude that a naive adiabatic quantum searching
process does not  produce an optimal algorithm neither verifies  
 step-by-step majorization. This property is observed as the two cumulants decrease
in time for some time steps, since there are wiggles which indicate that the system is evolving too fast to remain close enough to the ground state,
and thus not verifying step-by-step majorization along the flow in time. 

\subsubsection{Analysis of the local adiabatic evolution}

\begin{figure}
\centering
\includegraphics[angle=-90, width=0.7\textwidth]{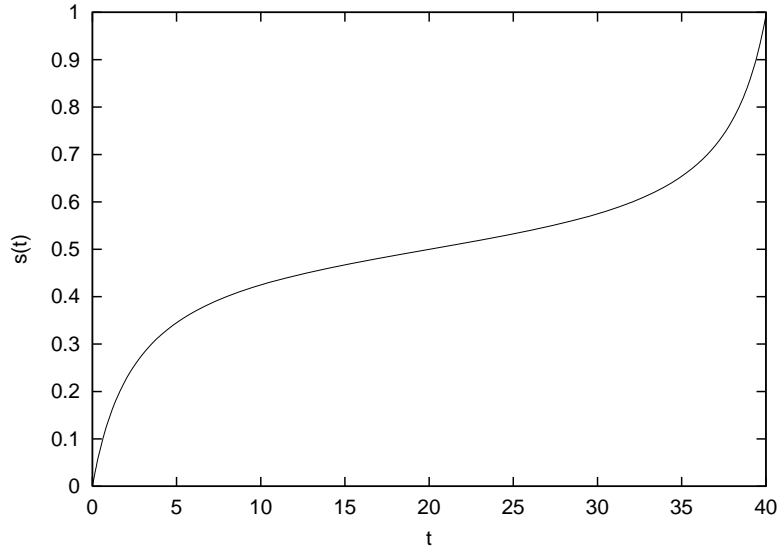}
\caption{Interpolating parameter $s(t)$ for quantum searching using local adiabatic evolution.}
\label{esefig}
\end{figure}

The preceding global adiabatic method can be improved if we apply
the adiabatic condition given in Eq.\ref{cond} of Chapter 4 locally.
That is, let us divide the interval $[0, T]$ into many
small subintervals and let us apply Eq.\ref{cond} to each one of these
subintervals individually. Taking the limit of the size of the
subintervals going to zero, we find that the adiabatic restriction
has to be fulfilled locally at each time $t$:

\begin{equation}
\frac{|\frac{dH_{1,0}}{dt}|}{g^2(t)} \leq \epsilon \qquad \forall t \ ,
\label{cond2}
\end{equation}
where $H_{1,0}$ is the Hamiltonian matrix element between the ground state and the first excited state and $g(t)$
 is the energy gap between these two states, everything given at $t$. 
This is a less demanding condition than Eq.\ref{cond}, and
means that the adiabaticity condition must be satisfied at each
infinitesimal time interval. It can be shown (see, for example,
\cite{Roland02}) that proceeding in this way the function $s(t)$
must have a precise form which is given by the relation

\begin{equation}
t = \frac{1}{2 \epsilon} \frac{N}{\sqrt{N-1}} \left(
\arctan({\sqrt{N-1}(2s-1)}) + \arctan({\sqrt{N-1}})\right) \ .
\label{te}
\end{equation}
We can observe this dependence in Fig.\ref{esefig}, in
the case of $\epsilon = 0.2$ and $N = 32$.
The local adiabatic
process implies that the smaller the energy gap between the
ground and first excited states is, the slower the rate at which the Hamiltonian changes.
With this information it can be proven \cite{Roland02,Dam01}
that the evolution
time for the algorithm to succeed with sufficiently high probability is, in
the limit $N \gg 1$,
\begin{equation}
T = \frac{\pi}{2 \epsilon}\sqrt{N} \ .
\label{time}
\end{equation}
Hence, in the case of local adiabatic evolution the computational
process takes $O(\sqrt{N})$ time, just as in Grover's quantum
searching algorithm, obtaining an square-root speed-up with
respect to the best classical searching.

\begin{figure}
\centering
\includegraphics[angle=-90, width=0.7\textwidth]{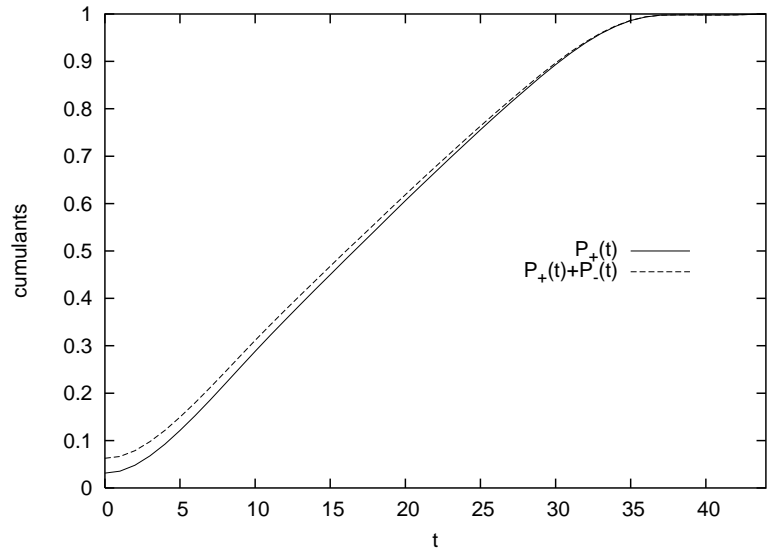}
\caption{Quantum searching using local adiabatic evolution with parameters
$\epsilon = 0.2$, $N = 32$ and $T = 44$.} 
\label{adi2}
\end{figure}

Defining $P_+(t)$ and $P_-(t)$ as before,
we can again restrict ourselves to the study of the two
non-trivial cumulants $P_+(t)$ and $P_+(t) + P_-(t)$ in order to
observe the evolution of majorization. We
have numerically solved the dynamical equations for $\epsilon = 0.2$ and $N
= 32$, and have found the evolution of the two quantities, which
is given in Fig.\ref{adi2}.
From the numerical analysis, it follows that a local adiabatic 
searching algorithm is not only optimal in time, but also verifies 
step-by-step majorization. 

\subsubsection{Analysis of slower global adiabatic evolutions}

Let us now consider global adiabatic
evolutions which are not necessarily tight in time, that is,
extremely slow time variations of the Hamiltonian, much 
slower than the minimum necessary for the adiabatic theorem to
hold. In the case we are dealing with, this implies the
consideration of the case in which $T > \frac{N}{\epsilon}$, that is, 
the adiabatic inequality from Eq.\ref{timmme} is not saturated. 

\begin{figure}
\centering
\includegraphics[angle=-90,
width=0.7\textwidth]{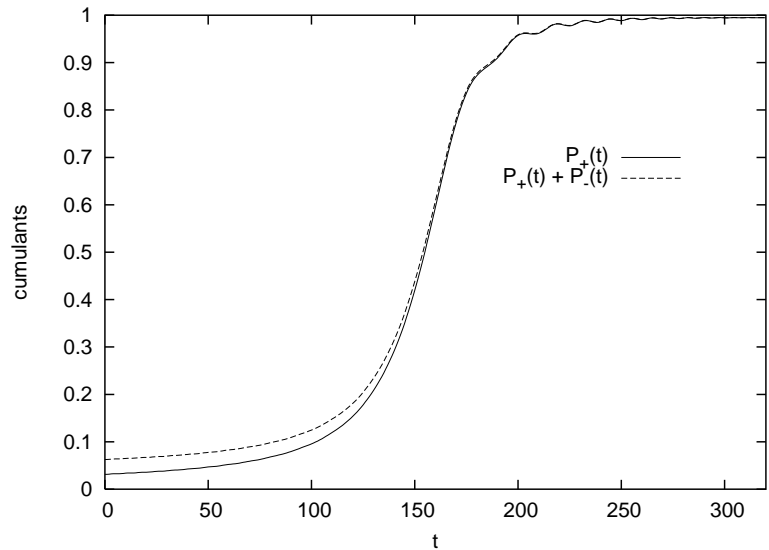}
\caption{Quantum searching using global adiabatic evolution with  parameters $\epsilon = 0.2$, $N = 32$, and
 $T = 320$.} 
\label{adi34}
\end{figure}

\begin{figure}
\centering
\includegraphics[angle=-90, width=0.7\textwidth]{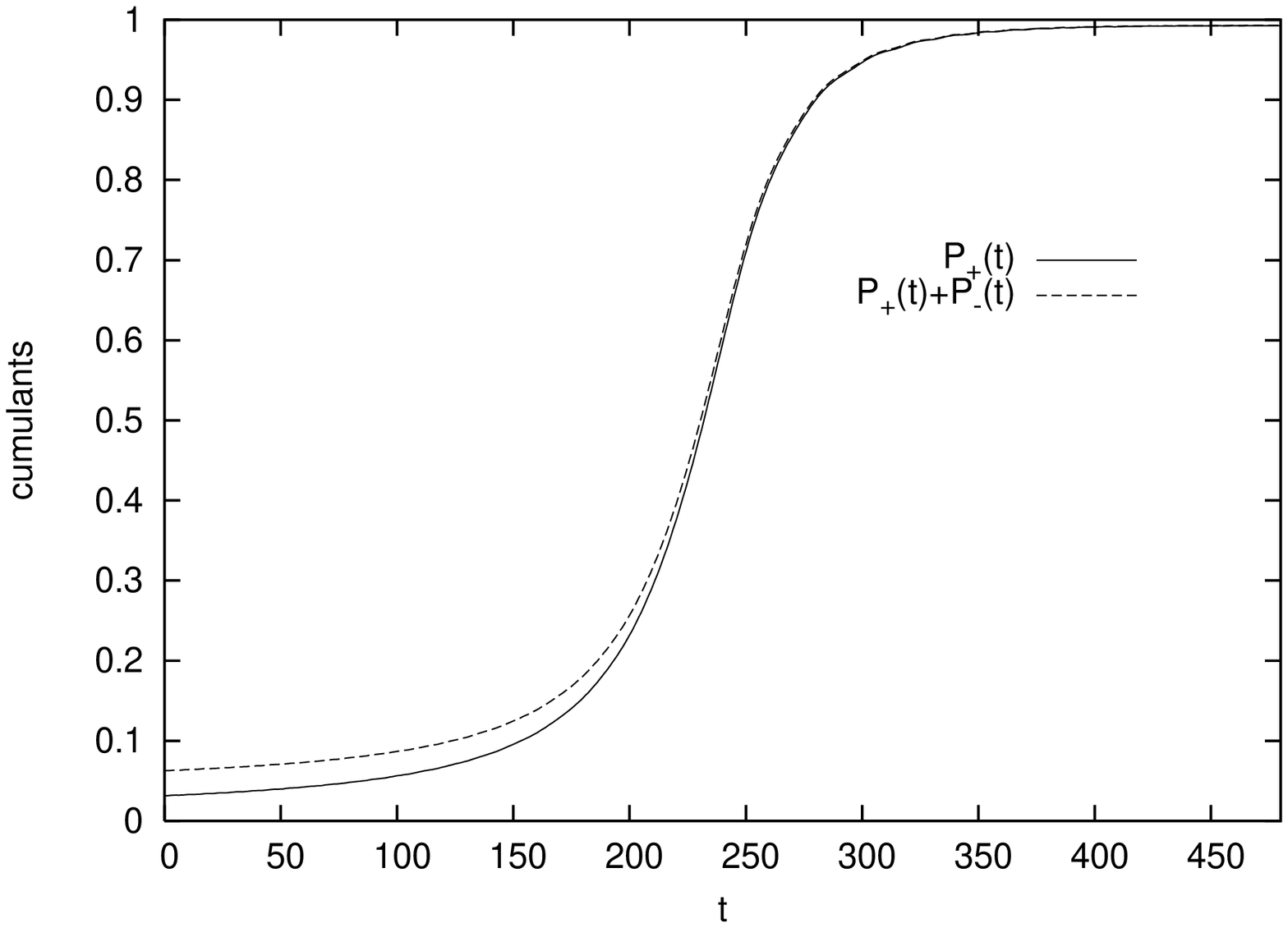}
\caption{Quantum searching using global adiabatic evolution with parameters $\epsilon = 0.2$, $N = 32$, and $T =
  480$.}
\label{adi35}
\end{figure}

We have again performed a numerical analysis for the time evolution
of the two non-trivial cumulants $P_+(t)$ and $P_+(t) + P_-(t)$, for $\epsilon = 0.2$, $N = 32$,
and $T = 320$ and $480$ (both cases bigger than $\frac{N}{\epsilon}
= 160$). The results are plotted in Fig.\ref{adi34} and Fig.\ref{adi35}.
From these two plots, we observe 
that a step-by-step majorization tends to appear as long as the
evolution of the Hamiltonian becomes slower and slower.
From a physical point of view, this means that the probability of ``jumping'' to the
first excited state decreases as long as the evolution is
performed at slower changing rates, thus satisfying better the
assumptions of the adiabatic theorem.
Step-by-step majorization may thus
appear in global adiabatic searching processes for a slow enough evolution
rate. 

\section{Majorization in a quantum walk algorithm 
with exponential speed-up}

The extension of classical random walks to the quantum world has
been widely studied, yielding  two different models of
quantum random walks, namely, those which operate in discrete time
by means of a ``coin operator'' \cite{Aha93, AAKV01, ABNVW01}
 and those  based on a
Hamiltonian evolution in continuous time \cite{Farhi98, Childs01, Childs02_3}.
Regarding the discrete-time model of quantum random walk, two
indicative algorithmic results have been found, namely, an
exponentially fast time when crossing the hypercube with respect to the
classical random walk \cite{Kempe05} and a quantum searching algorithm
achieving Grover's quadratic speed-up \cite{Shenvi03}. As a matter of fact, the first one of
these two results does not provide any algorithmic speed-up, as there exists
 a classical algorithm that solves the hitting problem in the
hypercube exponentially faster than the naive classical random walk,
that is, in a time $O({\rm poly}(\log_2{N}))$ where $N$ is the number of
nodes of the graph (see \cite{Kempe05,Kempe03}). Nevertheless, the second of these examples
shows algorithmic advantage with respect to any possible classical
strategy. The analysis of  the quantum random walk searching
algorithm
shows that  the quantum evolution can be understood as an (approximate)
rotation of the quantum state in a two-dimensional Hilbert space which
is exact in the limit of a very large database (see \cite{Shenvi03} for
details), resembling the original proposal of Grover's searching
algorithm which can be decomposed exactly in a two-dimensional Hilbert
space (see Eq.\ref{evol}). This rotational structure of the evolution implies again step-by-step
majorization when approaching the marked state, exactly in the same
way as the usual Grover's searching algorithm \cite{Grover96,LM02}.

Here we wish to restrict ourselves to the continuous-time 
model of quantum walk and analyze a proposed quantum
algorithm based on a quantum walk on continuous time solving a
classically hard problem \cite{Childs02_3}. We sketch the main ingredients 
of the problem setting and its efficient solution in terms of a quantum evolution (the interested reader is addressed to \cite{Childs02_3} for specific details). For a more generic review on quantum walks both in discrete and continuous time, see \cite{Kempe03}. 

\subsection{The exponentially fast quantum walk algorithm}

\begin{figure}
\centering
\includegraphics[angle=0, width=.5\textwidth]{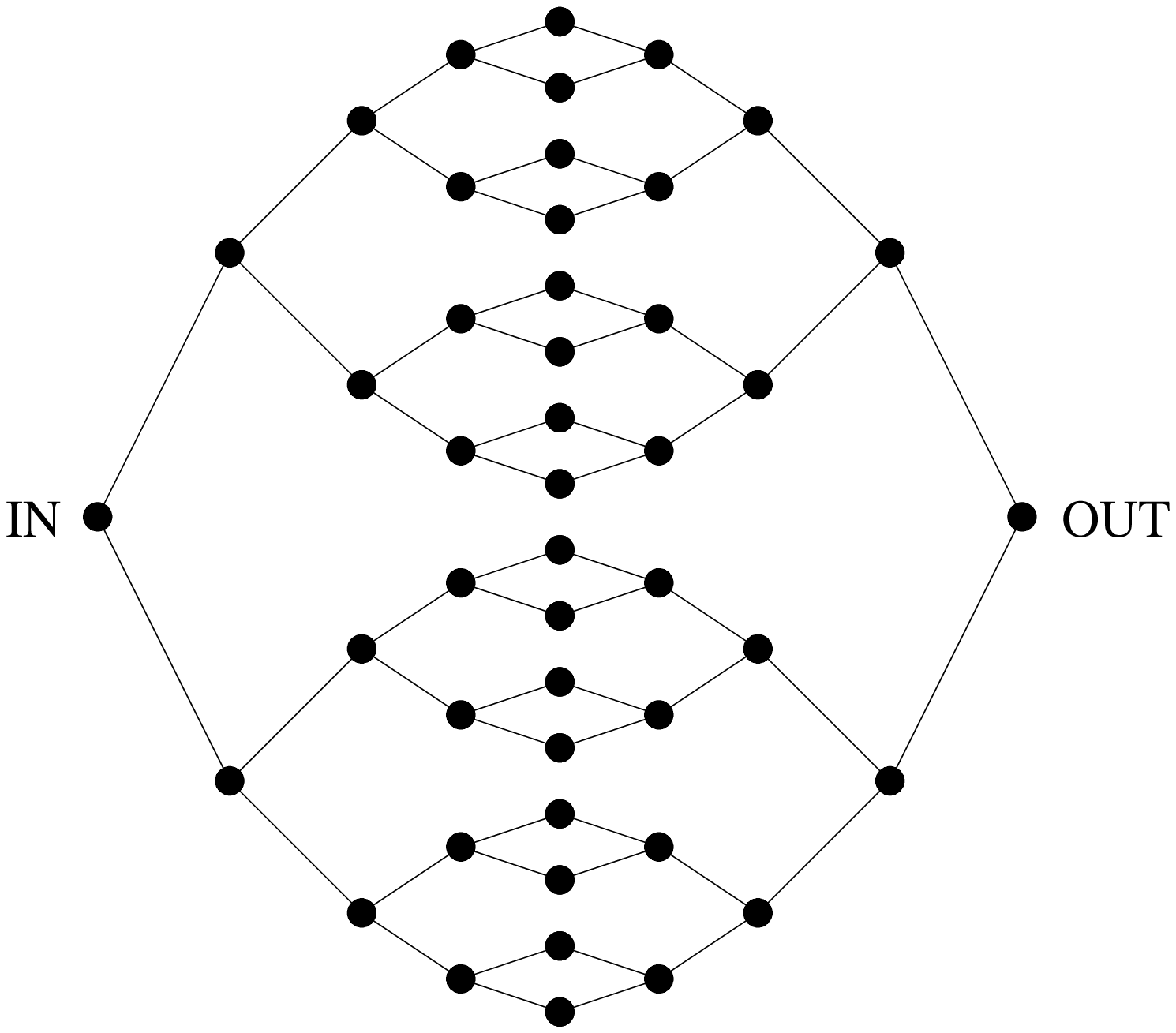}
\caption{A possible graph constructed from two binary trees with $n = 4$.}
\label{ge}
\end{figure}

\begin{figure}
\centering
\includegraphics[angle=0, width=.6\textwidth]{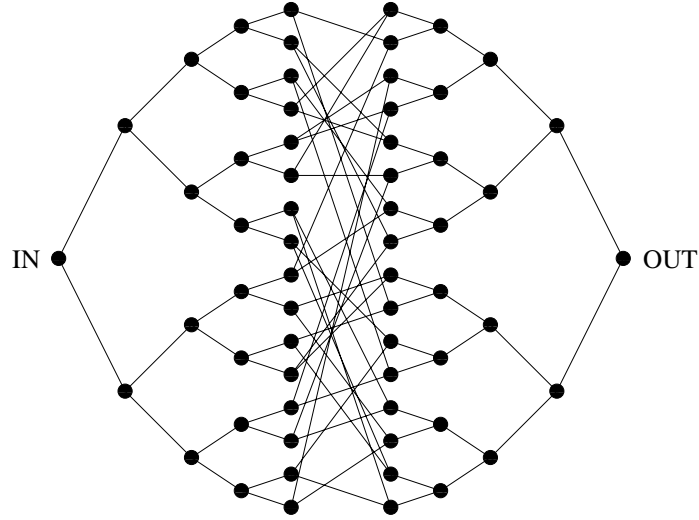}
\caption{An alternative graph constructed from two binary trees
with $n = 4$. Connection between the leaves is made through a
random cycle.} \label{geprime}
\end{figure}

The problem we wish to solve is defined by means of a graph built in the
following way (see \cite{Childs02_3}): suppose we are given two 
balanced binary trees of height $n$ with the $2^n$ leaves 
of the left tree identified with
the $2^n$ leaves of the right tree in a simple way, as shown in
Fig.\ref{ge}. A way of modifying such a graph is to connect
the leaves by a random cycle that alternates between the leaves of
the two trees, instead of identifying them directly. An example of
such a graph is shown in Fig.\ref{geprime}.

Suppose that the edges of such a graph are assigned a consistent
coloring (that is, not two edges incident in the same vertex have
the same color), and that the vertices are each one given a
different name (with a $2n$-bit string, so there are more possible
names than the ones assigned). We now define a black-box that
takes two inputs, a name $a$ given as a $2n$-bit string and a
color $c$, and acts in the following way: if the input name $a$
corresponds to a vertex that is incident with an edge of color
$c$, then the output corresponds to the name of the vertex joined
by that edge; if $a$ is not the name of a vertex or $a$ is the
name of a vertex but there is no incident edge of color $c$, the
output is the special $2n$-bit  string $(1,1,\ldots, 1)$, which is not
the name of any vertex.

Now, the problem we wish to solve reads as follows: 
given a black-box for a graph such as the one previously
described, and given the name of the {\rm IN} vertex, find out the
name of the {\rm OUT} vertex.

In \cite{Childs02_3} it was proven that no classical algorithm can
transverse a graph such as the one in Fig.\ref{geprime} in
polynomial time, given such a black-box. Furthermore, 
an explicit construction of
a quantum algorithm based on a continuous-time quantum walk on the
graph that succeeds in finding the solution for this oracular
problem in polynomial time was given.
The quantum algorithm of \cite{Childs02_3} for this problem  can be briefly summarized as
follows: consider the $(2n+2)$-dimensional subspace spanned by the
states

\begin{equation}
|{\rm col} \ j \rangle = \frac{1}{\sqrt{N_j}}\sum_{a \  \in \  {\rm column} \  j}|a\rangle \ ,
\label{column}
\end{equation}
where $N_j = 2^j$ if $0 \leq j \leq n$ and $N_j = 2^{2n+1-j}$ if
$n+1 \leq j \leq 2n+1$. We call this subspace the ``column
subspace'', and each state of the basis is an equally weighted sum
of the states corresponding to the vertices lying on each column
of the graph. We now define a Hamiltonian acting on this subspace
by the following non-zero matrix elements:

\begin{equation}
\langle {\rm col} \ (j+1) |H|{\rm col} \ j \rangle =
\langle {\rm col} \ j |H|{\rm col} \ (j+1)\rangle = 
\begin{cases}
1 \ , &  {\rm if}   \ 0 \le j \le n-1 \ , \ n+1 \le j \le 2n \\
2^{1/2} \ , & {\rm if} \ j = n \ . 
\end{cases}
\label{chamiltonian}
\end{equation}
The action of this Hamiltonian on the graph is nothing but
promoting transitions between adjoint vertices, so a quantum walk
on the graph (on the whole Hilbert space) generated by this
Hamiltonian is equivalent to a quantum walk on the line (on the
column subspace). Because of that, from now on we only focus our
attention on the quantum walk on the line generated by the
Hamiltonian from Eq.\ref{chamiltonian}. Moreover, it  can be
proven that given the structure of the graph in the form of a
black-box such as the one already described, our Hamiltonian can
be efficiently simulated by means of a quantum circuit \cite{Childs02_3}.

The quantum walk works as follows: at first the ``wave packet''
will be precisely localized at the IN vertex (the initial state
will be $|{\rm col} \ 0\rangle$). Due to the unitary time evolution driven by the Hamiltonian, it
will initially spread out through the different vertices at the
left hand side of the graph (those belonging to the left binary
tree), but after a short time (once half the graph has been
transversed) it will begin to spread through the vertices on the
right hand side, interfering constructively in the OUT vertex as
the time goes on. Physically, this is nothing but a wave
propagation. Should we wait longer, the wave packet would come
back to be localized at IN vertex and the process would similarly be repeated
again. Actually, due to the ``defect'' of the Hamiltonian in the
central vertices, it can be shown that the transmission through the
central columns is not perfect, but high enough for the
OUT node to be achieved with a very high probability in small computational time. In
\cite{Childs02_3} the authors prove that the succeeding time is
polynomial in $n$.

\subsection{Numerical results}

\begin{figure}
\centering
\includegraphics[angle=-90, width=0.7\textwidth]{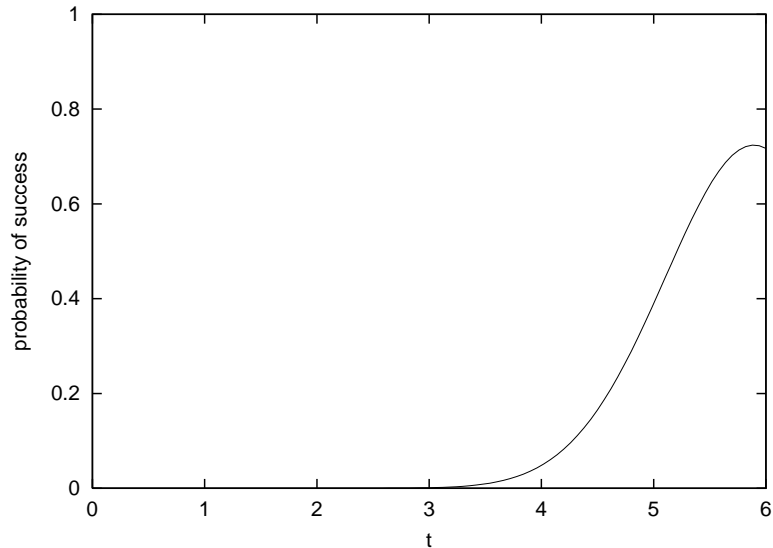}
\caption{Probability of finding the OUT node in the quantum walk algorithm, for $n=4$.}
\label{prob}
\end{figure}

We have numerically simulated this quantum walk for the particular
case of $n =4$, and have plotted the time evolution of the
probability of success in Fig.\ref{prob}. We observe that the
numerical result is in agreement with the prediction that the time
the algorithm takes in hitting the OUT node with high probability 
seems to be, at first sight, linear with the size of the system. 

\begin{figure}
\centering
\includegraphics[angle=-90, width=0.7\textwidth]{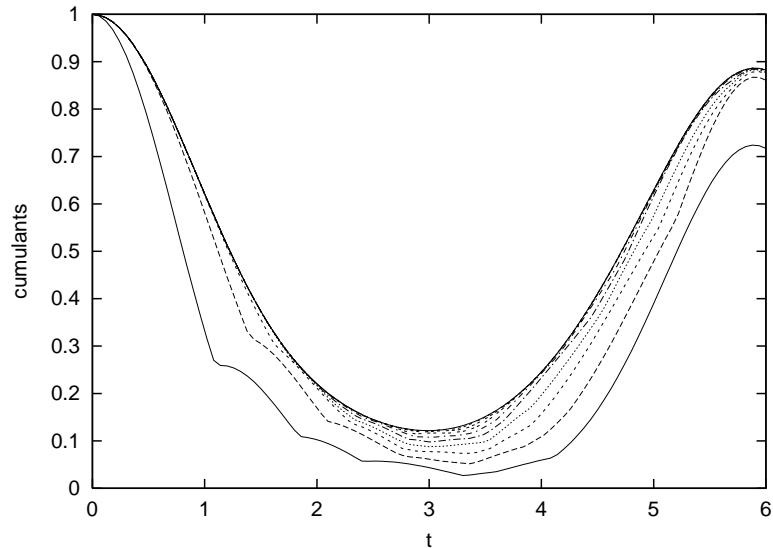}
\caption{Time evolution of the ten cumulants in the quantum walk algorithm when one node per column is considered, for $n=4$. The evolution follows a majorization cycle}
\label{cycle}
\end{figure}

\begin{figure}
\centering
\includegraphics[width=0.7\textwidth]{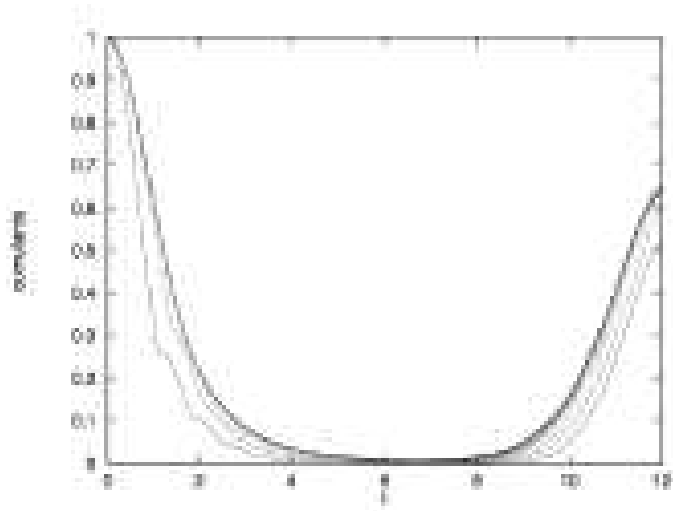}
\caption{Time evolution of the $22$ cumulants in the quantum walk algorithm when one node per column is considered, for $n=10$. The cumulants tend to collapse in the plot given the size of the graph. The evolution follows a majorization cycle.}
\label{cycle2}
\end{figure}

In order to analyze majorization, for
the case $n=4$ there are $62$ cumulants that can be computed from a set of $10$ non-trivial probabilities.
 This is so due to the fact that all the states of the
whole Hilbert space belonging to the same column always share the
same probability amplitude. The quantities to be considered
are then the probabilities of being at each column state
normalized by the number of nodes belonging to that column, that
is, the probability of being in one node of each column. In general, there are
then $2n+2$ different probabilities to be considered at each time
step. Given only these $10$ quantities, we were able to compute 
all of the $62$ cumulants corresponding to all the partial sums of sorted probabilities, according to Eq.\ref{deftwo} in Appendix A.
 In order to make the figures as clear as possible we have only plotted $10$ of these quantities in Fig.\ref{cycle}, 
 which correspond to the cumulants arising from the sorted probabilities when only one node per column is considered. 
 Our numerical simulations indicate that the rest of the cumulants  exhibit a behavior similar to that of the ones appearing in Fig.\ref{cycle} 
 and thus bring no further insight.  
We have also numerically simulated the algorithm in the case of a bigger graph, namely, in the case $n=10$. In this case there are $2n+2 = 22$ different probabilities to be considered at each time step. Proceeding in the same way than in the case $n=4$ (that is, not plotting all the cumulants, but the only the sorted sum of these $22$ probabilities), we obtain a similar behavior as in the case for $n=4$, as is shown in Fig.\ref{cycle2}.

Looking at the two plots, we conclude that the continuous time quantum walk 
 follows a step-by-step majorization
cycle all along the computation until it reaches the OUT
node. It is worth remarking as well that the time the algorithm spends
reversely majorizing the probability distribution is about half of the time
of the whole computation. The physical reason for this behavior
is clear, as this is the time the ``wave packet'' spends spreading
over the binary tree on the left hand side, thus leading to a
destructive interference part. Note that such a
destructive interference indeed strictly follows a step-by-step
reverse majorization of probabilities.
Furthermore, by combining Fig.\ref{prob} and Fig.\ref{cycle} we see 
that the raising of the probability of success is linked
to a step-by-step majorization. Physically, this is the part in
which the algorithm constructively interferes into the OUT node
once the wave packet is approximately in the right-hand-side binary
tree. We see that this constructive interference follows a
majorization arrow. Actually, the observed
majorization cycle is very similar to the one that we already found in the 
quantum phase-estimation algorithm,
 but in this case we have numerically checked
that the present cycle does not seem to follow the rules of 
natural majorization. Complementarily, we have also observed 
that the probability amplitudes follow the rule that
those belonging to even columns are real, while those belonging to 
odd columns are imaginary. 

\begin{figure}
\centering
\includegraphics[angle=-90, width=0.7\textwidth]{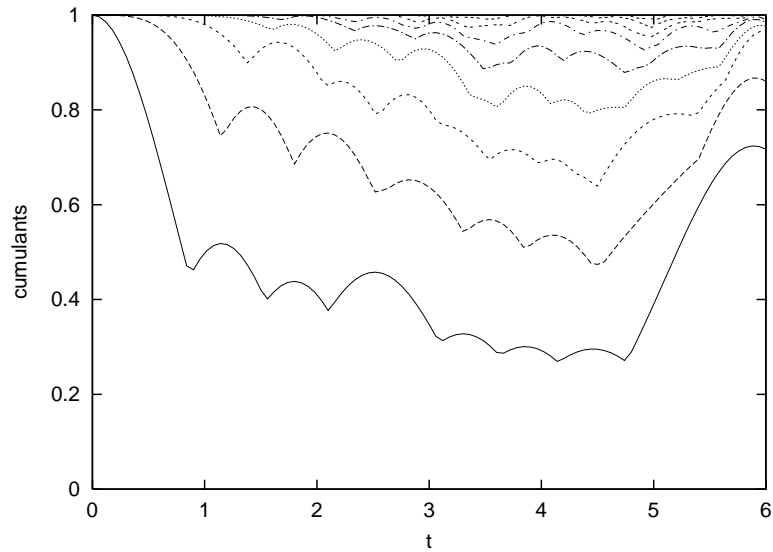}
\caption{Time evolution of the nine cumulants in the quantum walk algorithm when the column-measurement is considered, for $n=4$. No majorization cycle is present.}
\label{nocycle}
\end{figure}

\begin{figure}
\centering
\includegraphics[angle=-90, width=0.7\textwidth]{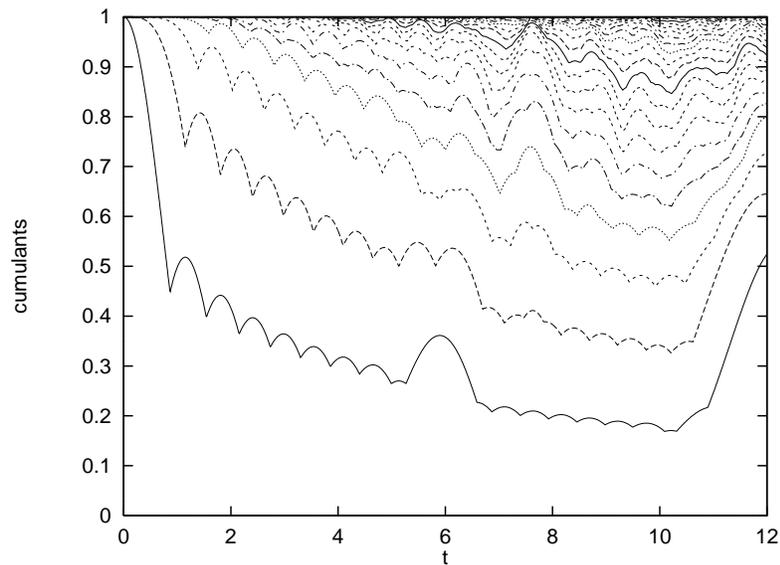}
\caption{Time evolution of the $21$ cumulants in the quantum walk algorithm when the column measurement is considered, for $n=10$. No majorization cycle is present.}
\label{nocycle2}
\end{figure}

The quantum random walk heavily exploits
the column structure of the problem. The register works
on a superposition of columns, that is of states belonging
to the same column with equal weight. It is then natural
to ask whether a step-by-step majorization cycle operates
also at the level of columns. The idea behind this
analysis corresponds to accept that the final measurement will
filter each one of the columns as a whole. The result of the measurement 
would correspond to determining a particular column. The point
here is to find to what extent the success of finding the
OUT state is related to the column structure of the algorithm.   
We have numerically considered the column amplitudes
 for $n=4$  and $n=10$ with a total of
 $9$ and $21$ cumulants to be calculated respectively from the sorted 
probabilities at each time step of being \emph{at each column} 
of the graph. In Fig.\ref{nocycle} and Fig.\ref{nocycle2} we plot our 
results, which show that there does not exist a majorization 
cycle when the final measurement is carried on columns. 
The conclusion is that deterministic quantum walks cleverly 
exploit the column subspace structure of the problem to achieve
step-by-step majorization on the individual states, but not on the individual columns.

\section{Conclusions of Chapter 6}

We have seen in this Chapter that majorization seems to appear in the fauna of quantum algorithms in a very specific way, namely, in such a way that some instances of efficient quantum algorithms seem to step-by-step majorize the probability distribution of the final outcomes all along the flow in time. In order to be precise:

\begin{itemize}
\item{We have proven that the quantum Fourier transform in quantum phase-estimation algorithms majorizes step-by-step the probability distribution of the final outcomes. This step-by-step majorization is seen to appear in a natural way from the absence of some interference terms in the unitary evolution, in contrast with what is 
found for Grover's quantum searching algorithm. The example of a quantum algorithm solving a hidden affine function problem also shows the same basic features than the quantum-phase estimation algorithm, which can be understood in terms of a majorization cycle along the complete time-evolution. However, Shor's quantum factoring algorithm, though being based on a variant of the quantum phase-estimation algorithm, does not globally obey step-by-step majorization on the whole set of relevant probabilities.}
\item{We have seen that step-by-step majorization in adiabatic quantum searching algorithms is heavily attached to the optimality of the interpolating path. Those paths which do not produce an optimal quantum search are seen to step-by-step majorize the probability of the final outcomes only if the change rate of the Hamiltonian is extremely slow. On the contrary, 
the optimal path producing a square-root speed-up directly obeys step-by-step majorization.}
\item{We have observed that there is a majorization cycle of the probabilities of the final outcomes in an exponentially fast  quantum walk algorithm solving a classically hard problem defined in terms of a non-trivial graph. This majorization cycle does not appear if alternative collective measurements are considered.} 
\end{itemize}

Our conclusion is that some broad families of quantum algorithms seem to have an underlying majorization structure in the way they proceed in order to get the desired solution to the  problem that they deal with. This behavior is somehow similar to the one of greedy algorithms in classical computation, which always evolve such that the probability of the ``winner'' increases at each time step. Majorization is, though, a far more severe condition, since it not only involves constraints on one single and specific probability, but on the complete probability distribution. In some sense, majorization seems to be a plausible candidate to look at in order to have a good understanding of the performance of a quantum algorithm, together with entanglement. How these two quantities behave along the computational evolution of a given quantum algorithm  may already provide a lot of information about its performance.

    \chapter{General conclusions and outlook}

The work presented in this thesis tries to bring together different fields of physics. We used tools from quantum information science to analyze problems in quantum field theory and condensed-matter physics in Chapters 1 and 3. Conformal field theory can in turn be useful to analyze problems in quantum information science, as we saw in Chapters 1 and 2. Moreover, quantum phase transitions and quantum algorithms are seen to be very much related, as we have seen in Chapters 4 and 5. Furthermore,  ideas related to the performance of some quantum algorithms were shown in Chapter 6 by using majorization theory. All in all, we have seen that the fields of quantum information science, condensed-matter physics, and quantum field theory have very much in common, and that their multidisciplinary intersection is useful.

Let us consider several future directions. First, the use of majorization theory and conformal field theory together with related techniques applied to a comprehension of both the irreversibility of renormalization group flows and the behavior of the single-copy entanglement in more than $(1+1)$ dimensions is something that remains to be done. Also, it is still a theoretical challenge to know whether adiabatic quantum algorithms can solve NP-complete problems in polynomial time or not, which in the end amounts to ask about the possibilities of quantum computation to solve the celebrated P$\ne$NP conjecture. Further analysis of adiabatic quantum algorithms could be done, for instance, by means of a parallelization of the local truncation scheme that we used in Chapter 5, or by means of non-local truncation schemes, adapted valence-bond ansatzs for the ground state wavefunction, or other related techniques. Indeed, classical numerical simulations using the ideas from Chapter 5 of some other quantum algorithms, like Shor's factoring quantum algorithm, could bring further insight both for the quantum algorithm and for the classical simulation technique itself. The big problem in quantum computation remains to be, yet, the design of new, useful and efficient quantum algorithms. Furthermore, from the many-body physics point of view, the challenge now is to perform reliable and accurate classical simulations of the properties of $(2+1)$-dimensional quantum many-body systems, for which new numerical techniques are beginning to be discovered. However, the tools developed so far do not apply to the study of critical fermionic systems in more than one spatial dimension, since some of these systems  break the entropic area-law scaling \cite{Wolf06, Gio06, Bart06}. A better understanding of these models, both from a theoretical and numerical point of view, together with a plausible numerical ansatz for their ground state wave function, remains as an open problem.
     
     \appendix
      
     \chapter{Majorization}

Majorization theory  deals with the notion of relative order of probability distributions. It was originally introduced 
within the fields of mathematical statistics and economics \cite{Muirhead1903, Hardy78, Marshall79, Bathia97}, 
and its basic idea relies on the comparison of two given probability distributions by means of a set of order relations to be
satisfied by their components. 

We now precisely define the notion of majorization \cite{Bathia97}. 
Let $\vec{x}$, $\vec{y}\in \mathbb{R}^{+N}$ be two normalized probability 
vectors, $\sum_{i = 1}^N x_i = \sum_{i=1}^N y_i = 1$. We say that distribution $\vec{y}$ 
majorizes distribution $\vec{x}$, written $\vec{x}\prec \vec{y}$, 
if and only if there exist a set of permutation matrices $P_k$ 
and probabilities $p_k \ge 0$, $\sum_k p_k=1$, such that 
\begin{equation}
\vec{x} = \sum_k p_k P_k \vec{y} \ .
\label{defone}
\end{equation}
Since, from the previous definition, $\vec{x}$ can be obtained by means 
of a probabilistic combination of permutations of $\vec{y}$, we get the
intuitive notion that probability distribution $\vec{x}$ is 
more disordered than probability distribution $\vec{y}$. This defines a partial order in the space
of probability distributions. 

There are two alternative equivalent definitions of majorization which turn out to be useful. The first one  reads as follows. We say that a given $N \times N$ matrix $D$ is doubly stochastic if it has non-negative
entries and each row and column adds up to $1$. Then, $\vec{y}$ majorizes $\vec{x}$ if and only if there is a doubly
stochastic matrix $D$ such that

\begin{equation}
\vec{x} = D \vec{y} \ .
\label{defthree}
\end{equation}

Notice that in Eq.\ref{defone}, $\sum_k p_k P_k \equiv D$ defines a doubly 
stochastic matrix, that is, $D$ has nonnegative entries and each row and 
column adds up to unity, thus satisfying Eq.\ref{defthree}. 

The third equivalent definition of majorization can be stated in terms 
of a set of inequalities between partial sums of the two distributions. Consider the 
components of the two probability vectors sorted in decreasing order. Then, $\vec{x} \prec \vec{y}$ if and only if

\begin{equation}
\sum_{i=1}^k x_i \leq \sum_{i=1}^k y_i \qquad k = 1, 2, \ldots , N -1\ .
\label{deftwo}
\end{equation}
All along this thesis, we refer to  these partial sums of sorted probabilities as \emph{cumulants}.

A powerful relation between majorization and any convex function $f$ over the
set of probability vectors states that 
\begin{equation}
\vec{x} \prec \vec{y} \Rightarrow f \left(\vec{x}\right) 
\le f\left(\vec{y}\right) \ .
\label{powerful}
\end{equation}
From this relation it follows that the Shannon entropy $H(\vec{z}) \equiv -\sum_{i=1}^N z_i \log_2{z_i}$ 
of a probability distribution $\vec{z} \in \mathbb{R}^N$ satisfies $H \left(\vec{x}\right) 
\ge H\left(\vec{y}\right)$ whenever $\vec{x} \prec \vec{y}$. Majorization is, therefore, a stronger notion of 
order for probability distributions that the one imposed by the entropy $H(\vec{z})$.

The connection between majorization and quantum mechanics can be established whenever a probability distribution appears. 
For instance, one could be interested in the majorization properties of the probability distribution arising from the spectrum of some
given reduced density matrix, as happens often in the field of quantum information science. For two reduced density operators $\rho$ and $\sigma$
with spectrums $\vec{\rho}$ and $\vec{\sigma}$, we say that $\rho \prec \sigma$ if and only if $\vec{\rho} \prec \vec{\sigma}$. This extends the notion of majorization
 to positive semi-definite operators by considering their normalized spectrum.

     \chapter{Some notions about conformal field theory}

The aim of this Appendix is to give a brief, non-technical and non-exhaustive idea about some of the basic concepts of conformal field theory. The interested reader is referred to the specific literature in the field for further details and developments (see for  example \cite{Ginsparg} and references therein).

Consider a metric $g_{\mu \nu}(x)$ of signature $(p,q)$ in a space of total dimension $D$, where $x$ stands for a given point of this space in some given coordinate system. Under a change of coordinates $x \rightarrow x'$, the metric transforms as $g^{\prime}_{\mu \nu}(x^{\prime}) =  \frac{\partial x^\alpha}{\partial x^{\prime \mu}} 
\frac{\partial x^\beta}{\partial x^{\prime \nu}} g_{\alpha \beta}(x)$, where sums are to be understood on repeated indices from now on. The conformal group in $D$ dimensions is, by definition, the subgroup of coordinate transformations that leave the metric invariant up to a local change of scale,
\begin{equation}
g_{\mu \nu}(x) \rightarrow g'_{\mu \nu}(x') = \Omega(x) g_{\mu \nu}(x) \ , 
\label{defCFT}
\end{equation}
where $\Omega(x)$ is a local dilatation factor. It is possible to exactly characterize the form of these transformations, which are given by the Poincar\'e group
\begin{eqnarray}
x& \rightarrow &x' = x + a \nonumber \\
x &\rightarrow &x' = \Lambda x \ \ \ \ \ \ (\Lambda \in {\rm SO}(p,q)) 
\end{eqnarray}
with $\Omega(x) = 1$, the dilatations
\begin{equation}
x \rightarrow x' = \lambda x 
\end{equation}
with $\Omega = \lambda^{-2}$, and the so-called special conformal transformations
\begin{equation}
x \rightarrow x' = \frac{x + bx^2}{1 + 2 b \cdot x + b^2x^2} 
\end{equation}
with $\Omega(x) = (1 + 2b \cdot x + b^2 x^2)^2$.
Conformal symmetry can then be understood as some generalization of scale symmetry. Those field theories defined in the continuum that are invariant under conformal transformations constitute the so-called \emph{conformal field theories}. 

Conformal symmetry is especially powerful in the case of $2$ dimensions, typically denoted as $(1+1)$, in the case of having one temporal and one spatial dimension. Given the coordinates of the plane $x^1$ and $x^2$, and defining new complex coordinates $z = x^1 + i x^2$ and $\bar{z} = x^1 - i x^2$ (respectively called holomorphic and antiholomorphic coordinates), conformal transformations in $2$ dimensions coincide with the set of analytic coordinate transformations in the plane 
\begin{eqnarray}
z &\rightarrow& f(z) \nonumber \\
\bar{z} &\rightarrow& \bar{f}(\bar{z}) \  ,
\end{eqnarray}
$f$ and $\bar{f}$ being analytic complex functions. Typically, it is useful to work with $z$ and $\bar{z}$ treated as independent variables, so that the physical condition $\bar{z} = z^*$ is left to be imposed at our convenience. The fact that conformal transformations in the plane precisely coincide with the group of analytic coordinate transformations is very notorious, since the number of generators of the conformal group in $2$ dimensions is then \emph{infinite}, which only happens for this number of dimensions. The behavior of conformally-invariant field theories in $2$ dimensions is, then, heavily constrained by the symmetry. 

In order to be more specific, assume that we are given a conformally-invariant quantum field theory in $D=2$. Those operator fields $\Phi(z,\bar{z})$ that transform under conformal transformations like 
\begin{equation}
\Phi(z,\bar{z}) \rightarrow \left( \frac{\partial f}{\partial z} \right)^h \left( \frac{\partial \bar{f}}{\partial \bar{z}} \right)^{\bar{h}} \Phi(f(z),\bar{f}(\bar{z})) \ , 
\end{equation}
with positive real $h$ and $\bar{h}$, are called primary fields of conformal weight $(h,\bar{h})$. Conformal symmetry imposes that the two-point correlation function of two primary fields $\langle \Phi_1(z_1,\bar{z}_1) \Phi_2(z_2,\bar{z}_2)\rangle $ must be
\begin{equation}
\langle \Phi_1(z_1,\bar{z}_1) \Phi_2(z_2,\bar{z}_2)\rangle = \frac{1}{z_{12}^{2h} \bar{z}_{12}^{2\bar{h}}}
\label{corrCFT}
\end{equation}
if $(h_1,\bar{h}_1) = (h_2,\bar{h}_2)$ and zero otherwise, where $z_{12} = z_1 - z_2$, $\bar{z}_{12} = \bar{z}_1 - \bar{z}_2$. Note that the decay of the correlation function in Eq.\ref{corrCFT} is algebraic, as is the typical situation of critical condensed-matter systems. This is not strange, since many critical quantum many-body systems can be understood at criticality as the regularization on a lattice of some given conformal field theory, as is the case, for example, of the critical Ising quantum spin chain \cite{Ginsparg,Rico2005}. Indeed, conformal symmetry imposes similar decaying laws for the two-point correlators in any number of dimensions. 

An important quantity which is to play a role is the \emph{stress-energy tensor} $T_{\mu \nu}(x)$, which can be always defined for any field theory.  For instance, for a free-bosonic quantum field theory defined in terms of a Lagrangian ${\mathcal L}$, the stress-energy tensor reads
\begin{equation}
T_{\mu \nu}(x) = \frac{\partial {\mathcal L}}{\partial (\partial^{\mu} \phi)}\partial_{\nu} \phi - {\mathcal L}g_{\mu \nu} \ ,
\end{equation}
where $\phi$ stands for the quantum field of the free boson. It can be seen that in two dimensions, the stress-energy tensor of a conformally-invariant quantum field theory has only two non-vanishing components, which are called $T(z)$ and $\bar{T}(\bar{z})$. An important property of a primary field $\Phi(w,\bar{w})$  is that its operator product expansion with the stress-energy tensor reads
\begin{eqnarray}
T(z) \Phi(w,\bar{w}) &=& \frac{h}{(z-w)^2}\Phi(w,\bar{w}) + \frac{1}{(z-w)}\partial_w \Phi(w,\bar{w}) + \cdots \nonumber \\
\bar{T}(\bar{z}) \Phi(w,\bar{w}) &=& \frac{\bar{h}}{(\bar{z}-\bar{w})^2}\Phi(w,\bar{w}) + \frac{1}{(\bar{z}-\bar{w})}\partial_{\bar{w}} \Phi(w,\bar{w}) + \cdots \ ,
\label{alterCFT}
\end{eqnarray}
which can be understood as an alternative definition of a primary field of conformal weight $(h,\bar{h})$. 

The stress-energy tensor is an example of a quantum field that is not primary. Computing its operator product expansion with itself, one gets
\begin{eqnarray}
T(z)T(w) &=& \frac{c/2}{(z-w)^4} + \frac{2}{(z-w)^2}T(w) + \frac{1}{(z-w)}\partial_w T(w) \nonumber \\
\bar{T}(\bar{z})\bar{T}(\bar{w}) &=& \frac{\bar{c}/2}{(\bar{z}-\bar{w})^4} + \frac{2}{(\bar{z}-\bar{w})^2}\bar{T}(\bar{w}) + \frac{1}{(\bar{z}-\bar{w})}\partial_{\bar{w}} \bar{T}(\bar{w}) \ ,
\end{eqnarray}
which clearly differs from Eq.\ref{alterCFT}. The above equations define the so-called holomorphic and antiholomorphic central charges $c$ and $\bar{c}$, which depend on the particular theory under consideration, much in the same way as the  conformal weights $(h,\bar{h})$ do. For example, for a free bosonic quantum field theory $c = \bar{c} = 1$, whereas for a free fermionic quantum field theory $c = \bar{c} = 1/2$. Yet, another property of the stress-energy tensor for conformally-invariant quantum field theories in $2$ dimensions is that it is possible to expand it in terms of modes as follows: 
\begin{eqnarray}
T(z) &=& \sum_{n \in \mathbb{Z}} z^{-n-2} L_n \nonumber \\
\bar{T}(\bar{z}) &=& \sum_{n \in \mathbb{Z}} \bar{z}^{-n-2} \bar{L}_n \ ,
\end{eqnarray} 
where the operators $L_n$ and $\bar{L}_n$ satisfy the commutation relations
\begin{equation}
[L_n,L_m] = (n-m)L_{n+m} + \frac{c}{12}(n^3-n)\delta_{n+m,0} \nonumber 
\end{equation}
\begin{equation}
[\bar{L}_n,\bar{L}_m] = (n-m) \bar{L}_{n+m} + \frac{\bar{c}}{12} (n^3-n) \delta_{n+m,0} \nonumber 
\end{equation}
\begin{equation}
[L_n,\bar{L}_m] = 0 \ .
\end{equation}
The above equations define two copies of an algebra which is called the Virasoro algebra. Every conformally-invariant quantum field theory determines a representation of this algebra, with some $c$ and $\bar{c}$. 

The construction of the Hilbert space for a conformal field theory in $2$ dimensions is very much related to the above operator algebra. Given a vacuum $|\Omega\rangle$ which is assumed to exist by hypothesis, the state 
\begin{equation}
|h, \bar{h}\rangle \equiv \Phi(0,0) |\Omega\rangle 
\end{equation}
created by a primary field $\Phi(z,\bar{z})$ of conformal weight $(h,\bar{h})$ satisfies
\begin{eqnarray}
L_0 |h,\bar{h}\rangle &=& h |h,\bar{h}\rangle \nonumber \\
\bar{L}_0 |h,\bar{h}\rangle &=& \bar{h} |h,\bar{h}\rangle \nonumber \\
L_n|h,\bar{h}\rangle &=& \bar{L}_m|h,\bar{h}\rangle = 0 \ \ \forall n, m > 0 \ .
\end{eqnarray}
Any state satisfying the above relations is called a highest-weight state. States of the form 
\begin{equation}
L_{-n_1} L_{-n_2} \cdots L_{-n_j} \bar{L}_{-m_1} \bar{L}_{-m_2} \cdots \bar{L}_{-m_k} |h,\bar{h}\rangle 
\end{equation}
are called descendant states, and are also eigenstates of $L_0$ and $\bar{L}_0$ with eigenvalues $h+n_1+n_2+\cdots+n_j$ and $\bar{h} + m_1 + m_2 + \cdots + m_k$ respectively. 
The full tower of eigenstates of $L_0$ and $\bar{L}_0$ constructed in this way is known as the Verma module. Therefore, the Hilbert space of a conformally-invariant quantum field theory in $2$ dimensions decomposes as the direct sum of Verma modules, the number of which depends only on the number of primary fields appearing in the theory.

     \chapter{Some notions about classical complexity theory}

In this Appendix our aim is to give some very basic notions and non-technical background on classical complexity theory. Excellent textbooks on this topic are those of Garey and Johnson \cite{Garey-Johnson} and Papadimitriou \cite{Papadimitriou}. A review on complexity theory, with extensions to quantum complexity theory, is given by Aharonov and Naveh in \cite{Aharonov02}. 

Let us begin with the following definition:

\bigskip

{\bf Definition C.1:}  {\it An alphabet $\Sigma$ is a set of symbols.}

\bigskip

We did not define the concept of \emph{symbol} since we believe its meaning to be clear from the context. Examples of alphabets are $\Sigma_1 \equiv \{ a, b, \ldots , z \}$, $\Sigma_2 \equiv \{ \alpha, \beta, \ldots , \omega \}$, and $\Sigma_3 \equiv \{0, 1\}$. The alphabet $\Sigma_3$, with only two symbols, is usually referred to as the \emph{binary alphabet}. 

\bigskip

{\bf Definition C.2:}  {\it A language $L$ over an alphabet $\Sigma$ is a set of strings of symbols from $\Sigma$.}

\bigskip

For instance, $L_1 \equiv \{jack, sam, daniel, tealc \}$ is a language over the alphabet $\Sigma_1$, and  
$L_2 \equiv \{ 010, 00010, 1001 \}$ is a language over the binary alphabet $\Sigma_3$. 

\bigskip

{\bf Definition C.3:}  {\it A decision problem is a problem for which the answer belongs to a binary alphabet.}

\bigskip

This is the kind of ``yes'' or  ``no'' problems. That is, questions of the type ``will the universe expand forever?'', or ``do I prefer chocolate or lemon ice-creams?'', but also questions like ``is the number $1761935875391$ the product of two or more primes?''.  An important part of the theory of computational complexity is built in terms of decision problems. More concretely, one has to decide whether a given string of symbols from an alphabet, called \emph{instance}, belongs to a certain language or not. From now on we shall always restrict ourselves to the binary alphabet, whose symbols are called \emph{bits}. 

Languages are classified in terms of \emph{complexity classes}, according to different criteria. We now define a complexity class that plays a major role in complexity theory:

\bigskip

{\bf Definition C.4:}  P {\it is the class of languages $L$ for which a deterministic Turing machine can decide in a time $O({\rm poly}(|x|)$ if an instance $x$ belongs to $L$ or not, $|x|$ being the number of bits of $x$.}

\bigskip

In the above definition, we understand that a \emph{deterministic Turing machine} is our classical model of computation. Usually, it is said that languages $L \in$ P can be \emph{decided} in polynomial time by a deterministic Turing machine. Intuitively, we understand that a language $L$ belongs to the complexity class P if there is an \emph{efficient} classical algorithm that allows to deterministically decide whether a given  instance $x$ belongs to $L$ or not, where by the term ``efficient'' we mean ``polynomial in the size of the instance''. Let us now define another important complexity class:

\bigskip

{\bf Definition C.5:}  NP {\it is the class of languages $L$ for which there exists a deterministic polynomial-time verifier $V$ such that
\begin{itemize}
\item{ $\forall x \in L$, there is a $y$ such that $|y| = {\rm poly}(|x|)$ and $V(x,y) = 1$, and}
\item{$\forall x \notin L$ and $\forall y$ such that  $|y| = {\rm poly}(|x|)$, $V(x,y) = 0$.}
\end{itemize}
}

\bigskip

Usually $y$ is referred to as the \emph{witness} or \emph{certificate}. Both the witness $y$ and the verifier $V$ help in deciding whether the instance $x$ belongs to $L$ or not. Let us clarify Definition C.5 by means of an example: let $L =$ COMPOSITE be the language of numbers that can be decomposed as a product of two or more primes. Let $x = 161$ be an instance of the decision problem ``does $x$ belong to COMPOSITE?''. A possible witness $y$ can be given by the two prime numbers $7$ and $23$, and the verifier $V$ can be a classical deterministic algorithm that performs the following check: $7 \times 23 = 161$. Notice then that if the instance $161$ belongs to COMPOSITE there is a witness $7$, $23$ such that the verifier can check that the instance belongs to the language. On the contrary, if we are given an instance that does not belong to COMPOSITE (for instance, $x = 17$), then there is no witness $y$ such that our verifier can check that $17$ is a product of two or more primes. In a way, the witness has to be thought of as the ``proposal of solution", and the verifier has to be considered as a classical algorithm that allows to deterministically and efficiently check whether the proposed solution to the specific instance is correct or not. This example shows that COMPOSITE $\in NP$, which in less mathematical words is commonly referred to as ``the problem of deciding whether a given number is the product of two or more primes is NP".  

Given the Definition C.5 of the NP complexity class, we can now define the following:

\bigskip

{\bf Definition C.6:}  NP-hard {\it is the class of languages $L$ such that the problem of deciding whether an instance $x'$ belongs or not to a language $L' \in$ NP can be efficiently reduced to the problem of deciding whether an instance $x$ belongs or not to $L$,  $\forall x'$ and $L' \in $ NP.}

\bigskip

In plain words, a problem is said to be NP-hard if \emph{all} the instances of \emph{all} the NP problems can be efficiently mapped to specific instances of the NP-hard problem. Therefore, if a language $L \in$ NP-hard can be decided by some deterministic classical algorithm, the same procedure can essentially be applied to decide all the languages in the complexity class NP, and ``solve all the NP problems". 

Let us now define the important concept of NP-complete:

\bigskip

{\bf Definition C.7:}  NP-complete {\it is the class of languages $L$ such that $L \in$} NP-hard {\it and $L \in$} NP. 

\bigskip

According to Definition C.7, NP-complete languages are those languages in NP such that being able to decide about one of them implies being able to decide about \emph{the whole} complexity class NP. An important example of an NP-complete language is 3-SAT. A possible instance of the 3-SAT decision problem is a boolean formula in conjunctive normal form over $n$ bits $\phi(x_1, x_2, \ldots , x_n) = C_1 \land C_2 \land \cdots \land C_m$, where $x_i$, $i = 1, 2, \ldots , n$, denotes the value of the bits, and $C_j$, $j = 1, 2, \ldots , m$, are the so-called \emph{clauses}. Each clause $C_j$ is built in the way $C_j = (\tilde{x}_{j,1} \lor \tilde{x}_{j,2} \lor \tilde{x}_{j,3})$, where $\tilde{x}_{j,\alpha}$ is a \emph{literal} for bit $\alpha$ of clause $j$, which can be any of the $n$ bit variables or its negation. The decision problem is properly defined by the following question: ``given an instance $\phi$ is there a string of $n$ bits $(y_1, y_2, \ldots , y_n)$ such that $\phi(y_1, y_2, \ldots , y_n) = 1$?", or equivalently, ``is there a string of $n$ bits 
$(y_1, y_2, \ldots , y_n)$ such that all the $m$ clauses are satisfied?". 

The proof of the NP-completeness of 3-SAT is one of the most relevant results in the field of complexity theory, and is due to the original work of Cook \cite{Cook71}. That proof opened the door to the discovery of many other NP-complete languages and, today, NP-complete languages (or problems) appear in many different fields of mathematics, physics, and science in general. Their relevance comes in part from the fact that they are at the heart of one of the most celebrated open questions in mathematics, which reads as follows:

\bigskip

{\bf Problem C.1:}  {\it Is} P $\ne$ NP {\it ?}

\bigskip

To determine the answer to the above question, it would be sufficient to prove that it is possible to deterministically decide some NP-complete language efficiently, and then P $=$ NP, or on the contrary to prove that it is impossible to efficiently and deterministically decide an NP-complete language, and therefore P $\ne$ NP. While the most accepted opinion is that P $\ne$ NP, it has been so far impossible to produce a precise and mathematical proof of this, neither of the opposite statement P $=$ NP. Indeed, Problem C.1 remains today as probably the most challenging open problem in computer science \cite{Papadimitriou}. 

Let us mention as well that the deterministic complexity classes P, NP, NP-hard and NP-complete can be further generalized if we consider classical probabilistic models of computation, the equivalent \emph{probabilistic} complexity classes being called BPP, MA, MA-hard and MA-complete. Furthermore, if the underlying computational model is a quantum computer, the corresponding generalized \emph{quantum} complexity classes are called BQP, QMA, QMA-hard and QMA-complete. The study of these classes is beyond the scope of this Appendix, and we refer the reader to \cite{Aharonov02} and references therein for further details on quantum complexity theory and its consequences for quantum computation.


  
\bibliography{bibliografia}
\bibliographystyle{unsrt}  
  

    \end{document}